\documentclass[12pt, twoside]{report}
\usepackage{latexsym, amssymb, a4, float, epsfig, graphicx, mathbbol}
\newcommand{\graphicwidth}{0.8\textwidth}

\newcommand{\bml}{\begin{mathletters} \baselineskip 10pt}
\newcommand{\eml}{\baselineskip 12pt \end{mathletters}}
\newcommand{\nn}{\nonumber}

\newcommand{\m}{{\scriptscriptstyle -}}
\newcommand{\p}{{\scriptscriptstyle +}}

\newcommand{\intl}{\int\limits_{-L}^L}

\newcommand{\om}{\omega}
\newcommand{\la}{\lambda}

\newcommand{\vi}{\varphi}
\newcommand{\bra}{\langle}
\newcommand{\ket}{\rangle}
\newcommand{\cond}{\bra 0 | \bar \psi \psi | 0 \ket}

\newcommand{\simgeq}{\scriptstyle{\stackrel{>}{\sim}}}
\newcommand{\lca}{\phi_{\mathrm{LC}}}
\newcommand{\bsa}{\chi_{\mathrm{BS}}}

\newcommand{\Proj}{\mathbb{P}}

\newcommand{\Pn}{\Proj_{\mbox{\scriptsize \sf N}}}
\newcommand{\sfn}{\mbox{\scriptsize \sf N}}
\newcommand{\Pnm}{\Proj_{\mbox{\scriptsize \sf N-1}}}

\newcommand{\sfrac}[2]{{\textstyle \frac{#1}{#2}}}
\newcommand{\pad}[2]{\frac{\partial #1}{\partial #2}}
\newcommand{\comm}[2]{\left[#1 \, , \, #2 \right] }
\newcommand{\vc}[1]{\mbox{\bf #1}}
\newcommand{\svc}[1]{\mbox{\footnotesize\bf #1}}
\newcommand{\vcg}[1]{\mbox{\boldmath$#1$}}
\newcommand{\svcg}[1]{\mbox{\footnotesize\boldmath$#1$}}
\newcommand{\pb}[2]{\left\{#1 \, , \, #2 \right\}}
\newcommand{\spinor}[2]{\left[ \begin{array}{c}
                               #1  \\  #2
                            \end{array} \right]}

\newcommand{\sgn}{{\mbox{sgn}}}
\newcommand{\tr}{\mbox{tr}}

\newcommand{\ad}{\; \mbox{ad}}
\newcommand{\e}{\mbox{e}}

\begin{document}

\title{\bf \Huge Light-Cone Dynamics of Particles and Fields}

\vspace{5cm}

\author{Thomas Heinzl\footnote{present address:
Friedrich-Schiller-Universit\"at Jena, Theoretisch-Physikalisches
Institut, Max-Wien-Platz 1, D-07743 Jena, e-mail: {\tt heinzl@tpi.uni-jena.de}} \\
        Universit\"at Regensburg \\
        Institut f\"ur Theoretische Physik \\
        93040 Regensburg \\
        Germany \\
        }
\date{}

\maketitle

\thispagestyle{empty}
\cleardoublepage

\begin{abstract}
  We review the foundations as well as a number of important
  applications of light-cone dynamics. Beginning with the peculiarities
  of relativistic particle dynamics we discuss the choice of a time
  parameter as the gauge fixing within reparametrization invariant
  dynamical systems. Including Poincar\'e invariance, we are naturally
  led to Dirac's forms of relativistic dynamics. Among these, the front
  form is our main focus as it is the basis for light-cone dynamics. We
  explain the peculiar features of the light-cone formulation such as
  boost and Galilei invariance or separation of relative and
  center-of-mass motion. Combining light-cone dynamics and field
  quantization leads to the introduction of light-cone quantum field
  theory. We show how the positivity of the kinematical longitudinal
  momentum implies the triviality of the light-cone vacuum. We point
  out that its special features make the light-cone formulation a unique
  framework to deal with bound states as few-body systems based on
  quantum field theory.  In a first application, we analyze spontaneous
  symmetry breaking for scalar field theory in 1+1 dimensions. The
  importance of modes with vanishing longitudinal momentum is
  elucidated. For fermionic field theories, we suggest to reconstruct
  vacuum properties, like chiral condensates, from the particle
  spectrum. The latter can be obtained by solving the light-cone
  Schr\"odinger equation as we explicitly demonstrate for the 't~Hooft and 
  Schwinger models. Finally, we make contact with phenomenology by
  calculating the pion wave function within the Nambu and Jona-Lasinio
  model. We are thus able to predict a number of observables like the
  pion charge and core radius, the r.m.s.~transverse momentum and the
  pion distribution amplitude. The latter turns out being not very
  different from the asymptotic one.
\end{abstract}

\setcounter{page}{4}
\newpage
\thispagestyle{empty}
\cleardoublepage
\renewcommand{\baselinestretch}{0.95}
\tableofcontents
\cleardoublepage


\renewcommand{\baselinestretch}{1}


\chapter{Introduction: Strong Interactions}

\section{Hadrons}

The word `hadron' originates from Greek where $\alpha \delta \! \rho o
\sigma$ means `big' or `strong'. It refers to all particles participating in
strong interactions. One can therefore say that hadrons are the physical
degrees of freedom in strong interaction physics. Everything we know
empirically in this area has been obtained by studying hadronic
processes and properties. A (very) incomplete list of hadrons is
provided by Table~\ref{T1}.

\begin{table}
\begin{center}
\renewcommand{\arraystretch}{1.2}
\caption{\label{T1} \textsl{Masses and quantum numbers of a few hadrons. The
  upper part consists of baryons ($B=1$), the lower part of mesons
  ($B=0$). Masses are given in MeV. }}
\vspace{.5cm}

\begin{tabular*}{\textwidth}[h]{ @{\extracolsep\fill} l c r c c c }
\hline\hline
particle & & mass   & isospin & spin & $B$ \\
\hline
proton  & $p$       &  938 & 1/2 & 1/2 & 1 \\
neutron & $n$       &  940 & 1/2 & 1/2 & 1 \\
delta   & $\Delta$  & 1236 & 3/2 & 3/2 & 1 \\
lambda  & $\Lambda$ & 1116 & 0   & 1/2 & 1 \\
\hline
pion    & $\pi$     &  138 & 1   &  0  & 0 \\
kaon    & $K$       &  495 & 1/2 &  0  & 0 \\
rho     & $\rho$    &  770 & 1   &  1  & 0 \\
\hline\hline
\end{tabular*}
\end{center}
\end{table}

There are more than a hundred additional hadrons which, together with
their properties can be found in the reviews of the particle data group
\cite{barnett:96}. One distinguishes between baryons having baryon
number $B$ = 1 and half-integer spin (fermions), and mesons with $B=0$
and integer spin (bosons). This great variety of hadrons has been
classified in the sixties within the quark model of Gell-Mann and Zweig
\cite{gell-mann:64a, zweig:64} which states that hadrons are grouped
into multiplets of the group $SU(3)_F$, where the `$F$' stands for the
quantum number `flavor'. (Anti-)\-Quarks are postulated as the
realizations of the fundamental (anti-)triplet of $SU(3)_F$. This
implies that quarks come in three flavors, `up' $u$, `down' $d$ and
`strange' $s$ (the anti-quarks being $\bar u$, $\bar d$ and $\bar s$),
which combine to give the quantum numbers of the hadrons. The quarks
were postulated to be fermions carrying charges which are multiples of a
third of the electron charge. Nowadays one knows a total number of $N_F
= 6$ quarks with the additional heavy flavors `charm' $c$, `bottom' $b$
and `top' $t$.  In the quark scheme, mesons consist of a quark and an
antiquark, (anti-)baryons of three (anti-)quarks. Thus, for example, one
finds the following octet of pseudo-scalar mesons with spin-parity
$J^\pi = 0^-$,
\begin{equation}
  \begin{array}{l c l c l c l }
    \pi^0 & = & \sqrt{1/2} (u \bar u - d \bar d) && K^0 & = & d \bar s \\
    \pi^- & = & d \bar u                         && K^+ & = & u \bar s \\
    \pi^+ & = & u \bar d                         && K^- & = & s \bar u \\
    \eta  & = & \sqrt{1/3} (u \bar u + d \bar d + s \bar s) && \bar K^0 &=&
    s \bar d 
  \end{array}
\end{equation}
This octet structure is reminiscent in the name `the eightfold way', the
title of a famous book by Gell-Mann and Ne'eman \cite{gell-mann:64b}.

It was soon realized thereafter, that the quark-flavor model is in
conflict with the Pauli principle. The baryon resonance $\Delta^{++} =
uuu$, with all quark spins pointing upwards, cannot have an
anti-symmetric wave function as it should as a three-fermion state. The
loophole is to introduce an additional quantum number called color, giving
rise to another symmetry group, $SU(3)_C$. Upon postulating that hadrons
should always appear as color singlets the Pauli principle is satisfied.

Using flavor and color symmetry, the wave function of mesons and
baryons can thus be written as
\begin{eqnarray}
  | M \ket &=& \frac{1}{\sqrt{N_C}} M_{ij} \delta_{ab} |q_i^a \bar q_j^b
  \ket \; ,   \\
  | B \ket &=& \frac{1}{2\sqrt{N_C}}B_{ijk} \epsilon_{abc} | q_i^a q_j^b
  q_k^c  \ket   \; .
\end{eqnarray}
Here, $N_C = 3$ is the number of colors (indices $a,b,c$), $M_{ij}$ and
$B_{ijk}$ are tensors in flavor space.  It is obvious that the baryon
state is anti-symmetric in color which solves the $\Delta^{++}$-puzzle.

If we include spin degrees of freedom and the spatial part of the hadron
wave function, the total hadronic state can be written symbolically as
\begin{equation}
  | \mathrm{hadron} \ket = | space \ket \otimes |spin \ket \otimes 
  | flavor \ket \otimes |   color \ket \; , 
\end{equation}
where the space, spin and flavor part is always symmetric, the color
part guaranteeing overall anti-symmetry. The non-spatial part, $|spin
\ket \otimes | flavor \ket \otimes | color \ket$, is governed by the
symmetry group $SU(2)_S \otimes SU(3)_F \otimes SU(3)_C$. For the
positively charged pion $\pi^+$ one finds, for instance, 
\begin{equation}
  | \pi^+ \ket = \frac{1}{\sqrt{2N_C}} ( | \bar d_{a\uparrow} u_{a
  \downarrow} \ket - | \bar d_{a \downarrow} u_{a \uparrow} \ket ) \; ,
\end{equation}
where the spatial part has been suppressed. The latter will, however,
become relevant in later chapters.

It should be pointed out, that the flavor symmetry is not realized in
nature as an exact symmetry, because the states in the $SU(3)_F$
multiplets have different masses. The color symmetry, on the other hand,
is unbroken.

If one calculates the magnetic moments ($\mu_i = e_i/2M_i$) of the 
hadrons $i =p$, $n$, $\Lambda$ by adding those of their quark
constituents, one can determine the {\em constituent masses} of 
the quarks
\begin{equation}
  \label{CONST_MASSES}
  M_u = M_d = 340 \; \mathrm{MeV} \; , \quad  M_s = 510 \; \mathrm{MeV} \; .
\end{equation}
The constituent quark masses thus roughly sum up to the total bound
state (hadron) mass.

At this point, the question arises whether quarks can really be `seen'
experimentally. The answer is both `yes' and `no'. On the one hand,
quarks have never been observed as free particles outside a hadron, a
property called {\em quark confinement}. On the other hand, the
deep-inelastic scattering (DIS) experiments beginning in the late
sixties have revealed a rich substructure of hadrons, in particular the
nucleon (Nobel prize 1990 for Friedman, Kendall, and Taylor). This
substructure consists of quasi-free, point-like constituents (`partons')
which carry the quantum number of quarks.

This large variety of facts about hadrons has to be explained by the
dynamical theory of strong interactions which is the subject of the next
section.

\section{Quantum Chromodynamics}

Twenty-five years after its inception
\cite{gross:73,weinberg:73,fritzsch:73}, Quantum Chromodynamics (QCD) is
generally accepted as {\em the} theory of strong interactions. It
describes the interaction of matter (quarks and antiquarks) with
intermediate gauge bosons named gluons. Like the other theories of the
fundamental interactions, QCD is a gauge theory; its symmetry group,
$SU(3)$ is associated with three `color' charges mediating the strong
interaction. As the gauge group $SU(3)$ is non-Abelian, the (eight)
gluons themselves carry color charge and therefore can interact with
each other\footnote{This explains the name `gluon' which pictorially
  expresses the idea that these particles act like glue when the bind
  among themselves (`glueballs') or force quarks to bind into hadrons.}.
Nevertheless, in a somewhat condensed notation, the QCD Hamiltonian
appears very similar to the one of Quantum Electrodynamics (QED). The
former is
\begin{equation}
  \label{HQCD}
  \mathcal{H}_\mathrm{QCD} = \sfrac{1}{2} (\vc{E}^2 + \vc{B}^2) + \psi^\dagger
  (\vcg{\alpha}\cdot \vc{D} + \beta \mathcal{M}) \psi \; ,
\end{equation}
where $\vc{D}$ denotes the covariant derivative (in the fundamental
representation), 
\begin{equation}
  \vc{D} = i \nabla - g \vc{A} \; 
\end{equation}
and $\mathcal{M} = diag(m_u, m_d , m_s, \ldots)$ the quark-flavor mass
matrix. If its mass eigenvalues were all equal, $SU(N_F)$ would be an
exact symmetry of QCD. However, as already stated, the only exact
(local) symmetry is its gauge invariance under $SU(3)_C$.     

Note that, in contrast to the Abelian gauge theory QED, the
chromo-electric and -magnetic fields $\vc{E}$ and $\vc{B}$ as well as
the gauge potentials $\vc{A}$ are non-commuting (color) matrices so that
there are implicit summations over color indices in (\ref{HQCD}). The
self-interaction of the gluon fields $\vc{A}$ becomes  evident upon
expanding the chromo-magnetic fields $\vc{B}$ in terms of the $\vc{A}$,
which, however, will not be done here as it can be found in any textbook
on QCD.

Like its older `relative' QED, QCD is a renormalizable relativistic
quantum field theory. Any infinities arising from the point-like (local)
nature of the interaction can therefore be consistently absorbed into a
redefinition of the physical parameters (masses, couplings).  As a
result, the strong coupling parameter $\alpha_s$ is not a constant but
is running with the typical momentum scale of the physical process under
consideration. The microscopic reason for this are vacuum polarization
effects: quarks screen and therefore weaken the color charge (analogous
to QED), whereas the self-interacting gluons {\em anti}-screen the color
charge which is the dominating effect. Unlike in QED, therefore, the
running coupling $\alpha_s (Q)$ of QCD is weak for high momentum
transfer $Q$ (small distances). This is the realm of `asymptotic
freedom' where perturbative methods work. For small momentum transfer
$Q$ (large distances), the coupling is large, perturbation theory breaks
down, and one has to utilize nonperturbative methods. Typical and
well-established values for $\alpha_s$ are \cite{barnett:96}
\begin{eqnarray}
  \alpha_s (M_Z) &=& \alpha_s (91.2 \mathrm{GeV}) = 0.118 \pm 0.006 \;
  , \\
  \alpha_s (M_\tau) &=& \alpha_s (1.78 \mathrm{GeV}) = 0.33 \pm 0.03 \; .
\end{eqnarray}
Thus, the nonperturbative domain is characterized by a maximum momentum
scale of approximately 1 GeV.  In some loose sense one can therefore speak of
two relevant phases of QCD, the weak coupling phase or perturbative QCD,
and the strong coupling phase (nonperturbative QCD).

Let us return to our first subject, the hadrons.  In principle, it is
quite clear, what a hadron is in QCD: it is an eigenstate of the QCD
Hamiltonian,
\begin{equation}
  \label{QCD_SEQ}
  H_{QCD} |\mathrm{hadron} \ket = M_H | \mathrm{hadron} \ket \; ,
\end{equation}
where $M_H$ denotes the hadron mass\footnote{ In writing down
  (\ref{QCD_SEQ}) we neglect any possible `fine splitting' due to
  electromagnetic or weak interactions.}. The question, of course, is,
whether the `QCD Schr\"odin\-ger equation' (\ref{QCD_SEQ}) can be solved.
If we consider a typical hadronic scale like, say, the nucleon radius, 
$R_N \simeq $ 1 fm, the associated energy scale is $Q \simeq \hbar c /R_N
\simeq $ 200 MeV. This number tells us that we are in the low-energy
regime which implies that the binding of quarks into hadrons is a
nonperturbative phenomenon. In other words, a perturbative solution of
the `QCD Schr\"odinger equation' will make no sense in general.

There are two main routes out of this dilemma. Firstly, one can try to
perform brute-force ab-initio calculations which involve sophisticated
computer simulations on the largest machines available. Technically, one
makes use of a space time discretization leading to a lattice
formulation of gauge theories. This program has been developed over more
than twenty years now, and has become the subject of entire textbooks
\cite{montvay:94}. 

Secondly, one can rely on a reputable tradition of physics, namely model
building. There is an abundance of hadron models on the market, the most
popular one being the constituent quark model and variants thereof. In
this model, one starts with a non-relativistic phenomenological
Hamiltonian of the form
\begin{equation}
  H = H_0 + V_{\mbox{\footnotesize conf}} \; ,
\end{equation}
with an ad-hoc confining potential $V_{\mbox{\footnotesize conf}}$ which
typically is proportional to the inter-quark distance $r$ (or $r^2$).
The Hamiltonian describes the dynamics of two or three constituent
quarks with masses $M_q \, \simgeq $ 300 MeV like in (\ref{CONST_MASSES})
which are treated as parameters. The main virtue of the model consists
in its rather accurate reproduction of the hadron masses (`spectroscopy'). An
exhaustive compendium of results can be found in the recent reviews
\cite{lucha:91,mukherjee:93,chakrabarty:94}.

However, as any model, also the constituent quark model has its
shortcomings. First of all, it works where it is not supposed to work.
A hadron size of $R_H \simeq $ 1~fm leads to an average constituent
momentum of $\bra p \ket \simeq $ 200~MeV which is of the order of the
constituent mass $M_q$. Therefore, hadrons are {\em relativistic} bound
states and a non-relativistic treatment is not appropriate. Furthermore,
the relation of the model with QCD is rather unclear. In QCD, as in any
relativistic quantum field theory, one expects a bound state like
e.g.~the pion to be of the form
\begin{equation}
  | \pi \ket \sim \psi_2 | q \bar q \ket + \psi_3 | q \bar q g \ket +
    \psi_4 | q \bar q q \bar q \ket + \ldots \; .
\end{equation}
This means that hadrons are states containing an infinite number of
quarks and gluons, which is consistent with the results of the DIS
experiments where, with growing resolution $Q^2$, an increasing number
of partons is observed.  This confirms that there are non-vanishing
amplitudes $\psi_2$, $\psi_3$ ... to find two quarks, two quarks and a
gluon ..., in general to find an arbitrary number of quarks and gluons
in a hadron.  

Another flaw of the model is that the constituent masses $M_q$ do not
coincide with the masses in the QCD Hamiltonian (\ref{HQCD}). This is
qualitatively explained by the idea that the constituent quarks are {\em
  effective} quarks (quasi-particles) with effective masses originating
from `medium effects' due to the QCD vacuum. A quantitative derivation
of this mass generation within QCD is to a large extent still missing.

A more fundamental drawback of the constituent quark model is the
following: it does not take into account all symmetries of QCD. This is
particularly true for one prominent symmetry which is believed to govern
most of the physics of light hadrons, namely chiral symmetry. This is
our next topic.

\section{Chiral Symmetry in QCD}

\begin{table}
\begin{center}
\renewcommand{\arraystretch}{1.2}
\caption{\label{T2} \textsl{The presently observed quark flavors. $Q/e$ is the
  electric charge in units of the electron charge. The (scale
  dependent!) quark masses are given for a scale of 1 GeV.}}
\vspace{.5cm}

\begin{tabular*}{\textwidth}[h]{ @{\extracolsep\fill} l c c r c }
\hline\hline
flavor & & $Q/e$ & mass & \\
\hline
down    & $d$ & $-1/3$ & 10  & MeV \\
up      & $u$ & $+2/3$ & 5   & MeV \\
strange & $s$ & $-1/3$ & 150 & MeV \\
\hline
charm   & $c$ & $+2/3$ & 1.5 & GeV \\
bottom  & $b$ & $-1/3$ & 5.1 & GeV \\
top     & $t$ & $+2/3$ & 180 & GeV \\
\hline\hline
\end{tabular*}
\end{center}
\end{table}

If we have a look at Table \ref{T2}, which provides a list of all quark
flavors, we realize that there are large differences in the quark masses
as they appear in the QCD Hamiltonian. In particular, there is a
hierarchy,
\begin{equation}
  \underbrace{m_u , m_d \ll m_s}_{\mbox{light quarks}} \ll \underbrace{m_c, m_b,
  m_t}_{\mbox{heavy quarks}} \; .
\end{equation}
As the masses of heavy and light quarks are separated by the same scale
as the perturbative and nonperturbative regime, one expects different
physics associated with those two kinds of quarks. This expectation
turns out to be true. The physics of heavy quarks is governed by a
symmetry called `heavy quark symmetry' leading to a very successful
`heavy quark effective theory' \cite{neubert:94}. The physics of light
quarks, on the other hand, is governed by chiral symmetry which we are
now going to explain.

Let us write the QCD Hamiltonian in the following way,
\begin{equation}
  \label{QCD_HAM}
  H_{\mathrm{QCD}} = H_\chi + \bar \psi \mathcal{M} \psi \; , 
\end{equation} 
where we recall the mass matrix for the light flavors, $\mathcal{M} =
diag(m_u , m_d, m_s)$. To a good approximation, one can set $\mathcal{M}
= 0$. In this case, the QCD Hamiltonian $H_\chi$ is invariant under the
symmetry group $SU(3)_R \otimes SU(3)_L$, the chiral flavor group. Under
the action of this group, the left and right handed quarks independently
undergo a chiral rotation. Due to Noether's theorem, there are sixteen
conserved quantities, eight vector charges and, more important for us,
eight pseudo-scalars, the chiral charges $Q_5^a$ satisfying
\begin{equation}
  [Q_5^a , H_\chi ] = 0 \; .
\end{equation}
This states both that the chiral charges are conserved, and that
$H_\chi$ is chirally invariant. Under parity, $Q_5^a \to - Q_5^a$. Now,
if $|A \ket$ is an eigenstate of $H_\chi$, so is $Q_5^a |A \ket$ with
the same eigenvalue. Thus, one expects (nearly) degenerate parity
doublets in nature, which, however, do not exist empirically. The only
explanation for this phenomenon is that chiral symmetry is spontaneously
broken. In contradistinction to the Hamiltonian, the QCD ground state
(the vacuum) is {\em not} chirally invariant,
\begin{equation}
  Q_5^a | 0 \ket \ne 0 \; . 
\end{equation}
For this reason, there must exist a non-vanishing vacuum expectation
value, the quark condensate,
\begin{equation}
  \bra \bar \psi \psi \ket = \bra \bar \psi_R \psi_L + \bar \psi_L
  \psi_R \ket  \; .
\end{equation}
This condensate is not invariant (it mixes left and right) and therefore
serves as an order parameter of the symmetry breaking. Note that in QCD
the quark condensate is a renormalization scale dependent quantity. A
recent estimate can be found in \cite{dosch:98}, with a numerical value,
\begin{equation}
  \cond (1 \; \mathrm{GeV}) \simeq (-229 \; \mathrm{MeV})^3 \; .
\end{equation}
In terms of the full quark propagator,
\begin{equation}
  S(p) =  \frac{p \!\!\!/ + M(p)}{p^2 - M^2(p)} \; ,
\end{equation}
where we have allowed for a momentum dependent (or `running') mass
$M(p)$, the quark condensate is given by 
\begin{equation}
  \cond = - i \int \frac{d^4 p}{(2\pi)^4} \tr S(p) = - 4i N_C \int \frac{d^4
    p}{(2\pi)^4} \frac{M(p)}{p^2   - M^2 (p)} \; .
\end{equation}
We thus see that the involved Dirac trace yields a non-vanishing
condensate only if the effective quark mass $M(p)$ is non-zero.  This
links the existence of a quark condensate to the mechanism of dynamical
mass generation. In this way we have found  another argument that the
bare quarks appearing in the QCD Hamiltonian (\ref{QCD_HAM}) indeed
acquire constituent masses. Of course we are still lacking a microscopic
mechanism for that.

Goldstone's theorem \cite{goldstone:61} now states that for any symmetry
generator which does not leave the vacuum invariant, there must exist a
massless boson with the quantum number of this generator. This results
in the prediction that in massless QCD one should have an octet of
massless pseudo-scalar mesons. In reality one finds what is listed in
Table~\ref{T3}.

\begin{table}
\begin{center}
\renewcommand{\arraystretch}{1.2}
\caption{\protect\label{T3} \textsl{Masses of the  pseudo-scalar octet
    mesons (in MeV).}} 
\vspace{.5cm}

\begin{tabular*}{\textwidth}[h]{ @{\extracolsep\fill} l  c c c c c }
\hline\hline
meson & $\pi^0$ & $\pi^\pm$ & $K^0$, $\bar K^0$ & $K^\pm$ & $\eta$ \\
\hline
mass  & 135 & 140 & 500 & 494 & 549 \\
\hline\hline
\end{tabular*}
\end{center}
\end{table}

The non-vanishing masses of these mesons are interpreted as stemming
from the non-vanishing quark masses in the QCD Hamiltonian which break
chiral symmetry explicitly. They can be consistently viewed as
perturbations within an effective field theory dubbed `chiral
perturbation theory', which has become one of the standard tools to make
quantitative predictions for low-energy strong-interaction physics
\cite{holstein:95}.

It should be stressed that chiral symmetry has nothing to say about the
mechanism of confinement which presumably is a totally different story.
This is also reflected within a QCD based derivation of chiral symmetry
breaking, the instanton model \cite{diakonov:95}. This model explains
many facts of low-energy hadronic physics but is known not to yield
confinement. It is therefore possible that confinement is not
particularly relevant for the understanding of hadron structure
\cite{diakonov:95}.

\section{Outline}

This report is divided into two main parts, devoted to the foundations and
applications of light-cone dynamics, respectively. In the first part, we
begin with some general remarks on relativistic dynamics (Chapter~2). As
a paradigm example we discuss the free relativistic particle which is
the prototype of a reparametrization invariant system.  We show that the
choice of the time parameter is not unique as this corresponds to a
gauge fixing, the purpose of which is to get rid of the reparametrization
redundancies. By considering the stability subgroups of the Poincar\'e
group, we find that there are essentially three reasonable choices of
`time' for a relativistic system, corresponding to Dirac's instant, point
and front form, respectively. 

The latter choice is the basis of light-cone dynamics which we analyze
in detail in  Chapter~3. We find some unique features like a maximal
number of kinematical Poincar\'e generators, boost invariance of
relative coordinates and light-cone wave functions and separation of
center-of-mass and relative motion. Many of these features are
reminiscent of what is known from non-relativistic physics. 

Chapter~4 is devoted to light-cone \emph{field} quantization. We show
how the Poincar\'e generators are defined in this case and utilize
Schwinger's quantum action principle to derive the canonical
commutators. The relation between equal-time commutators, the field
equations and their solutions for different initial and/or boundary
conditions is clarified. It is found that light-cone field theories,
being of first order in the velocity, generally are constrained systems
which require a special treatment. We use a method of phase space
reduction due to Faddeev and Jackiw to take this into account. This
leads us naturally to the construction of Fock space.

In Chapter~5 we discuss what might be viewed as the most spectacular
feature of light-cone quantum field theory, the triviality of the
vacuum. We show how this property comes about and illustrate the
difference to standard instant-form quantization for some very simple model
field theories.

We conclude the presentation of the foundations with Chapter~6, where we
introduce the notion of light-cone wave functions. Their special
properties make them useful tools for investigating hadronic physics. We
show how light-cone wave functions of hadrons can be obtained by solving
the light-cone Schr\"odinger equation.
 
The first application we will study in Part~II is the phenomenon of
spontaneous symmetry breaking for scalar field theory (Chapter~7), in a
setting which can be viewed as a simplified variant of the Higgs
mechanism. We analyze how a non-vanishing vacuum expectation value can
arise in a trivial vacuum. We find that the vacuum properties are
carried by the field mode carrying vanishing longitudinal momentum. This
mode satisfies a complicated constraint equation which we solve both
perturbatively and nonperturbatively.

As the methods of Chapter~7 do not work for fermionic fields, we have to
consider alternatives which is the subject of Chapter~8. We use the particle
spectrum of  model field theories in 1+1 dimensions to reconstruct their
vacuum properties. The spectrum is obtained by solving the light-cone
Schr\"odinger equation. This is done here with unprecedented accuracy,
which is made possible through the use of computer algebraic methods
yielding analytic solutions even in high orders.

In Chapter~8 we finally make contact with phenomenology. We calculate
the light-cone wave function of the pion within the Nambu and
Jona-Lasinio model. This model is known to provide a good description of
spontaneous chiral symmetry breaking, as it is governed by the same
symmetry group as low-energy QCD. With the pion wave function at hand we
predict a number of observables like the pion charge and core radius,
the electromagnetic form factor and the r.m.s.~transverse momentum. We
conclude with a calculation of the pion distribution amplitude which is
important for the study of exclusive processes in QCD.

\newpage
\pagestyle{empty}
\cleardoublepage

\vspace*{7cm}

\begin{center}
   
  \Huge \textbf{Part I}\\[1cm]
  \textbf{Foundations}

\end{center}
\addcontentsline{toc}{part}{I Foundations}

\cleardoublepage

\pagestyle{headings}

\chapter{Relativistic Particle Dynamics}

The nature of elementary particles calls for a synthesis of relativity
and quantum mechanics. The necessity of a quantum treatment is quite
evident in view of the microscopic scales involved which are several
orders of magnitude smaller than in atomic physics. These very scales,
however, also require a relativistic formulation. As we have seen
already, a typical hadronic scale of 1 fm, for instance, corresponds to
momenta of the order of $p \sim \hbar c / 1$ fm $\simeq 200$ MeV. For
particles with masses of the order of 1 GeV and below this implies sizable
velocities of 0.2 $c$ and larger.

It turns out that the task of combining the principles of quantum
mechanics and relativity is not a straightforward one. One can neither
simply extend ordinary quantum mechanics to include relativistic physics
nor quantize relativistic mechanics using the ordinary correspondence
rules. Nevertheless, Dirac and others have succeeded in formulating what
is called ``relativistic quantum mechanics'', which has become a subject
of text books since \cite{bjorken:64, yndurain:96}. It should, however,
be pointed out that this formulation, which is based on the concept of
single-particle wave-functions and equations, is not really consistent.
It does not correctly account for relativistic causality (retardation
effects etc.)  and the existence of anti-particles. As a result, one has
to struggle with issues like the Klein paradox \cite{bjorken:64,
  yndurain:96}, the definition of position operators \cite{newton:49}
and the like.

The solution to these problems is provided by quantum {\em field}
theory, with an inherently correct description of anti-particles that
entails relativistic causality. In contrast to single-particle wave
mechanics, quantum field theory is a (relativistic) many body
formulation that necessarily involves (anti-)particle creation and
annihilation. Physical particle states are typically a superposition of
an infinite number of ``bare'' states, as any particle has a finite
probability to emit or absorb other particles at any moment of time.
With the advent of QCD, however, a conceptual difficulty concerning this
many-particle picture has appeared and already been discussed in the
introduction. At low energy or momentum transfer, hadrons, the bound
states of QCD, are reasonably described as bound states of two or three
constituent quarks and thus as {\em few-body} systems. It is rather
unclear how such a constituent picture can arise in a quantum field
theory such as QCD.

The physically motivated desire to describe hadrons as bound states of a
small, fixed number of  constituents is our rationale to go back
and reanalyze the relation between Hamiltonian quantum mechanics and
relativistic quantum field theory.

Quite generally, bound states are obtained by solving the Schr\"odinger
equation,
\begin{equation}
  i \hbar \pad{}{\tau} |\psi(\tau) \ket = H | \psi(\tau) \ket \; , 
\end{equation}
for {\em stationary states},  
\begin{equation}
  |\psi(\tau) \ket = e^{-i E \tau} | \psi(\tau = 0) \ket .  
\end{equation}
This leads to the bound-state equation
\begin{equation}
  H |\psi(\tau = 0) \ket = E | \psi(\tau = 0) \ket \; ,
\end{equation}
where $E$ is the bound state energy. We would like to make this
Hamiltonian formalism consistent with the requirements of relativity.
It is, however, obvious from the outset that this procedure is not
manifestly covariant as it singles out a time $\tau$ (and an energy $E$,
respectively).  Furthermore, it is not even clear what the time $\tau$
really is as it does not have an invariant meaning.

\section{The Free Relativistic Particle}

To see what is involved it is sufficient to consider the classical
dynamics of a free relativistic particle. We want to find the associated
canonical formulation as a basis for subsequent quantization. We will
proceed by analogy with the treatment of classical free strings which is
described in a number of textbooks \cite{green:87, polyakov:87}.
Accordingly, the relativistic point particle may be viewed as an
infinitely short string.  

We recall that the action for a relativistic particle is essentially
given by the arc length of its trajectory
\begin{equation}
  \label{REL_ACT1}
  S = - m s_{12} \equiv - m \int_1^2 ds  \; .
\end{equation}
This action\footnote{We work in natural units, $\hbar = c = 1$.} is a
Lorentz scalar as
\begin{equation}
  ds = \sqrt{g_{\mu \nu} x^\mu x^\nu} \; 
\end{equation}
is the (infinitesimal) invariant distance. We can rewrite the action
(\ref{REL_ACT1}) as 
\begin{equation}
  \label{REL_ACT2}
  S = -m \int_1^2 ds \, \sqrt{\dot x_\mu \dot x^\mu} \equiv 
  \int_1^2 ds \, L(s) \; ,
\end{equation}
in order to introduce a Lagrangian $L(s)$ and the four velocity $\dot
x^\mu \equiv dx^\mu/ds $. The latter obeys 
\begin{equation}
  \label{V2}
  \dot x^2 \equiv \dot x_\mu \dot x^\mu = 1 \; , 
\end{equation}
as the arc length provides a natural parametrization. Thus, $\dot x^\mu$
is a time-like vector, and we assume in addition that it points into the
future, $\dot x^0 > 0$. In this way we guarantee relativistic causality
ensuring that a real particle passing through a point $A$ will always
propagate into the future light cone based at $A$.

We proceed with the canonical formalism by calculating the canonical
momenta as
\begin{equation}
  \label{CAN_MOM1}
  p^\mu = - \pad{L}{\dot x_\mu} = m \dot x^\mu \; .
\end{equation}
These are not independent, as can be seen by calculating the square
using (\ref{V2}),
\begin{equation}
  p^2 = m^2 \dot x^2 = m^2 \; , 
\end{equation}
which, of course, is the usual mass-shell constraint. This constraint
indicates that the Lagrangian $L(s)$ defined in (\ref{REL_ACT2}) is {\em
  singular}, which means that its Hessian $W^{\mu \nu}$ with respect to
the velocities,
\begin{equation}
  W^{\mu \nu} = \frac{\partial^2 L}{\partial \dot x_\mu \partial \dot
  x_\nu} = - \frac{m}{\sqrt{\dot x^2}} \left( g^{\mu \nu} - \frac{\dot
  x^\mu \dot x^\nu}{\dot x^2} \right) =  - m \left( g^{\mu \nu} - \dot
  x^\mu \dot x^\nu \right)\; ,  
\end{equation}
is degenerate. It has a zero mode given by the velocity itself,
\begin{equation}
  \label{SING_LAG}
  W^{\mu \nu} \dot x_\nu = 0 \; . 
\end{equation}
The Lagrangian being singular implies that the velocities cannot  be
uniquely expressed in terms of the canonical momenta. This, however, is
not obvious from (\ref{CAN_MOM1}), as we can easily solve for the
velocities, 
\begin{equation}
  \dot x^\mu = p^\mu /m \; .
\end{equation}
But if one now calculates the canonical Hamiltonian,
\begin{equation}
  \label{VAN_HAM}
  H_c = - p_\mu \dot x^\mu - L = -m \dot x^2 + m \dot x^2 = 0 \; ,
\end{equation}
one finds that it is vanishing! It therefore seems that we do not have a
generator for the  time evolution of our dynamical system. In the following,
we will analyze the reasons for this peculiar finding.

First of all we note that the Lagrangian is homogeneous of first degree
in the velocity,    
\begin{equation}
  L (\alpha \dot x^\mu) = \alpha L(\dot x^\mu) \; .
\end{equation}
Thus, under a  reparametrization of the world-line,
\begin{equation}
  \label{REPAR} 
  s \mapsto s^\prime \; , \quad x^\mu (s) \mapsto x^\mu \big(s^\prime (s)\big) \; ,
\end{equation}
where the mapping $s \mapsto s^\prime$ is one-to-one with $ds^\prime /
ds > 0$ (orientation conserving), the Lagrangian changes according to
\begin{equation}
  L(dx^\mu/ds) = L\Big( (dx^\mu/ds^\prime) (ds^\prime/ds) \Big) = (ds^\prime/ds)
  L(dx^\mu/ds^\prime)\; .
\end{equation}
This is sufficient to guarantee  that the action is invariant under
(\ref{REPAR}), that is, {\em reparametrization invariant}, 
\begin{equation}
  S =  \int_{s_1}^{s_2}ds \, L(dx^\mu/ds) = 
  \int_{s_1^\prime}^{s_2^\prime}ds^\prime \, \frac{ds}{ds^\prime}\frac{ds^\prime}{ds}
  L(dx^\mu/ds^\prime) \equiv S^\prime \; , 
\end{equation}
if the endpoints remain unchanged, $s_{1,2} = s_{1,2}^\prime$. On the
other hand, $L$ is homogeneous of the first degree if and only if
Euler's formula holds, namely
\begin{equation}
  \label{EULER}
  L = \pad{L}{\dot x^\mu} x^\mu = - p_\mu \dot x^\mu \; .
\end{equation}
This is exactly the statement (\ref{VAN_HAM}), the vanishing of the
Hamiltonian. Furthermore, if we differentiate (\ref{EULER}) with respect
to $\dot x^\mu$, we recover (\ref{SING_LAG}) expressing the singular
nature of the Lagrangian. Summarizing, we have found the general result
\cite{hanson:76,sundermeyer:82} that if a Lagrangian is homogeneous of
degree one in the velocities, the action is reparametrization invariant,
and the Hamiltonian vanishes. In this case, the momenta are homogeneous
of degree zero, which renders the Lagrangian singular.

In order to properly incorporate the reparametrization invariance we
chose an arbitrary parametrization, $\tau \mapsto x^\mu (\tau)$, so that
the infinitesimal arc length becomes
\begin{equation}
  \label{METRIC}
  ds^2 = \left(\frac{ds}{d\tau}\right)^2 d\tau^2 \equiv h(\tau) d\tau^2
  \; , 
\end{equation}
introducing a world-line metric (or einbein) $h(\tau)$. In what follows
we let the over-dot denote the derivative with respect to $\tau$,
e.g.~$\dot{f} = df/d\tau$.  From (\ref{METRIC}) we get that
\begin{equation}
  \label{V_SCALE}
  \dot x^2 = \left(\frac{ds}{d\tau}\right)^2 = h(\tau) \; ,
\end{equation}
so that, according to (\ref{V2}),  $h(\tau)=1$ corresponds to $d\tau =
ds$. Therefore, the world-line metric $h$ provides a `length' scale for
the four-velocity.  We write the action as
\begin{equation}
  \label{TAU_ACT}
  S = -m \int_1^2 d\tau \, \sqrt{\dot x^2} = -m \int_1^2 d\tau \,
  \sqrt{h(\tau)} \equiv \int_1^2 d\tau \, L(\tau) \; .
\end{equation}
The canonical momenta are
\begin{equation}
  \label{CAN_MOM2}
  \pi^\mu \equiv - \pad{L}{\dot x_\mu} = m \frac{\dot x^\mu}{\sqrt{\dot x^2}} =
  m \frac{\dot x^\mu}{\sqrt{h(\tau)}} = \frac{m}{\sqrt{h(\tau)}} \,
  \frac{dx^\mu}{ds} \frac{ds}{d\tau} = m \frac{dx^\mu}{ds}  \equiv p^\mu
  \; .
\end{equation}
We note that they are independent of the world-line metric $h$, so that,
upon squaring (\ref{CAN_MOM2}) one again finds the mass shell constraint
$\pi^2 = p^2 = m^2$. However, if we want to express the velocities in
terms of the momenta, we get
\begin{equation}
  \label{X_DOT}
  \dot x^\mu = \sqrt{h} \, p^\mu / m \; .
\end{equation}

Therefore, it is exactly the world-line metric $h$ that expresses the
arbitrariness involved when solving for the velocities. Unlike for the
momenta, the scale for the velocities is not fixed, unless one chooses a
particular world-line metric as in (\ref{V2}).

By the same arguments as before, the canonical Hamiltonian still
vanishes. The Lagrangian one-form thus coincides with the canonical
one-form,
\begin{equation}
  L(\tau) d\tau = - p_\mu dx^\mu \; .
\end{equation}
From the latter we read off the non-vanishing Poisson brackets,
\begin{equation}
  \label{PB_COV1}
  \pb{x^\mu}{p^\nu} = -g^{\mu\nu} \; , 
\end{equation}
which will be needed in a moment.  To proceed we utilize the
Dirac-Bergmann algorithm for singular systems
\cite{dirac:50,dirac:64,bergmann:56,anderson:51} (for monographs, see
\cite{hanson:76,sundermeyer:82,gitman:90,henneaux:92}) by adding the
constraint
\begin{equation}
  \label{CONSTRAINT}
  \theta \equiv p^2 - m^2 
\end{equation}
to the (vanishing) canonical Hamiltonian via a Lagrange multiplier
$\lambda$. This yields the primary Hamiltonian
\begin{equation}
  H_p = \lambda (p^2 - m^2) \; .
\end{equation}
The constraint (\ref{CONSTRAINT}) is the only one arising and has vanishing
Poisson bracket with itself whereupon it is  called {\em first
  class}. Technically, this implies that demanding consistency,
i.e.~conservation of the constraint in time $\tau$, does not lead to an
equation determining the Lagrange multiplier. Physically, this means
that the time development of the system is not completely
determined. Such a behavior is typical for systems with a gauge or
redundancy symmetry, an example of which is the reparametrization
invariance discussed above. The redundancy consists in the fact that a
single trajectory (world-line) can be described by an infinite number of
different parametrizations.  The physical objects, the trajectories, are therefore
equivalence classes obtained by identifying all reparametrizations. An
infinitesimal reparametrization is given by 
\begin{eqnarray}
  \tau^\prime (\tau) &=& \tau + \delta \tau (\tau) \; , \\
  \delta x^\mu (\tau) &=& x^\mu (\tau + \delta \tau ) - x^\mu (\tau) =
  \dot x^\mu \delta \tau \label{INF_REP1} \; .  
\end{eqnarray}
In systems with redundancy symmetries, these symmetries are generated by
the first class constraints. A familiar example is provided by gauge
fields (like in electromagnetism), where the first class constraints are
represented by Gau\ss's law, the generator of (time-independent) gauge
transformations. For the case at hand, we find using (\ref{PB_COV1}),
\begin{equation}
  \label{INF_REP2}
  \delta x^\mu = \pb{x^\mu}{\delta G} \equiv \pb{x^\mu}{\theta \delta \epsilon} = 
  - 2 p^\mu \delta \epsilon = - 2 m  
  \frac{\dot x^\mu}{\sqrt{h}} \delta\epsilon \; .  
\end{equation}
Identifying $\delta \tau \equiv (2m/ \sqrt{h}) \delta\epsilon$ to
account for the different dimensionalities, (\ref{INF_REP1}) and
(\ref{INF_REP2}) coincide. Thus, the reparametrization (\ref{INF_REP1})
is indeed generated by the first-class constraint $\theta$ of
(\ref{CONSTRAINT}). The Hamiltonian equations of motion become
\begin{eqnarray}
  \dot x^\mu &=& \pb{x^\mu}{H_p} = - 2 \lambda p^\mu \; , \label{HAM_EQ1}\\
  \dot p^\mu &=& \pb{p^\mu}{H_p} = 0 \; . \label{HAM_EQ2}
\end{eqnarray}
It is important to note at this point that the equation for $\dot x^\mu$
contains the Lagrange multiplier $\lambda$ which is undetermined so
far. This makes the whole dynamics of the system undetermined. If we
compare (\ref{HAM_EQ1}) with (\ref{X_DOT}), we find that the
arbitrariness is again encoded in the world-line metric as the
multiplier is
\begin{equation}
  \label{LH}
  \lambda = - \sqrt{h}/2m \; .
\end{equation}
To overcome the problem of determining $\lambda$ or $h$, one has to {\em
  fix a gauge} by choosing an auxiliary condition,
\begin{equation}
  \label{GF_GEN}
  \chi (x ;\tau) = 0 \; .
\end{equation}
Demanding consistency, i.e.~conservation of the gauge fixing condition
in time,
\begin{equation}
  \label{LM_EQ}
  \dot \chi = \pad{\chi}{\tau} + \pb{\chi}{H_p} = \pad{\chi}{\tau}+
  \lambda \pb{\chi}{\theta} = \pad{\chi}{\tau} - 2
  \lambda p^\mu \pad{\chi}{x^\mu}  = 0 \; ,
\end{equation}
we obtain the desired equation determining the Lagrange multiplier
$\lambda$,
\begin{equation}
  \label{LM_GEN}
  \lambda = - \pb{\chi}{\theta}^{-1} \, \pad{\chi}{\tau} =  
  \frac{1}{2 (p  \cdot \partial)  \chi} \, \pad{\chi}{\tau} \; .
\end{equation}
We see that it is crucial for the determination of $\lambda$ that the
gauge fixing $\chi$ depends \emph{explicitly} on the time parameter
$\tau$ and on at least one of the $x^\mu$. This is necessary for
rendering the Poisson bracket $\pb{\chi}{\theta}$ nonvanishing. The
analogue of this condition in the gauge theory context is well known: 
it corresponds to having a nonvanishing Faddeev-Popov matrix. 
For the relativistic particle, one can say that the gauge condition
amounts to defining a time parameter $\tau$ in terms of the coordinates
$x^\mu$. This suggests that $\chi$ will in general be of the form
\begin{equation}
  \label{GF_TAU}
  \chi(x ; \tau) = \tau - F(x) \; ,
\end{equation}
stating that $\tau$ is some function $F$ of the coordinates, $\tau =
F(x)$. In this case, (\ref{LM_GEN}) becomes
\begin{equation}
  \label{LM_DET}
  \lambda =   - \frac{1}{2 (p \cdot \partial)  F(x)} \; .
\end{equation}
The quantity $(p \cdot \partial) F(x)$ is nothing but the projection of
the four-momentum $p^\mu$ onto the normal $N^\mu \equiv \partial^\mu F$
of the hypersurface $\tau = const$. The knowledge of the Lagrange
multiplier (or the einbein $h$) makes the dynamics uniquely determined,
as (\ref{HAM_EQ1}) leads to
\begin{equation}
  \label{XDOT_DET}
  \dot x^\mu = - 2 \lambda p^\mu = \frac{p^\mu}{N \cdot p} \; . 
\end{equation}
Squaring this expression or using (\ref{LH}), the world-line metric is
found to be
\begin{equation}
  \label{LH_DET}
  h (\tau) = 4 m^2 \lambda^2 = \frac{m^2}{(N \cdot p)^2} =
  \frac{p^2}{(N \cdot p)^2} \; .
\end{equation}
It thus measures the normal component of the momentum (squared) compared
to $p^2 = m^2$.
The most common choice for $\tau$ is the Galileian time $t$ of, say, the
observer rest frame,
\begin{equation}
  \chi = x^0 - \tau \equiv t - \tau = 0 \; .
\end{equation}
This gauge choice corresponds to a particular choice for the world-line
metric $h$, given by 
\begin{equation}
  \label{METRIC_T}
  ds^2 = dt^2 - dx^i dx^i = (1 - v^2)dt^2 = \frac{1}{\gamma^2} dt^2 \equiv
  h(t) dt^2 
  \; ,
\end{equation}
with $v^i = dx^i /dt$. In this case, the Lagrange multiplier as derived
from (\ref{LM_GEN}) becomes
\begin{equation}
  \label{LM_T}
  \lambda = -\frac{1}{N \cdot p} = - \frac{1}{2 p^0} \; .
\end{equation}
From (\ref{HAM_EQ1}) and (\ref{XDOT_DET}) we get
\begin{equation}
  \dot x^\mu = - 2 \lambda p^\mu = p^\mu/p^0 \; ,
\end{equation}
This yields  the four-velocity squared
\begin{equation}
  \label{4V_SQUARED}
  \dot x^2 = \frac{p^2}{p_0^2}= \frac{m^2}{p_0^2} = \frac{1}{\gamma^2}
  \equiv h(t)
  \; ,
\end{equation}
in accordance with the result (\ref{METRIC_T}) for the world-line
metric. The Hamiltonian generating the $t$-evolution is found by solving
the mass-shell constraint $\theta$ for $p^0 = N \cdot p$,
\begin{equation}
  H \equiv p^0 = \sqrt{\vc{p}^2 + m^2} \; .
\end{equation}
It generates the equations of motion
\begin{eqnarray}
  \dot x^i &=& \pb{x^i}{H} = \frac{p^i}{p^0} \; , \\
  \dot p^i &=& \pb{p^i}{H} = 0 \; ,
\end{eqnarray}
where the first one is consistent with (\ref{XDOT_DET}) so that the
second one coincides with the Euler-Lagrange equation following from
(\ref{TAU_ACT}). The Poisson brackets above are interpreted as being
taken at \emph{equal time} $t$.

Altogether, the gauge-fixed dynamics is well-determined as can be seen
from the (unique) solution
\begin{equation}
  x^i (t) = \frac{p^i}{p^0} t + x^i (0) = v^i t + x^i (0) \; ,
\end{equation}
which corresponds to  linear motion with constant velocity.

We have thus seen that it is the choice of time which makes the dynamics
unambiguous. This problem and its solution are  actually typical for
reparametrization invariant systems, the most important example probably
being general relativity. There, the problem of choosing a natural time
variable is the main obstacle for developing a quantum theory of gravity
(for a thorough discussion, see \cite{isham:91}). 
 
In the remainder of this section we will perform a general analysis of
the possible choices of gauge fixings, that is, of different time
parameters $\tau$. The choice of a particular time variable corresponds
to a foliation of Minkowski space into hypersurfaces of equal time,
$\tau = const$, which in general are three-dimensional objects, and the
time direction `orthogonal' to them. The time development thus
continuously evolves the hypersurface $\Sigma_0: \tau = \tau_0$ into
$\Sigma_1: \tau = \tau_1 > \tau_0$. Put differently, initial conditions
provided on $\Sigma_0$ together with the dynamical equations (being
differential equations in $\tau$) determine the state of the dynamical
system on $\Sigma_1$. 

The gauge fixing (\ref{GF_TAU}) expresses the time parameter in terms of
the coordinates $x^\mu$. These are orthogonal coordinates which are
essentially tailored to fit the common choice of time, $\tau = t$. On
the other hand, we have explicitly seen within this particular example,
that the gauge (or time) choice corresponds to an associated choice of
the world-line metric. It is the latter, on which we now want to
concentrate. Consider some arbitrary coordinates $\xi^\alpha =
\xi^\alpha (x)$, which may be curvilinear. We imagine that the $\xi^i$,
$i=1,2,3$, parametrize the three-dimensional hypersurface $\Sigma$, so
that the remaining one, $\xi^0$, represents the time variable,
i.e.~$\tau = \xi^0$. Then we have
\begin{equation}
  \label{H_METRIC}
  ds^2 = g_{\mu\nu}dx^\mu dx^\nu =
  g_{\mu\nu}\pad{x^\mu}{\xi^\alpha} \pad{x^\nu}{\xi^\beta} d\xi^\alpha
  d\xi^\beta \equiv h_{\alpha\beta}(\xi) d\xi^\alpha d\xi^\beta \; . 
\end{equation}
Introducing a vierbein $e^\mu_{\;\alpha} (\xi)$ the metric
$h_{\alpha\beta}(\xi)$ is alternatively given by
\begin{equation}
  h_{\alpha\beta} (\xi) = g_{\mu\nu} \, e^\mu_{\;\; \alpha} (\xi) e^\nu_{\;\;
  \beta} (\xi) \; .
\end{equation}
This transformation is well known from general relativity, where it
corresponds to the transformation from a local inertial frame described
by the flat metric $g_{\mu\nu}$ to a non-inertial frame with
coordinate dependent metric $h_{\alpha\beta}(\xi)$. For our purposes we
write this metric in a (3+1)-notation as follows, 
\begin{equation}
  \label{(3+1)_METRIK}
  h_{\alpha\beta} = \left( \begin{array}{cc}
                            h_{00} & h_{0i} \\
                            h_{i0} & h_{ij} 
                      \end{array}        \right) \; .
\end{equation}
Of particular interest is the component $h_{00}$, which  explicitly
reads 
\begin{equation}
  h_{00} = g_{\mu\nu} \pad{x^\mu}{\xi^0} \pad{x^\nu}{\xi^0} = g_{\mu\nu}
  e^\mu_{\;\; 0} e^\nu_{\;\; 0} \equiv n^2 \; ,
\end{equation}
where we have defined the unit vector in $\xi^0$-direction
\begin{equation}
  n^\mu = \pad{x^\mu}{\xi^0} =  e^\mu_{\;\; 0} \; .
\end{equation}
It is related to the normal vector $N^\mu$ via
\begin{equation}
  n \cdot N = e^\mu_{\;\; 0} e^{\;\; 0}_\mu = \pad{\xi^0}{x^\mu} \pad
  {x^\mu}{\xi^0} = 1 \; .
\end{equation}
The $h_{ij}$ are the metric components associated with the
hypersurface. The invariant distance element thus becomes (setting
$h_{0i} \equiv h_i$),
\begin{eqnarray}
  ds^2 &=& h_{00} d\xi^0 d\xi^0 + 2 h_{0i} d\xi^0 d\xi^i + h_{ij} d\xi^i
  d\xi^j \; , \nn \\
  &=& \left( n^2 + 2 h_i \frac{d\xi^i}{d\tau} +
  h_{ij}\frac{d\xi^i}{d\tau} \frac{d\xi^j}{d\tau} \right) d\tau^2 \; , 
\end{eqnarray}
where, in the second step, we have used that $\xi^0 = \tau$. The
quantities $w^i \equiv d\xi^i/d\tau$ are the velocities expressed in the
new coordinates, so that the world-line metric can be written as
\begin{equation}
  h(\tau) = n^2 + 2 h_i w^i + h_{ij} w^i w^j \; .
\end{equation}
For $\tau \equiv \xi^0 \equiv t$ we find $n = N = (1, \vc{0})$, $N^2 =
1$, $h_i = 0$, $w^i = v^i$ and $h_{ij} = -\delta_{ij}$ leading to the
world-line metric $h(t) = 1 - v^2$ of (\ref{METRIC_T},
\ref{4V_SQUARED}).  The dynamics of the $\xi^i$ is easily found as
follows,
\begin{equation}
  \dot \xi^i = \partial_\mu \xi^i \dot x^\mu = \partial_\mu \xi^i
  \frac{p^\mu}{N \cdot p} = \frac{p^\mu e_\mu^{\;\; i}}{N \cdot p} \equiv
  \frac{\pi^i}{N \cdot p} \; ,
\end{equation}
where we have introduced the momenta $\pi^i$ canonically conjugate to
$\xi^i$, satisfying
\begin{equation}
  \pb{\xi^i}{\pi^j} = h^{ij}(\xi) \; .
\end{equation}

\section{Dirac's Forms of Relativistic Dynamics}

At this point one needs a satisfactory criterion in order to decide
which choices for the time variable (or the world-line metric $h$) are
sensible. To this end it is not sufficient to consider only the
$\tau$-development and the associated generator of time translations
(i.e.~the Hamiltonian). Instead , one has to refer to the full
Poincar\'e group to be able to guarantee full relativistic invariance.
The generators of the Poincar\'e group are the four momenta $P^\mu$ and
the six operators $M^{\mu \nu}$ which combine the angular momenta and
boosts according to
\begin{eqnarray}
  L^i &=& \frac{1}{2}\epsilon^{ijk}M^{jk} \; , \label{LK1} \\
  K^i &=& M^{0i} \; , \label{LK2}
\end{eqnarray}
with $i$, $j$, $k$ = 1,2,3. These generators are elements of the
Poincar\'e algebra which is defined by the Poisson bracket relations,
\begin{eqnarray}
  \label{PC_ALGEBRA}
  \pb{P^\mu}{P^\nu}  &=& 0 \; , \nn \\
  \pb{M^{\mu \nu}}{P^{\rho}}  &=& g^{\nu \rho}P^\mu - g^{\mu \rho}
  P^\nu  \; , \\
  \pb{ M^{\mu \nu}}{M^{\rho \sigma}}  &=& g^{\mu \sigma} M^{\nu \rho} -
  g^{\mu \rho} M^{\nu \sigma} - g^{\nu \sigma} M^{\mu \rho} + g^{\nu
  \rho} M^{\mu \sigma} \; . \nn
\end{eqnarray}
It is well known that the momenta $P^\mu$ generate space-time
translations and the $M^{\mu \nu}$ rotations and Lorentz boosts,
cf.~(\ref{LK1}, \ref{LK2}). In the following we will only consider
proper and orthochronous Lorentz transformations, i.e.~we exclude space
and time reflections.

Any Poincar\'e invariant dynamical theory describing e.g.~the
interaction of particles should provide a particular realization of the
Poincar\'e algebra. For this purpose, the Poincar\'e generators are
constructed out of the fundamental dynamical variables like positions,
momenta, spins etc. An elementary realization of (\ref{PC_ALGEBRA}) is
given as follows. Choose the space-time point $x^\mu$ and its conjugate
momentum $p^\mu$ as canonical variables, i.e.~adopt (\ref{PB_COV1}),
\begin{equation}
  \label{PB_COV2}
  \pb{x^\mu}{p^\nu}  = -  g^{\mu \nu} \; .
\end{equation}
The Poincar\'e generators are then found to be
\begin{equation}
  \label{ELEM_PC_GEN}
  P^\mu = p^\mu \; , \quad M^{\mu \nu} = x^\mu p^\nu - x^\nu p^\mu
\end{equation}
as is easily confirmed by checking (\ref{PC_ALGEBRA}) using
(\ref{PB_COV2}).  An infinitesimal Poincar\'e transformation is thus
generated by
\begin{equation}
  \delta G = - \sfrac{1}{2}\delta\omega_{\mu\nu} M^{\mu\nu} 
  + \delta a_\mu P^\mu \; ,
\end{equation}
in the following way,
\begin{equation}
  \delta x^\mu = \pb{x^\mu}{\delta G} = \delta \omega^{\mu\nu} x_\nu +
  \delta a^\mu \; , \quad \delta\omega^{\mu\nu} = - \delta\omega^{\nu\mu} \; .  
\end{equation}
The action of the Poincar\'e group on some function $F(x)$ is thus
given by
\begin{equation}
  \label{PC_ACTION}
  \delta F = \pb{F}{\delta G} = \partial^\mu F \, \delta a_\mu -
  \sfrac{1}{2}  (x^\mu
  \partial^\nu - x^\nu \partial^\mu)F \, \delta \omega_{\mu\nu} \; .
\end{equation}
Though the realization (\ref{ELEM_PC_GEN}) is covariant, it has several
shortcomings. It does not describe any interaction; for several
particles the generators are simply the sum of the single particle
generators. This point, however, is of minor importance to us, and will
only be touched upon at the end of the next chapter. The solution of the
problem, as already mentioned in the introduction to this chapter, is
the framework of local quantum field theory. More importantly, the
representation (\ref{ELEM_PC_GEN}) does not take into account the
mass-shell constraint, $p^2 = m^2$, which we already know to guarantee
relativistic causality as it generates the dynamics.

To remedy the situation we proceed as before by choosing a `time'
variable $\tau$, i.e.~a foliation of space time into essentially
space-like hypersurfaces $\Sigma$ with time-like or light-like normals.
$\Sigma$ should be chosen in such a way that it intersects all possible
world-lines once and only once.  Apart from this necessary consistency
with causality this foliation appears quite arbitrary. However, given a
particular foliation one can ask the question which of the Poincar\'e
generators will leave the hypersurface $\Sigma$ invariant. The set of
all such generators defines a subgroup of the Poincar\'e group called
the stability group $G_\Sigma$ of $\Sigma$. The associated generators
are called {\em kinematical}, the others {\em dynamical}. The latter map
$\Sigma$ onto another hypersurface $\Sigma^\prime$ and thus involve the
development in $\tau$. One thus expects that the dynamical generators
will depend on the Hamiltonian (and, therefore, the interaction) which,
by definition, is a dynamical quantity.

It is clear, however, that the stability group corresponding to a
particular foliation will be empty if the associated hypersurface looks
very irregular and thus does not have a high degree of symmetry. One
therefore demands in addition that the stability group acts transitively
on $\Sigma$: any two points on $\Sigma$ can be connected by a
transformation from $G_\Sigma$. With this additional requirement there 
are exactly five inequivalent classes of hypersurfaces
\cite{leutwyler:78} which are listed in Table \ref{T4}.

\begin{table}
\begin{center}
\renewcommand{\arraystretch}{1.3}
\caption{\label{T4} \textsl{All possible choices of hypersurfaces $\Sigma$:
  $\tau = const$ with transitive action of the  stability group
  $G_\Sigma$. $d$ denotes the dimension of $G_\Sigma$, that is, the
  number of kinematical Poincar\'e generators, $\vc{x}_\perp \equiv (x^1,
  x^2)$.}}
\vspace{.5cm}

\begin{tabular*}{\textwidth}[h]{ @{\extracolsep\fill} l  l  l  c }
\hline\hline
name        & $\Sigma$        & $\tau$          & $d$ \\
\hline
instant     & $x^0 = 0$       & $t$             & 6 \\
light front & $x^0 + x^3 = 0$ & $t + x^3/c = 0$ & 7 \\ 
hyperboloid & $x_0^2 - \vc{x}^2 = a^2 > 0$, $x^0 > 0$ & $(t^2 - \vc{x}^2/c^2 -              a^2/c^2)^{1/2}$ & 6 \\
hyperboloid & $x_0^2 - \vc{x}_\perp^2 = a^2 > 0$, $x^0 > 0$ & $(t^2 -
              \vc{x}_\perp^2 /c^2 -  a^2/c^2)^{1/2}$ & 4 \\
hyperboloid & $x_0^2 - x_1^2  = a^2 > 0$, $x^0 > 0$ & $(t^2 -
              x_1^2 )/c^2 -  a^2/c^2)^{1/2}$ & 4 \\
\hline\hline
\end{tabular*}
\end{center}
\end{table}

The first three choices have been found by Dirac in his seminal 1949
paper \cite{dirac:49} on `forms of relativistic dynamics'.  He called
the associated forms the `instant', `front' and `point' forms,
respectively. These are the most important choices as the other two
forms have a rather small stability group and thus are not very useful.
We have only listed them for the sake of completeness.

It is important to note that for all forms one has $\lim_{c \to \infty}
\tau = t$, which means that in the non-relativistic case there is only
one possible foliation leading to the absolute Galileian time $t$. This
is consistent with the fact that there is no limiting velocity in this
case implying that particle trajectories can have arbitrary slope
(tangent vector). Therefore, the hypersurface $\Sigma_{nr}: t = const$
is the only one intersecting all possible world-lines.

To decide which of the Poincar\'e generators are kinematical we use the
general formula (\ref{PC_ACTION}) describing their action. Imagine that
$\Sigma$ is given in the form $\Sigma: \tau = F(x)$ as in
Table~\ref{T4}. If $P^\mu$ or $M^{\mu\nu}$ are kinematical for some
$\mu$ or $\nu$, then, for these particular superscripts, we must have
\begin{equation}
  \label{KINEMATIC}
  \partial^\mu F = 0 \; , \quad (x^\mu \partial^\nu - x^\nu
  \partial^\mu)F = 0 \; ,
\end{equation}
saying that the gradient and the `angular derivative' of $F$ have to
vanish.  Let us now discuss the different forms in more detail along
these lines.

\subsection*{The Instant Form}

The choice of Galileian time $\tau = t$ is of course the most common one
also in the relativistic case, and we have discussed it briefly in the
preceding subsection. To complete this discussion, we construct the
associated representation of the Poincar\'e generators on $\Sigma: t
=0$.  The idea is to explicitly saturate the constraint $p^2 = m^2$ by
eliminating the variable conjugate to $\tau = t$, i.e.~$p^0 = N \cdot p
= \sqrt{\vc{p}^2 + m^2}$, and setting $x^0 = 0$ in (\ref{ELEM_PC_GEN}).
Alternatively, one can follow Dirac \cite{dirac:49} and add the
constraint to {\em any} of the Poincar\'e generators,
\begin{eqnarray}
  P^\mu &=& p^\mu + \lambda^\mu (p^2 - m^2) \; , \label{P_MU} \\
  M^{\mu\nu} &=& x^\mu p^\nu + x^\nu p^\mu + \lambda^{\mu\nu}(p^2 - m^2)
  \; . \label{M_MUNU}
\end{eqnarray}
Now we demand that the right-hand-sides should be independent of
$p^0$. Differentiating the right-hand-side of (\ref{P_MU}) with respect
to $p^0$ one finds
\begin{equation}
  1 + 2 \lambda^0 p^0 = 0 \; ,
\end{equation}
which is exactly the consistency condition (\ref{LM_EQ}) for the case at
hand. Thus $\lambda^0$ is determined as $\lambda^0 = -1/2 p^0$ which
coincides with (\ref{LM_T}) upon identifying $\lambda^0 \equiv \lambda$.
The multipliers $\lambda^i$ turn out to be zero.  The same procedure
applied to the $M^{\mu\nu}$ of (\ref{M_MUNU}) yields as the nonvanishing
Lagrange multipliers the $\lambda^{0i}$ with
\begin{equation}
  \lambda^{0i} = - \frac{x^i}{2 p^0} \; .
\end{equation}
Altogether we obtain the following (3+1)-representation of the Poincar\'e
generators, 
\begin{equation}
  \label{PC_ALG_IF}
  \begin{array}{l c l l c l l}
    P^i & = & p^i \; ,     & M^{ij} & = & x^i p^j - x^j p^i \; , &  \\
    P^0 & = & \omega_p \; , & M^{i0} & = & x^i \omega_p \; , &
    \omega_p  \equiv (p^i p^i + m^2)^{1/2} \; . 
  \end{array} 
\end{equation}
This result is as expected: Compared to (\ref{ELEM_PC_GEN}), $p^0$ has
been replaced by $\omega_p$, and $x^0$ has been set to zero. It should,
however, be pointed out that for non-Cartesian coordinates the
construction of the Poincar\'e generators is less straightforward, as
will be seen in a moment. 
  
As a general rule we note that only the generators containing $p^0$ in
the representation (\ref{P_MU}, \ref{M_MUNU}) lead to non-vanishing
Lagrange multipliers. These are just the dynamical generators which are,
for the instant form, given by the ordinary Hamiltonian $P^0$
and the boost generators $M^{i0}$. In agreement with (\ref{KINEMATIC}),
one has 
\begin{equation}
  \partial^i F = (x^i \partial^j - x^j \partial^i)F = 0 \; ,
\end{equation}
so that $\Sigma$ is both translationally and rotationally invariant
confirming that the dimension of its stability group is six
(cf.~Table~\ref{T4}). On the other hand,
\begin{eqnarray}
  \partial^0 F = 1 &\ne& 0 \; , \\
  (x^0 \partial^i - x^i \partial^j )F = - x^i &\ne& 0 \; ,
\end{eqnarray}
from which we read off that, apart from the Hamiltonian, also the boosts
are dynamical, i.e., $\Sigma$ is not boost invariant. The latter fact
is, of course, well known because the boosts  mix space and time. Under a
boost along the $\vc{n}$-direction with velocity $\vc{v}$ ($\vc{n} =
\vc{v}/v$), $t$ transforms as
\begin{equation}
  \label{BOOST}
  t \to t^\prime = t \cosh \omega + (\vc{n} \cdot \vc{x}) \sinh \omega \; ,  
\end{equation}
where $\omega$ is the rapidity, defined through $\tanh \omega = v$. From
(\ref{BOOST}) it is evident that the hypersurface $\Sigma: t=0$ is not
boost invariant. 

In obtaining the representation (\ref{PC_ALG_IF}), we make the
Poincar\'e algebra compatible with the instant-form constraint $x^0 =
0$. An elementary calculation, using $\pb{x^i}{p^j} = \delta^{ij}$,
indeed shows that the generators (\ref{PC_ALG_IF}) really obey the
bracket relations (\ref{PC_ALGEBRA}). We have already seen that the
Hamiltonian $P^0 = \omega_p$ generates the correct dynamics.

\subsection*{The Point Form}

As might have been expected, the situation is somewhat more complicated
for the point form. The hypersurfaces of equal $\tau = (x^2 -
a^2)^{1/2}$ are hyperboloids implying that curvilinear coordinates are
involved. For the following discussion we chose $a=0$ for
simplicity. Note, however, that in this case the hypersurface $\Sigma:
\tau = 0$ is the light-cone which is not transitive and has a smaller
stability group of dimension four. We therefore assume $\tau > 0$ in what
follows. 

It is natural to introduce hyperbolic coordinates defined via
\begin{eqnarray}
  x^0 &=& \tau \cosh \omega \; , \nn \\ 
  x^1 &=& \tau \sinh \omega \sin \theta \cos \phi \; , \nn \\
  x^2 &=& \tau \sinh \omega \sin \theta \sin \phi \; , \nn \\
  x^3 &=& \tau \sinh \omega \cos \theta \; . \label{HYP_COORD}
\end{eqnarray}
The coordinates can thus be written in the compact form
\begin{equation}
  x^\mu = \tau n^\mu = \tau (\cosh \omega , \vc{e}_r \sinh \omega) \; ,
\end{equation}
where $\vc{e}_r$ is the radial unit vector in $\mathbb{R}^3$. The new
coordinates are a straightforward generalization of the
(three-dimensional) polar coordinates. Using the general formula
(\ref{H_METRIC}), they lead to a world-line metric
\begin{equation}
  h(\tau) = 1 - \tau^2 \left( \frac{d\omega}{d\tau} \right)^2 -
  \tau^2 \sinh^2 \omega \left(\frac{d\theta}{d\tau} \right)^2 - \tau^2
  \sinh^2 \omega \, \sin^2 \theta \left(\frac{d\phi}{d\tau}\right)^2  \; .
\end{equation}
The normal vector on $\Sigma$ is $N^\mu = n^\mu = x^\mu / \tau$ which
can be interpreted as a four-dimensional radial vector. The Lagrange
multiplier is therefore given by
\begin{equation}
  \lambda = - \frac{1}{2 N \cdot p} = - \frac{\tau}{2 p \cdot x} 
   \; ,
\end{equation}
so that the equation of motion for the $x^\mu$ becomes
\begin{equation}
  \dot x^\mu = \frac{\tau}{p \cdot x} \, p^\mu \; .
\end{equation}
The Hamiltonian, that is, the generator of $\tau$-development, is 
\begin{equation}
  H = N \cdot p = x \cdot p / \tau \; .
\end{equation}
Obviously, we cannot set $\tau$ equal to zero.  Note that this 
Hamiltonian does \emph{not} belong to the Poincar\'e generators as it is
essentially the dilatation operator generating scale transformations
\cite{fubini:73}. Its explicit $\tau$-dependence indicates that the
dynamical system is not scale-invariant which is obvious as we consider
a massive particle.

Let us analyze the Poincar\'e generators. To decide which of them are
dynamical, we calculate with $\tau = F(x) = \sqrt{x^2}$,
\begin{eqnarray}
  \partial^\mu F = N^\mu = \frac{x^\mu}{\tau}  &\ne& 0\; , \\
  (x^\mu \partial^\nu - x^\nu \partial^\mu) F = \frac{1}{\tau}(x^\mu
  x^\nu - x^\nu x^\mu)F &=& 0 \; .
\end{eqnarray}
Not unexpectedly, all momenta are dynamical, but Lorentz transformations
are kinematical. This is obvious from the fact that $\tau$ is a Lorentz
scalar rendering $\Sigma$ boost and rotation invariant. The stability
group $G_\Sigma$ thus has again dimension six.

Let us finally mention that the point form, though already introduced by
Dirac, has only rarely been used in physical applications
\cite{peres:68,fubini:73,sommerfield:74,gromes:74}. The reason is, of
course, the curvilinear nature of $\Sigma$ which particularly hampers
the quantization.

We are left with the front form which is at the basis of the present
work. We reserve the next chapter for its investigation. 

\chapter{The Front Form}

\section{Generalities}

\begin{figure}
   \caption{\label{fig-front} \textsl{The hypersurface $\Sigma : x^\p =
        0$ defining the front form. It is a null-plane tangential to the
        light-cone, $x^2 =0$.}}
   \begin{center} 
     \vspace{1cm}
     \includegraphics[scale=0.8]{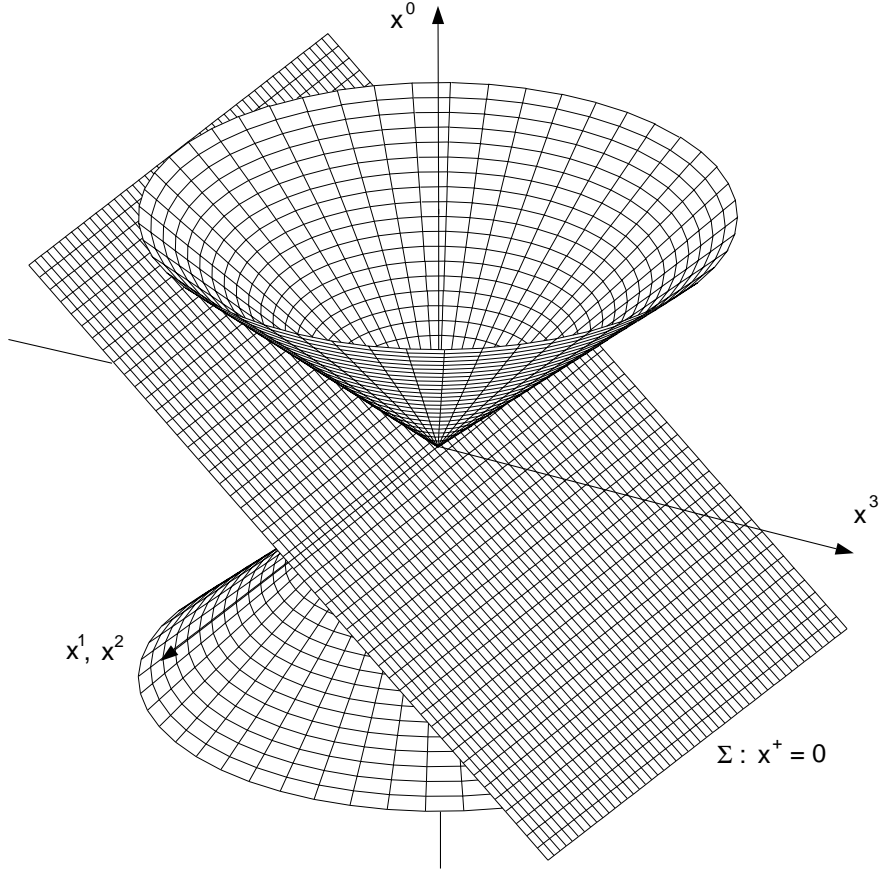}
   \end{center}    
\end{figure}

For an arbitrary four-vector $a$ we perform the following transformation
to {\em light-cone} coordinates,
\begin{equation}
  (a^0, a^1, a^2 , a^3 ) \mapsto (a^\p , a^1 , a^2 , a^\m) \; ,
\end{equation}
where we have defined
\begin{equation}
  a^\p = a^0 + a^3 \; , \quad a^\m = a^0 - a^3 \; .
\end{equation}
We also introduce the transverse vector part of $a$ as
\begin{equation}
  \vc{a}_\perp = (a^1 , a^2) \; .
\end{equation}
The metric tensor becomes
\begin{equation}
  g_{\mu\nu} = \left( \begin{array}{rrrr}
                         0   &  0 &  0 & 1/2 \\ 
                         0   & -1 &  0 &  0  \\
                         0   &  0 & -1 &  0  \\
                        1/2  &  0 &  0 &  0
                      \end{array} \right) 
\end{equation}
The entries 1/2 imply a slightly unusual scalar product
\begin{equation}
  \label{SCAL_PROD}
  a \cdot b = g_{\mu\nu}a^\mu b^\nu = \sfrac{1}{2}a^\p b^\m + \sfrac{1}{2} a^\m b^\p
  - a^i b^i \; , \quad i = 1,2 \; .
\end{equation}
According to Table \ref{T4}, the front form is defined by choosing the
hypersurface $\Sigma: x^\p = 0$, which is a plane tangent to the
light-cone. It can equivalently be viewed as the wave front of a plane
light wave traveling towards the positive $z$-direction. Therefore,
$\Sigma$ is also called a \emph{light-front}. It corresponds to a gauge fixing
\begin{equation}
  \label{N_FRONT}
  \chi (x , \tau) = \tau - N \cdot x \; , \quad N = (1, 0,0, -1) \; ,
  \quad N^2 = 0 \; ,  
\end{equation}
where $N$ has been written in ordinary coordinates. We see that $N^\p =
N^0 + N^3 = 0$ which implies that the normal $N$ to the hypersurface
lies {\em within} the hypersurface \cite{neville:71a,rohrlich:71}. As
$N$ is a light-like or null vector, $\Sigma$ is often referred to as a
\emph{null-plane} \cite{neville:71a, coester:92}. We have depicted the
front-form hypersurface $\Sigma$ together with the light-cone in
Fig.~\ref{fig-front}.

The unit vector in $x^\p$-direction is another null-vector,
\begin{equation}
  n^\mu = \pad{x^\mu}{x^\p} = \sfrac{1}{2} (1 , 0 , 0, 1) \; ,
\end{equation}
so that $n \cdot N = 1$ as it should.  Given the scalar product
(\ref{SCAL_PROD}), we infer the invariant distance element
\begin{equation}
  ds^2 = g_{\mu\nu}dx^\mu dx^\nu = dx^\p dx^\m - dx^i dx^i
  = \left( \frac{dx^\m}{dx^\p} - \frac{dx^i}{dx^\p} \frac{dx^i}{dx^\p} 
  \right) dx^\p dx^\p
\end{equation}
from which the metric $h$ can be read off as
\begin{equation}
  h(x^\p) = \dot x^\m - \dot x^i \dot x^i  \equiv
  v^\m - \vc{v}_\perp^2 \; .
\end{equation}
Within the Hamiltonian formulation we find that $(p \cdot \partial) \chi
= N \cdot p = p^\p$, so that, from (\ref{LM_DET}), the Lagrange
multiplier $\lambda$ becomes
\begin{equation}
  \lambda = -1/2p^\p \; ,
\end{equation}
implying the equation of motion $\dot x^\mu = p^\mu / p^\p$. Note that
$p^\p$ must be non-zero. The equations of motion are generated by the
Hamiltonian
\begin{equation}
  \label{FF_HAM}
  H = n \cdot p = p^\m / 2 = \frac{p_\perp^2 + m^2}{2p^\p} \; ,
\end{equation}
by means of the Poisson brackets  
\begin{equation}
  \dot x^\m = \pb{x^\m}{H} = \frac{p^\m}{p^\p} \; , \quad
  \dot x^i = \pb{x^i}{H} = \frac{p^i}{p^\p} \; .
\end{equation}
and
\begin{equation}
  \dot p^\p = \pb{p^\p}{H} = 0 \; , \quad 
  \dot p^i = \pb{p^i}{H} = 0 \; .
\end{equation}
The solution is
\begin{eqnarray}
  x^\m (x^\p) &=& \frac{p^\m}{p^\p}x^\p + x^\m (0) = v^\m x^\p + x^\m
  (0) \; \\
  x^i (x^\p) &=& \frac{p^i}{p^\p} x^\p + x^i (0) = v^i x^\p + x^i (0) \; ,
\end{eqnarray}
and again corresponds to linear motion with constant (front-form)
velocity. Note that the Hamiltonian (\ref{FF_HAM}) is \emph{not} the
normal projection $N \cdot p$ of the momentum, because $N \cdot p$ lies
within $\Sigma$ and thus corresponds to a kinematical direction.  As for
the instant form, the light-cone representation of the Poincar\'e
generators can be obtained by solving the constraint $p^2 = m^2$ for
$p^-$, inserting the result into the elementary representation
(\ref{ELEM_PC_GEN}) of the generators and setting $x^\p = 0$.  The
kinematical generators are
\begin{equation}
 P^i = p^i \; , \quad P^\p = p^\p \; , \quad M^{\p i} = - x^i p^\p \; ,
 \quad M^{\p \m} = - x^\m p^\p \; , \quad M^{12} = x^1 p^2 - x^2 p^1 \; .  
\end{equation}
The dynamical ones (or Hamiltonians) are given by 
\begin{equation}
  P^\m = \frac{p_\perp^2 + m^2}{p^\p} \; , \quad M^{\m i} = x^\m
  p^i - x^i p^\m \; .
\end{equation}
We thus have found \emph{seven} kinematical generators, so that the
front form leads to the largest stability group. Furthermore, it is
important to note that the Hamiltonian $P^\m$ does not contain a square
root as already pointed out by Dirac. However, there is a singularity at
$p^\p = 0$ which therefore is a somewhat peculiar point. In the gauge
theory language, it corresponds to a vanishing Faddeev-Popov `matrix'.
We will quite frequently be concerned with this issue in what follows.

\section{Lorentz Boosts}

Consider the following boost in $z$-direction with rapidity $\omega$
written in instant-form coordinates,
\begin{eqnarray}
  t &\to&  t \cosh \omega + z \sinh \omega \; , \\
  z &\to&  t \sinh \omega + z \cosh \omega \; .
\end{eqnarray}
As stated before, such a boost mixes space and time coordinates $z$ and
$t$, respectively. However, if we add and subtract these equations, we
obtain the action of the boost for the front form,
\begin{eqnarray}
  x^\p &\to& e^\omega x^\p \; , \label{XP_BOOST}\\
  x^\m &\to& e^{-\omega} x^\m \; . \label{XM_BOOST}
\end{eqnarray}
We thus find the important result that a boost in $z$-direction does not
mix light-cone space and time but rather rescales the coordinates! Note
that $x^\p$ and $x^\m$ are rescaled inversely with respect to each
other. The scaling factor can be written as
\begin{equation}
  e^\omega = \sqrt{\frac{1-v}{1+v}} \; ,
\end{equation}
if the rapidity $\omega$ is defined in the usual manner in terms of the
boost velocity $v$, $\tanh \omega = v$. One should note in particular,
that one has the fixed point hypersurface $\Sigma: x^\p = 0$ which is
mapped onto itself, so that the relevant generator, $M^{\p\m}= 2 M^{30}=
- 2 K^3$, is kinematical. However, we see explicitly that this is not
true for $x^\p \ne 0$, where we get a rescaling. Stated differently, the
transformation to light-cone coordinates diagonalizes the boosts in
$z$-direction. Therefore, the behavior under such boosts becomes
especially simple. A pedagogical discussion of some elementary
applications can be found in \cite{parker:70a}.

In our discussion of the boosts we are still left with the other two
kinematical generators, the $M^{\p i}$. These act on the transverse
coordinates as
\begin{equation}
  \label{DELTA_X_I}
  \delta x^i = \pb{x^i}{M^{\p j}} \delta \beta^j = x^\p \delta \beta^i  = 0 \; , 
\end{equation}
where the last identity holds on $\Sigma$. Translated to
instant-form coordinates, $M^{\p i}$ is a combination of boosts and
angular momenta,
\begin{equation}
  M^{\p i} = K^i + \epsilon^{ij} L^j \; .
\end{equation}
The finite transformation corresponding to (\ref{DELTA_X_I}) is found
via exponentiation. In general, if the infinitesimal transformation is
written as
\begin{equation}
  \delta_B A = \pb{A}{B} \equiv i \ad B (A) \; ,
\end{equation}
the finite transformation is given by
\begin{equation}
  A^\prime = \exp(i \ad B ) (A) \; .
\end{equation}
From (\ref{DELTA_X_I}) we thus obtain
\begin{equation}
  \label{X_I_BOOST}
  x^i \to x^i + x^\p \beta^i \; ,
\end{equation}
which coincides with the infinitesimal transformation because the
exponential series terminates after the term linear in the boost
parameter $\beta^i$. Again,  on $\Sigma$ the $x^i$ remain unchanged.

We are actually more interested in the transformation properties of the
momenta, as these, being Poincar\'e generators, are more fundamental
quantities than the coordinates, in particular in the quantum theory
\cite{leutwyler:78}. As $P^\mu$ transforms as a four-vector we just have
to replace $x^\mu$ by $P^\mu$ in the boost transformations
(\ref{XP_BOOST}, \ref{XM_BOOST}, \ref{X_I_BOOST}) and obtain,
\begin{eqnarray}
  P^\p &\to& e^\omega P^\p \; , \label{P_BOOST} \\
  P^\m &\to& e^{-\omega} P^\m \; , \label{M_BOOST} \\
  P^i  &\to& P^i + \beta^i P^\p  \label{T_BOOST}\; . 
\end{eqnarray}
We remark that $P^\p = 0$ is a fixed point under longitudinal boosts. In
quantum field theory, it corresponds to the vacuum.  Note further, that
in the third identity describing the action of $M^{\p i}$, longitudinal
and transverse momenta (which are both kinematical) get mixed.

We can now ask the question how to boost from $(P^\p , P^i)$ to momenta
$(Q^\p , Q^i)$. This can be done by fixing the boost parameters $\omega$
and $\beta^i$ as
\begin{equation}
  \omega = - \log \frac{Q^\p}{P^\p} \; , \quad \beta^i = \frac{Q^i -
  P^i}{P^\p} \; .
\end{equation}
Obviously, this is only possible for $P^\p \ne 0$. We emphasize that in
the construction above there is no dynamics involved. For the quantum
theory, this means that we can build states of arbitrary light-cone
momenta with very little effort. All we have to do is applying some
kinematical boost operators. The simple behavior of light-cone momenta
under boosts will be important for the discussion of bound states. The
same is true for the subject of the next section.

\section{Galilei Subgroups and Consequences}
 
Let us analyze the Poincar\'e algebra in light-cone coordinates in more
detail. We are particularly interested in the Poisson bracket relations
of the seven generators $P^\mu$, $M^{12}$, $M^{\p i}$,  

\parbox{12cm}{
\begin{eqnarray*}
  \pb{M^{12}}{M^{\p i}} &=& \epsilon^{ij} M^{\p j} \; , \\
  \pb{M^{12}}{P^i}     &=& \epsilon^{ij} P^j \; , \\
  \pb{M^{\p i}}{P^\m}   &=& - 2 P^i \; , \\
  \pb{M^{\p i}}{P^j}    &=& - \delta^{ij} P^\p
\end{eqnarray*}}
\hfill
\parbox{1.5cm}{\begin{eqnarray}\label{GALILEI_SUB}\end{eqnarray}}
All other brackets of the seven generators vanish. Consider now the
two-dimensio\-nal Galilei group. Its generators (for a free particle of
mass $m$) are: two momenta $k^i$, one angular momentum $L =
\epsilon^{ij}x^i k^j$, two Galilei boosts $G^i = m x^i$, the Hamiltonian
$H = k^i k^i /2m$ and the mass $m$, which is the Casimir generator. Upon
using $\pb{x^i}{k^j} = \delta^{ij}$ and identifying $P^i \leftrightarrow
k^i$, $M^{12} \leftrightarrow L$, $M^{\p i} \leftrightarrow -2 G^i$,
$P^\p \leftrightarrow 2m$ and $P^\m \leftrightarrow H$, one easily finds
that (\ref{GALILEI_SUB}) forms a subalgebra of the Poincar\'e algebra
which is isomorphic to the Lie algebra of the two-dimensional Galilei
group. (A second isomorphic subalgebra is obtained via identifying $M^{\m
  i} \leftrightarrow 2 G^i$ and exchanging $P^\p$ with $P^\m$.) The
first two identities in (\ref{GALILEI_SUB}), for instance, state that
$M^{\p i}$ and $P^i$ transform as ordinary two-dimensional vectors. 

One thus expects that light-cone kinematics will partly  show a
non-relativistic behavior which is associated with the transverse
dimensions and governed by the two-dimensional Galilei group. 

A very important example where these features are profitably at work is
provided by a system of many particles which, for the time being, will
be assumed as non-interacting. Let the $i^{th}$ particle  have mass $m_i$
and light-cone four momentum 
\begin{equation}
  p_i = (p_i^\p , \vc{p}_{\perp i} , p_i^\m) \; .
\end{equation}
As the particles are free, the total four momentum is given by the sum
of the individual momenta, 
\begin{equation}
  \label{P_SUM}
  P = \sum_i p_i \; .
\end{equation}
Note that both the individual particles and the total system are  on-shell, 
\begin{eqnarray}
  p_i^\m &=& \frac{p_{\perp i}^2 + m_i^2}{p_i^\p} \; , \\
  P^\m &=& \frac{P_\perp^2 + M_0^2}{P^\p} \; ,
\end{eqnarray}
where $M_0$ denotes the total mass. We introduce {\em relative} momentum
coordinates $x_i$ and $\vc{k}_{\perp i}$ via
\begin{eqnarray}
  p_i^\p &\equiv& x_i P^\p \; , \\
  \vc{p}_{\perp i} &\equiv& x_i \vc{P}_\perp + \vc{k}_{\perp i} \; .
\end{eqnarray}
Thus, $x_i$ and $\vc{k}_{\perp i}$ denote the longitudinal momentum
fraction and the relative transverse momentum of the $i^{th}$ particle,
respectively. Comparing with (\ref{P_SUM}) we note that these variables
have to obey the constraints
\begin{equation}
  \label{RELMOM_SUM}
  \sum_i x_i = 1 \; , \quad   \sum_i \vc{k}_{\perp i} = 0 \; .
\end{equation}
A particularly important property of  the relative momenta is their boost
invariance. To show this we calculate, using (\ref{P_BOOST}),
\begin{equation}
  \label{X_INV}
  x_i^\prime = e^\omega p_i^\p / e^\omega P^\p = x_i \; .
\end{equation}
From this and (\ref{T_BOOST}) we find in addition
\begin{equation}
  \vc{k}_{\perp i}^\prime = \vc{p}_{\perp i}^\prime - x_i
  \vc{P}_{\perp}^\prime 
  = \vc{p}_{\perp i}+ \vcg{\beta} p_i^\p - x_i (\vc{P}_\perp
  + \vcg{\beta}P^\p ) = \vc{k}_{\perp i} \; ,
\end{equation}
which indeed proves the frame independence of $x_i$ and $\vc{k}_{\perp
  i}$.

Let us calculate the total light-cone energy of the system in terms of
the relative coordinates. Making use of the constraints
(\ref{RELMOM_SUM}), we obtain
\begin{eqnarray}
  P^\m &=& \sum_i p_i^\m = \sum_i \frac{p_{\perp i}^2 +
  m_i^2}{p_i^\p} 
  = \sum_i \frac{(x_i \vc{P}_\perp + \vc{k}_{\perp i})^2 + m_i^2}{x_i
  P^\p} \nn \\
  &=& \frac{1}{P^\p}\left( P_\perp^2 + \sum_i \frac{k_{\perp i}^2 +
  m_i^2}{x_i} \right) \equiv P_\mathrm{CM}^\m  + P_\mathrm{r}^\m 
  \label{SEP_HAM}\; .
\end{eqnarray}
This is another important result: the light-cone Hamiltonian $P^\m$
separates into a center-of-mass term,
\begin{equation}
  \label{P_CM}
  P_\mathrm{CM}^\m = P_\perp^2 / P^\p \; ,
\end{equation}
and a term containing only the relative coordinates,
\begin{equation}
  \label{P_REL}
  P_\mathrm{r}^\m = \frac{1}{P^\p}\left( \sum_i \frac{k_{\perp i}^2 +
  m_i^2}{x_i} \right) \equiv \frac{M_0^2}{P^\p} \; .
\end{equation}
The second identity, which states that $P_r^\m$ is essentially the free
invariant mass squared, follows upon multiplying (\ref{SEP_HAM}) by
$P^\p$,
\begin{equation}
  \label{FIMS}
  P^\p P_r^\m = P^\p P^\m - P_\perp^2 = M_0^2 = \sum_i \frac{k_{\perp i}^2 +
  m_i^2}{x_i} \; . 
\end{equation}
To simplify things even more, one often goes to the `transverse rest
frame'  where $\vc{P}_\perp$ and therefore the center-of-mass
Hamiltonian $P_\mathrm{CM}^\m$ from (\ref{P_CM}) vanish. 
 
Summarizing we note that the special behavior under boosts together
with the transverse Galilei invariance leads to frame independent
relative coordinates and a separation of the center-of-mass motion
reminiscent of ordinary non-relati\-vis\-tic physics. This is at variance
with the instant form, where the appearance of the notorious square root
in the energy, $P^0 = (\vc{P}^2 + M_0^2)^{1/2}$, prohibits a similar
separation of  variables.

For a system of interacting particles, we add an interaction $W$ to the
free invariant mass squared defined in (\ref{FIMS}),
\begin{equation}
  \label{P2_INT}
  P^2 = M^2 = M_0^2 + W \; .  
\end{equation}
The total light-cone  energy $P^\m$ is thus no longer given by the sum of
the individual energies, but rather by
\begin{equation}
  P^\m = \frac{P_\perp^2 + M^2}{P^\p} = \sum_i \frac{p_{\perp i}^2 +
  m_i^2}{p_i^\p} + W/ P^\p \; . 
\end{equation}
Though we will not enter an intensive discussion of quantization at
this point, we note that, when translated to the quantum theory,
(\ref{P2_INT}) gives rise to the following light-cone Schr\"odinger
equation, $(\hat P^2 - M^2) |\psi \ket = 0$, or 
\begin{equation}
  \label{IF_SEQ}
  (M^2 - \hat M_0^2) | \psi \ket = \hat W |\psi \ket \; , 
\end{equation}
where $M^2$ now denotes the mass-squared eigenvalue. In later chapters
we will derive and solve such equations within quantum field theory.

Of course, in the interacting case, all dynamical Poincar\'e generators
differ from their free counterpart by some `potential' like $W$. This
has already been pointed out by Dirac \cite{dirac:49}, who also stated
that finding potentials which are consistent with the commutation
relations of the Poincar\'e algebra is the ``real difficulty in the
construction of a theory of a relativistic dynamical system'' with a
fixed number of particles. 

It has turned out, however, that Poincar\'e invariance alone is not
sufficient to guarantee a reasonable Hamiltonian formulation. There are
no-go theorems both for the instant \cite{leutwyler:65} and the front
form \cite{jaen:84}, which state that the inclusion of any potential
into the Poincar\'e generators, even if consistent with the commutation
relations, spoils relativistic {\em covariance}. The latter is a
stronger requirement as it enforces particular transformation laws for
the particle trajectories (world-lines). Thus, covariance imposes rather
severe restrictions on the dynamical system \cite{leutwyler:78}.

The  physical reason for these problems is  that potentials imply an
instantaneous interaction-at-a-distance which is in conflict with the
existence of a limiting velocity and retardation effects. Relativistic
causality is thus violated. This is equivalently obvious from the fact
that a \emph{fixed} number of particles is in conflict with the necessity of
particle creation and annihilation and the appearance of
anti-particles. 

Nevertheless, with considerable effort, it is possible to construct
dynamical quantum systems with a fixed number of constituents which are
consistent with the requirements of Poincar\'e invariance and
relativistic covariance \cite{leutwyler:78,sokolov:79,coester:82}.

At this point one might finally ask the question whether the different
forms of relativistic dynamics are physically equivalent. From the point
of view that different time choices correspond to different gauge
fixings it is clear that equivalence must hold.  After all, we are just
dealing with different coordinate systems. People have tried to make
this equivalence more explicit by working with coordinates which
smoothly interpolate between the instant and the front form
\cite{prokhvatilov:88,prokhvatilov:89,lenz:91,hornbostel:92}. In the
context of relativistic quantum mechanics, it has been shown that the
Poincar\'e generators for different forms are unitarily equivalent
\cite{sokolov:79}.

We are, however, more interested in what might be called a `top-down
approach'. Our aim is to describe few-body systems not within quantum
mechanics but quantum field theory to which we now turn.

\chapter{Light-Cone Quantization of Fields}

\section{Construction of the Poincar\'e Generators}

We want to derive the representation of the Poincar\'e generators within
field theory and their dependence on the hypersurface $\Sigma$ chosen to
define the time evolution. To this end we follow \cite{fubini:73} and
describe the hypersurface mathematically through the equation
\begin{equation}
  \Sigma:  F(x) = \tau  \; .
\end{equation}
The surface element on $\Sigma$ is implicitly defined via
\begin{equation}
  \int_\Sigma d\sigma _\mu u(x) = \int d^4 x \, N^\mu  
  \delta(F(x) - \tau) u(x) \; ,
\end{equation}
where, as before, $N^\mu = \partial^\mu F(x)$ is the normal on $\Sigma$
and $u$ some arbitrary function. We will write this expression
symbolically as
\begin{equation}
  \label{SURF_EL} 
  d\sigma _\mu = d^4 x \, N^\mu  \delta(F(x) - \tau) \; .
\end{equation}
The central object of this section will be the energy-momentum tensor, 
\begin{equation}
  T^{\mu \nu} = \pad{\mathcal{L}}{(\partial_\mu \phi)} \partial^\nu \phi -
  g^{\mu \nu} \mathcal{L} \; ,
\end{equation}
with $\mathcal{L}$ being the Lagrangian which depends on fields that are
collectively denoted by $\phi$. With the help of the energy-momentum
tensor we can define a generator
\begin{equation}
  \label{GENA}
  A[f] = \int_\Sigma d \sigma_\mu f_\nu (x) T^{\mu \nu}(x) \; ,
\end{equation}
where $A$ and $f$ can be tensorial quantities carrying dummy indices
$\alpha$, $\beta, \ldots$ which we have suppressed.  $A[f]$ generates the
infinitesimal transformations
\begin{eqnarray}
  \delta_f B(x) &=& f_\mu (x) \, \partial^\mu B (x) \; , \\
  \delta_f x^\mu &=& f^\mu (x) \; ,
\end{eqnarray}
where $f$ is now understood as being infinitesimal. In the same way as
for a finite number of degrees of freedom, the generator $A$ is called
kinematical, if it leaves $\Sigma$ invariant, that is,
\begin{equation}
  \label{AKIN}
  \delta_f F = f_\mu \partial^\mu F = 0 \; .
\end{equation}
Otherwise, $A$ is  dynamical. With the energy-momentum tensor
$T^{\mu\nu}$ at hand we can easily show that kinematical generators are
interaction independent. We decompose $T^{\mu \nu}$, 
\begin{equation}
  T^{\mu \nu} = T_0^{\mu \nu} - g^{\mu \nu} \mathcal{L}_{\mathrm{int}}
  \; ,
\end{equation}
into a free part $T_0^{\mu \nu} = T^{\mu \nu}(g= 0)$, $g$ denoting the
coupling, and an interacting part (we exclude the case of derivative
coupling). If $A$ is kinematical, we have from (\ref{SURF_EL},
\ref{GENA}, \ref{AKIN}),
\begin{equation}
  A_{\mathrm{int}}[f] = - \int d^4 x \, \delta(F -
  \tau) \mathcal{L}_{\mathrm{int}} f^\mu \partial_\mu F
    = - \int d^4 x \, \delta(F -\tau) \mathcal{L}_{\mathrm{int}}
  \delta_f F  = 0 \; ,
\end{equation}
which indeed shows that $A$ does not depend on the interaction. Of
course, we are particularly interested in the Poincar\'e generators,
$P^\alpha$ and $M^{\alpha \beta}$. The momenta $P^\alpha$ and Lorentz
transformations $M^{\mu\nu}$ correspond to the choices $f_\mu^\alpha =
g_\mu^\alpha$ and $f_\mu^{\alpha \beta} = x^\alpha g_\mu^\beta - x^\beta
g_\mu^\alpha$, respectively so that, from (\ref{GENA}), they are given
in terms of $T^{\mu\nu}$ as
\begin{eqnarray}
  P^\alpha &=& \int_\Sigma d\sigma_\mu T^{\mu\alpha} \; , \label{QFT_POINC_P}\\
  M^{\alpha\beta} &=& \int_\Sigma d\sigma_\mu (x^\alpha T^{\mu\beta} -
  x^\beta T^{\mu\alpha}) \; .\label{QFT_POINC_M}
\end{eqnarray}
From (\ref{AKIN}) it is easily seen that the Poincar\'e generators
defined in (\ref{QFT_POINC_P}, \ref{QFT_POINC_M}) act on $F(x) = \tau$
exactly as described in (\ref{PC_ACTION}). The remarks of Chapter~2 on
the kinematical or dynamical nature of the generators in the different
forms are therefore equally valid in quantum field theory. Let us
discuss the three main forms of relativistic dynamics explicitly.

\subsection*{The Instant Form}

We recall the hypersurface of equal time,  
\begin{equation}
  \Sigma: F(x) \equiv N \cdot x \equiv  x^0 = \tau \; ,   
\end{equation}
which leads  to a surface element
\begin{equation}
  d \sigma^\mu = d^4 x \, N^\mu \delta(x^0 - \tau) \; , \quad N^\mu =
  (1, \vc{0}) \; . 
\end{equation}
Using (\ref{QFT_POINC_P}, \ref{QFT_POINC_M}), the Poincar\'e generators are obtained as
\begin{eqnarray}
  P^\mu &=&  \int_\Sigma d^3 x \, T^{0 \mu} \; ,   \\
  M^{\mu\nu} &=& \int_\Sigma d^3 x \, \big(x^\mu T^{0 \nu} - x^\nu
  T^{0 \mu}\big) \; .
\end{eqnarray}

\subsection*{The Front Form}

Hypersurface and surface element are given by
\begin{equation}
  \Sigma: F(x) \equiv N \cdot x \equiv x^\p = \tau \; , \quad
  d\sigma^\mu = d^4 x \, N^\mu \, \delta(x^\p - \tau)  \; ,
\end{equation}
where $N$ is the light-like four-vector $N = (1,0,0,-1)$, $N^2 = 0$ of
(\ref{N_FRONT}).  In terms of $T^{\mu\nu}$, the Poincar\'e generators are
\begin{eqnarray}
  P^\mu &=& \sfrac{1}{2} \int_\Sigma dx^\m d^2 x_\perp \, T^{\p \mu} \; ,
  \label{FF_P} \\
  M^{\mu\nu} &=& \sfrac{1}{2} \int_\Sigma dx^\m d^2 x_\perp \, 
  \big(x^\mu T^{\p \nu} - x^\nu  T^{\p \mu}\big) \; . \label{FF_M}
\end{eqnarray}
The somewhat peculiar factor 1/2 is the Jacobian which arises upon
transforming to light-cone coordinates.

\subsection*{The Point Form}

As we have already seen, the point form is slightly more involved. The
hypersurface of equal time,
\begin{equation}
  \Sigma: F(x) = \sqrt{x^2} = \tau > 0 \; , \quad x^2 > 0 \; ,
\end{equation}
has the radial vector $N^\mu = x^\mu/\tau$ as its normal. In terms of
the hyperbolic coordinates (\ref{HYP_COORD}), the Poincar\'e generators
become  
\begin{eqnarray}
  P^\mu &=& \tau^2 \int_\Sigma d\omega \, d\theta \, d\phi \, \sinh^2
  \omega \, \sin   \theta \, x_\alpha T^{\mu\alpha} \; , \\
  M^{\mu\nu} &=& \tau^2 \int_\Sigma d\omega \, d\theta \, d\phi \,
  \sinh^2 \omega \,  \sin   \theta \, x_\alpha \big(x^\mu T^{\alpha\nu} - x^\nu T^{\alpha\mu}\big) \; .  
\end{eqnarray}
Note again that the limit $\tau \to 0$ does not make sense.

\section{Schwinger's (Quantum) Action Principle}

Our next task is to actually quantize the fields on the hypersurfaces
$\Sigma: \tau = F(x)$ of equal time $\tau$. There is more than one
possibility to do so, and we will explain a few of these. We begin with
a method that is essentially due to Schwinger \cite{schwinger:51,
  schwinger:53a,schwinger:53b}. We define a four-momentum density
\begin{equation}
  \Pi^\mu = \pad{\mathcal{L}}{(\partial_\mu \phi)} \; ,
\end{equation}
so that the energy-momentum tensor $T^{\mu\nu}$ can be written as 
\begin{equation}
  T^{\mu\nu} = \Pi^\mu \partial^\nu \phi - g^{\mu\nu} \mathcal{L} \; .
\end{equation}
In some sense, this can be viewed as a covariant generalization of the
usual Legendre transformation between Hamiltonian and Lagrangian. Using
the normal $N^\mu$ of the hypersurface $\Sigma$, we define the canonical
momentum (density) as the projection of $\Pi^\mu$,
\begin{equation}
  \pi \equiv N \cdot \Pi \; .
\end{equation}
Schwinger's action principle states that upon variation the action $S =
\int d^d x \, \mathcal{L}$ changes at most by a surface term which (if
$\Sigma$ is not varied, i.e.~$\delta x^\mu = 0$) is given by
\begin{equation}
  \delta G (\tau) = \int_\Sigma d\sigma_\mu \, \Pi^\mu \, \delta\phi =
  \int d^d x \, \delta   (\tau - F) \, \pi \,  \delta \phi \;  
\end{equation}
taken between the initial and final hypersurfaces $\Sigma_1$ and
$\Sigma_2$.  The quantity $\delta G$ is interpreted as the generator of
field transformations, so that we have 
\begin{equation}
  \label{DELTA_G_SL}
  \delta \phi = \pb{\phi}{\delta G} \; ,
\end{equation}
in case that $\Sigma$ is entirely space-like (with time-like normal)
\cite{schwinger:51, schwinger:53a}.  We note in passing that the
generator $\delta G$ is a field theoretic generalization of the
canonical one-form used in analytical mechanics.  In the last expression
we have been a bit sloppy with the space-time arguments, so let us
improve on this. As in the preceding chapter, we parametrize space-time
with foliation coordinates $\xi^\mu$, with $\xi^0 = \tau$ and the
$\xi^i$ living on $\Sigma$. Then we calculate
\begin{eqnarray}
  \delta \phi (\xi) &=& \pb{\phi (\xi)}{\delta G (\tau)} \nn \\
  &=& \int d\xi^{\prime 0} \int d^{d-1} \xi^\prime \, J(\xi^\prime)
  \delta (\tau - \xi^{\prime 0})
  \pb{\phi(\xi)}{\pi(\xi^\prime)} \delta \phi (\xi^\prime) \nn \\
  &=& \int_\Sigma d^{d-1} \xi^\prime \, J(\xi^\prime)
  \pb{\phi(\xi)}{\pi (\xi^\prime)} 
  \delta \phi (\xi^\prime) \Big|_{\xi^0 = \xi^{\prime 0} = \tau} \; ,
\end{eqnarray}
where $J(\xi)$ denotes the Jacobian associated with the change of
variables $x^\mu \to \xi^\alpha$. From the last expressions we infer
that the canonical equal-time bracket must be
\begin{equation}
  \label{GEN_BAS_BRA} 
  \pb{\phi (\xi)}{\pi (\xi^\prime)}_{\xi , \xi^\prime \in \Sigma} = 
  \pb{\phi (\xi)}{\pi (\xi^\prime)}_{\xi^0 = \xi^{\prime 0} = \tau} =
  J^{-1}({\vcg{\xi}}) \, \delta^{d-1}(\vcg{\xi} - \vcg{\xi^\prime}) \; .
\end{equation}
As usual, commutators are abstracted from the basic bracket by invoking
Dirac's correspondence principle, that is, by replacing the bracket by
$i$ times the commutator. For arbitrary classical observables, $A$, $B$,
this means explicitly,
\begin{eqnarray}
  [\hat A , \hat B] = i \widehat{\pb{A}{B}} \; ,
\end{eqnarray}
With minor modifications, the approach leading to (\ref{GEN_BAS_BRA})
can also be used for the front form, resulting in what is called
light-cone or light-front quantization. The canonical light-cone
momentum is
\begin{equation}
  \pi = N \cdot \Pi = N \cdot \partial \phi = \partial^\p \phi \equiv 2
  \pad{}{x^\m} \phi \; ,
\end{equation}
which is peculiar to the extent that it does not involve a (light-cone)
time derivative. Therefore, $\pi$ is a dependent quantity which does not
provide additional information, being known on $\Sigma$ when the field is
known there, so that $\pi$ is merely an abbreviation for $\partial^\p
\phi$. Again, the reason is that the normal $N^\mu$ of the null-plane
$\Sigma$ lies within $\Sigma$. As a result, $\phi$ and $\partial^\p
\phi$ have to be treated on the same footing within Schwinger's approach
which leads to an additional factor 1/2 compared to (\ref{DELTA_G_SL}),
\begin{equation}
  \label{DELTA_PHI_FF}
  \sfrac{1}{2} \delta \phi = \pb{\phi}{\delta G} \; ,
\end{equation}
with a front-form generator
\begin{equation}
  \label{DELTA_G_FF}
  \delta G (x^\p) = \sfrac{1}{2} \int_\Sigma dx^\m d^2 x_\perp
  \partial^\p \phi \, \delta \phi \; .
\end{equation}
The appearance of the peculiar factor 1/2 in (\ref{DELTA_PHI_FF}) has
been discussed at length by Schwinger in \cite{schwinger:53b} (see also
\cite{chang:73a}). This factor cancels the light-cone Jacobian $J = 1/2$
in (\ref{DELTA_G_FF}), so that we are left with the equal-time
commutator (for $d$ = 4),
\begin{equation}
  \label{NON_FUND_COMM}
  [\phi (x), \pi (y)]_{x^\p = y^\p = \tau} = i \delta (x^\m -
  y^\m) \delta^2 (\vc{x}_\perp - \vc{y}_\perp) 
\end{equation}
As the independent quantities are the fields themselves, we invert the
derivative $\partial^\p$ and obtain the more fundamental commutator
\begin{equation}
  \label{FUND_COMM}
  [\phi (x) , \phi (y)]_{x^\p = y^\p = \tau} = - \sfrac{i}{4}\sgn(x^\m -
  y^\m) \delta^2 (\vc{x}_\perp - \vc{y}_\perp) \; .
\end{equation}
In deriving (\ref{FUND_COMM}) we have chosen the anti-symmetric Green
function $\sgn(x^\m)$ satisfying
\begin{equation}
  \pad{}{x^\m} \sgn (x^\m) = 2 \delta (x^\m) \; ,
\end{equation}
so that (\ref{NON_FUND_COMM}) is re-obtained upon differentiating
(\ref{FUND_COMM}) with respect to $y^\m$.  We will see later that the
\emph{field} commutator (\ref{FUND_COMM}) can be directly obtained from
Schwinger's method. Before that, however, we will study the relation
between the choice of initializing hypersurfaces, the problem of field
quantization and the solutions of the dynamical equations.

\section{Quantization as  an Initial/Boundary-Value Problem}

As a prototype field theory we consider a massive scalar field in 1+1
dimensions. Its dynamics is encoded in the action
\begin{equation}
  \label{SCAL_ACTION}
  S[\phi] = \int d^2 x \, \mathcal{L} =  \int d^2 x \left( \sfrac{1}{2}
  \partial_\mu \phi \partial^\mu
  \phi - \sfrac{1}{2} m^2 \phi^2 - \mathcal{V}[\phi] \right)\; ,
\end{equation}
where $\mathcal{V}$ is some interaction term like e.g.~$\lambda \phi^4$
and $\mathcal{L} = \mathcal{L}_0 + \mathcal{V}$. By varying the free
action in the standard way we obtain
\begin{eqnarray}
  \delta S = \int_{\partial M} d\sigma_\mu
  \Pi^\mu  \delta \phi + \int_M \left[ 
  \pad{\mathcal{L}_0}{\phi} - \partial_\mu \pad{\mathcal{L}_0}{(\partial_\mu
  \phi)} \right] \delta \phi \; .
\end{eqnarray}
If we do not vary on the boundary of our integration region $M$, $\delta
\phi|_{\partial M} = 0$, the surface term in $\delta S$ vanishes and we
end up with the (massive) Klein-Gordon equation in 1+1 dimensions,
\begin{equation}
  \label{KG}
  (\Box + m^2) \phi = 0 \; .
\end{equation}
We are going to solve this equation by specifying initial and/or
boundary conditions for the scalar field $\phi$ on different
hypersurfaces $\Sigma$.  In particular we will clarify the relation
between the associated initial value problems and the determination of
`equal-time' commutators.

It may look rather trivial to consider just the free theory, but this is
not really true.  Let us analyze what quantization of a field theory
means in the light of the different forms of relativistic dynamics.  One
specifies canonical commutators like $[\phi (x), \phi (0)]_{x \in
  \Sigma}$, where the equation for the hypersurface $\Sigma$: $\tau =
const$ defines an evolution parameter $\tau$ and the notion of `equal
time' as already pointed out several times.  $\Sigma$ should be chosen
such that the interaction does not change the commutator on it, i.e.~the
latter is a kinematical quantity.

Now, if $\phi$ is a {\it free} field, the commutator,
\begin{equation}
 [\phi (x) , \phi(0)] = i \Delta (x) \; ,
\end{equation}
is exactly known: it is the Pauli-Jordan or Schwinger function $\Delta$
which is a special solution of the Klein-Gordon equation (\ref{KG}). It
can be obtained directly from the action in a covariant manner as a
Peierls bracket \cite{peierls:52,dewitt:83}. Alternatively, one can find it
from its Fourier representation which yields  \cite{heinzl:94a}
\begin{eqnarray}
  \label{PJ}
  \Delta (x) &=& - \frac{i}{2\pi} \int d^2 p \, \delta (p^2 - m^2) \,
  \sgn(p^0) e^{- i p \cdot x} \nn \\
  &=& - \sfrac{1}{2} \sgn(x^0) \, \theta (x^2) \, J_0 (m \sqrt{x^2}) \nn \\
  &=& - \sfrac{1}{4} \big[ \sgn(x^\p) + \sgn(x^\m) \big] \; J_0 (m
  \sqrt{x^\p x^\m}) \; , 
\end{eqnarray}
We note that it is antisymmetric, $\Delta (x) = - \Delta (-x)$, and
vanishes outside the light-cone, i.e.~for $x^2 < 0$.

If $\phi$ is an {\em interacting} field one still has some information
from causality.  For space-like hypersurfaces (instant and point form)
the equal-time commutator vanishes like in the free field case,
\begin{equation}
[\phi (x) , \phi (0)]_{x^2 < 0} = 0 \; . 
\end{equation}
This expresses the fact that fields which are separated by a space-like
distance cannot communicate with each other.  For the front form with
hypersurface $\Sigma: x^\p = 0$, the situation is slightly different. In
1+1 dimensions, $\Sigma$ is the light-cone and therefore entirely
light-like. In higher dimensions, $\Sigma$ still contains light-like
directions namely where $x^\m = \vc{x}_\perp = 0$. This can also be seen
from the (free) commutator (\ref{FUND_COMM}) which does not vanish at
these points corresponding to a light-like separation. It may therefore
well be that the interaction changes the light-cone behavior of the
commutator. This is actually seen in deep-inelastic scattering, where
the assumption of free-field commutation relations leads to the parton
model and scaling \cite{cheng:84}, which, however, is violated in the
interacting theory, i.e.~QCD (though only weakly).

Concluding, we can say that, if the fields are separated by a space-like
distance, the free field `equal-time' commutators are boundary
conditions also for the interacting theory and thus of sizable interest.

\subsection{The Cauchy Problem}

Before discussing the case of the front form, we would like to recall
that the conventional quantization on a space-like surface (based on the
instant form) corresponds to a Cauchy problem: if one specifies the
field $\phi$ and its time derivative $\dot \phi$ on $\Sigma: x^0 = 0$,
the solution of the Klein-Gordon equation is uniquely determined as
\cite{schweber:61,neville:71a}
\begin{equation}
  \label{KG_SOL}
  \phi    (x)    =   \int_{y^0    =0}    dy^1    \Big[\phi(y)
  \frac{\stackrel{\leftrightarrow}{\partial}}{\partial  y^0} \Delta  (x-y)
  \Big] \; . 
\end{equation}
This solution expresses the well-known fact that the creation and
annihilation operators in the Fock expansion of the quantum field are
given in terms of the field \emph{and} the velocity on $\Sigma$,
\begin{equation}
  \label{FOCK_IF}
  a (p^1) = \int dx^1 e^{-i p^1 x^1} \Big[ \omega_p \, \phi(x^0 = 0 , x^1) + i
  \dot \phi (x^0 = 0 , x^1) \Big] \; ,
\end{equation}
with $\omega_p = (p_1^2 + m^2)^{1/2}$. If we compare (\ref{KG_SOL}) with
the two independent canonical commutators
\begin{eqnarray}
  \Big [\phi   (x),  \phi  (0) \Big]_{x^0   =  0}  &=&  i  \Delta 
  (x)\vert_{x^0 = 0} = 0 \; , \nn \\  
  \Big [\dot \phi   (x),  \phi  (0) \Big]_{x^0   =  0}  &=&  i \dot \Delta 
  (x)\vert_{x^0 = 0} = -i \delta (x^1) \; ,  \label{CAUCHY_DATA}
\end{eqnarray}  
we see that they can be viewed as Cauchy data for the Pauli-Jordan
function $\Delta$.  Using the data (\ref{CAUCHY_DATA}) in formula
(\ref{KG_SOL}) we obtain a consistency condition,
\begin{equation}
  [\phi  (x), \phi (0)] = \int_{y^0 = 0} dy^1 [\phi  (y), \phi (0)] 
  \frac{\stackrel{\leftrightarrow}{\partial}}{\partial  y^0}
  \Delta (x-y) \; , 
\end{equation}
which  expresses  the commutator  for arbitrary  time differences
$x^0$ in terms of its initial values given by (\ref{CAUCHY_DATA}).

\subsection{The  Characteristic Initial-Value Problem} 

In the following we will perform an analogous discussion for the
hypersurfaces $\Sigma: x^\pm = const$, which, in $d+1+1$, constitute the
light-cone. In Dirac's classification, the light-cone corresponds to a
degenerate point form with parameter $a =0$. One thus does not have
transitivity as points on different `legs' of the cone are not related
by a kinematical operation. The Hamiltonians are $P^\p$ and $P^\m$, so
that a Hamiltonian formulation appears a little bit strange and will
involve two time parameters. We will not pursue this issue here. A
Lagrangian formulation, however, is no problem; it just leads to the
Klein-Gordon equation. Let us discuss how to solve it using initial
data on $\Sigma$.

The light-fronts $x^\pm = 0$ are characteristics of the Klein-Gordon equation
\cite{domokos:71}.  Therefore, one is dealing with a characteristic
initial-value problem \cite{courant:62,myint-u:87}: if one specifies the
field $\phi$ on $x^+ = x_0^+$ {\it and} $x^- = x_0^-$, say $\phi (x^+,
x_0^- ) = f (x^+)$ and $\phi (x_0^+, x^-) = g(x^-)$, the solution of the
Klein-Gordon equation is \cite{neville:71a,rohrlich:71}
\begin{equation}
  \label{KG_SOL_CHAR}
  \phi       (x)       =      \int_{x_0^+}^\infty
  dy^+  f(y^+)  \frac{\stackrel{\leftrightarrow}{\partial}}{\partial  y^+}
  \Delta (x-y) \bigg\vert_{y^- = x_0^-} \, + 
  \int_{x_0^-}^\infty            dy^-           g(y^-)
  \frac{\stackrel{\leftrightarrow}{\partial}}{\partial  y^-} \Delta  (x-y)
  \bigg\vert_{y^+ = x_0^+} \; .  
\end{equation}
This amounts to quantization on {\it two} characteristics, $x^\pm = 0$,
i.e., in $d=1+1$, really \emph{on} the light cone, $x^2 = 0$.  The
following two independent commutators,
\begin{equation}
  [\phi (x), \phi (0)]_{x^\pm = 0} = i \Delta (x)\vert_{x^\pm  = 
  0} = - \sfrac{i}{4} \sgn (x^\mp) \; , 
\end{equation}
are then characteristic data for the Pauli-Jordan function. Again this
can be cast in the form of a consistency condition,
\begin{eqnarray}
  \label{CONSIST_CHAR}
  [\phi (x), \phi (0)] &=&  \int_0^\infty
  dy^+           [\phi           (y),           \phi           (0)]
  \frac{\stackrel{\leftrightarrow}{\partial}}{\partial  y^+} \Delta  (x-y)
  \bigg\vert_{y^- = 0} \, + \nn  \\
  &+&   \int_{0}^\infty    dy^-    [\phi    (y),    \phi    (0)]
  \frac{\stackrel{\leftrightarrow}{\partial}}{\partial  y^-} \Delta  (x-y)
  \bigg\vert_{y^+ = 0} \; , 
\end{eqnarray}
which expresses the commutator in terms of its characteristic initial
values.

It turns out that, in case the field $\phi$ is massless, the above
quantization procedure is the only consistent one (in $d$=1+1), if one
wants to use light-like hypersurfaces \cite{bogolubov:90}.

\subsection{The Hamiltonian Formulation}

In this subsection we want to analyze whether a (Hamiltonian)
quantization on a {\it single} light-front is possible in the massive
case. Note that in the instant form with its corresponding Cauchy
problem everything is quite clear. The data for $\phi$ and $\dot \phi$
translate into data for $\phi$ and the momentum $\pi$, which is
independent of $\phi$. We already know that within the front form there
is no independent momentum as the normal of a null-plane lies within the
null-plane. If one does not want to use a second light-front in order to
avoid an awkward Hamiltonian formulation with two times, it seems that
one only needs half of the data as compared to the Cauchy or the
characteristic initial-value problem \cite{chang:73a}. This seems strange
and, indeed, is not the case as we will see.

Let us write the Klein-Gordon equation (\ref{KG}) in terms of light-cone
coordinates
\begin{equation}
  (\partial^\p \partial^\m + m^2) \phi = 0 \; , 
\end{equation}
and solve for $\dot \phi \equiv \partial \phi / \partial x^\p$ (which
amounts to inverting $\partial^\p = 2 \partial/\partial x^\m$). This
gives
\begin{equation}
  \label{EL_INIT}
  \dot \phi (x^+, x^-) = -\frac{m^2}{4}  \int dy^- G (x^-,  y^-) 
  \phi (x^+ , y^-)  \; , 
\end{equation}
where  the  Green  function  $G$ is an inverse  of the derivative
$\partial/\partial   x^-$,
\begin{equation}
  \label{GREEN_DEF}
  \pad{}{x^\m} G(x^\m , y^\m) = \delta (x^\m - y^\m) \; . 
\end{equation}
The Hamiltonian equation of motion is given by the Poisson bracket with
$H = \sfrac{1}{2} m^2 \int dx^\m \phi^2$,
\begin{equation} 
  \label{HAM_INIT}
  \dot  \phi  (x^+ , x^-) =  \frac{m^2}{2}  \int  dy^- \pb{\phi 
  (x^+,   x^-)}{\phi(x^+   ,   y^-)}   \phi   (x^+   ,   y^-)   
\end{equation}
with a presently unknown bracket of the fields $\phi$. Clearly,
Euler-Lagrange and Hamiltonian equation of motion, (\ref{EL_INIT}) and
(\ref{HAM_INIT}) have to be  equivalent. This is realized  if one identifies
\begin{equation}
  \pb{\phi  (x^+, x^-)}{\phi  (x^+ , y^-)} \equiv -
  \sfrac{1}{2}G(x^- , y^-) \; , \label{BAS_BRACK}
\end{equation}
Upon quantization, this should, of course, coincide with
(\ref{FUND_COMM}) specialized to $d= 1+1$. To make the whole procedure
of solving the equations of motion unambiguous, we need an initial
condition for $\phi$ on $\Sigma$ and a \emph{further} condition which
makes the inversion of $\partial^\p$ unique. Thus, as announced, we need
again \emph{two} conditions.
 
After quantization the basic bracket (\ref{BAS_BRACK}) becomes the free
field commutator according to the correspondence principle.  The free
field commutator on the light front $x^+ = 0$ thus essentially is the
Green function $G$. In momentum space, $G$ is given by (a regularization
of) the expression $1/p^+$.  In this sense, quantization on $x^+ = 0$
amounts to inversion of the light-cone momentum $p^+$ (viewed as a
distribution). It is well known that there is no unique inverse (or
regularization) of $1/p^\p$ \cite{gelfand:64}. The possible
regularizations differ by terms proportional to $\delta(p^\p)$ which
correspond to homogeneous solutions of (\ref{GREEN_DEF}). To solve this
problem, one has to specify additional conditions with respect to the
variable $x^-$ which is conjugate to~$p^+$.

It turns out that data on a second characteristic, say $\phi (x^+, x^- =
x_0^-) = f (x^+)$, do not work: they do not lead to an antisymmetric
Green function in (\ref{EL_INIT}) which is needed for the commutator.
Therefore one does not obtain equivalence between Euler-Lagrange and
Hamiltonian equations of motion, which is a severe inconsistency.  As a
result, there is no Hamiltonian formulation with data on {\it two}
characteristics and a {\it single} light-cone Hamiltonian.

There exists, however, an alternative. We follow the philosophy of an
approach called `discretized light-cone quantization'
\cite{maskawa:76,pauli:85a,pauli:85b} and enclose the system in a finite volume,
$-L \le x^- \le L$.  Upon demanding periodic boundary conditions for the
field,
\begin{equation}
  \label{PBC}
  \phi(x^\p , x^\m = -L) = \phi(x^\p , x^\m = L) \; ,
\end{equation}
which are to hold for all light-cone times $x^\p$, the longitudinal
momenta become discrete,
\begin{equation}
  k_n^\p = 2 \pi n /L \; , \quad n \in \mathbb{Z} \; .
\end{equation}
As the spatial boundaries are not separated by a space-like but a
light-like distance, one has to check whether the boundary conditions
(\ref{PBC}) are consistent with the equation of motion and thus with
relativistic causality. This will be done in the next subsection.
Assuming this consistency, the Euler-Lagrange equation (\ref{EL_INIT})
becomes
\begin{equation}
  \dot \phi (x^+,  x^-) = -\frac{m^2}{4}  \intl  dy^- G_p (x^- , 
  y^-) \phi (x^+, y^-)  \; , 
\end{equation}
where $G_p$ is uniquely determined as the periodic sign function,
\begin{equation}
  G_p (x^- , y^-) = \frac{1}{2}  \sgn (x^- - y^-) - \frac{x^- - 
  y^-}{2L} \; , 
\end{equation}
so that there is no arbitrariness left. From (\ref{BAS_BRACK}) we obtain a
unique and consistent commutator,
\begin{equation}
  \label{LC_COMM}
  [\phi (x), \phi (0)]_{x^+ = 0} = - \frac{i}{2} G_p (x^-) = -\frac{i}{2} \Big[
  \sfrac{1}{2}  \sgn (x^- - y^-) - \frac{x^- -   y^-}{2L} \Big]   \; .
\end{equation}
Apart from the second term, which is a finite volume correction, this
coincides with (\ref{FUND_COMM}).  We are left with checking the
consistency of the procedure above.

\subsection{The Initial-Boundary-Value Problem}

The attempt  to solve the Klein-Gordon equation with periodic boundary
conditions in $x^\m$ poses an  initial-boundary-value problem which is
explicitly  defined through the conditions
\begin{eqnarray}
  \phi (x^+ = 0, x^-) &=& g (x^-) \; , \label{INIT_DATA}\\
  \phi (x^+, x^- =L) &=& \phi (x^+, x^- = -L)  \label{BC}\; , 
\end{eqnarray}
where (\ref{INIT_DATA}) represents the initial data and (\ref{BC}) the
boundary conditions. The unique solution of the Klein-Gordon equation
obeying these conditions is obtained via Fourier transformation,
\begin{equation}
  \label{SOL_IBVP}
  \phi (x^+,  x^-)  = \sum_{n  \ne 0} g_n e^{-i k_n^+ x^- /2 - i 
  \hat k_n^- x^+/2} \equiv \sum_{n  \ne 0} g_n e^{-i \hat k_n \cdot 
  x} \; . 
\end{equation}
The on-shell energy is $k_n^\m = m^2/k_n^\p = m^2 L /2\pi n$, and the
$g_n$ are the Fourier modes of the function $g(x^-)$ specified in
(\ref{INIT_DATA}), i.e.
\begin{equation}
  \label{G_N}
  g_n = \frac{1}{2L} \intl dx^\m \, g(x^\m) \, e^{in\pi x^\m /L}
\end{equation}
Investigating the Klein-Gordon equation in momentum space one finds that
the zero mode $g_0$ is vanishing (see also Section~4.4). It is therefore
already omitted in (\ref{SOL_IBVP}). From this solution it is obvious
that data on a second characteristic are not necessary. In the quantum
theory, (\ref{G_N}) corresponds to the peculiar fact that the light-cone
Fock operators are expressible in terms of the field alone; the velocity
is not needed \cite{leutwyler:70,chang:73a}.

Another consistency check is to demand periodic boundary conditions for
the solution (\ref{KG_SOL_CHAR}) of the characteristic initial-value
problem.  Together with the data $f(x^+)$ this would over-determine the
problem, which can only be avoided, if $f$ is expressible in terms of
$g$. This is indeed the case,
\begin{equation}
  \label{F(X)+}
  f(x^+) = \frac{1}{2\pi  i} \int_{c-i\infty}^{c+i  \infty}  dp 
  \, \e^{px^+} \frac{H_L[g,p]}{1 - \exp (-m^2 L/2p)} \; , 
\end{equation}
where $H_L[g, p]$ is a complicated but known functional of the data $g$
\cite{heinzl:94a}.

In addition, let us check the consistency of the solution
(\ref{HAM_INIT}) for an important example, the periodic Pauli-Jordan
function,
\begin{equation}
  \Delta_L (x^+ , x^-) \equiv -\sum_{n \ne 0} \frac{i}{4\pi  n} e^{-i 
  \hat k_n^- x^+ /2 - in\pi x^- /L} \; .
\end{equation}
In analogy with (\ref{CONSIST_CHAR}) this leads to the light-cone
commutators
\begin{eqnarray}
  \Big[\phi   (x),   \phi   (0)\Big]_{x^+   =  0}   &=&   -\frac{i}{2}
  \Big[\frac{1}{2}  \sgn (x^-)  - \frac{x^-}{2L}\Big]  \equiv  g
  (x^-)  \; , \label{CIVP_COMM} \\ 
  \Big[\phi (x), \phi (0)\Big]_{x^- = -L} &=& \frac{1}{4\pi}  \sum_{n \ne 0}
  \frac{(-1)^n}{n}  \exp \Big(-i  \frac{m^2  L}{4\pi n} x^+ \Big)
  \equiv f(x^+) \; , 
\end{eqnarray}
where (\ref{CIVP_COMM}) coincides with (\ref{LC_COMM}), as expected.
Due to the periodic boundary conditions the second commutator given by
$f(x^+)$ is not independent but satisfies exactly (\ref{F(X)+}) with $g$
given by (\ref{CIVP_COMM}).  As in the case of the Cauchy problem and
the characteristic initial-value problem one can write down a
consistency condition,
\begin{equation}
  [\phi  (x) , \phi (0)]  = \intl  dy^-  [\phi  (y) , \phi  (0)]
  \left.  \pad{}{y^-} \Delta_L (x-y) \right\vert_{y^+ = 0} \; , 
\end{equation}
which expresses the commutator through its value on {\it one}
characteristic ($y^+ = 0$) and the {\it periodic} Pauli-Jordan function.

\subsection{Fermions}

For fermions, light-cone quantization is slightly more involved because
not all of the fermion field components are independent. This is easily
seen by considering the free massive Dirac equation (again, for
simplicity, in 1+1 dimensions),
\begin{equation}
  \label{DEQ}
  (i \partial \!\!\!/ - m ) \psi = 0 \; .
\end{equation}
If we write the two-spinor $\psi$ in terms of a lower and upper
component as 
\begin{equation}
  \psi = \left( \begin{array}{c} 
                    \psi_1 \\
                    \psi_2 
                \end{array} \right) \; ,
\end{equation}
the Dirac equation (\ref{DEQ}) splits up into
\begin{eqnarray}
  \partial^\m \psi_1 &=& -i m \psi_2 \; , \label{DEQ_P1} \\
  \partial^\p \psi_2 &=& -i m \psi_1 \; . \label{DEQ_P2}
\end{eqnarray}
The important point to note is that (\ref{DEQ_P2}) represents a constraint as
$\partial^\p$ is a spatial rather than a temporal derivative. It should
be solved for the dependent lower component $\psi_2$ in terms of the
independent upper component $\psi_1$, 
\begin{equation}
  \label{PSI_SOL}
  \psi_2 = -i m (\partial^\p)^{-1} \psi_1 \; .
\end{equation}
Obviously, at this point, the ubiquitous inverse of $\partial^\p$ arises
again, and we have to make sure that it becomes well defined. Like for
the scalar field, the mentioned ambiguity affects the quantization
procedure. The covariant anti-commutator of two fermion fields with
arbitrary space-time separation $x$ is given in terms of an invariant
function $S$,
\begin{equation}
  \label{COV_AC}
  \pb{\psi(x)}{\bar \psi(0)} = i S(x) \; ,
\end{equation}
where  $S$ can be inferred from the Pauli-Jordan function as
\begin{equation}
  \label{S(X)}
  S(x) = (i \partial \!\!\!/ + m) \Delta (x) \; .
\end{equation}
From this representation it is obvious that $S$ obeys the Dirac equation
(as it must), as $\Delta$ is a solution of the Klein-Gordon equation. We
thus expect that $S$ will also contain dependent degrees of freedom. To
make these explicit, we introduce a two-by-two matrix notation and write
(\ref{COV_AC}) in the form
\begin{equation}
  \label{S_IJ}
  \pb{\psi_i (x)}{\psi_j^\dagger (0)} = i S_{ij}(x) \; , \quad i,j = 1,2
  \; .
\end{equation}
According to (\ref{S(X)}), the components $iS_{ij}$ are 
\begin{eqnarray}
  i S_{11} (x) &=& - \partial^\p \Delta (x) \; , \\
  i S_{12} (x) = i S_{21} (x) &=& i m \Delta (x) \; , \\
  i S_{22} (x) &=& - \partial^\m \Delta (x) \; .
\end{eqnarray}
Comparing the second and the third identity we learn that $S_{22}$ is a
dependent quantity satisfying the constraint (\ref{DEQ_P2}), 
\begin{eqnarray}
  \label{S_CONSTR}
  \partial^\p S_{22} = -i m S_{12} \; .
\end{eqnarray}
Of course, taking into account (\ref{S_IJ}), this is consistent with the
fact that $\psi_2$ is the dependent field component. There is, however,
one subtlety that deserves a separate analysis. If one performs the
derivatives in (\ref{S(X)}) and writes the Pauli-Jordan function
(\ref{PJ}) as
\begin{equation}
  \Delta (x) \equiv D(x) J_0 (m \sqrt{x^2}) \; ,
\end{equation}
the $iS_{ij}$ are explicitly found to be
\begin{eqnarray}
  i S_{11} (x) &=& \delta(x^\m) + m D(x) \sqrt{x^\p / x^\m} J_1 (m
  \sqrt{x^2}) \; , \\
  i S_{12} (x) &=& i m D(x) J_0 (m \sqrt{x^2}) \; , \\
  i S_{22} (x) &=& \delta(x^\p) + m D(x) \sqrt{x^\m / x^\p} J_1 (m
  \sqrt{x^2}) \; . 
\end{eqnarray}
There are several remarks in order. First of all, the first identity
provides the canonical anti-commutator of the independent field
components,
\begin{equation}
  \pb{\psi_1(x)}{\psi_1^\dagger (0)}_{x^\p = 0} = i S_{11}(x^\p = 0 ,
  x^\m) = \delta (x^\m) \; .
\end{equation}
In addition, we note that $S_{11}$ and $S_{22}$ are related by an
exchange of $x^\p$ and $x^\m$. Going back to instant-form coordinates,
this amounts to the spatial inversion $x^1 \to -x^1$, that is, a {\em
  parity} transformation. Obviously, as parity exchanges light-cone
space and time, it is a dynamical and therefore complicated
transformation. Furthermore, the leading light-cone singularities of the
$S_{ii}$ are mass independent, as is well known from, say, the
space-time analysis of deep-inelastic scattering \cite{jackiw:72}. In
particular, we have
\begin{equation}
  \label{SM0}
  \lim_{m \to 0} i S_{22} (x) \equiv i S_{22}^0 = i S_{22} (x^\m = 0) 
  = \delta (x^\p) \; 
\end{equation}
On the other hand, if we formally solve (\ref{S_CONSTR}) for $S_{22}$,
we obtain
\begin{equation}
  \label{S22_SOL}
  S_{22} (x) = - i m (\partial^\p)^{-1} S_{12} \; , 
\end{equation}
which, at first glance, seems to be in conflict with (\ref{SM0}) upon
performing the limit $m \to 0$. One might be tempted to add the
homogeneous solution $S_{22}^0$ to the right-hand-side of
(\ref{S22_SOL}). This would imply that $S_{22}$ is not entirely
determined in terms of $S_{12}$, and one would have to provide
$S_{22}(x^\p, x^\m = 0)$ as a boundary condition for \emph{all
  light-cone times} $x^\p$. Analogous arguments apply to the relation
between $\psi_2$ and $\psi_1$. In the language of the preceding
subsections this would correspond to a characteristic initial-value
problem.  Fortunately, these difficulties can be circumvented, if one
makes use of parity, as we are now going to show.

We assume knowledge of the field $\psi_1$ at light-cone time $x^\p
=0$. For arbitrary times $x^\p$ it is then given by
\begin{equation}
  \label{PSI1_FOCK}
  \psi_1 (x) = \frac{1}{\sqrt{2\pi}}\int dk^\p \, \left[ b(k^\p) e^{-i \hat k
  \cdot x} + d^\dagger (k^\p) e^{i \hat k \cdot x} \right] \; ,
\end{equation}
where
\begin{equation}
  \hat k \cdot x = \sfrac{1}{2}k^\p x^\m + \sfrac{1}{2}\hat k^\m x^\p \;
  , \quad \hat k^\m = m^2 / k^\p \; ,
\end{equation}
and the Fock operators obey the anti-commutation relations
\begin{equation}
  \pb{b(k^\p)}{b^\dagger (q^\p)} = \pb{d(k^\p)}{d^\dagger (q^\p)} =
  \delta (k^\p - q^\p) \; .
\end{equation}
Plugging (\ref{PSI1_FOCK}) into (\ref{PSI_SOL}), the dependent field becomes
\begin{equation}
  \psi_2 (x) = \frac{m}{\sqrt{2\pi}} \int \frac{dk^\p}{k^\p} 
  \left[ b(k^\p) e^{-i \hat k  \cdot x} + d^\dagger (k^\p) e^{i \hat k
  \cdot x} \right] \; ,
\end{equation}
from which one infers the anti-commutator
\begin{equation}
  i S_{22} (x) = \frac{m^2}{2\pi} \int \frac{dk^\p}{(k^\p)^2} [e^{-i
  \hat k  \cdot x} + e^{i \hat k  \cdot x}] \; .
\end{equation}
Note the $m^2$ in front of the integral which is consistent with
(\ref{S22_SOL}) as $S_{12}$ is of order $m$.  The trick is now to invoke
parity by substituting $k^\p = m^2 / k^\m$ with $k^\m$ as the new
integration variable. This absorbs the $m^2$ in front of the integral
and results in
\begin{equation}
  i S_{22} (x) = \int \frac{dk^\m}{2\pi} \exp(- im^2 x^\m / k^\m - i
  k^\m x^\p) \; .
\end{equation}
Evaluating this for $x^\m = 0$, we finally  obtain
\begin{equation}  
  i S_{22} (x^\p , x^\m = 0) = \pb{\psi_2 (x)}{\psi_2 (0)}_{x^\m = 0} =
  \delta(x^\p) \; .
\end{equation}
We reemphasize that this result was obtained by only assuming initial
data (in terms of the usual Fock expansion) for the independent field
component $\psi_1$, and parity. Data on the second characteristic are
once more unnecessary.

Altogether, we have thus shown that, for massive bosons and fermions in
space-time dimension $d$ = 1+1, quantization on a {\it single} light
front $x^+ = 0$ is possible. However, there are special requirements to
be met. In the bosonic case, periodic boundary conditions in the spatial
direction lead to a consistent and simple light-cone Hamiltonian
formulation.  Data on a second characteristic are not necessary.  The
result justifies the approach of discretized light-cone quantization
\cite{maskawa:76,pauli:85a,pauli:85b}.  For fermions, we have used
parity as an additional input.

For $d$ = 3+1 analogous arguments apply because the solution of the
Klein-Gordon equation is still of the form (\ref{KG_SOL_CHAR}) with just
additional integrations over the transverse dimensions
\cite{neville:71a,rohrlich:71}.

\section{First-Order Systems}

In this section we will encounter yet another method of Hamiltonian
reduction, i.e.~of identifying and quantizing the physical degrees of
freedom. It will serve us to construct the light-cone Fock space. We
will restrict our attention to the bosonic case.  According to
(\ref{SCAL_ACTION}), the Lagrangian of a scalar field in 1+1 dimensions,
when written in terms of light-cone coordinates, takes the following
form,
\begin{equation}
  \label{LC_PHI_LAG}
  L [\phi] = \sfrac{1}{2}\int  dx^\m \left( \sfrac{1}{2}\partial^+ \phi
  \, \partial^- \phi  - \sfrac{1}{2} m^2 \phi^2 - \mathcal{V}[\phi]
  \right)   \; .
\end{equation}
The important thing to note is the fact that the Lagrangian is {\em
  linear} in the light-cone velocity $\partial^- \phi = 2\partial
\phi/\partial x^\p$. This means that the appropriate method for
quantization is the one of Faddeev and Jackiw for first order systems
\cite{faddeev:88} called ``(Constrained) Quantization Without Tears'' by
Jackiw \cite{jackiw:93}. It avoids many of the technicalities of the
Dirac-Bergmann formalism and is in general more economic. It reduces
phase-space right from the beginning as there are no `primary
constraints' introduced. The method is essentially equivalent to
Schwinger's action principle, especially in the form presented by
Schwinger in \cite{schwinger:53b}.

The approach becomes particularly transparent for the case at hand if we
again discretize the system in momentum space
\cite{maskawa:76,pauli:85a,pauli:85b} by enclosing it in a finite
spatial volume, $-L \le x^\m \le L$. We proceed by Fourier expanding the
classical free field as in (\ref{SOL_IBVP}),
\begin{equation}
  \label{FOCK_EXP1}
  \phi(x^\p , x^\m) = a_0 + \sum_{n \ne 0} g_n
  e^{-i (k_n^\p x^\m / 2  \, + \, \hat k_n^\m x^\p / 2)} \; .
\end{equation}
Note that we have allowed for a zero-momentum mode $a_0$ in the
expansion (\ref{FOCK_EXP1}). In what follows we will show that it
actually vanishes. If we define $a_n \equiv (4 \pi |n|)^{-1/2}$, separate
positive and negative modes in (\ref{FOCK_EXP1}), and use the fact that
$a_{-n} = a_n^*$ (as $\phi$ is real), the summation can be restricted to
run over only positive modes. At light-cone time $x^\p = 0$ the field
$\phi$ thus becomes
\begin{equation}
  \phi(x^\p = 0 , x^\m) = a_0 + \sum_{n > 0} \frac{1}{\sqrt{4\pi n}} \Big(a_n
  e^{-i n \pi x^\m / L } + a_n^* e^{i n \pi x^\m / L }\Big) \; , 
\end{equation}
Plugging this into the free part of the Lagrangian we get (discarding a
total time derivative) 
\begin{equation}
  \label{SCAL_LAG}
  L_0 [a_n ,a_0 ] = -i \sum_{n>0} a_n \dot a_n^*  - m^2 L a_0^2 -\sum_{n>0}
  \frac{m^2 L}{4 \pi n}a_n^* a_n \equiv -i \sum_{n>0} a_n \dot a_n^* - H
  \; ,
\end{equation}
with $H$ denoting the Hamiltonian. The Euler-Lagrange equations are
\begin{eqnarray}
  -i \dot a_n + \frac{m^2 L}{4\pi n}a_n &=& 0 \; ,\label{EL_SCAL1} \\ 
  2 m^2 L a_0 &=& 0 \; ,  \label{EL_SCAL2}
\end{eqnarray}
where we have defined $\dot a_n = \partial a_n / \partial x^\p$. The
first equation, (\ref{EL_SCAL1}), is just the free Klein-Gordon equation
which can be seen upon multiplying by $k_n^\p$. The second
identity, (\ref{EL_SCAL2}), is non-dynamical and thus a {\em constraint}
which states that a free field does not have a zero mode, $a_0 = 0$. 

The method of Faddeev and Jackiw is based on equivalence of the
Euler-Lagrange and Hamiltonian equations of motion. The latter are
\begin{equation}
  \label{HAM_SCAL1}
  \dot a_n = \pb{a_n}{H} =  \sum_{k>0} \frac{m^2 L}{4 \pi k}
  \pb{a_n}{a_k^*} a_k  \; , 
\end{equation}
Obviously, this coincides with (\ref{EL_SCAL1}), if the canonical bracket
is
\begin{eqnarray}
  \label{PB_SCAL}
  \pb{a_k}{a_n^*} = -i \delta_{kn} \; .
\end{eqnarray}
The constraint (\ref{EL_SCAL2}) is obtained by differentiating the
Hamiltonian, 
\begin{equation}
  \label{HAM_SCAL2}
  \pad{H}{a_0} =  2m^2 L a_0 = 0 \; . 
\end{equation}
Let us briefly show that the approach presented above is equivalent to
Schwinger's \cite{schwinger:53b}. From (\ref{SCAL_LAG}) we read off a
generator
\begin{equation}
  \delta G = -i \sum_{n>0} a_n \delta a_n^*
\end{equation}
effecting the transformation
\begin{equation}
  \delta a_n^* = \pb{a_n^*}{\delta G} = -i \sum_{k>0} \pb{a_n^*}{a_k}
  \delta a_k^* \; ,
\end{equation}
which in turn implies the canonical bracket (\ref{PB_SCAL}). 

Quantization is performed as usual by employing the correspondence
principle, so that, from (\ref{PB_SCAL}), the elementary commutator
is given by
\begin{eqnarray}
  [a_m , a_n^\dagger] = \delta_{mn} \; .
\end{eqnarray}
The Fock space expansion for the (free) scalar field $\phi$ thus becomes
\begin{eqnarray}
  \label{FOCK_EXP2}
  \phi(x^\p = 0 , x^\m) = \sum_{n > 0} \frac{1}{\sqrt{4\pi n}} \Big(a_n
  e^{-i n \pi x^\m / L } + a_n^\dagger e^{i n \pi x^\m / L }\Big) \; , 
\end{eqnarray}
Note that the Fock `measure' $1/\sqrt{4\pi n}$ does not involve any
scale like the mass $m$ or the volume $L$. This is at variance with the
analogous expansion in the instant form which reads
\begin{equation}
  \label{IF_FOCK}
  \phi(x, t=0) = \frac{1}{\sqrt{2L}} \sum_n \frac{1}{\sqrt{2 (k_n^2
  + m^2)}}   \left( a_n e^{i k_n x} + a_n^\dagger e^{-i k_n x} \right) \; ,
\end{equation}
where $-L \le x \le L$, $k_n = \pi n /L$, and $[a_n , a_m^\dagger] =
\delta_{mn}$. Obviously, $\phi$ \emph{does} depend on $m$ and $L$. We
will discuss some consequences of this difference in the next chapter.

We can use the result (\ref{FOCK_EXP2}) to calculate the free field
commutator at equal light-cone time $x^\p$,
\begin{eqnarray}
  \label{COMM_SCAL}
  [\phi(x) , \phi(0)]_{x^\p = 0} = \sum_{n \ne 0}\frac{1}{4 \pi n} e^{-i n \pi x^\m
  / L} = -\frac{i}{2} \left[ \frac{1}{2}\sgn (x^\m) - \frac{x^-}{2L} \right] \; . 
\end{eqnarray}
which coincides with (\ref{FUND_COMM}), (\ref{LC_COMM}) and
(\ref{CIVP_COMM}).  The commutator (\ref{COMM_SCAL}) has originally been
obtained in \cite{maskawa:76} using the Dirac-Bergmann algorithm for
constrained systems. The Faddeev-Jackiw method, however, is much more
economic and transparent. In particular, it makes clear that the basic
canonical variables of a light-cone field theory are the Fock operators
or their classical counterparts. The $a_n$ with, say, $-N \le n \le N$
in (\ref{FOCK_EXP1}) can be viewed as defining a $(2N + 1)$-dimensional
phase space. A phase space, however, should have even dimension. This is
accomplished by choosing a polarization in terms of positions and
momenta, here $a_n$ and $a_n^\dagger$, with $n > 0$, and by the
vanishing of the zero mode, $a_0 = 0$. We will later see that this
vanishing is a peculiarity of the free theory.

\chapter{The Light-Cone Vacuum}

\section{Basics}

One of the basic axioms of quantum field theory states that the spectrum
of the four-momentum operator is contained within the closure of the
forward light-cone (in momentum space) \cite{streater:64, bogolubov:75}.
The four-momentum $P^\mu$ of any physical, that is, observable particle
thus obeys
\begin{equation}
  \label{SPECTRUM_COND}
  P^2 \ge 0 \; , \quad P^0 \ge 0 \; ,
\end{equation}
which is, of course, consistent with the mass-shell constraint, $p^2 =
m^2$. In other words, there are no tachyons having imaginary mass ($m^2 <
0$). The tip of the cone, the point $P^2 = P^0 = 0$, corresponds to the
vacuum. From the spectrum condition (\ref{SPECTRUM_COND}) we infer that
\begin{equation}
  P_0^2 - P_3^2 \ge P_\perp^2 \ge 0 \quad \mathrm{or} \quad P^0 \ge |P^3| \; . 
\end{equation}
This implies for the longitudinal light-cone momentum,
\begin{equation}
  P^\p = P^0 + P^3 \ge |P^3| + P^3 \ge 0 \; .
\end{equation}
We thus have the important kinematical constraint that physical states
must have non-negative longitudinal momentum,
\begin{equation}
  \bra \mathrm{phys} | P^\p | \mathrm{phys} \ket \ge 0 \; . 
\end{equation}
The spectrum of $P^\p$ is thus bounded from below. Due to Lorentz
invariance, the vacuum $| 0 \ket$ must have vanishing four-momentum, and
in particular
\begin{equation}
  P^\p | 0 \ket = 0 \; .
\end{equation}
Therefore, the vacuum is an eigenstate of $P^\p$ with the lowest
possible eigenvalue, namely zero. We will be interested in the
phenomenon of spontaneous symmetry breaking, i.e.~in the question
whether---roughly speaking---the vacuum is degenerate. We are thus going
to analyze whether there is another state, $| p^\p = 0 \ket$, having the
same eigenvalue, $p^\p = 0$, as the vacuum. If so, it must be possible
to create this state from the vacuum with some operator $U$,
\begin{equation}
  | p^\p = 0 \ket = U | 0 \ket \; , 
\end{equation}
where $U$ must not produce any longitudinal momentum. Note that within
ordinary quantization such a construction is straightforward and quite
common, for example, in BCS theory. A state with vanishing
three-momentum can be obtained via
\begin{equation}
  \label{BCS}
  | \vc{p} = 0 \ket = \int d^3 k f(\vc{k}) a^\dagger (\vc{k}) a^\dagger
    (-\vc{k}) | 0 \ket  \; ,
\end{equation}
where $f$ is an arbitrary wave function. Evidently, the contributions
from modes with positive and negative momenta cancel each other. It is
obvious as well, that within light-cone quantization things must be
different as there cannot be an analogous cancelation for the
longitudinal momenta which are always non-negative. Instead, one could
imagine something like
\begin{equation}
  \label{LC_BCS}
  |p^\p = \vc{p}_\perp = 0 \ket = \int_0^\infty dk^\p \! \int d^2
   k_\perp \, 
   f(\vc{k}_\perp) \delta(k^\p) a^\dagger (k^\p , \vc{k}_\perp)
   a^\dagger(k^\p, - \vc{k}_\perp) | 0 \ket \; .
\end{equation}
The problem thus boils down to the question whether there are Fock
operators carrying light-cone momentum $k^\p = 0$. As we have seen in
the preceding section there are no such operators, and a construction
like  (\ref{LC_BCS}) is impossible. 

The only remaining possibility is that, if $U$ contains a creation
operator $a^\dagger (k^\p > 0)$ carrying longitudinal momentum $k^\p \ne
0$, there must be annihilators that annihilate exactly the same amount
$k^\p$ of momentum. Thus, after Wick ordering, $U$ must have the general
form
\begin{eqnarray}
  \label{U}
  U = \bra 0 | U | 0 \ket &+& \int\limits_{k^\p > 0} dk^\p \, f_2 (k^\p)
  a^\dagger (k^\p) a(k^\p)  \nn \\
  &+& \int\limits_{p^\p > 0} dp^\p \int\limits_{k^\p > 0} dk^\p \, 
  f_3 (k^\p , p^\p )
  a^\dagger (p^\p + k^\p) a(p^\p) a(k^\p) \nn \\
  &+& \int\limits_{p^\p > 0} dp^\p \int\limits_{k^\p > 0} dk^\p \, 
  \tilde f_3 (k^\p , p^\p )
  a^\dagger (p^\p ) a^\dagger (k^\p) a(k^\p + p^\p) \nn \\
  &+&  \ldots \; .  
\end{eqnarray}
It follows that the light-cone vacuum $|0\ket$ is an eigenstate of $U$,
\begin{equation}
  U | 0 \ket = \bra 0 | U | 0 \ket | 0 \ket \; .
\end{equation}
As we only deal with rays in Hilbert space, the action of $U$ on the
vacuum does not create a state distinct from the vacuum. This result is
actually rather general. Any quantity that is obtained by integrating
some functional of the fields over longitudinal space, i.e.,
\begin{equation}
  F[\phi] = \int dx^\m \mathcal{F}[\phi] \; ,
\end{equation}
is of the form (\ref{U}), because the integration can be viewed as a
projection onto the longitudinal momentum sector $k^\p = 0$. The most
important examples for such quantities are the Poincar\'e generators, as
is obvious from the representations (\ref{FF_P}, \ref{FF_M}).  This
implies in particular that the trivial light-cone vacuum is an
eigenstate of the fully interacting light-cone Hamiltonian $P^\m$,
\begin{equation}
  P^\m | 0 \ket \equiv 2 H | 0 \ket = \bra 0 | P^\m | 0 \ket | 0 \ket \; . 
\end{equation}
This can be seen alternatively by considering
\begin{equation}
  P^\p P^\m |0 \ket = P^\m P^\p | 0 \ket = 0 \; ,
\end{equation}
which says that $P^\m | 0 \ket$ is a state with $k^\p = 0$, so that
$P^\m$ must have a Fock representation like $U$ in (\ref{U}).

The actual value of $\bra 0 | P^\m | 0 \ket$ is not important at this
point as it only defines the zero of light-cone energy.  Note that
within the instant form the Fock or trivial vacuum is {\em not} an
eigenstate of the full Hamiltonian as the latter usually contains terms
with only creation operators where positive and negative three-momenta
compensate to zero as in (\ref{BCS}). The instant-form vacuum thus is
unstable under time evolution.  Such a vacuum, a typical example of which
is provided by (\ref{BCS}), is called `nontrivial'.

In the next section we perform an explicit comparison of  nontrivial
instant-form vacua and their trivial light-cone counterparts.

\section{Trivial versus Nontrivial Vacua}

Let us discuss some simple field theoretic models which are exactly
solvable so that one can explicitly determine the ground state and the
particle spectrum built on top of it.

\subsection{Constant Source}

Consider first a massive scalar field $\phi$ in $d = 1+1$ coupled to a
constant external source $\rho$. We solve this model field theory both
in the instant and front form. Using coordinates interpolating between
the two forms the same model has been studied by Hornbostel
\cite{hornbostel:92}.  We work in a finite spatial volume to circumvent
any possible infrared problems. The Lagrangian density is
\begin{equation}
  \label{SOURCE_LAG}
  \mathcal{L} = \sfrac{1}{2}\partial_\mu \phi \partial^\mu \phi -
  \sfrac{1}{2} m^2 \phi^2 + \rho \phi  = \mathcal{L}_0 + \rho \phi \; .
\end{equation}
Note that the external source explicitly breaks the reflection symmetry
$\phi \to - \phi$ of the free Lagrangian $\mathcal{L}$.  Upon completing
the square, the Lagrangian (\ref{SOURCE_LAG}) becomes
\begin{equation}
  \label{SOURCE_LAG2} 
  \mathcal{L} = \sfrac{1}{2}\partial_\mu \phi \partial^\mu \phi -
  \sfrac{1}{2} m^2 (\phi - \rho/m^2)^2 + \rho^2 /2 m^2 \; .
\end{equation}
The minimum of the (classical) potential is thus located at $\phi = \rho
/m^2 \equiv \phi_0$, which, in the quantum theory, corresponds to a
non-vanishing vacuum expectation value,
\begin{equation}
  \bra \Omega |\phi | \Omega \ket = \phi_0 \; .
\end{equation}
The state $|\Omega \ket$ denotes a `non-trivial' vacuum in the following
sense. If we expand the field $\phi$ in Fock space like in (\ref{IF_FOCK}),
\begin{equation}
  \phi(x, t=0) = \frac{1}{\sqrt{2L}} \sum_n \frac{1}{\sqrt{2 \omega_n}}
  \left( A_n e^{i k_n x} + A_n^\dagger e^{-i k_n x} \right) \; ,
\end{equation}
where $-L \le x \le L$, $k_n = \pi n/ L$,  $\omega_n^2 = k_n^2 +
m^2$ and 
\begin{equation}
  \label{FOCK_COMM}
  [A_n , A_m^\dagger] = \delta_{mn} \; ,
\end{equation}
we must have 
\begin{equation}
  A_n | \Omega \ket \ne 0 
\end{equation}
for at least one $n$. The vacuum state $|\Omega \ket$ must therefore
contain particles corresponding to the $A$-modes of $\phi$. If we denote
the `trivial' vacuum, i.e.~the one annihilated by the $A_n$, as usual by
$| 0 \ket$, the two vacua must be related as in (\ref{BCS}) via
\begin{equation}
  \label{VAC_REL}
  | \Omega \ket = U[A_n , A_n^\dagger] \,  | 0 \ket \; .
\end{equation}
Let us determine the explicit form of $U$.  To this end we note that the
system under consideration is a field theoretic version of the shifted
harmonic oscillator. If we define a translated field
\begin{equation}
  \varphi \equiv \phi - \phi_0 \; ,
\end{equation}
this does not have an expectation value in $|\Omega \ket$, as
\begin{equation}
  \bra \Omega | \varphi | \Omega \ket = \bra \Omega | \phi | \Omega \ket
  - \phi_0 = 0 \; .
\end{equation}
Thus, from the Fock expansion of  $\varphi$ 
\begin{equation}
  \varphi(x, t=0) = \frac{1}{\sqrt{2L}} \sum_n \frac{1}{\sqrt{2 \omega_n}}
  \left( a_n e^{i k_n x} + a_n^\dagger e^{-i k_n x} \right) \; ,
\end{equation}
we infer that the $a_n$ annihilate the non-trivial vacuum $|\Omega \ket$
and thus correspond to quasi-particle operators built on the ground
state $|\Omega \ket$. Put differently, $\varphi$ can be viewed as a free
field with vacuum $|\Omega \ket$, as is also obvious from the
Klein-Gordon equation,
\begin{equation}
  (\Box + m^2)\phi = (\Box + m^2) \varphi + \rho = \rho \; ,
\end{equation}
which implies that $\varphi$ obeys a \emph{free} equation of motion.
The relation between the $A$- and $a$-modes is easily found. As $\phi$
and $\varphi$ are related by a translation, we must have
\begin{equation}
  \label{USHIFT}
  \phi = U^\dagger \varphi U \; ,
\end{equation}
where $U$ is the translation operator
\begin{equation}
  \label{U_PI}
  U  = \exp (-i \rho \Pi / m^2 ) = \exp (- i \phi_0 \Pi ) \; . 
\end{equation}
Accordingly, the operator $U$ from (\ref{VAC_REL}) is determined in
terms of the momentum $\Pi$ conjugate to $\phi$, which is the integral
over the momentum density 
\begin{equation}
  \pi (x, t=0) = \frac{-i}{\sqrt{2L}} \sum_n \sqrt{ \omega_n / 2}
  \left( A_n e^{i k_n x} - A_n^\dagger e^{-i k_n x} \right) \; .
\end{equation}
The momentum $\Pi$ is thus essentially the zero mode of the density
$\pi$, 
\begin{equation}
  \Pi (t=0) = \intl dx \, \pi(x, t=0) = i \sqrt{mL}\; (A_0^\dagger - A_0)
  \; .
\end{equation}
Putting everything together we obtain the explicit representation of $U$
in terms of Fock operators,
\begin{equation}
  \label{U_DET}
  U [A_n , A_n^\dagger] = \exp[\sqrt{mL} \, \phi_0 \, (A_0^\dagger - A_0)] \; .
\end{equation}
Note that this quantity is entirely determined by the zero momentum
modes $A_0^\dagger$ and $A_0$. Plugging (\ref{U_DET}) into
(\ref{VAC_REL}), one finds that $|\Omega \ket$ is a coherent state of zero
momentum modes,
\begin{equation}
  |\Omega \ket = \exp \big(- mL \phi_0^2 / 2\big) \exp \Big(\sqrt{mL} \phi_0
   A_0^\dagger\Big) \; | 0 \ket \; ,
\end{equation}
where we have used the Baker-Campbell-Hausdorff formula to factorize
$U$. Being a coherent state, $| \Omega \ket$ must be an eigenstate of
the annihilators $A_n$. This is confirmed by a simple calculation
yielding
\begin{equation}
  A_{n\ne 0} |\Omega \ket = 0 \; , \quad A_0 |\Omega \ket = \sqrt{mL} \phi_0
  \; |\Omega \ket \; .
\end{equation}
The non-vanishing vacuum expectation value thus arises from the action
of the zero mode operators,
\begin{equation}
  \label{IF_VEV}
  \bra \Omega | \phi | \Omega \ket = \frac{1}{2L} \intl dx \, \bra \Omega
  | \phi | \Omega \ket =  \frac{1}{2\sqrt{mL}} \bra
  \Omega | A_0 + A_0^\dagger | \Omega \ket = \phi_0 \; .
\end{equation}
The quasi-particle operators are obtained by the unitary transformation
(\ref{USHIFT}),
\begin{eqnarray}
  \label{FOCK_SHIFT}
  a_n &=& U A_n U^\dagger = A_n \; , \qquad \qquad (n \ne 0)  \nn \\
  a_0 &=& U A_0 U^\dagger = A_0 - \sqrt{mL} \phi_0 \; ,  
\end{eqnarray}
As might have been expected, only the zero modes get shifted through the
action of $U$, namely by the $c$-number $- \sqrt{mL} \, \phi_0$. This
transformation is a particular example of a canonical (or Bogolubov)
transformation. It conserves the commutation relations
(\ref{FOCK_COMM}), i.e.
\begin{equation}
  [a_n , a_m^\dagger] = \delta_{mn} \; ,
\end{equation}
and diagonalizes the Hamiltonian, as is shown by the following
calculation,   
\begin{eqnarray}
  H &=& \frac{1}{2} \intl dx \, \left[ \pi^2 + \phi \bigg(
  - \frac{\partial^2}{\partial x^2} + m^2 \bigg) \phi \right] - \intl dx \,
  \rho \phi \nn \\
  &=& \sum_n \omega_n A_n^\dagger A_n - \frac{\rho}{m}\sqrt{mL} (A_0 +
  A_0^\dagger) \nn \\
  &=& \sum_n \omega_n a_n^\dagger a_n - (2L) \frac{\rho^2}{2m^2}  \; .
\end{eqnarray}
Obviously, the Hamiltonian becomes diagonal in the $a$-modes. In
addition, we  find a vacuum energy 
\begin{equation}
  E_{\mathrm{vac}} \equiv \bra \Omega | H | \Omega \ket = -
  (2L) \frac{\rho^2}{2m^2}  \; , 
\end{equation}
which is consistent with the shift in energy density in
(\ref{SOURCE_LAG2}). We thus have solved the problem entirely: the
Hamiltonian is diagonal and its ground state as well as the excited
states are determined. The entire effect of adding a constant source is
a rearrangement of the vacuum state accompanied by a lowering of the
ground state energy. The spectrum of excited states which is determined
by the mass $m$ (i.e.~the curvature of the oscillator potential) remains
unchanged.

Let us see now whether we can do the same within light-cone
quantization. Again, we enclose the system in a finite spatial box, $-L
\le x^\m \le L$ and demand periodic boundary conditions in $x^\m$. We
perform the same decomposition of the fields,
\begin{equation}
  \label{LC_DECOMP}
  \phi = \varphi + \phi_0 \; ,
\end{equation}
where $\phi_0$ represents the vacuum expectation value, and look for the
translation operator $U$ relating $\phi$ and $\varphi$. The momentum
(density) conjugate to $\phi$ is
\begin{equation}
  \pi  \equiv \partial^\p \phi \; .
\end{equation}
We know from the discussion of the initial-value problem that $\pi$ does
not provide any new information: knowledge of $\phi$ implies knowledge
of $\pi$ as $\partial^\p$ is a spatial derivative. Its zero mode
is therefore determined through the spatial boundary conditions
according to
\begin{equation}
  \label{PI=0}
  \Pi = \intl dx^\m \, \pi(x^\m) = 2 \intl dx^\m \, \pad{}{x^\m}
  \phi(x^\m) = 2 \big[ \phi (x^\m = L) - \phi (x^\m = -L) \big] = 0 \; ,
\end{equation}
and we note that it is vanishing. This has far-reaching
consequences. First of all we note that $U$ is the identity,
\begin{equation}
  U = \exp(-i \phi_0 \Pi) = \Eins \; ,
\end{equation}
so that the vacuum does not change under action of $U$, 
\begin{equation}
  |\Omega \ket = | 0 \ket \; ,
\end{equation}
nor do the Fock operators. How can this observation be reconciled with
the decomposition (\ref{LC_DECOMP}), i.e.~the existence of a
non-vanishing vacuum expectation value $\phi_0$ ? To this end we
consider the Klein-Gordon equation for $\phi$,
\begin{equation}
  (\Box + m^2)\phi = (\partial^\p \partial^\m + m^2) \phi = \rho \; ,
\end{equation}
where the external source $\rho$ provides an inhomogeneous term.  Upon
integrating this over $x^\m$ the derivative term vanishes as in
(\ref{PI=0}), and we are left with
\begin{equation}
  \frac{1}{2L} \intl dx^\m \, \phi (x^\m) = \rho / m^2 \equiv \phi_0 \; .
\end{equation}
which is nothing but a straightforward generalization of
(\ref{EL_SCAL2}).  The zero mode of $\phi$ thus coincides with the
vacuum expectation value $\phi_0$. It is a $c$-number, entirely
determined by the source (or `interaction') $\rho$.  This is at variance
with ordinary quantization where we have
\begin{equation}
  \frac{1}{2L} \intl dx \,  \phi (x) = \frac{1}{2\sqrt{mL}}(A_0 +
  A_0^\dagger) \ne \phi_0 \; .
\end{equation}
Here, the zero mode is {\em operator valued} and yields the vacuum
expectation value only after acting on the non-trivial vacuum
$|\Omega\ket$ as is evident from (\ref{IF_VEV}).

The decomposition (\ref{LC_DECOMP}), when written in Fock space, becomes
\begin{equation}
  \phi (x^\m) = \sum_{n > 0} \frac{1}{4\pi n}\Big( a_n e^{i k_n^\p x^\m / 2}
  +  a_n^\dagger e^{-i k_n^\p x^\m / 2} \Big) + \phi_0 \; . 
\end{equation}
Obviously, the Fock operators in $\phi$ and $\varphi$ are the same, and
the zero mode of $\varphi$, which is a free field, vanishes in
accordance with (\ref{EL_SCAL2}).  As a result we conclude that, in
order to determine the vacuum expectation value of $\phi$, we have to
calculate its zero mode which is constrained by the Euler-Lagrange
equation of motion. This will be discussed exhaustively in Chapter~7.

\subsection{Mass Shift: Bosons}

Let us consider two free scalar fields, $\phi_0$ and $\phi_1$ with
different masses, $m_0$ and $m_1$ \cite{stern:94}. We stay in $d=1+1$
but go over to a continuum formulation. The Heisenberg fields
obey the Klein-Gordon equations 
\begin{equation}
  (\Box + m_i^2) \, \phi_i = 0  \; , \quad i = 0, 1 \; , 
\end{equation}
and have the instant-form Fock expansions 
\begin{equation}
  \label{DIFFMASS_FOCK_BOS}
  \phi (x, t) = \int \frac{dk}{2\pi} \frac{1}{\sqrt{2\omega_i (k)}} \Big[
  a_i (k) \, e^{-i k_i \cdot x} + a_i^\dagger (k) \, e^{i k_i \cdot x} \Big] \; ,
\end{equation}
with $\omega_i (k) \equiv (k^2 + m_i^2)^{1/2}$, $k_i \cdot x \equiv
\omega_i (k) t - kx$, and the commutation relations
\begin{equation}
[a_i (k), a_i^\dagger (p)] = 2 \pi \delta (k-p) \; . 
\end{equation}
It is important to note that the instant-form fields even at $t=0$
(i.e.~the Schr\"odinger fields) depend on the masses $m_i$ both
explicitly via $\omega_i$ and implicitly via the Fock operators $a_i$
and $a_i^\dagger$. Accordingly, also the (Fock) vacua defined through
\begin{equation}
  \label{MASS_VAC}
  a_i (k) | \Omega_i \ket = 0 \; ,
\end{equation}
are mass dependent. This can be seen explicitly in the functional
Schr\"odinger picture \cite{jackiw:87} in which the states are
functionals $\Psi [\varphi] \equiv \bra \varphi | \Psi \ket$ of the
classical field $\varphi$, and the momentum operator $\pi(x)$ acts as
the functional derivative $-i \delta / \delta \varphi (x)$. In this very
intuitive formulation of quantum field theory, the free field vacua
(\ref{MASS_VAC}) are represented by Gaussian functionals
\cite{jackiw:87}, 
\begin{eqnarray}
  \bra \varphi | \Omega_i \ket &=& \mathrm{det}^{1/4}(\omega_i / \pi) \,
  \exp
  \left[ - \sfrac{1}{2} \int dx dy \, \varphi (x) \, \omega_i (x - y)
  \, \varphi (y) \right] \nn  \\
  &=& \mathrm{det}^{1/4}(\omega_i / \pi) \, \exp
  \left[ - \sfrac{1}{2} \int \frac{dk}{2\pi} \, \varphi (-k) \, \omega_i (k)
  \, \varphi (k) \right] \; ,
\end{eqnarray}
where the mass dependence resides in the `covariances' $\omega_i$. If we
rewrite the Klein-Gordon equation for $\phi_1$ as
\begin{equation}
   (\Box + m_0^2) \phi_1 = - (m_1^2 - m_0^2) \phi_1 \equiv \delta m^2
   \phi_1 \; ,
\end{equation}
we can equivalently view $\phi_1$ as an interacting field with the
interaction given by the Hamiltonian
\begin{equation}
  H_\mathrm{int} \equiv \sfrac{1}{2} \int dx \, \delta m^2 \phi^2 \; .
\end{equation}
The field $\phi_1$ in the interaction picture, and the Heisenberg field
$\phi_0$ thus satisfy the same free field equation of motion. This
implies that the associated fields in the Schr\"odinger picture are the
same. If we choose $t=0$ as the time where the different dynamical
pictures coincide, we have the common initial conditions,
\begin{equation}
  \label{COMM_INIT_IF}
  \phi_0 (x , t=0) = \phi_1 (x , t=0) \; ,
\end{equation}
and analogously for the momenta $\pi_i$.  This relates the Fock
operators according to
\begin{equation}
  \label{BOG_1}
  a_1 (k) = \alpha_k \, a_0 (k) + \beta_k \, a_0^\dagger (-k) \; , 
\end{equation}
with the coefficients $\alpha_k$ and $\beta_k$ given by
\begin{eqnarray}
  \alpha_k &=& \sfrac{1}{2} \left( \sqrt{\omega_1 (k)/ \omega_0 (k)} + \sqrt{\omega_0
  (k)  / \omega_1 (k)} \right) \; ,  \\
  \beta_k &=& \sfrac{1}{2} \left( \sqrt{\omega_1 (k) / \omega_0 (k)} - \sqrt{\omega_0
   (k)/ \omega_1 (k)} \right) \; . 
\end{eqnarray}
As in the preceding subsection, we can interpret the $a_1 (k)$ as the
annihilators of quasi- or dressed particles which by some (unspecified)
mechanism acquire an effective mass $m_1 \ne m_0$.  Using the fact that
$\alpha_k^2  - \beta_k^2 = 1$, one can define an angle $\theta_k$
such that
\begin{equation}
  \cosh \theta_k \equiv \alpha_k \; , \quad  \sinh \theta_k =
  \beta_k \; .
\end{equation}
Therefore, the transformation (\ref{BOG_1}) can be rewritten as 
\begin{equation}
  a_1 (k) =  a_0 (k) \cosh \theta_k + a_0^\dagger (-k) \sinh \theta_k \; , 
\end{equation}
which is the standard Bogolubov transformation for bosons. As this
transformation is unitary, it can equivalently be written with the help
of a unitary operator $U$,
\begin{equation}
  a_1 (k) = U  a_0 (k) U^\dagger  \; , 
\end{equation}
given by 
\begin{equation}
  U \equiv U \big(\theta_k \big) = \exp  \int
  \frac{dk}{4 \pi}  \, \theta_k \left[ a_0 (k) a_0 (-k)  - a_0^\dagger (k)
  a_0^\dagger (-k) \right] \; .
\end{equation}
The vacuum $|\Omega_1 \ket$ is thus found to be
\begin{equation}
  | \Omega_1 \ket = U | \Omega_0 \ket = \exp  \int
  \frac{dk}{4 \pi}  \, \theta_k \left[ a_0 (k) a_0 (-k)  - a_0^\dagger (k)
  a_0^\dagger (-k) \right] | \Omega_0 \ket \; . 
\end{equation}
In quantum optics, such a state is called a `squeezed' state
\cite{vogel:94}. It corresponds to a harmonic oscillator where the
parabola representing the quasi-particle potential is `squeezed'
compared to the parabola of the particles ($a_0$-modes). It is exactly
this `squeezing' which leads to a different curvature at the minimum
thus implying a shift in the mass, $m_0 \to m$. From the point of view
of the particles, the vacuum $| \Omega_1 \ket$ is non-trivial, as $a_0 |
\Omega_1 \ket \ne 0$.

Again, within light-cone quantization, the situation is drastically
different. We have seen that the initial field at $x^\p = 0$ does not
involve any scale, in particular no mass scale. As a consequence, both
the light-cone Fock operators and the light-cone vacuum are mass
independent \cite{leutwyler:70,schlieder:72,rohrlich:72}. Heuristically,
one can argue that the front form is equivalent to a formulation in the
`infinite-momentum frame' \cite{chang:69b} which is obtained from the
instant form through the limit of a Lorentz boost with velocity $v \to
c$ \cite{susskind:68}. In this frame, all masses can be neglected
compared to the longitudinal momentum, and thus should only play a minor
role \cite{rohrlich:72}.

Technically, for the case at hand, the initial condition analogous to
(\ref{COMM_INIT_IF}),
\begin{equation}
  \phi_1 (x^\m, x^\p = 0) = \phi_0 (x^\m, x^\p = 0) \; , 
\end{equation}
implies that the $a_0$- and $a_1$-modes coincide so that $U$ becomes the
identity. Once again we thus find that the light-cone vacuum is not
changed by the interaction, but remains trivial.

\subsection{Mass Shift: Fermions}

Consider finally the same situation for fermions, however, now in $d=3+1$
\cite{heinzl:88}.  Our starting point are the two Dirac equations
\begin{equation}
  (i \partial \!\!\!/ - m_i) \, \psi_i = 0 \; , \quad i = 0, 1 \; .
\end{equation}
We expand the Heisenberg fields as
\begin{equation}
  \label{DIFFMASS_FOCK_FER}
  \psi_i (x) = \sum_\lambda \int \frac{d^3 k}{(2 \pi)^{3/2}}
  \frac{1}{\sqrt{2 \omega_i}} \left[ b_i (\vc{k}, \lambda) u_i (\vc{k},
  \lambda ) e^{-i k_i \cdot x} + d_i^\dagger (\vc{k} , \lambda) v_i
  (\vc{k} , \lambda) e^{i k_i \cdot x} \right] \; ,
\end{equation}
with $\omega_i$ and $k_i$ as in (\ref{DIFFMASS_FOCK_BOS}). The
elementary spinors $u$ and $v$ are those of Itzykson and Zuber
\cite{itzykson:80}, multiplied by a factor $(2m_i)^{1/2}$ to allow for a
straightforward massless limit. The Fock operators obey the canonical
anti-commutation relations,
\begin{equation}
  \left\{ b_i (\vc{k}, \lambda) , b_i^\dagger (\vc{p}, \lambda) \right\}
  = \left\{ d_i (\vc{k}, \lambda) , d_i^\dagger (\vc{p}, \lambda)
  \right\} = \delta_{\lambda \lambda^\prime} \, \delta^3 (\vc{k} - \vc{p})
  \; .
\end{equation}
As before, the Fock operators with subscripts $i = 0$ and $i=1$
(corresponding to bare and dressed particles, respectively) are
connected by a Bogolubov  transformation,
\begin{equation}
  b_1 (\vc{k} , \lambda) = U b_0 (\vc{k} , \lambda) U^\dagger \; , \quad  
  d_1 (\vc{k} , \lambda) = U d_0 (\vc{k} , \lambda) U^\dagger \; , 
\end{equation}
which conserves the canonical anti-commutation relations.  Like in the
bosonic case, $U$ is found from the common initial condition,
\begin{equation}
  \psi_0(x , t=0) = \psi_1 (x , t=0) \; .
\end{equation}
A somewhat lengthy calculation involving a number of spinor matrix
elements finally yields \cite{heinzl:88}
\begin{eqnarray}
  b_1 (\vc{k} , \lambda) &=& \alpha_k \, b_0 (\vc{k} , \lambda) +
  \beta_k \, d_0^\dagger
  (-\vc{k} , \lambda) \; ,  \label{BOG_FERM1} \\
  d_1 (\vc{k} , \lambda) &=& \alpha_k \, d_0 (\vc{k} , \lambda) -
  \beta_k \, b_0^\dagger
  (- \vc{k} , \lambda) \; ,\label{BOG_FERM2}
\end{eqnarray}
with  coefficients
\begin{eqnarray}
  \alpha_k &=& \sqrt{\sfrac{1}{2}(1 + \gamma_k)} \; , \\
  \beta_k  &=& \sqrt{\sfrac{1}{2}(1 - \gamma_k)} \; , \\
  \gamma_k &=& \frac{m_0 \, m_1 + k^2}{\omega_0 \, \omega_1}  \label{GAMMA_K} \; .
\end{eqnarray}
Obviously, we have $\alpha_k^2 + \beta_k^2 = 1$, so that this time there
is an angle $\theta_k$ allowing to write
\begin{eqnarray}
  b_1 (\vc{k} , \lambda) &=&  b_0 (\vc{k} , \lambda) \cos \theta_k +  d_0^\dagger 
  (- \vc{k} , \lambda) \sin \theta_k \; ,   \\
  d_1 (\vc{k} , \lambda) &=&  d_0 (\vc{k} , \lambda) \cos \theta_k  - b_0^\dagger
  (- \vc{k} , \lambda) \sin \theta_k \; .
\end{eqnarray}
The unitary operator $U$ is found to be
\begin{equation}
  U = \exp \left\{ - \sum_{\lambda}  \int d^3 k \,
  \theta_k  \left[
  b_0 (- \vc{k} , \lambda) d_0 (\vc{k} , \lambda) + b_0^\dagger (\vc{k} ,
  \lambda) d_0^\dagger   (- \vc{k} , \lambda) \right] \right\} \; ,
\end{equation}
defining the nontrivial vacuum $|\Omega_1 \ket = U | \Omega_0 \ket $,
with $b_0|\Omega_1 \ket$, $d_0|\Omega_1 \ket \ne 0$. 

Upon initializing the fields on the null-plane $x^\p = 0$, however,
things again become very different. The fermion fields $\psi_i$ are
decomposed into $\psi_i = \psi_{i, +} + \psi_{i , -}$, where only the 
$\psi_{i, +}$  are independent fields (see Section~4.3.5 and
App.~A). These obey a Klein-Gordon equation,
\begin{equation}
  i \partial^\m \psi_{i, +} = (i \partial^\p)^{-1}(- \partial_\perp^2 +
  m_i^2) \, \psi_{i, +} \; , \quad i = 0, 1 \; ,
\end{equation}
which is obtained after eliminating the $\psi_{i , -}$. The
(independent) Heisenberg fields have the Fock expansion
\begin{equation}
  \psi_{i , +}(x) = \sum_\lambda
  \int_0^{\infty}\frac{dk^\p}{ k^\p}\int \frac{d^2k_{\perp}}{16 \pi^3} \,
  \left[b_i(\vcg{k},\lambda) u_{i,+}(\vcg{k}, \lambda) e^{- i k_i \cdot x} +
  d_i^{\dagger} (\vcg{k},\lambda) v_{i,+}(\vcg{k}, \lambda) e^{i k_i \cdot x}
  \right]  
\end{equation}
where $\vcg{k} = (k^\p , \vc{k}_\perp)$ and 
\begin{equation}
  k_i \cdot x = \sfrac{1}{2}   \, \hat {k}_i^\m x^\p + \sfrac{1}{2} \, k^\p x^\m
  - \vc{k}_\perp \cdot  \vc{x}_\perp
  \equiv \sfrac{1}{2} \, \hat {k}_i^\m x^\p + \vcg{k} \cdot \vcg{x}
\end{equation}
with the $\hat {k}_i^\m$ being on shell, $\hat {k}_i^\m = (k_\perp^2 +
m_i^2)/k^\p$ .  Again, it is crucial to note that both the Fock
operators and the spinors $u_{i,+}$ and $v_{i,+}$ are mass independent
(see App.~A). As a result, the common initial condition,
\begin{equation}
  \psi_{0,+} (\vcg{x}, x^\p =0) = \psi_{1,+} (\vcg{x}, x^\p =0) \; ,
\end{equation}
leads to a trivial Bogolubov transformation with $U = \Eins$
\cite{heinzl:88,dietmaier:89}, like in the bosonic case . The light-cone
vacuum, therefore, is stable under mass generation and the same for both
fields $\psi_{i,+}$, namely the trivial Fock vacuum. This is
corroborated by taking the infinite-momentum limit of the coefficient
$\gamma_k$,
\begin{equation}
  \lim_{|\svc{k}| \to \infty} \gamma_k = 1 \; ,
\end{equation}
which also makes the Bogolubov transformation (\ref{BOG_FERM1},
\ref{BOG_FERM2}) trivial. In Chapters~8 and~9 we will discuss how one
can obtain a fermion condensate in the trivial light-cone vacuum (see
also \cite{dietmaier:89}).

\chapter{Light-Cone Wave Functions}

In this chapter we collect some basic facts about the eigenvalue problem
of the light-cone Hamiltonian, or, in other words, about the light-cone
Schr\"odinger equation and its solutions, the light-cone wave functions.
We follow the discussion of Brodsky and Lepage \cite{brodsky:89} and
also use their conventions. Accordingly, this chapter contains only few
original results, but we think it worthwhile to include to make our
presentation more self-contained.

\section{Definition}

Let us first stick to a discrete notation and, for the time being, stay
in 1+1 dimensions. We thus have a Fock basis of states
\begin{eqnarray}
  && | n \ket = a_n^\dagger | 0 \ket \; , \nn \\
  && | m , n \ket = a_m^\dagger a_n^\dagger |0 \ket \; , \nn \\
  && \vdots \nn \\
  && | n_1 , \ldots , n_N \ket = a_{n_1}^\dagger \ldots a_{n_N}^\dagger | 0 \ket
  \; . \label{FOCK_BASIS}
\end{eqnarray}
This leads to a completeness relation defining the unit operator in Fock
space, 
\begin{eqnarray}
  \Eins &=& | 0 \ket \bra 0 | + \sum_{n > 0} | n \ket \bra n | +
  \sfrac{1}{2} \sum_{m,n > 0} | m,n \ket \bra m,n | + \ldots \nn \\ 
  &=&  | 0 \ket \bra 0 | + \sum_{N = 1}^\infty \frac{1}{N!} \sum_{n_1,
  \ldots , n_N >0} |n_1 , \ldots , n_N \ket \bra n_1 , \ldots , n_N | \; .
\end{eqnarray}
An arbitrary state $|\psi \ket$ can thus be expanded as
\begin{equation}
  |\psi  \ket = \sum_{n>0} \bra n | \psi \ket | n \ket +
   \sfrac{1}{2} \sum_{m,n > 0} \bra m,n | \psi \ket | m,n \ket + \ldots
   \; .
\end{equation}
The sums are such that the longitudinal momenta in each Fock sector add
up to the total longitudinal momentum of $|\psi \ket$.  Note that the
vacuum state does not contribute as it is orthogonal to any particle
state, $\bra 0 | \psi \ket = 0$. The normalization of this state is
obtained as
\begin{eqnarray}
  \bra \psi | \psi \ket &=& \sum_{n>0} |\bra n | \psi \ket|^2 +
  \sfrac{1}{2} \sum_{m ,n >0} |\bra m ,n | \psi \ket|^2 + \ldots \nn \\
  &=& \sum_{N=1}^\infty \frac{1}{N!} \sum_{n_1 , \ldots , n_N > 0} |\bra
  n_1 , \ldots , n_N | \psi \ket|^2 \; .
\end{eqnarray}
Let us assume that the state $|\psi \ket$ corresponds to a bound-state
obeying the light-cone Schr\"odinger equation (\ref{IF_SEQ}),
\begin{equation}
  (M^2 - \hat M_0^2) | \psi \ket = \hat W | \psi \ket \; .
\end{equation}
We want to project this equation onto the different Fock sectors. For
this we need the eigenvalues of the free invariant mass squared when
applied to an $N$-particle state
\begin{equation}
  | N \ket \equiv | n_1 , \ldots n_N \ket \; .
\end{equation}
We find that $\hat M_0^2$ is diagonal as expected,
\begin{equation}
  \bra N |\hat M_0^2 |N^\prime \ket = \bra N |P^\p \sum_{i=1}^N
  \frac{m^2}{k_{n_i}^+} | N^\prime \ket \equiv \sum_{i=1}^N
  \frac{m^2}{x_i} \bra N | N^\prime \ket \equiv M_{N}^2 \bra N |
  N^\prime \ket \; .
\end{equation}
The light-cone Schr\"odinger equation thus becomes a system of coupled
eigenvalue equations,
\begin{equation}
  \label{GEN_LCBSE} 
  \left \lceil  \begin{array}{cl}
                  (M^2 - M_1^2) & \bra l | \psi \ket \\
                  (M^2 - M_2^2) & \bra k l | \psi \ket \\
                  \vdots & 
               \end{array} \right \rceil = 
  \left \lceil \begin{array}{r@{}c@{}l r@{}c@{}l c}
                \bra l | & W & | m \ket & \bra l | & W & | m n \ket &
                \ldots  \\
                \bra k l |& W & | m \ket & \bra k l |& W & | m 
                n \ket & \ldots  \\
                & \vdots & & & \vdots & & \ddots 
              \end{array} \right \rceil 
  \left \lceil \begin{array}{c}
                \bra l | \psi \ket \\
                \bra k l | \psi \ket \\
                \vdots    
             \end{array} \right \rceil \; . 
\end{equation}
Clearly, this represents an infinite number of equations which in
general will prove impossible to solve unless the matrix is very sparse
and/or the matrix elements are small. The former condition is usually
fulfilled as the interaction $W$ in the light-cone Hamiltonian typically
changes particle number at most by two\footnote{Note that terms with
  only creation operators are forbidden by $k^\p$-conservation.  Still,
  in 1+1 dimensions, things can become messy as interactions with
  polynomials of arbitrary powers in $\phi$ are allowed
  \cite{zinn-justin:96}}. Assuming the matrix elements to be small
amounts to dealing with a perturbative situation. This will be true for
non-relativistic bound states of heavy constituents, but not for light
hadrons which we are mainly interested in. We shall, however, encounter
situations where the magnitude of the amplitudes decreases enormously
with the particle number $N$, so that it is a good approximation to
restrict to the lowest Fock sectors. In instant form field theory this
has long been known as the Tamm-Dancoff method
\cite{tamm:45,dancoff:50}.

Let us turn to the more realistic case of 3+1 dimensions in a continuum
formulation.  In the discussion of the Klein-Gordon equation we have
seen that the solutions are of the form
\begin{equation}
  \phi (x) = \int \frac{d^4 p}{2 \pi^4} \, 2\pi \, \delta(p^2 - m^2) \theta
  (p^\p) e^{-i p \cdot x} \chi (p) 
  = \int_0^\infty \frac{dp^\p}{16 \pi^3 p^\p} \int d^2 p_\perp \, e^{- i
  \hat p \cdot x} \chi (\hat p)\; ,   
\end{equation}
where $\hat p$ denotes the on-shell four-momentum with $\hat p^2 =
m^2$. From this expression we read off that the invariant normalization
of a momentum eigenstate $|P^\p, \vc{P}_\perp \ket \equiv | \vcg{P}
\ket $ is given by
\begin{equation}
  \label{MOM_NORM}
  \bra \vcg{P} | \vcg{K} \ket = 16 \pi^3 P^\p
  \delta^3 (\vcg{P} - \vcg{K})  \; .
\end{equation}
We already know that the bare Fock vacuum is an eigenstate of the
interacting Hamiltonian. It thus serves as an appropriate ground state
on top of which we can build a reasonable Fock expansion. If we
specialize immediately to the case of QCD, we are left with the Fock
basis states
\begin{eqnarray}
  && |0 \ket \; , \nn \\
  && | q \bar q : \vcg{k}_i , \alpha_i \ket = b^\dagger
  (\vcg{k}_1 , \alpha_1) d^\dagger (\vcg{k}_2 , \alpha_2)
  |0 \ket \; , \\
  && | q \bar q g : \vcg{k}_i , \alpha_i \ket = b^\dagger
  (\vcg{k}_1 , \alpha_1) d^\dagger (\vcg{k}_2 , \alpha_2)
  a^\dagger (\vcg{k}_3 , \alpha_3) |0 \ket \; , \nn \\
  && \vdots
\end{eqnarray}
In these expressions, $b^\dagger$, $d^\dagger$ and $a^\dagger$ create
quarks $q$, antiquarks $\bar q$ and gluons $g$ with momenta $\vcg{k}_i$
from the trivial vacuum $|0 \ket$. The $\alpha_i$ denote all other
relevant quantum numbers, like helicity, flavor and color.

In a more condensed notation we can thus describe, say, a pion with
momentum $\vcg{P} = (P^\p, \vc{P}_\perp )$, as
\begin{equation}
  \label{GEN_PI_WF}
  | \pi (\vcg{P}) \ket = \sum_{n , \lambda_i} \int \overline{\prod_i}
    \frac{dx_i}{\sqrt{x_i}} \frac{d^2 k_{\perp i}}{16\pi^3} \psi_{n/\pi} 
    (x_i, \vc{k}_{\perp i}, \lambda_i ) \Big| n: x_i
    P^\p , x_i \vc{P}_\perp + \vc{k}_{\perp i} ,
    \lambda_i \Big \rangle  \; ,
\end{equation}
where we have suppressed all discrete quantum numbers apart from the
helicities $\lambda_i$. The integration measure takes care of the
constraints (\ref{RELMOM_SUM}) which the relative momenta in each
Fock state (labeled by $n$) have to obey,
\begin{eqnarray}
  \overline{\prod_i} dx_i &\equiv& \prod_i dx_i \, \delta \bigg(1 - \sum_j
  x_j \bigg)  \; , \\ \relax
  \overline{\prod_i} d^2 k_{\perp i} &\equiv& 16 \pi^3 \prod_i   
  d^2k_{\perp i} \, \delta^2 \bigg(\sum_j \vc{k}_{\perp j} 
  \bigg) \; . 
\end{eqnarray}
As a mnemonic rule, we note that any measure factor $d^2 k_{\perp i}$ is
always accompanied by $1 / 16\pi^3$.  

The most important quantities in (\ref{GEN_PI_WF}) are the
\emph{light-cone wave functions}
\begin{equation}
  \psi_{n/\pi} (x_i, \vc{k}_{\perp i}) \equiv \bra  n: x_i P^\p , x_i
  \vc{P}_\perp +  \vc{k}_{\perp i} , \lambda_i |\pi (\vcg{P}) \ket
  \; , 
\end{equation}
which are the amplitudes to find $n$ constituents with relative momenta
$p_i^\p = x_i P^\p$, $\vc{p}_{\perp i} = x_i \vc{P}_\perp +
\vc{k}_{\perp i}$ and helicities $\lambda_i$ in the pion. Due to the
separation properties of the light-cone Hamiltonian the wave functions
do not depend on the total momentum $\vcg{P}$ of the pion. Applying
(\ref{MOM_NORM}) to the pion state (\ref{GEN_PI_WF}), we obtain the
normalization condition
\begin{equation}
  \label{LCWF_NORM}
  \sum_{n , \lambda_i} \int \overline{\prod_i}  dx_i  \frac{d^2 k_{\perp
  i}}{16\pi^3}   |\psi_{n/\pi} (x_i,
  \vc{k}_{\perp i} , \lambda_i) |^2 = 1 \; .
\end{equation}
The light-cone bound-state equation for the pion is a straightforward
generalization of (\ref{GEN_LCBSE}),  
\begin{equation}
  \label{PI_LCBSE} 
  \left \lceil  \begin{array}{cl}
                  (M^2 - M_{q \bar q}^2) & \bra q \bar q| \pi \ket  \\
                  (M^2 - M_{q \bar q g}^2) & \bra q \bar q g | \pi \ket \\
                  \vdots & 
               \end{array} \right \rceil = 
  \left \lceil \begin{array}{r@{}c@{}l r@{}c@{}l c}
                \bra q \bar q | & W & | q \bar q \ket & \bra q \bar q |
                & W & | q \bar q g \ket &
                \ldots  \\
                \bra q \bar q g |& W & | q \bar q \ket & \bra q \bar q g
                | & W & | q \bar q g \ket & \ldots  \\
                & \vdots & & & \vdots & & \ddots 
              \end{array} \right \rceil 
  \left \lceil \begin{array}{c}
                \bra q \bar q | \pi \ket \\
                \bra q \bar q g | \pi \ket \\
                \vdots    
             \end{array} \right \rceil \; . 
\end{equation}
If a constituent picture for the pion were true, its wave function would
be entirely given by the projection $\bra q \bar q| \pi \ket$ onto the
valence state. All the higher Fock contributions would vanish and the
unitarity sum (\ref{LCWF_NORM}) would simply reduce to 
\begin{equation}
  \label{VAL_NORM}
  \sum_{\lambda \lambda^\prime}  \int_0^1 dx \int \frac{d^2
  k_\perp}{16\pi^3}   |\psi_{q \bar q/\pi} (x,
  \vc{k}_{\perp }, \lambda , \lambda^\prime ) |^2 = 1 \; .
\end{equation}
We will later discuss a model where this is indeed a good approximation
to reality.

\section{Properties}

Let us rewrite the light-cone bound-state equation (\ref{GEN_LCBSE}) by
collecting all light-cone wave functions $\psi_n = \bra n | \psi \ket$
into a  vector $\Psi$, 
\begin{equation}
  \label{PSI}
  \Psi  = \frac{W \Psi}{M^2 - M_0^2} \; .
\end{equation}
From this expression it is obvious that all light-cone wave functions
tend to vanish whenever the denominator
\begin{equation}
  \epsilon \equiv M^2 - M_0^2 = M^2 - \Big( \sum_i p_i \Big)^2 =  
  M^2 - \sum_i \frac{k_{\perp i}^2 +  m_i^2}{x_i} 
\end{equation}
becomes very large. We know from Section 3.3 that this quantity measures
how far off energy shell the total system, i.e.~the bound-state is, 
\begin{equation}
  P^\m - \sum_{i}p_i^\m = \epsilon/P^\p \; .
\end{equation}
For this reason, $\epsilon$ is sometimes called the `off-shellness'
\cite{namyslowski:85,lavelle:87}.  We thus learn from (\ref{PSI}) that
there is only a small overlap of the bound-state with Fock states that
are far off shell. This implies the limiting behavior
\begin{equation}
  \label{WF_BC}
  \psi(x_i , \vc{k}_{\perp i} , \lambda_i) \to 0 \quad \mathrm{for} \;
  x_i \to 0 \; , k_{\perp i}^2 \to \infty \; .
\end{equation}
These boundary conditions are related to the self-adjointness of the
light-cone Hamiltonian and to the finiteness of its matrix elements.
Analogous criteria have been used recently to relate wave functions for
different Fock states $n$ (via ``ladder relations'',
\cite{antonuccio:97a}) and to analyze the divergence structure of
light-cone perturbation theory \cite{burkardt:98}.

Suppressing spin, flavor and color degrees of freedom, a light-cone
wave function will be a scalar function $\phi(x_i , \vc{k}_{\perp i})$ of
the parameter $\epsilon$. This is used for building models, the most
common one being to assume a Gaussian behavior
\cite{terentev:76,dziembowski:88,jaus:90,ji:92}, 
\begin{equation}
  \phi(x_i , \vc{k}_{\perp i}) = N \exp(-|\epsilon|/\beta^2) \; ,
\end{equation}
where $\beta$ measures the size of the wave function in momentum space.
Note, however, that a Gaussian ansatz is in conflict with perturbation
theory which is the appropriate tool to study the high-$\vc{k}_\perp$
behavior and indicates a power decay of the {\em renormalized} wave
function (up to possible logarithms, \cite{brodsky:89}). For the
unrenormalized wave functions the boundary conditions (\ref{WF_BC}) are
violated unless one uses a cutoff as a regulator.

As the off-shellness $\epsilon$ is the most important quantity
characterizing a light-cone wave function let us have a closer look by
specializing to the simplest possible system, namely two bound particles
of equal mass. One can think of this, for instance,  as the valence
wave function of the pion. The off-shellness becomes
\begin{equation}
  \label{2P_OFFS}
  \epsilon = M^2 - \frac{k_\perp^2 + m^2}{x (1-x)} 
  = - \frac{1}{x(1-x)} \Big[ M^2 \big(x - \sfrac{1}{2}\big)^2 
  + \underbrace{\frac{4m^2 -
  M^2}{4}}_{\ge 0} + k_\perp^2 \Big] \; .
\end{equation}
The second term on the right-hand-side is positive because, for a bound
state, the binding energy,
\begin{equation}
  \label{EB}
  E = M - 2m \; , 
\end{equation}
is negative so that $2m > M$. As a result, the off-shellness is always
negative. Only for free particles it is zero, because {\em all} momentum
components (including the energy) sum up to the total momentum. In this
case, each individual term in (\ref{2P_OFFS}) vanishes,
\begin{equation}
  M = 2m \; , \quad x = \sfrac{1}{2} \; , \quad \vc{k}_\perp = 0 \; .
\end{equation}
It follows that the light-cone wave function of a free two-particle system
(with equal-mass constituents) is of the form
\begin{equation}
  \phi (x , \vc{k}_\perp ) \sim \delta (x - 1/2) \, \delta^2 (\vc{k}_\perp)
  \; .
\end{equation}
One expects that for weak binding, in particular in the non-relativistic
case, the wave functions will not look too differently, and thus will be
highly peaked around $ x= 1/2$ (in the equal mass case) and
$\vc{k}_\perp = 0$. Let us check this explicitly. We go to the particle
rest frame with $\vc{P} = 0$ or $P^\p = P^\m = M$ and $\vc{P}_\perp =
0$, implying $\vc{p}_{\perp i} = \vc{k}_{\perp i}$. In this frame, the
non-relativistic limit is defined by the following inequalities for the
constituent masses and momenta (in ordinary instant-form coordinates),
\begin{equation}
  \label{HIERARCHY}
  p_i^0 - m_i \simeq \frac{\vc{p}_i^2}{2 m_i}   \ll |\vc{p}_i| 
  \ll m_i \; .
\end{equation}
The prototype systems in this class are of hydrogen type where we have
for binding energy and r.m.s.~momentum,
\begin{eqnarray}
  |E| &=& \bra \frac{\vc{p}^2}{2m} \ket = \frac{m\alpha^2}{2} \; , \\
  \bra p \ket &=& m \alpha \; ,
\end{eqnarray}
with $\alpha = 1/137$ the fine structure constant and $m$ the reduced
mass. In this case, the hierarchy (\ref{HIERARCHY})  becomes
\begin{equation}
  \frac{\alpha^2}{2} \ll \alpha \ll 1 \; ,
\end{equation}
which is fulfilled to a very good extent in view of the smallness of
$\alpha$. 

Consider now the longitudinal momentum of the $i^{th}$ constituent,
\begin{equation}
  p_i^\p = p_i^0 + p_i^3 \simeq m_i + \frac{\vc{p}_i^2}{2m_i} + p_i^3 
  = x_i P^\p =   x_i M \; .
\end{equation}
We thus find that we should replace $p_i^3$ in instant-form
non-relativistic wave functions by
\begin{equation}
  \label{REPLACE}
  p_i^3 = x_i M - m_i \; ,
\end{equation}
where we neglect terms of order $\vc{p}_i^2 / m_i$.  Let us analyze the
consequences for the off-shellness. The latter is in ordinary
coordinates
\begin{equation}
  \label{IF_OFFS}
  \epsilon = M^2 - M_0^2 = (M + \sum_i p_i^0)(M - \sum_i p_i^0) \; .
\end{equation}
We thus need 
\begin{equation}
  \sum_i p_i^0 \simeq \sum_i m_i + \sum_i \frac{\vc{p}^2}{2m_i} = (M -
  E) + \sum_i \frac{\vc{p}^2}{2m_i} \; ,
\end{equation}
with $E = M - \sum m_i$, so that the off-shellness (\ref{IF_OFFS})
becomes
\begin{equation}
  \epsilon \simeq 2M \bigg(E - \sum_i \vc{p}_i^2 / 2m_i \bigg) \simeq 2M \left[ E -
  \sum_i \frac{k_{\perp i}^2 + (Mx_i - m_i)^2}{2m_i} \right] \; ,
\end{equation}
where we have performed the replacement (\ref{REPLACE}) in the second
identity. The light-cone wave functions will be peaked where the
off-shellness is small, that is, for
\begin{equation}
  x_i = m_i / M \, , \quad \mathrm{and} \quad \vc{k}_{\perp i} = 0 \; ,
\end{equation}
as expected from the non-interacting case. 

As an explicit example we consider the ground state wave function of
positronium, which in ordinary momentum space is given by the Fourier
transform of $\psi(r) = N \exp(-m \alpha r)$. Performing (\ref{REPLACE})
once more and normalizing to one, we obtain
\begin{equation}
  \label{PSI_POS}
  \psi (x , \vc{k}_\perp) = \sqrt{\gamma^3 / \pi} \frac{8\pi
  \gamma}{\Big[k_\perp^2 + (xM - m_e)^2 + \gamma^2 \Big]^2} \; , 
\end{equation}
where $m_e$ is the electron mass and $\gamma = m \alpha$ the
r.m.s.~momentum. The result (\ref{PSI_POS}) is valid for small momenta,
i.e.~when $k_\perp^2, (xM - m_e)^2 \ll m_e^2$. It is obvious from
(\ref{PSI_POS}) that the positronium wave function is sharply peaked
around $x = m_e /M \simeq 1/2$ and $k_\perp^2 = 0$.



 

\newpage
\pagestyle{empty}
\cleardoublepage

\vspace*{7cm}

\begin{center}
   
  \Huge \textbf{Part II}\\[1cm]
  \textbf{Applications}

\end{center}

\addcontentsline{toc}{part}{II Applications}

\cleardoublepage

\pagestyle{headings}

\chapter{Zero Modes and the Light-Cone Vacuum}


Our first application will not (yet) deal with light-cone wave functions
but rather with the issue of Chapter~5, the triviality of the light-cone
vacuum. We will, however, be concerned with a non-trivial example of an
interacting field theory, namely $\phi^4$-theory in 1+1 dimensions. As
before we will address the question how a non-vanishing vacuum
expectation value of the field $\phi$ can arise in the light-cone
vacuum. As a further motivation we note that the dynamical system under
consideration may be viewed as a low-dimensional caricature of the Higgs
sector of the standard model which is also given by a $\phi^4$-type
scalar field theory. 

Historically, it was Weinberg in 1966, who realized that within
``old-fashioned" Hamiltonian perturbation theory the infinite-momentum
limit of many diagrams, in particular vacuum diagrams, is vanishing
\cite{weinberg:66a}. This limit was later on shown to be (at least
perturbatively) equivalent to light-cone quantization
\cite{susskind:68,chang:69b}. Within the context of current algebra
\cite{fubini:65,dashen:66,alfaro:73} the success of this limit was
traced to the fact that light-like charges always annihilate the vacuum,
irrespective of whether they are conserved or not \cite{leutwyler:70}.
Thus, Coleman's theorem \cite{coleman:66}, ``the symmetry of the vacuum
is the symmetry of the world", does not apply \cite{heinzl:96b}.  As we
have seen in Chapter~5, the Fock vacuum, i.e.~the ground state of the
free Hamiltonian, is stable under interaction.  In the same manner as
the light-like charges, the fully interacting Hamiltonian annihilates
the vacuum, the formal reason being that both quantities conserve
longitudinal momentum and thus are of the form (\ref{U}).

With the advent of QCD as the theory of strong interactions, however,
people began to feel somewhat uneasy.  There was (and is) growing
evidence, that many of the phenomenological aspects of hadron physics,
like confinement and chiral symmetry breaking are related to the
non-trivial features of the QCD vacuum (within standard instant-form
quantization). Let us only mention features like quark- and gluon
condensates, instantons, monopole condensation etc., which all indicate
that the vacuum is densely populated by non-trivial quantum
fluctuations, which furthermore are not accessible to perturbation
theory.

The concern that arose at these points can be put into the question: can
the existence of these large vacuum fluctuations be reconciled with the
triviality of the light-front vacuum?  For QCD, the answer to this
question is not (yet) known. For model field theories, some answers have
already been found. Not surprisingly, it is the zero modes that are to
some extent the carriers of non-trivial vacuum properties. One generally
distinguishes between constrained and unconstrained zero modes. The
latter typically arise in (low-dimensional) gauge theories
\cite{heinzl:91b,heinzl:92b,kalloniatis:93,kalloniatis:94a,kalloniatis:94c,
  mccartor:94a,mccartor:97a}. The former we have already met in the
discussion of the constant source (Subsection~5.2.1). For
$\phi^4$-theory in 1+1 dimensions, it has been shown that the zero modes
are responsible for spontaneous symmetry breaking of the reflection
symmetry $\phi \to -\phi$ \cite{heinzl:92c,robertson:93,bender:93,
  pinsky:94,pinsky:95}. Let us review the basic features of what is
going on \cite{heinzl:96a}.

\section{The Constrained Zero Mode}

We recall that the Lagrangian of a scalar field theory in 1+1 dimensions, 
written in terms of light-cone coordinates, cf.~(\ref{LC_PHI_LAG}),
\begin{equation}
  \label{LFLag}
  L[\phi,  \partial^\m  \phi]  = \sfrac{1}{2} \int  dx^{\m}
  \sfrac{1}{2}   \partial^{\p}  \phi \,
  \partial^\m \phi - U [\phi] \; , 
\end{equation}
is linear in the velocity $\partial^\m \phi$. According to the
discussion of Chapter~4, the elementary bracket between the fields is
essentially given by the inverse of the spatial derivative, $\partial^\p
= 2 \partial/\partial x^\m $.  In order to uniquely define this inverse,
one has to specify the domain of $\partial^\m$ and boundary conditions.
As before, we enclose our spatial variable $x^{\m}$ in a box, $-L \le
x^{\m} \le L$, and impose periodic boundary conditions on our fields. We
know that in this case the operator $\partial^{\p}$ has zero modes,
namely all spatially constant functions. To deal with this aspect, we
split our field $\phi$ into a zero mode $\omega$ and its complement
$\varphi$,
\begin{eqnarray}
  \label{OMVI}
  \phi (x^{\p} , x^{\m}) &=& \omega (x^{\p}) + \varphi  (x^{\p} , x^{\m} ) \;
  , \\ 
  \omega (x^{\p})  &\equiv&  \frac{1}{2L}  \intl dx^{\m} \phi (x^{\p} ,
  x^{\m}) \; , 
\end{eqnarray}
such that $\varphi$ does \emph{not} have a zero mode,
\begin{equation}
  \intl dx^{\m} \varphi (x^{\p} , x^{\m} ) = 0 \; .
\end{equation}
The  Lagrangian   (\ref{LFLag})   can  then be  rewritten  as
\begin{equation}
  L[\varphi,   \omega]   =  \intl   dx^{\m}   \sfrac{1}{2}   \varphi
  (-2 \partial_{\m}) \dot \varphi - H [\varphi, \omega] \; .
\end{equation}
For brevity, we use the derivative $\partial_{\m} = \partial / \partial
x^\m$ and $\dot \varphi \equiv \partial \varphi / \partial x^\p$. Note
that $\omega$ does not appear in the kinetic term.  The Hamiltonian $H
\equiv P^\m$ coincides with the potential $U$ from (\ref{LFLag}) after
the replacement (\ref{OMVI}) has been performed.  We thus find that the
zero mode $\omega$ is constrained by its `equation of motion',
\begin{equation}
  \label{ZMC}
  \theta \equiv \frac{\delta H}{\delta \omega} = 0 \; . 
\end{equation}
The basic bracket can be read off from the kinetic term in the
Lagrangian (\`{a} la Schwinger or Faddeev/Jackiw, cf.~Chapter~4),
\begin{equation}
  \label{COMMPHI}
  \{\varphi  (x^{\p} , x^{\m} ) , \varphi (x^{\p} , y^{\m} ) \} \equiv  -
  \frac{1}{2}\langle  x^\m \vert  \partial_{\m}^{-1}  \vert  y^{\m}
  \rangle  = - \frac{1}{2} \left[ \frac{1}{2} \sgn (x^{\m} - y^{\m} )  -
  \frac{x^{\m} - y^{\m}}{2L} \right]\; , 
\end{equation}
which, of course, coincides with (\ref{LC_COMM}) upon quantization via
the correspondence principle.

As is well known, not all classical observables can be quantized
unambiguously due to possible operator ordering problems
\cite{weyl:27,moyal:49,hillery:84,grosche:93}.  Such problems do not
arise for the field $\varphi$ and the bracket (\ref{COMMPHI}), where the
field-independent right-hand-side leads to a $c$-number commutator.  The
constraint (\ref{ZMC}), however, implies a functional dependence of the
zero mode $\omega$ on $\varphi$ and thus a non-vanishing commutator of
$\om$ with $\vi$.  This can be explicitly verified by calculating the
associated Dirac bracket within the Dirac-Bergmann algorithm
\cite{heinzl:91a,heinzl:91c,heinzl:92a}.  For the quantum theory, it
results in an ordering ambiguity with respect to $\om$ and $\vi$
\cite{heinzl:92c,pinsky:94}.  Therefore, a definite ordering has to be
prescribed.  We chose Weyl (or symmetric) ordering \cite{weyl:27,lee:81}
which is explicitly hermitian.  Using this prescription, one finds the
quantum Hamiltonian for light-cone $\phi^4_{1+1}$-theory,
\begin{eqnarray} 
  \label{HAM}
  H   =   \intl   dx^\m \!\!\! &\Bigg(& \!\!\!\! \frac{1}{2} m^2 \vi^2 +
  \frac{\lambda}{4!} \vi^4\Bigg) + \nonumber \\
  +   \intl    dx^\m  \!\!\!\! &\Bigg[& \!\!\!\! \frac{1}{2}    m^2   \om^2    +
  \frac{\lambda}{4!}  \Big( \om^4  + \om\vi^3  + \vi \om \vi^2  +
  \vi^2 \om \vi + \vi^3 \om + \nonumber \\
  && + \, \om^2 \vi^2 + \vi^2 \om^2 + \om \vi \om \vi + \vi \om \vi \om 
  + \om \vi^2 \om + \vi \om^2 \vi \Big) \Bigg] \; . 
\end{eqnarray}
Note that we have chosen the sign of the mass term(s) in such a way that
there is no spontaneous symmetry breaking at tree level as the classical
potential does not have any turning points.  With this Hamiltonian,
equation (\ref{ZMC}) for the constrained zero mode $\omega$ reads
explicitly
\begin{equation} 
  \label{TH1} 
  \theta = \frac{\delta H}{\delta \om}   =    m^2    \om   +   
  \frac{\lambda}{3!}\om^3    +
  \frac{\lambda}{3!}  \frac{1}{2L} \intl dx^\m \left [\vi^3 + \vi^2
  \om + \vi \om \vi + \om \vi^2 \right] = 0 \; .  
\end{equation}
This is nothing but the zero mode of the Euler-Lagrange equation of motion for
the total field $\phi$ decomposed into $\om$ and $\vi$
\cite{maskawa:76,wittman:89,pinsky:94}.  The remainder of this chapter
is concerned with different approaches to solve this equation for
$\omega$.

\section{Perturbative Solution}

To obtain  a perturbative  solution  for $\om$ we expand  it in a
power series in $\lambda$,
\begin{equation}  
  \om  \equiv \sum_{n=0}^{\infty} \la^{n}\om_{n}
  \label{SERIES_PERT}
\end{equation}
Inserting this into (\ref{TH1}) determines the coefficients $\om_n$
recursively. For the first three we find
\begin{eqnarray}
  \!\!\om_0 \!\!&=& \!\! 0 \; , \\
  \!\!\om_{1} \!\!  &=&  \!\! -\frac{1}{6m^2}\frac{1}{2L} \intl  \!  dx^\m   \vi^3(x)\,   ,
  \label{OM1} \\
  \!\!\om_{2}  \!\! &=& \!\! \frac{1}{36m^4}   \frac{1}{(2L)^2}  \intl
  \!\! dx^\m  dy^\m  \Big[ 
  \vi^2(x)\vi^3(y) \! + \! \vi(x)\vi^3(y)\vi(x) \!+ \! \vi^3(y)\vi^2(x)\Big]
\end{eqnarray}
All higher orders may be obtained similarly.  Unfortunately, however, we
have not been able to find a closed formula for $\om_n$ which would
allow for summing up the whole series (\ref{SERIES_PERT}).

If we expand the quantum field $\vi$ in terms of Fock operators,
cf.~(\ref{FOCK_EXP2}), 
\begin{equation}
  \label{FOCK}
  \vi(x) = \sum_{n=1}^{\infty}  \frac{1}{\sqrt{4\pi  n}}\Big [a_{n}
  e^{-ik_{n}^{+} x^{-}/2} + a_{n}^{\dagger} e^{ik_{n}^{+}x^{-}/2} \Big]
  \; ,
\end{equation}
with the discretized longitudinal momentum $k^\p_n = 2\pi n/L$, one
notes that the vacuum expectation value of $\om$ is zero to all orders
in $\la$, since $\om_n$ contains an {\it odd} number of Fock operators,
i.e.~$\bra 0 \vert \om_n \vert 0 \ket = 0$. Thus
\begin{equation}
  \bra 0 \vert \om \vert  0 \ket = \sum_{n=0}^\infty  \la^n  \bra 0
  \vert \om_n \vert 0 \ket = 0 \; . 
\end{equation}
This seems to imply that a non-vanishing vacuum expectation value for
$\om$ can only arise non-perturbatively and must be non-analytic in the
coupling $\la$.  We will discuss this issue in the next sections and
continue for the time being within the framework of perturbation theory.
In particular, we will study the effect of the zero mode $\om$ on the
mass renormalization. To this end, we split up the Hamiltonian $H$ from
(\ref{HAM}) into two pieces,
\begin{equation} 
  H = H_0 + H_\om \; ,
\end{equation}
where $H_0$ is independent of $\om$,
\begin{equation}
  \label{HZERO}
  H_0 = \intl dx^\m \left(  \frac{1}{2}  m^2 \vi^2 + \frac{\la}{4!}
  \vi^4 \right) \; , 
\end{equation}
and $H_\om$ is the $\om$-dependent interaction given by the last two
lines in (\ref{HAM}). $H_\omega$ can be simplified using the constraint
equation (\ref{TH1}),
\begin{eqnarray}
  \label{HOM}
  H_\om &=& H_\om - \frac{L}{2}\left(\om \, \theta + \theta \, \om \right) 
  \nonumber \\
  &=& \frac{\la}{4!} \intl  dx^\m
  \left( -\om^4 + \vi \om \vi^2 + \vi^2 \om \vi 
  + \vi \om^2 \vi - \om \vi^2 \om \right) \,. 
\end{eqnarray}
Because $\om$  is of order  $\la$,  the zero mode dependent  part 
$H_\om$  is of order $\la^2$. Explicitly, we find, neglecting terms of
order $\lambda^3$, 
\begin{eqnarray} 
  \label{CORR}
  H_\om  &=& \frac{\la}{4!}  \intl dx^\m \left ( \vi \om \vi^2 + \vi^2
  \om \vi \right)   \nonumber\\
  &=& - \frac{\la^2}{144m^2} \frac{1}{2L} \intl dx^\m dy^\m \left[ \vi(x) \vi^3(y)
  \vi^2(x) + \vi^2(x) \vi^3(y) \vi(x) \right]   ,
\end{eqnarray}
where we have used the first order term (\ref{OM1}). This induces
a mass shift of order $\la^2$ which is given by \cite{heinzl:92a} (see
also  \cite{robertson:93})  
\begin{equation}
  \label{SHIFT}
  \delta  m_n^2 \equiv \frac{2\pi  n}{L} \Big( \bra n \vert H_\om
  \vert n \ket - \bra 0 \vert H_\om \vert 0 \ket \Big) \; . 
\end{equation}
Here, $\vert n \ket \equiv a_n^\dagger \vert 0 \ket$ denotes a
one-particle state of longitudinal momentum $k^\p_n = 2\pi n/L$. We
have subtracted the constant vacuum energy, $\bra 0 \vert H_\om
\vert 0 \ket$, which diverges linearly with the volume. This is
sufficient to render the mass shift finite.  After inserting the
Fock-expansion (\ref{FOCK}) into (\ref{CORR}) one obtains for
(\ref{SHIFT})
\begin{eqnarray}
  \label{SHIFTRES}
  \delta m_n^2 &=& -\frac{\la^2}{6m^2} \frac{L}{(4\pi)^3} \frac{1}{n}
  \bigg[ \sum_{m=1}^{n-1} \frac{1}{(n-m)m} +
  4 \sum_{m=1}^{\infty} \frac{1} {(n+m)m} \bigg] 
  \frac{2\pi n}{L} \nonumber\\
  &=& -\frac{\la^2}{6m^2} \frac{1}{16 \pi^2 n}  \Big[
  3\gamma + \Psi(n) + 2\Psi(1+n) \Big]   < 0 \;,
\end{eqnarray}
where $\Psi$ denotes the Digamma-function \cite{abramowitz:70}, and
$\gamma = - \Psi (1) = 0.577216\ldots$ is Euler's constant.  The
question now is, whether the result (\ref{SHIFTRES}) is a finite size
effect which vanishes in the infinite-volume limit.  The latter is
obtained by replacing
\begin{equation}
  \sum_{n=1}^N   f(n)   \to   \lim_{L   \to \infty}   \frac{L}{2\pi}
  \int\limits_{2\pi/L}^{2\pi N/L} dk^\p f(k^\p L / 2 \pi ) \;  
\end{equation}
in such a way that $k^\p_n$ approaches a finite limit $k^\p$, i.e.~$n,L
\to \infty$ with $n/L$ finite.  For the case at hand, this amounts to
replacing the Digamma-function by its asymptotics \cite{abramowitz:70}
\begin{equation}
  \label{PSI_ASYMPT}
  \Psi(z) \simeq \mbox{log}(z) - \frac{1}{2z} + O(1/z^2) \; ,
  \quad (z \rightarrow \infty \quad   
  \mbox{in} \quad \vert\mbox{arg}(z) \vert  < \pi)  \,.
\end{equation}
Denoting the infinite-volume limit of the mass shift (\ref{SHIFT}) by
$\delta m^2$, we finally obtain
\begin{equation}
   \delta m^2 = -\frac{\la^2}{m^2}  \frac{1}{16\pi
  k^{\p}}    \lim_{L\to\infty}     \frac{1}{L}    \bigg[
  \mbox{log}(k^{\p} L/2\pi) + \gamma + O(1/k^\p L) \bigg] = 0 \,.
\end{equation}
The vanishing of this expression implies that the zero mode induced
second order mass shift $\delta m_n^2$ is indeed a finite size effect.
We do not have a general proof that this is also true for higher orders
in $\la$.  However, it seems to be plausible that a {\it single} mode
like $\om$ constitutes a ``set of measure zero'' within the infinite
number of modes as long as one applies perturbation theory. From the
theory of condensation, however, it is well known that single modes,
especially those with vanishing momentum, can significantly alter the
perturbative results.  To analyze this possibility one clearly has to
use non-perturbative methods. This will be done in the next sections.

\section{Mean-Field Ansatz}

As $\phi^4_{1+1}$-theory is super-renormalizable, the number of
divergent diagrams is finite, namely one: the tadpole resulting from a
self-contraction of two fields at the same space-time point.  Therefore,
renormalization can, at least within perturbation theory, be done via
normal-ordering, which is nothing but making use of Wick's theorem:
expanding the appearing powers of $\phi$ in a sum of normal-ordered
terms with more and more self-contractions, one separates the finite
term (with no contraction) from the divergent ones (with at least one
contraction).  The latter terms then are just the negative of the
required counter-terms. For conventional $\phi^4$-theory one finds
\begin{eqnarray}
  \label{WICK}
  \frac{1}{2}m^2 \phi^2 + \frac{\la}{4!} \phi^4 &=& 
  \frac{1}{2}m^2 \Big( :\! \phi^2 \!\! : + \, T \Big) + 
  \frac{\la}{4!} \Big( :\! \phi^4 \!\! : +\, 6\, T :\! \phi^2 \!\! : +\, 3\, T^2
  \Big) \nonumber \\
  &=& \frac{1}{2} \left( m^2 + \frac{\la}{2} T\right) :\! \phi^2 \!\! : +\, 
  \frac{\la}{4!} :\! \phi^4 \!\! : +\, \frac{m^2}{2} T + \frac{\la}{8} \, T^2 \; ,
\end{eqnarray}
where  
\begin{equation}
  T   \equiv    \bra   \phi^2    \ket   =   \int    \frac{dk}{4\pi}
  \frac{1}{\sqrt{k^2 + m^2}} 
\label{ET-TADPOLE} 
\end{equation}
formally   denotes   the   logarithmically    divergent   tadpole
contribution  (in  $d$ = 1+1)  which  coincides  with  the vacuum
expectation value  of $\phi^2$ or a self-contraction of the field.

It is obvious  that only mass and vacuum energy get renormalized.
The  renormalized   Hamiltonian   is  obtained   by  adding   the
counter-terms
\begin{equation}
  \delta{\cal H} \equiv - \frac{\la}{4} T : \phi^2 :
 - \frac{1}{2}m^2  T  - \frac{\la}{8}  T^2 \; . 
  \label{COUNTER}
\end{equation}
If the vacuum expectation value of $\phi^2$ is taken in a Fock vacuum
corresponding to the bare  mass $m$, the latter coincides with the 
renormalized mass to order $\lambda$. 

This  rather  trivial   renormalization   procedure   cannot   be
straightforwardly  extended to light-cone field theory, simply because we
do not  know  the  Hamiltonian!   The  zero mode $\om$  is a complicated
functional of the Fock operators $a_n, a_n^\dagger$, which has to
be  found   from  the  constraint   (\ref{TH1})   before   we  can
normal-order. There is, however, a way around this obstacle, such
that  an  {\it  exact}  knowledge  of  $\om$  is not  needed  for
renormalization.

To this end we use the following general ansatz for $\om$,
\begin{eqnarray}
  \label{MFA}
  \om &=& \om_0  + \sum_{n>0}  \om_n  a_{n}^\dagger  a_n +
  \sum_{m,n >0} \om_{mn} a_{m}^\dagger a_{n}^\dagger a_{m+n} +
  \sum_{m,n>0} \om_{mn}^* a_{m+n}^\dagger a_m a_n + \nonumber \\
  &+&  \sum_{l,m,n>0}  \om_{lmn} \, a^\dagger_{l+m+n}  a_l  a_m a_n +
  \sum_{l,m,n>0}  \om_{lmn}^* \, a_l^\dagger  a_m^\dagger a_n^\dagger
  a_{l+m+n} + \nonumber \\
  &+& \sum_{k,l,m,n>0}  \delta_{k+l,  m+n} \,  \om_{klmn} \, a_k^\dagger
  a_l^\dagger a_m a_n + \dots  \; ,
\end{eqnarray}
where $\omega_0$, $\omega_n$, ... are $c$-numbers to be
determined. Obviously, $\omega_0$ is the vacuum expectation value of the
operator $\omega$,
\begin{equation}
  \omega_0 = \bra 0 | \omega | 0 \ket \; ,
\end{equation}
whereas the coefficients $\omega_n$ etc.~contribute to matrix elements
of $\omega$ in higher Fock sectors.  The ansatz (\ref{MFA}) is hermitian
and takes care of the fact that $\om$ cannot transfer any momentum.
Thus, it is a discrete analogue of (\ref{U}) and can be understood as a
Wick expansion written in the opposite of the usual order: the first
term $\om_0$ is the sum of all contractions, the second the sum of all
contractions but one and so on.  Accordingly, each individual term in
the expansion is a normal-ordered operator. Plugging the ansatz
(\ref{MFA}) into $H_\omega$ one notes that the latter may be viewed as
an effective Hamiltonian where additional operators $\varphi^n$, $n \ge
2$ have been induced by the zero mode. This is reminiscent of a
renormalization group analysis in which the same operators, being all
marginal, are generated by the renormalization group flow \cite{amit:84,
  zinn-justin:96}. In what follows, we will, however, truncate the
ansatz (\ref{MFA}) after the second term ($\sim a_n^\dagger a_n$) which
only introduces additional $\varphi^2$ contributions and thus
corresponds to a mean-field treatment.

Inserting the ansatz (\ref{MFA}) into (\ref{HAM}) and (\ref{TH1}) we
obtain for the constraint and the Hamiltonian
\begin{eqnarray}
  \theta  &=&  \theta_0  + \sum_{n>0}  \theta_n  a_n^\dagger  a_n +
  \ldots \; , 
  \label{MFT} \\
  H &=& H_0 + \sum_{n>0} H_n a_n^\dagger a_n + \ldots \; , 
  \label{MFHAM}
\end{eqnarray}
where,   in  accordance   with  the  truncation   of  our  ansatz
(\ref{MFA}),  we have omitted terms containing more than two Fock
operators.    The  coefficients   $H_0$,  $H_n$  and  $\theta_0$,
$\theta_n$   are  functions   of  $\om_0$  and  $\om_n$.    Thus,
(\ref{MFHAM})  is an effective one-body or mean-field  Hamiltonian
describing the influence of the zero mode $\om$.  Explicitly,  one finds
for the coefficients of the constraint,
\begin{eqnarray}
  \label{T0MF} 
  \theta_0   &=&  \left(m^2   +  \frac{\la}{2}   T  \right)  \om_0  +
  \frac{\la}{3!}  \left( \om_0^3 + \sum_{n>0} \frac{\om_n}{4\pi  n}
  \right) \; , \\
  \label{TH2}
  \theta_n  &=&  \left(  m^2  +  \frac{\la}{2}  T \right)  \om_n  +
  \frac{\la}{3!} \left( \om_n^3 + 3 \om_0 \om_n^2 + 3 \om_0^2 \om_n
  + \frac{6\om_0}{4\pi  n}  + \frac{6\om_n}{4\pi  n}  \right)   ,
\end{eqnarray}
and of the Hamiltonian (scaled by 2$L$)
\begin{eqnarray}
  \label{E0MF}
  \frac{H_0}{2L}  &=&  \frac{1}{2}\left(  m^2  +  \frac{\la}{2}  T  \right)
  \om_0^2  +  \frac{\la}{4!}  \left(  \om_0^4  + 4  \sum_{n>0}
  \frac{\om_0 \om_n}{4\pi n} + \sum_{n>0} \frac{\om_n^2}{4\pi  n}
  \right) \nn \\ 
  &+& \frac{m^2}{2} T + \frac{\la}{8} T^2 , \\
  \label{ENERGIES1}
  \frac{H_n}{2L} &=& \frac{1}{2}\left( m^2 + \frac{\la}{2} T \right) \left(
  \om_n^2 + 2\om_0 \om_n + \frac{1}{2\pi  n} \right) + \nonumber \\
  &+&  \frac{\la}{4!}   \left(  (\om_0  +  \om_n)^4   -  \om_0^4  +
  \frac{3}{\pi n} (\om_0 + \om_n)^2 + \frac{\om_n^2}{2\pi  n} \om_n
  \sum_{m>0}  \frac{\om_m}{\pi  m} \right)  \; .  
\end{eqnarray}
$T$  now denotes the light-cone tadpole (in discretized form)
\begin{equation}
  T= \bra \varphi^2 \ket = \sum_{n>0}\frac{1}{4\pi n} \; ,
\end{equation}
which is mass independent in contrast to (\ref{ET-TADPOLE}) (unless one
regularizes with a mass dependent cutoff, cf.~Chapter~9).  Note that the
one-particle matrix elements of $\theta$ and $H$ are given as the {\it
  sum} of two coefficients,
\begin{eqnarray}
  \bra n |\theta | n \ket &=&  \theta_0 + \theta_n  = \left( m^2 + 
  \frac{\la}{2}  T \right) (\om_0 + \om_n) \nn \\
  &+&  \frac{\la}{3!}  \left[  (\om_0  + \om_n)^3  + 6 \frac{\om_0  +
  \om_n}{4\pi  n} + \sum_{k>0}  \frac{\om_k}{4\pi  k} \right] ,
  \label{TNMF} 
\end{eqnarray}
and
\begin{eqnarray}
  \bra n | H/2L | n \ket &=& (H_0 + H_n)/2L = \nonumber \\
  &=& \frac{1}{2} \left( m^2 + \frac{\la}{2} T \right) (\om_0 +
  \om_n)^2  +  \frac{1}{2}  \left(  m^2  + \frac{\la}{2}  T \right)
  \frac{1}{2\pi n} + \nonumber \\
  &+&  \frac{\la}{4!}  \Bigg[  (\om_0  + \om_n)^4  + (12 \om_0^2  +
  24\om_0 \om_n + 14 \om_n^2)  \frac{1}{4\pi  n} + \nonumber \\
  && \quad\quad +\: 4(\om_0 + \om_n )
  \sum_{k>0} \frac{\om_k}{4\pi  k} + \sum_{k>0} \frac{\om_k^2}{4\pi
  k} \Bigg] + \nonumber \\
  &+& \frac{m^2}{2} T + \frac{\la}{8} T^2 \; . 
\label{HNMF}
\end{eqnarray}
From  the  divergence  structure  above  it  is  clear  that  all
coefficients can be made finite by adding the counter-term
\begin{equation}
  \delta H/2L = - \frac{\la}{4}  T \sum_{n=1}^\infty  \frac{1}{2\pi n}
  a_n^\dagger  a_n - \frac{\la}{4}  T \om^2  - \frac{1}{2}  m^2 T -
  \frac{\la}{8} T^2 \; , 
  \label{COUNTER2}
\end{equation} 
which can be obtained  from (\ref{COUNTER})  by integrating  over
$x^\m$  and decomposing  the field.  The renormalization  is thus
standard,  i.e.~performed by normal-ordering and formally achieved
by setting $T=0$ in the expressions above.

It is convenient to rescale the coefficients
\begin{eqnarray}
  \om_0 &\to& \frac{\om_0}{\sqrt{4\pi}} \; , \\
  \om_n &\to& \frac{\om_n}{\sqrt{4\pi}} \; , 
\end{eqnarray}
and define a dimensionless coupling $g$ as
\begin{equation} 
  g \equiv \frac{\la}{24\pi m^2} > 0 \; ,
\end{equation}
such that (\ref{T0MF} - \ref{ENERGIES1}) become
\begin{eqnarray}
  \!\!H_0/2L &=& \frac{m^2}{4\pi} \Bigg[\frac{1}{2} \om_0^2+
   \frac{g}{4}\left( \om_0^4 + 4 \om_0 \sum_{n=1}^{\infty}
   {\frac{\om_n}{n}} +    \sum_{n=1}^{\infty} {\frac{\om_n^2}{n}} 
   \right) \Bigg]  \; , \\
\label{ENERGIES2}
 \!\! H_n/2L &=&  \! \frac{m^2}{4\pi} \Bigg[ \frac{1}{2}
    \left( \om_n^2 + 2\om_0 \om_n +\frac{2}{n} \right) + \nonumber  \\ 
    \!\!&+& \! \frac{g}{4}\bigg[  \left(  \om_0 +\om_n  \right)^4-\om_0^4  +
    \frac{12}{n}       \left(      \om_0+\om_n       \right)^2      +
    4\om_n\sum_{k=1}^{\infty}{\frac{\om_k}{k}}+2\frac{\om_n^2}{n}
    \bigg] \! 
    \Bigg] 
\end{eqnarray}    
and
\begin{eqnarray}
  \label{ANSVAC}
  \om_0 &+& g \left( \om_0^3 + \sum_{n=1}^{\infty} {\frac{\om_n}{n}}  
  \right) = 0  \; , \\
  \label{ANSPART} 
  \om_n &+& g \left( \om_n^3 + 3\om_0 \om_n^2 + 3\om_0^2 \om_n + 
  \frac{6}{n} \left( \om_0 + \om_n \right)  \right) = 0 \;.
\end{eqnarray}
This system of equations has a trivial solution
\begin{equation}
        \om_0 = \om_n = 0 \;,
\end{equation}
corresponding to the symmetric phase with vanishing vacuum expectation
value of the field. Note further, that $\omega_n = 0$ \emph{implies}
$\omega_0 = 0$, which means that it is the \emph{operator} part of
$\omega$ that drives a possible phase transition. In the latter case,
where $\omega_n \ne 0$, there are non-trivial solutions of
(\ref{ANSVAC}, \ref{ANSPART}) which, however, cannot be obtained
exactly.  If we assume that there is a critical coupling where the field
starts to develop a vacuum expectation value, then, very close to this
coupling, $\om_0$ should be small so that we can expand $\om_n$ as a
power series in $\om_0$ \cite{heinzl:92c,heinzl:96a},
\begin{equation}
  \label{OMN}
  \om_n  =  \alpha_n  (g) \, \om_0 + \beta_n (g) \, \om_0^3  + O(\om_0^5)
  \; .  
\end{equation}
Plugging this into (\ref{ANSPART}), the expansion coefficients are
determined recursively as,
\begin{eqnarray}
  \alpha_n &=& - \frac{6g}{n + 6g} \: , \\
  \beta_n &=& - \frac{ng}{n + 6g} \alpha_n (3 + 3\alpha_n + \alpha_n^2) \nn \\
  &=& \frac{n}{6} \left( \frac{6g}{n+6g} \right)^2 \left[ 3 -
  \frac{18g}{n + 6g} + \left( \frac{6g}{n + 6g}  \right)^2 \right] \; .
\end{eqnarray}
Inserting  $\omega_n$  into  (\ref{ANSVAC}),  one  obtains  $\om_0$  as  a
function of the coupling $g$ via
\begin{equation}
  \label{OM_DET}
  a_0 (g) + a_2 (g) \, \omega_0^2 \equiv 1  +  g\sum_{n>0}
  \frac{\alpha_n}{n} 
  +  g \,   \left(  1 + \sum_{n>0} \frac{\beta_n}{n} \right) \om_0^2 = 0 \; .
\end{equation}
The coefficients $a_0$ and $a_2$ can be evaluated analytically with the
result 
\begin{eqnarray}
  \!\! a_0 (g) \!&=& \!\! 1 - g [\Psi (1 + 6g) + \gamma] \; , \label{A_0} \\
  \!\!a_2 (g) \!&=& \!\! g \Big[ 1 + \! 18 g^2 \left[ \Psi^{(1)} (1 + 6g) +
  \! 3 g   \Psi^{(2)} (1 + 6g) + \! 2 g^2 \Psi^{(3)} (1 + 6g) \right] \!
  \! \Big]
  \label{A_2} 
\end{eqnarray}
where the $\Psi^{(n)}$ are Polygamma functions \cite{abramowitz:70}.  As
$\beta_n$ and therefore $a_2$ are positive, equation (\ref{OM_DET})
develops two real solutions for $\omega_0$ (with opposite sign) if
\begin{equation}
  a_0 (g)  \le 0  \; .
  \label{SSBCOND}
\end{equation}
The critical  coupling  $g_c$  is determined  if equality  holds.
Using the explicit form of $a_0$ from (\ref{A_0}), one finds
\begin{equation}
  a_0 (g_c) = 1 - g_c [\Psi (1+6g_c) + \gamma] = 0 \; ,
  \label{GCRIT}
\end{equation}
with a numerical value for the critical coupling of 
\begin{equation}
  g_c = 0.53070059\ldots \; ,
\end{equation}
which   in   terms   of   the   original   parameters   is  (cf.~\cite{xu:95})
\begin{equation}
  \la_c = 24 \pi g_c \, m^2 \simeq 40.0 \, m^2 \; .
\end{equation}
Above this coupling, the vacuum expectation value $\om_0$ is
non-vanishing and acquires one of two possible signs so that the
reflection symmetry is spontaneously broken.

\begin{figure}
  \begin{center}
    \caption{\label{fig-omeg} \textsl{ Behavior of the rescaled 
      VEV $\om_0$ close to the critical coupling $g_c$. 
      Graphs {\rm 1},{\rm 2} and {\rm 3} refer to the expansion 
      {\rm (\ref{OMN})} of $\om_n$
      in $\om_0$ up to third, fifth and seventh order, respectively.}}
    \vspace{1cm}
    \includegraphics*[scale=0.7]{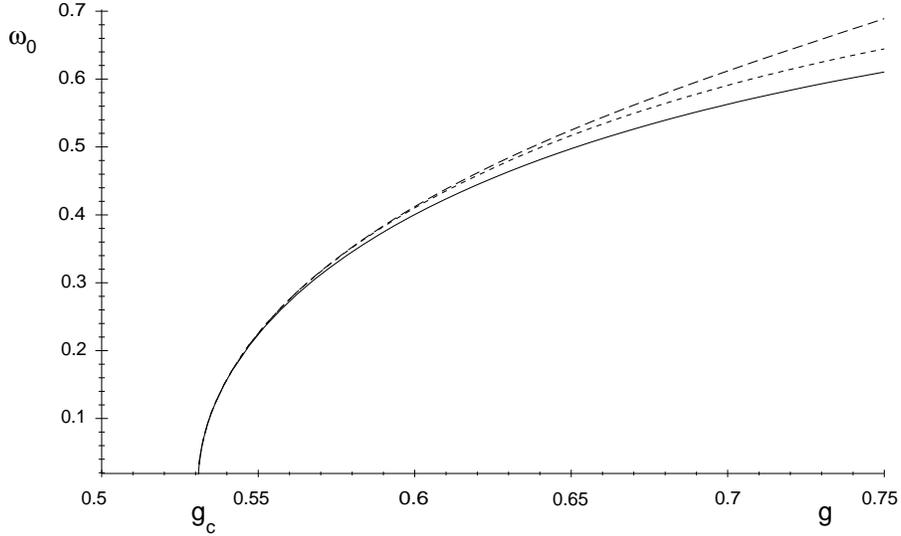}    
  \end{center} 
\end{figure}

It is very illuminating to visualize the discussion above in terms of an
effective potential $V_\mathrm{eff}$ \cite{heinzl:92c,simbuerger:92}. To
this end we interpret (\ref{OM_DET}), which determines the vacuum
expectation value $\omega_0$, as the equation localizing the minima of
$V_\mathrm{eff}$. The latter is thus obtained via integration,
\begin{equation}
  \label{V_EFF}
  V_\mathrm{eff} [\Omega] = \sfrac{1}{2} a_0 (g) \, \Omega^2 + \sfrac{1}{4}
  a_2 (g) \, \Omega^4 \; .
\end{equation}
The integration constant has been chosen such that $V_\mathrm{eff}
[\Omega = 0] = 0$. At the critical coupling, $g_c$, the coefficient
$a_0$ of $\Omega^2$ changes its sign from positive to negative.
Therefore, below $g_c$, the effective potential has no turning points
and its minimum is located at $\omega_0 = 0$ (symmetric phase). On the
other hand, above $g_c$, $V_\mathrm{eff}$ develops two non-trivial
minima at $\pm \omega_0$ (broken phase). As the order parameter
$\omega_0$ changes continuously, the phase transition is second order.
Obviously, $V_\mathrm{eff}$ has precisely the form of a Landau-Ginzburg
free energy which corroborates our mean-field interpretation. The final
confirmation is provided by the calculation of the critical exponents.
In a thermodynamic/magnetic analogy we view $g$ as a temperature,
$\Omega$ as a magnetization, and $\partial V_\mathrm{eff}/ \partial
\Omega$ as an external magnetic field. The coefficient $a_0$ is then an
inverse susceptibility which, near the critical coupling, should behave
like $a_0 \sim (g - g_c)^\gamma$. (Evidently, the exponent $\gamma$ has
nothing to do with Euler's constant.)  From (\ref{A_0}) and
(\ref{GCRIT}) we obtain via Taylor expansion around $g_c$,
\begin{equation}
  \label{EXP_GAMMA}
  a_0 (g) \simeq (g - g_c) \; a_0^\prime (g_c) \; ,
\end{equation}
yielding the expected mean field result, $\gamma = 1$. The order
parameter goes like $\omega \sim (g - g_c)^\beta$. Using (\ref{OM_DET})
and (\ref{EXP_GAMMA}), we find
\begin{equation}
  \label{OM_VEV}
  \omega_0 = \pm \, \big[ - a_0 (g)/a_2 (g) \big]^{1/2} \simeq \pm \, (g -
  g_c)^{1/2} \; 
  \big[ - a_0^\prime (g_c)/ a_2 (g_c) \big]^{1/2} \; , 
\end{equation}
i.e.~a mean field exponent $\beta = 1/2$. The dependence of the rescaled
vacuum expectation value $\om_0$ on $g$ is plotted in
Fig.~\ref{fig-omeg}. From both (\ref{OM_VEV}) and the figure it is
evident that $\omega_0$ is non-analytic in the coupling $g$.

The last exponent, $\delta$, is defined through the behavior of the
magnetic field at $g_c$, $\partial V_\mathrm{eff}/ \partial \Omega \sim
\Omega^\delta$. We calculate
\begin{equation}
  \left. \partial V_\mathrm{eff}/ \partial \Omega \right|_{g = g_c} =
  a_2 (g_c) \, \Omega^3 \; ,  
\end{equation}
so that $\delta = 3$. As a cross-check, we note that our critical
exponents obey the correct scaling law \cite{amit:84},
\begin{equation}
  \delta = 1 + \frac{\gamma}{\beta} \; .
\end{equation}
At this point it is instructive to make contact with perturbation
theory. For this purpose, we expand $a_0(g)$ as given in (\ref{A_0}) in
powers of $g$ using the series representation \cite{abramowitz:70}
\begin{equation}
  \label{PSI_EXP}
  \Psi (1 + z) = - \gamma + \sum_{n=2}^\infty (-1)^n \zeta (n) z^{n-1}
  \; ,
\end{equation}
from which we read off that $\Psi (1+z)$ is the generating function of
the sequence $\zeta(n)$, $\zeta$ being the Riemann zeta function. The
convergence region of the right-hand-side of (\ref{PSI_EXP}) is $|z| <
1$. For $a_0 (g)$ this yields,
\begin{equation}
  \label{A_0_PERT}
  a_0 (g) = 1 - 6 \, \zeta (2) \, g^2 + 36 \, \zeta (3) \, g^3 + O(g^4) \; ,
\end{equation}
which converges for $g < 1/6$. Keeping just the first two terms, one can
find a critical coupling perturbatively \cite{heinzl:92c,simbuerger:92,heinzl:91d}
via
\begin{equation}
  a_0 (g_{cp}) = 1 - 6 \, \zeta (2) \, g_{cp}^2 = 0 \; ,
\end{equation}
leading to 
\begin{equation}
  \label{G_CP}
  g_{cp} = 1/ \pi = 0.3183098... > 1/6
\end{equation}
which is beyond the radius of convergence of (\ref{A_0_PERT}). If we
include the third order in $g$, the equation $a_0(g) = 0$ does no longer
have a positive root like $g_{cp}$. We conclude that in the present
framework, perturbation theory is not applicable for studying the
critical behavior, and the value (\ref{G_CP}) for $g_{cp}$ has to be
abandoned. In other words, upon using perturbation theory one stays in the
symmetric phase with $\omega_0$ being zero, in agreement with the
discussion of Section~7.2.

Finally, from the one-particle energies (\ref{ENERGIES2}), we calculate
the mass-shift induced by the non-perturbative zero mode $\om$ as given
by the ansatz (\ref{MFA}),
\begin{eqnarray}
  \delta m_n^2 &=& \! \!P^\p_n P^\m_n - m^2 = \frac{2 \pi  n}{L} H_n - m^2
  = m^2 \Bigg[ \frac{n}{2}  \left( \om_n^2 + 2\om_0 \om_n \right) +
  \nonumber \\ 
  &+& \!\! \frac{ng}{4}\Bigg( \!\!\left( \om_0 +\om_n \right)^4-\om_0^4 +
  \frac{12}{n}       \left(      \om_0+\om_n       \right)^2      +
  4\om_n\sum_{k=1}^{\infty}{\frac{\om_k}{k}}+2\frac{\om_n^2}{n}
  \Bigg) \Bigg]  
\end{eqnarray}
Using the expression  (\ref{OMN})  for $\om_n$ this becomes (near
the critical coupling)
\begin{eqnarray}
  \delta  m_n^2  &=&  3g \, m^2  \om_0^2   \Bigg\{  1  -  2  \left[1  -
  \frac{g}{1+6g/n} \bigg(\Psi(1+6g) + \gamma \bigg) \right] + 
  \nonumber \\
  &+& 6g \frac{n + 7g}{(n+ 6g)^2} \Bigg\} \; .
\end{eqnarray}
In contrast  to the  perturbative  result  (\ref{SHIFTRES}), this
expression  is {\it not} vanishing  in the continuum  limit,  
\begin{equation}
  \delta m^2 =  3g \, m^2  \om_0^2 \Big[ 1 - 2 a_0 (g) \Big] \; . 
\end{equation}
Due to (\ref{SSBCOND}) and (\ref{GCRIT}), $a_0 (g)$ is a small
(negative) quantity.  The mass shift induced by $\om$ is thus positive.
We want to emphasize that the zero mode $\om$ has a non-trivial
influence on the spectrum.

\section{The Tamm-Dancoff Approximation}

\subsection{General Remarks}

The Tamm-Dancoff approximation was originally designed within the
conventional formulation of quantum field theory \cite{tamm:45}. The
principle of the method was to enormously reduce the particle number to
a finite (and small) one.  Due to the complicated many-body structure of
the equal-time vacuum, however, the approximation failed to lead to
quantitative results and was abandoned thereafter.  Although it has been
noted rather early that light-cone field theory might be better suited
for a Tamm-Dancoff approximation \cite{thooft:74} , it was not until
recently that people began to start systematic investigations
\cite{harindranath:90}.  The hope was (and is), of course, that many of
the problems of the original Tamm-Dancoff approximation can be avoided
due to the simplicity of the light-cone vacuum.  Specifically, people
first tried to calculate bound states for 1+1 dimensional field theories
where the method turned out to work extremely well (see Chapter~8 and
\cite{thooft:74,bergknoff:77,perry:91,harindranath:91a,
  harindranath:92}). Considerably more effort is needed in 3+1
dimensions \cite{mustaki:92,glazek:93a}, the essential problem to solve
being renormalization within a Hamiltonian formulation
\cite{glazek:93b,perry:94a,wilson:94,sugihara:96,harada:97b}, as the
structure of counter-terms as well as their required number is unclear.
Furthermore, dynamical symmetries like rotational invariance or parity
are violated in general \cite{ji:94}. Nevertheless, the first
phenomenological applications have appeared recently
\cite{brisudova:97,jones:97a,jones:97b}.

In the following we will use the model at hand to shed some light on the
issue of renormalization, in particular on the construction of the
counter-terms needed.  These could, in principle, be different for
different particle number sectors ("sector-dependent renormalization"
\cite{harindranath:90}). For a super-renormalizable theory like
$\phi^4_{1+1}$, which conventionally is renormalizable by
normal-ordering, this would be a rather undesired feature since it would
complicate the renormalization procedure enormously.  Therefore, in the
following, we will try to keep the renormalization as simple as
possible. To this end, we attempt to incorporate the normal-ordering
prescription into the Tamm-Dancoff approximation.  First we need some
definitions.  The projection operators on the lowest particle number
states are
\begin{eqnarray}
  P_0 &=& \vert 0 \ket \bra 0 \vert \\
  P_1 &=& \sum_{n>0} \vert n \ket \bra n \vert \\
  P_2 &=& \frac{1}{2}  \sum_{n,  m > 0} \vert  n,m\ket  \bra  m, n \vert \\
  \nonumber         &\vdots&\\  
  P_N &=& \frac{1}{N!} \sum_{n_1,\ldots,n_N > 0}
  \vert n_1,\ldots,n_N \ket  \bra n_1,\ldots,n_N  \vert
  \; ,
\end{eqnarray}
with the Fock basis states defined in (\ref{FOCK_BASIS}).  Most
important to us will be the Tamm-Dancoff projector
\begin{equation}
  \Pn \equiv \sum_{\alpha=1}^{\mbox{
  \scriptsize \sf N}} P_\alpha \; , 
\end{equation}
which projects  onto the direct sum of all sectors  with particle
number  less  than or equal  to {\sf N}.  Before  we perform  any
detailed  calculation,  let  us  make a  few  remarks  about  the
relativistic invariance of the Tamm-Dancoff approximation \cite{schmidt:95}.

Let ${\cal  P}$ denote  the Poincar\'e  group.   In $d$ space-time
dimensions the number of Poincar\'e generators is
\begin{equation}
  \label{DIM-P-GROUP}
  \mbox{dim} \, {\cal P} = \frac{d(d+1)}{2} \equiv D\; .
\end{equation}
If the Poincar\'e  generators are $G_0, G_1, \ldots G_{D-1}$, the
Tamm-Dancoff approximation is relativistically invariant if
\begin{equation}
  [\Pn G_i \Pn , \Pn G_k \Pn] = \Pn [G_i , G_k ] \Pn \; .
\end{equation}
It is easy to see, that this expression  can only hold if at most
one of the Poincar\'e generators fails to commute with $\Pn$.  So
we must have,  e.g. 
\begin{eqnarray}
  [G_0, \Pn] &\not=& 0 \; , \\
  \lbrack G_k , \Pn] &=& 0 \; , \; \;  k = 1, \ldots D-1 \; . 
  \label{INV}
\end{eqnarray}
The last identity implies that the $G_k$ conserves particle number and
therefore must be kinematical, i.e.~interaction independent.  Thus,
$G_0$ must be the only dynamical generator.  Clearly, this can only
happen if the dimension of the stability group (of kinematical
generators) is $D-1$.  There is only one single case where this
requirement is met, namely the front form in $d$ = 1+1, which is the
basis of the present discussion.  Here, according to
(\ref{DIM-P-GROUP}), there are only three generators; the dynamical
generator is $G_0 \equiv P^-$, the light-cone energy; the kinematical
generators are the momentum $P^+$ and the (longitudinal) boost $M^{+-}$.
We think that this unique feature is one of the reasons, why the
Tamm-Dancoff approximation (or more generally: Fock space truncation
methods) work so well for light-cone field theories in $d$ = 1+1.

Encouraged by this observations we continue and analyze the impact of
the Tamm-Dancoff approximation on the quantum nature of the scalar
field.  This will be important for the issue of normal-ordering.  We
imagine that any operator ${\cal O}$ can be built from the elementary
Fock operators $a_n$, $a_n^\dagger$, no matter how complicated its form
may be.  Thus, we {\it define} the $N$-particle Tamm-Dancoff
approximation by the replacement
\begin{eqnarray}
  a_n &\to& \Pn a_n \Pn \; , \label{NPLFTD1}\\
  a_n^\dagger &\to& \Pn a_n^\dagger \Pn \label{NPLFTD2} \; . 
\end{eqnarray}
In this way, however, one is mutilating  the quantum structure of
the theory.  This can be seen by calculating  the $N$-particle
Tamm-Dancoff approximation  of the elementary commutator,
\begin{equation}
  [a_m , a_n^\dagger]_{\sfn} \equiv  [\Pn a_m
  \Pn , \Pn a_n^\dagger \Pn] \; . 
\end{equation}
A somewhat tedious calculation  leads to the final result
\cite{heinzl:96a} 
\begin{equation} 
  \label{TDCOMM2} 
  [a_m  , a_n^\dagger  ]_{\sfn} = \delta_{mn}
  \Pnm - P_N a_n^\dagger a_m P_N \; . 
\end{equation}
Taking matrix elements of this expression one readily sees that the
correct result ($\delta_{mn}$) for the commutator within $N$-particle
Tamm-Dancoff approximation is reproduced only up to the ($N$-1)-particle
sector. The additional term on the right-hand-side of (\ref{TDCOMM2}) is
acting in the $N$-particle subspace only.  In other words, matrix
elements of expressions involving the elementary commutator, which are
calculated within $N$-particle Tamm-Dancoff approximation, should not be
trusted beyond the ($N$-1)-particle sector. This will be relevant for
the problem of normal-ordering to be discussed shortly.

To  be a little  bit  more  explicit,  we list  the  lowest  order
expressions for (\ref{TDCOMM2})
\begin{eqnarray}
  [a_m , a_{n}^{\dagger}  ]_1 &=& \delta_{mn} \vert 0 \ket \bra 0 \vert
  - \vert n \ket \bra m \vert \; , \\
  \lbrack a_m , a_{n}^{\dagger} ]_2 &=& \delta_{mn} \Big[ \vert 0 \ket \bra 0
  \vert + \sum_{l>0} \vert l \ket \bra l \vert \Big] - \sum_{l>0}
  \vert l,n \ket \bra m,l \vert \; .
\end{eqnarray}
What are now the implications of all that for the renormalization, in
particular mass renormalization?  As can be seen from (\ref{WICK}) and
(\ref{COUNTER}), the latter is encoded in the normal-ordering
prescription of the expression
\begin{eqnarray} 
  \label{NORD}
  \frac{1}{2L} \intl dx^\m \vi^2 (x) &=& \frac{1}{2L} \intl dx^\m :
  \vi (x)^2 : + T  \nonumber \\
  &=&  \sum_{n>0}  \frac{1}{2\pi  n} a_n^\dagger  a_n  + \sum_{n>0}
  \frac{1}{4\pi n} [a_n, a_n^\dagger] \; .
\end{eqnarray}
Normal-ordering  thus  amounts  to splitting  off  the  divergent
tadpole  contribution  $T$, which, in the Fock space language, is
given  by an elementary  commutator  (or contraction).   Thus the
remarks  above,  leading  to  (\ref{TDCOMM2}),   apply.   Let  us
calculate the $N$-particle Tamm-Dancoff approximation  of (\ref{NORD}),
\begin{eqnarray}
  \frac{1}{2L}   \intl   dx^\m   \vi^2   (x)  \stackrel{\mbox{\tiny
  $N$PTDA}}{\simeq}    &=&   \sum_{n>0}   \frac{1}{2\pi   n}   \Pnm
  a_n^\dagger  \Pnm a_n \Pnm + \Pnm T - \nonumber \\ 
  &-& \sum_{n>0} \frac{1}{4\pi n} P_N a_n^\dagger a_n P_N \; . 
\end{eqnarray}
The same  remarks  as for (\ref{TDCOMM2})  are in order.   Matrix
elements  of the last  expression  should  not be expected  to be
consistent beyond the ($N$-1)-particle  sector.  Furthermore, one
should  note that, as $\vi^2$  is a one-body  operator,  there is
only {\it one} commutator (or contraction)  involved in the above
normal-ordering.  For a $k$-body operator we therefore conjecture
that  its  renormalization   will  be  correct  only  up  to  the
($N-k$)-particle  sector.  For  example,  in  order  to  get  the
renormalization  of the two-body  operator  $\vi^4$ correct up to
the one-particle sector, a three-particle Tamm-Dancoff approximation  
will be needed.

\subsection{One-Particle Tamm-Dancoff Approximation}

The one-particle light-cone Tamm-Dancoff approximation is defined by the
replacements
\begin{eqnarray}
  a_n &\simeq&  \Proj_1 a_n \Proj_1  = \vert 0 \ket \bra n \vert \\
  a_n^\dagger  &\simeq&  \Proj_1 a_n^\dagger\Proj_1  = \vert n \ket
  \bra 0 \vert \; . 
\end{eqnarray}
From our general results we expect that within one-particle light-front
Tamm-Dancoff approximation we will get a consistent renormalization of 
the one-body operator $\vi^2$ in the vacuum (zero-particle) sector  only.

To solve for the zero mode $\om$ we make the following Tamm-Dancoff ansatz
\begin{equation} 
  \om_{TD} = c_0 \vert 0 \ket \bra 0 \vert +  \sum_{n>0}  c_n \vert n
  \ket \bra n \vert \; .
  \label{OMTD1}
\end{equation}
We similarly  expand the constraint  $\theta$ and the Hamiltonian
$H$ 
\begin{eqnarray}
  \theta   &\simeq&  \bar\theta_0  \vert  0  \ket  \bra  0  \vert  +
  \sum_{n>0} \bar\theta_n \vert n \ket \bra n \vert \; , \\
  H &\simeq&  \bar H_0 \vert 0 \ket \bra 0 \vert + \sum_{n>0}  \bar
  H_n \vert n\ket \bra n \vert \; ,
\end{eqnarray}
where   the  bars  simply   indicate   the  distinction   of  the
coefficients   above  from  those  of  the  mean field ansatz  (\ref{MFT})  and
(\ref{MFHAM}).   The coefficients  can be found  as functions  of
$c_0$,  $c_n$ upon inserting  the ansatz  (\ref{OMTD1})  into the
constraint (\ref{TH1}) and the Hamiltonian (\ref{HAM}) yielding
\begin{eqnarray}
  \bar \theta_0   &=&  \left(m^2   +  \frac{\la}{3}   T  \right)  c_0  +
  \frac{\la}{3!}   \left(c_0^3  +  \sum_{n>0}  \frac{c_n}{4\pi   n}
  \right) \; , \label{T0} \\ 
  \bar \theta_n  &=&  \left(  m^2 + \frac{\la}{12  \pi n} \right)  c_n +
  \frac{\la}{3!} \left( c_n^3 + \frac{c_0}{4\pi n} \right) \; ,
\end{eqnarray}
and
\begin{eqnarray}
  \frac{\bar H_0}{2L} &=& \frac{1}{2} \left( m^2 + \frac{\la}{4}  T \right) c_0^2 +
  \frac{\la}{4!} \left( c_0^4 + 2 \sum_{n>0} \frac{c_0c_n}{4\pi  n}
  + \sum_{n>0}  \frac{c_n^2}{4\pi  n} \right)  \nn \\
  &+&   \frac{m^2}{2}  T + \frac{\la}{4!} T^2 , \label{H0} \\
  \frac{\bar H_n}{2L}  &=& \frac{1}{2}  \left( m^2 + \frac{\la}{16 \pi n} \right)
  c_n^2 + \frac{1}{2} \left( m^2 + \frac{\la}{12} T \right)
  \frac{1}{4\pi n} \nn \\
  &+& \frac{\la}{4!}  \left ( c_n^4 + \frac{c_0  c_n}{2\pi n} +
  \frac{c_0^2}{4\pi n} \right) \!\!.
\end{eqnarray}
At a first look, these expressions for the coefficients appear to be a
disaster: the infinities in form of the tadpole $T$ do not appear
systematically, the mass gets renormalized in the vacuum sector only,
but differently for the constraint $\theta$ and the Hamiltonian $H$.
Both expressions differ from the standard expression $m^2 + \la T/2$.
There is also a divergent contribution from the $\vi^4$ term to $\bar
H_0$ which differs from the usual $\la T^2 /8$ of (\ref{WICK}).  As
stated above, we do not believe the coefficients $\bar \theta_n$ and
$\bar H_n$ to be correct within one-particle light-cone Tamm-Dancoff
approximation.  They will be discussed in the next subsection, when we
go to higher order.

The way to remedy the situation (for the coefficients $\bar \theta_0$
and $\bar H_0$) is the following.  We insist, firstly, on the standard
mass renormalization, $m^2 + \la T/2$, however, according to our general
discussion, in the vacuum sector only.  Secondly, we do not assume that
the coefficients $c_0$ and $c_n$ are independent, and use this freedom
to {\it redefine} them in the following way
\begin{eqnarray}
  c_0 &\equiv& \om_0 \; , \\
  c_n &\equiv& \om_0 + \om_n \; .
  \label{REDEF1}
\end{eqnarray}
Inserting this into (\ref{T0}) and (\ref{H0}) one finds
\begin{eqnarray}
  \bar \theta_0  &=&  \left(  m^2  +  \frac{\la}{2}   T  \right)\om_0  +
  \frac{\la}{3!}  \left( \om_0^3 + \sum_{n>0} \frac{\om_n}{4\pi  n}
  \right) \; ,  \label{1PT0} \\
  \bar H_0/2L &=& \frac{1}{2} \left( m^2 + \frac{\la}{2} T \right) \om_0^2 +
  \frac{\la}{4!}   \left(   \om_0^4   +   4\sum_{n>0}   \frac{\om_0
  \om_n}{4\pi  n}  + \sum_{n>0}  \frac{\om_n^2}{4\pi  n} \right)  +
  \nonumber \\
  &+& \frac{m^2}{2} T + \frac{\la}{4!} T^2 \; . \label{1PH0}
\end{eqnarray}
Remarkably, the simple redefinition (\ref{REDEF1}) has led to the
desired results.  The mass renormalization is standard and the same for
$\theta_0$ and $H_0$.  The divergences thus can be made to vanish by
adding the counter-term (\ref{COUNTER2}).  Both equations (\ref{1PT0})
and (\ref{1PH0}) coincide with the lowest order results from the
mean-field ansatz (\ref{T0MF}) and (\ref{E0MF}) (up to the constant
$\vi^4$-contribution to $H_0$ given by $\la T^2/4!$).  Note that there
are no two-body ($T^2$) contributions to the constraint.  This is
obviously true to all orders, so the renormalization of $\theta$ is
slightly simpler than that of $H$, namely just mass renormalization.

The coincidence  with the mean field results  is not accidental.   If one
calculates  the lowest  order matrix  elements  of the mean field 
ansatz (\ref{MFA}), one finds
\begin{eqnarray}
  \om_0 = \bra 0 | \om_{MF} | 0 \ket &=& \bra 0 | \om_{TD} | 0 \ket = c_0
  \; , \\
  \om_0 + \om_n = \bra n | \om_{MF} | n \ket &=& \bra n | \om_{TD} | n
  \ket = c_n \; .
\end{eqnarray}
Analogous   relations   hold  for  the  matrix  elements  of  the
constraint and the Hamiltonian,
\begin{eqnarray}
  \bra 0 | \theta | 0 \ket &=&  \theta_0 = \bar \theta_0 \; , \\
  \bra 0 | H | 0 \ket &=& H_0 = \bar H_0 \; , \\
  \bra  n  |  \theta  | n \ket  &=&  \theta_0  + \theta_n  = \bar
  \theta_n  \;  , \\  
  \bra  n | H | n \ket  &=&  H_0 + H_n = \bar H_n \; .
\end{eqnarray}
Thus, after the redefinition (\ref{REDEF1}), the zero- and one-particle
matrix elements of $\om$ calculated within the mean field ansatz and
Tamm-Dancoff approximation coincide.  As the renormalization within mean
field ansatz was conventional and straightforward, it is not too
surprising that the behavior of the redefined Tamm-Dancoff approximation
under renormalization gets improved. This will be another guideline in
the following.

\subsection{Two-Particle Tamm-Dancoff Approximation}

If we now go one  step  further  and  include  also  two-particle
states via
\begin{eqnarray}
  a_n &\simeq&  \Proj_2  a_n \Proj_2  = \vert 0 \ket \bra n \vert +
  \sum_{m>0} \vert m \ket \bra m,n \vert \\
  a_n^\dagger  &\simeq& \Proj_2 a_n^\dagger  \Proj_2 = \vert n \ket
  \bra 0 \vert + \sum_{m>0}  \vert n , m \ket \bra m \vert
  \; , 
\end{eqnarray}
we should further improve our renormalization program.  We expect
a consistent  renormalization  of  $\vi^2$-contributions  in  the
vacuum- and one-particle sector, and of $\vi^4$-contributions  in
the vacuum sector. The extended ansatz for $\om$ becomes
\begin{eqnarray}
  \om_{TD} &=& c_0 | 0 \ket \bra 0 | + \sum_{n>0} c_n | n \ket \bra n
  | + \nonumber \\
  &+&  \sfrac{1}{2}  \sum_{m,n>0}  c_{mn} \, | m,n \ket  \bra  m+n | +
  \sfrac{1}{2}  \sum_{m,n>0}  c_{mn}^* \, |  m+n  \ket  \bra  m,n  | +
  \nonumber \\ 
  &+& \sfrac{1}{4} \sum_{k,l,m,n>0}  \delta_{k+l,  m+n} \, c_{klmn}
  \, | k,l \ket \bra m,n| \; . 
\label{OMTD2}
\end{eqnarray}
It is now very plausible (though we cannot prove it a priori) that a
consistent renormalization requires a redefinition also of the
coefficients $c_{mn}$, $c_{mn}^*$, and $c_{klmn}$.  To proceed as
before, we would need the matrix elements $\bar \theta_0$, $\bar
\theta_n$, $\bar H_0$, and $\bar H_n$ with all two-particle
contributions in order to just get a consistent renormalization up to
the one-particle sector.  This is very tedious and inefficient, as we
are {\it only} interested in the {\it divergent} contributions from
higher order terms to lower order matrix elements.  Fortunately, there
is an alternative: we simply demand that the matrix elements of
$\om_{TD}$ and $\om_{MF}$ coincide also in the two-particle sector.
This gives us the desired redefinitions, namely
\begin{eqnarray}
  c_{mn} &=& 2 \om_{mn} \; , \\
  c_{mn}^* &=& 2 \om_{mn}^* \; , \\
  c_{klmn}  &=& (\om_0 + \om_n + \om_n)  (\delta_{km}\delta_{ln}  +
  \delta_{kn}\delta_{lm} ) + 4 \, \delta_{k+l, m+n}\, \om_{klmn} \; .
\end{eqnarray}
Thus,  essentially,   only  the  two-particle   matrix   elements
$c_{klmn}$ get redefined. Expression (\ref{OMTD2}) becomes
\begin{eqnarray}
  \om_{TD} &=& \om_0 | 0 \ket \bra 0 | + \sum_{n>0} (\om_0 + \om_n)
  | n \ket \bra n | + \nonumber \\ 
  &+& \sum_{m,n>0}  \om_{mn}  | m,n \ket \bra  m+n | + \sum_{m,n>0}
  \om_{mn}^* | m+n \ket \bra m,n | + \nonumber \\
  &+& \sfrac{1}{2}\sum_{m,n>0}  (\om_0 + \om_m + \om_n)
  | m,n \ket \bra m,n | +  \sum_{k,l,m,n>0} \delta_{k+l,
  m+n} \, \om_{klmn} \, | k,l \ket \bra m,n| \; . \nn
\end{eqnarray}
The diagonal two-particle term in the last line above contributes to the
equations determining the coefficients $\om_0$ and $\om_n$ and crucially
alters the renormalization behavior.  In \cite{pinsky:94}, it was noted
that our mean field ansatz amounts to including two-particle matrix
elements, and it is just these terms that we have now explicitly
displayed. Omitting them, as was done in \cite{pinsky:94}, leads to an
uncanceled tadpole contribution and thus to a spurious logarithmic
divergence.

If we neglect all terms containing $\om_{mn}$, $\om_{mn}^*$, and
$\om_{klmn}$, which are not of interest within two-particle Tamm-Dancoff
approximation, we find for the constraint
\begin{eqnarray}
  \!\!\bar\theta_0  \!&=&  \! \bigg(  m^2  +  \frac{\la}{2}   T  \bigg)\om_0  +
  \frac{\la}{3!}  \left( \om_0^3 + \sum_{n>0} \frac{\om_n}{4\pi  n}
  \right) \; ,  
  \label{2PT0} \\
  \!\!\bar\theta_n \!&=& \! \bigg( m^2 + \frac{\la}{2}  T \bigg) (\om_0 + \om_n)
  +  \frac{\la}{3!}  \left[  (\om_0  + \om_n)^3  + 6 \frac{\om_0  +
  \om_n}{4\pi  n} + \sum_{k>0}  \frac{\om_k}{4\pi  k} \right]  \; ,
  \label{2PTN}
\end{eqnarray}
and for the Hamiltonian
\begin{eqnarray}
  \bar H_0/2L &=& \frac{1}{2} \left( m^2 + \frac{\la}{2} T \right) \om_0^2 +
  \frac{\la}{4!}   \left[   \om_0^4   +   4\sum_{n>0}   \frac{\om_0
  \om_n}{4\pi  n}  + \sum_{n>0}  \frac{\om_n^2}{4\pi  n} \right]  +
  \nonumber \\ &+& \frac{m^2}{2} T + \frac{\la}{8} T^2 \; , 
  \label{2PH0} \\
  \bar H_n/2L &=& \frac{1}{2} \left( m^2 + \frac{\la}{2} T \right) (\om_0 +
  \om_n)^2  +  \frac{1}{2}  \left(  m^2  + \frac{\la}{3}  T \right)
  \frac{1}{2\pi n} + \nonumber \\
  &+&  \frac{\la}{4!}  \Bigg[  (\om_0  + \om_n)^4  + \Big[12 (\om_0  +
  \om_n)^2 + 2 \om_n^2 \Big]  \frac{1}{4\pi  n} + \nonumber \\
  && \quad\quad +\: 4(\om_0 + \om_n )
  \sum_{k>0} \frac{\om_k}{4\pi  k} + \sum_{k>0} \frac{\om_k^2}{4\pi
  k} \Bigg] + \nonumber \\
  &+& \frac{m^2}{2} T + \frac{\la}{4!} T^2 \; . 
  \label{2PHN}
\end{eqnarray}
Several remarks are in order.  As $\theta$ does not contain two-body
components, the renormalization is correct up to the one-particle
sector.  (\ref{2PT0}) and (\ref{2PTN}) thus coincide with (\ref{T0MF})
and (\ref{TNMF}). The coefficient $\theta_0$ is not even changed by
including the two-particle contributions as can be seen by comparing
with (\ref{1PT0}).  In the vacuum coefficient $\bar H_0$, the mass
renormalization (due to the $\vi^2$ contributions) {\it and} the vacuum
energy $m^2 T/2 + \la T^2 /8$ (with the $T^2$ contribution stemming from
the $\vi^4$-term) are correct, as expected.  So $\bar H_0$ is
consistently renormalized.  In the one-particle coefficient $\bar H_n$,
which should be compared with (\ref{HNMF}), only the mass
renormalization in the $\om$-sector is correct, as this is due to
one-body contributions like $\om^2\vi^2$.  As anticipated, mass
renormalization and vacuum energy stemming from the $\vi^4$-term differ
from the correct values by numerical factors. To get these correctly,
one would have to perform a three-particle Tamm-Dancoff approximation.
Presumably, this would only change the coefficients of the divergent
terms, whereas the coefficients of the finite terms would remain the
same. This would then be analogous to the change in $\bar H_0$ by going
from one-particle Tamm-Dancoff approximation (\ref{1PH0}) to
two-particle Tamm-Dancoff approximation (\ref{2PH0}).

Summarizing, we can say that, in order to obtain a consistent
renormalization within a $N$-particle light-cone Tamm-Dancoff
approximation, one has to include contributions from ($N+1$)-particle
matrix elements by appropriately redefining the coefficients in the
Tamm-Dancoff ansatz.  In this way, the Fock ansatz method (\ref{MFA})
and the Tamm-Dancoff approximation become completely equivalent.

\section{Discussion}

In this chapter we have analyzed the vacuum structure of light-cone
$\phi^4_{1+1}$-theory by comparing different methods of solving for the
constrained zero mode of the field operator.  Within perturbation
theory, the zero mode induces a second order mass correction which is
vanishing in the infinite volume limit.  We believe that to all orders
in perturbation theory the zero mode only induces finite size effects,
although we do not have a general proof.

We have presented two non-perturbative methods to obtain a solution for
the zero mode.  An ansatz in terms of an increasing number of Fock
operators, which we have truncated after the one-body term, seems to be
the most economic procedure.  With considerably more efforts, exactly
the same results can be obtained within a light-cone Tamm-Dancoff
approximation, if the renormalization procedure is properly designed.

With either method we find a non-vanishing vacuum expectation value
$\phi_c$ of the field if the coupling $\la$ exceeds a critical value of
$\la_c \simeq 40 \, m^2$, implying spontaneous breakdown of the
reflection symmetry $\phi \to - \phi$.  As the vacuum expectation value
changes continuously, the associated phase transition is of second
order, which has been rigorously established for the model at hand
\cite{griffiths:73}.  The order parameter $\phi_c$, for instance, shows
a square-root behavior as a function of the coupling, so that the
associated critical exponent is $\beta = 1/2$.  The critical behavior is
thus of mean field type, which is only qualitatively correct, as the
$\phi^4_{1+1}$ model is in the universality class of the two-dimensional
Ising model; thus, $\beta$ should be $1/8$.  Because the calculations
tend to become very messy, it is difficult to say, whether this
shortcoming can be removed if one extends our approximations to higher
orders. It is, however, plausible that the critical behavior will
change if more terms in the ansatz (\ref{MFA}) are retained as these
correspond to marginal operators $\varphi^n$ with $n > 2$. This issue,
of course, deserves further investigations.

Another problem we have to face is the absence of any volume dependence
of the phase transition.  We have been working in a finite spatial
volume of length $2L$, the length scale $L$, however, which does not
appear in the Fock measure, drops out of the equation (\ref{GCRIT})
determining the critical coupling.  On the other hand, there cannot be a
phase transition in a finite volume due to topological fluctuations
(kinks and anti-kinks) which have non-vanishing statistical weight for
$L<\infty$.  We have not incorporated these fluctuations by our choice
of (periodic) boundary conditions, so it is perhaps not too astonishing
that we do not obtain a volume dependence. It should, in addition, be
pointed out that the light-cone spatial volume, $-L \le x^\m \le L$, is of a
very peculiar nature. It corresponds to a light-like Minkowski region, 
the \emph{invariant}  length of which vanishes. This may give rise to
features that are not present if the volume is space-like \cite{lenz:91}. 

A better understanding of this problem is certainly desirable. As a
first step towards doing better, a pure continuum formulation of the
problem has been developed recently that leads to a slightly different
critical coupling \cite{grange:97}, namely
\begin{equation}
  \label{GC_GRANGE}
  g_c = 0.69812669\ldots \; . 
\end{equation}
Note that in \cite{grange:97} a different coupling normalization,
$\tilde g \equiv 6g$, is used so that $\tilde g_c = 4.18876012\ldots$.
The value (\ref{GC_GRANGE}) can be straightforwardly obtained from our
equation (\ref{GCRIT}) if the continuum limit is performed as in
Section~7.2, i.e.~via replacing the Digamma function by its asymptotics
(\ref{PSI_ASYMPT}). Instead of (\ref{GCRIT}), we find,
\begin{equation}
  1 - g_c \log(6g_c) = 0 \; , 
\end{equation}
which is solved by (\ref{GC_GRANGE}). This agreement of two independent
calculations provides additional confidence in our method. We also point
out that the critical coupling (\ref{GC_GRANGE}) is not too different
from the one obtained within sophisticated renormalization group
treatments \cite{zinn-justin:96}, which, using our normalization, is
given by $g_{c,RG} = 0.84 \pm 0.11$ \cite{grange:97}.


\chapter{'t~Hooft and Schwinger Model}

In the last chapter we have seen that zero modes carry at least some of
the vacuum structure for a \emph{scalar} field theory. Unfortunately,
the methods used there cannot be straightforwardly generalized to the
case of fermionic fields. There, a non-trivial vacuum shows up through
the appearance of a fermion condensate, which in QCD, for example,
signals spontaneous chiral symmetry breaking, as we have discussed in
the introduction,. The point now is, that a fermion condensate is
bilinear in the fields, and we do not expect fermionic zero modes to
play any particular role in the formation of the condensate. This is
also supported by the fact that a single fermionic zero mode cannot
condense as it does not have vacuum quantum numbers.  We will therefore
follow a different route in this chapter and try to reconstruct vacuum
properties from the particle spectrum. This is where light-cone wave
functions enter the stage.

\section{Prelude: Fermion Condensates}

Assume that we have a symmetry which is explicitly broken so that the
associated current is not conserved,
\begin{equation}
  \partial_\mu j^\mu (x) = A (x) \; .
\end{equation}
With the help of the corresponding  Ward identity \cite{coleman:71} one
can derive the following formula for an arbitrary operator $B$,
\begin{equation}
  \label{MASTER}
  \bra 0 | \delta  B / \delta  \alpha | 0 \ket = -i \int d^{\,4}  x
  \bra 0 | T A(x) B (0) | 0 \ket = - \sum_n \frac{\bra 0 | A(0) |
  n \ket \bra n | B(0) | 0 \ket}{m_n^2} \; .
\end{equation}
Here, $\delta B / \delta \alpha$  denotes the change of $B$ under
the symmetry transformation,  and in the last step a complete set
of states $ |n \ket$, each of mass $m_n$, has been inserted.  For
the case of chiral symmetry, choosing
\begin{equation} 
  A = B = 2 m \bar \psi i \gamma_5 \psi \; ,
\end{equation}
one finds for a single quark flavor of mass $m$
\begin{equation} 
  \label{COND_GEN}
  \bra  0 |\bar  \psi \psi | 0 \ket = -m \sum_n  \frac{|\bra  0 |
  \bar \psi i \gamma_5 \psi | n \ket|^2}{m_n^2} \; .
\end{equation}
In  \cite{chibisov:95a,burkardt:96d}, this expression was used as a
definition for the quark condensate in the `t~Hooft model (see the next
sections).  From the above derivation, however, it is clear that
(\ref{COND_GEN}) holds quite generally.  Using, for instance, the PCAC
relation for the axial vector current, $j_5^\mu$,
\begin{equation}
  \partial_\mu j_5^\mu (x) = f_\pi m_\pi^2 \, \pi (x) \; ,
  \label{PCAC}
\end{equation}
where  $\pi (x)$ is an interpolating  pion field and $f_\pi$  the
pion decay constant,  (\ref{MASTER})  involves the pion two-point
function and is easily evaluated with the result
\begin{equation}
  f_\pi^2  m_\pi^2  = - 4 m \bra 0 | \bar \psi \psi | 0 \ket \; .
  \label{GOR1}
\end{equation}
This is the famous Gell-Mann-Oakes-Renner relation \cite{gell-mann:68}
(to lowest order in the quark mass $m$ \cite{gasser:84}), which relates
the QCD parameters, $m$, the current quark mass, and the fermion
condensate to the observables $f_\pi$ and $m_\pi$.  We have written
everything in terms of bare quantities since the right-hand side does
not change under renormalization \cite{pascual:84}.  Thus, $m_\pi$ in
(\ref{GOR1}) is the physical pion mass.

The Gell-Mann-Oakes-Renner relation can of course also be derived from
(\ref{COND_GEN}) if one replaces
\begin{equation}
  \bar \psi (x) i \gamma_5 \psi (x) = \frac{1}{2m} f_\pi m_\pi^2 \,
  \pi (x) \; , 
\end{equation}
and assumes that, for small $m_\pi$, the sum is saturated  by the
pion.

In any case, we want to stress the fact that in (\ref{COND_GEN}) and
(\ref{GOR1}) above a vacuum quantity, the condensate, is expressed in
terms of the particle spectrum.  So, once the spectrum is known, after,
say, diagonalizing the light-cone Hamiltonian by one of the various methods on
the market, we can translate back properties of the spectrum
into properties of the vacuum.  In view of that, we suggest to
deemphasize the role of the vacuum, which is natural to the extent that
most of its properties are not directly observable. This is particularly
true within the light-cone framework, where the vacuum state seems to decouple
completely from the particle states.  Similar ideas have been put
forward long ago, in the context of chiral symmetry in the (light-cone)
parton model, by Susskind et al.~\cite{casher:71}: ``In this framework the
spontaneous symmetry breakdown must be attributed to the properties of
the hadron's wave function and not to the vacuum'' \cite{casher:74}. A related
point of view has also been taken more recently in \cite{lenz:91}.

Before  we pursue  the program  just  outlined  we would  like to
remark  that  the  `master  equation'  (\ref{MASTER})  cannot  be
derived by strictly  sticking  to the light-cone framework.   To this end
note that the first  term in (\ref{MASTER})  can be written  with
the help of the charge
\begin{equation}
  Q(x^0) = \int d^{\,3} x \, j^0 (x^0 , \vc{x} ) \; ,
\end{equation}
the generator of the symmetry, as
\begin{equation}
  \bra  0  |  \delta  B  /  \delta  \alpha  | 0 \ket  = -i  \bra  0
  |\comm{B}{Q} | 0 \ket \; ,
  \label{CHARGECOMM}
\end{equation}
where the commutator  is evaluated at equal time $x^0 = 0$.  This
expression  cannot be directly translated into the light-cone language by
replacing   the   ordinary   charge   $Q(x^0)$   by  the   light-cone  or
``light-like'' charge,
\begin{equation}
  Q(x^\p) = \int dx^\m d^2x_\perp \, j^\p (x^\p, x^\m, \vc{x}_\perp
  ) \; ,
\end{equation}
and evaluating the commutator at equal {\em light-cone} time.  The
reason for this is a peculiar property of light-like charges which was
briefly mentioned in the introduction to Chapter~7: they annihilate the
vacuum, irrespective of whether they generate a symmetry or not
\cite{leutwyler:70,jersak:68}, which is in accordance with the
triviality of the light-cone vacuum.  Thus, the right-hand side of
(\ref{CHARGECOMM}), evaluated on a null-plane, is always zero.  Hence,
that part of the operator $\delta B / \delta \alpha$ having a
non-vanishing vacuum expectation value cannot be obtained by an
infinitesimal transformation generated by the light-like charge $Q$.
For example, in the light-cone sigma model, the relation
\begin{equation}
  \comm{\pi}{Q_5} = -i \sigma \; 
  \label{SIGMA-COMM}
\end{equation}
does only hold for those modes of the field operators, $\pi$ and
$\sigma$, having non-vanishing light-cone three-momenta, $(p^\p ,
\vc{p}_\perp ) \ne 0$.  These non-zero modes do not have a vacuum
expectation value.  Thus, the vacuum expectation value of
(\ref{SIGMA-COMM}) consistently vanishes on both sides, as it should.

The moral is that we have to {\em assume} the validity of the identity
(\ref{MASTER}) for any possible choice of quantization hypersurface, in
particular for null planes.  Accordingly, we can use a complete set of
eigenstates of the {\em light-cone} Hamiltonian in the last term of
(\ref{MASTER}), as was done in \cite{chibisov:95a,burkardt:96d}. This
amounts to solving the light-cone Schr\"odinger equation for the
corresponding light-cone wave functions.

In the following, we will test the ideas outlined above for two
particular models due to  't~Hooft and Schwinger, respectively. Our
presentation is a slightly modified version of \cite{harada:98}.

\section{State of the Art}

The number of papers on the Schwinger \cite{schwinger:62} and 't~Hooft
\cite{thooft:74,thooft:75} models is legion -- for a twofold reason.
On the one hand, the two models are particularly simple and become,
for special choices of parameters, even exactly solvable.  On the
other hand, despite their simplicity, both models contain interesting
physics analogous or similar to properties of gauge theories in higher
dimensions.  The models therefore serve as interesting theoretical
laboratories for studying phenomena like confinement and chiral
symmetry breaking, to mention only the most prominent ones.  For
recent reviews on the two models we refer the reader to \cite{adam:97e}
for the Schwinger model and to \cite{abdalla:96} for the 't~Hooft model and
generalizations thereof.

While confinement (or charge screening) is realized in the same way in
both models, via a linearly rising Coulomb potential, the second
feature, chiral symmetry breaking, is not. In the 't~Hooft model, 1+1
dimensional QCD in the limit of large $N_C$, chiral symmetry is `almost'
spontaneously broken \cite{witten:78,zhitnitsky:86} and a {\em massless}
bound state arises in the chiral limit of vanishing quark mass
\cite{thooft:74,thooft:75}.  In the massive Schwinger model, QED in $d$ =
1+1 \cite{coleman:75a,coleman:76}, chiral symmetry is anomalously broken,
and the contribution of the anomaly to the mass gap survives the chiral
limit yielding the free, {\em massive} boson of the massless model.
This boson becomes interacting in the massive model and again can be
viewed as a bound state of fermionic constituents.  For the sake of
brevity we will call the lowest bound state of both models the `pion'
(although, in the real world of $d$ = 3+1, the Schwinger model boson
rather corresponds to the $\eta^\prime$-meson).  In this chapter we will try to
determine its mass and (light-cone) wave function as accurately as
possible. We mention in passing that the Schwinger model chiral anomaly
is closely related to the appearance of a vacuum $\theta$-angle
parameter \cite{lowenstein:71,kogut:75} which also affects the particle
spectrum \cite{coleman:76}.  Throughout this chapter, however, we implicitly
assume that $\theta$ is set to zero.

The calculation of bound state masses and wave functions for the two
models at hand has a long history, beginning with the exact solution of
the massless Schwinger model \cite{schwinger:62}.  For non-vanishing
`electron' mass, $m$, the model is no longer exactly solvable and one
has to resort to approximations.  One of them is to assume the mass $m$
being small and expand around the massless solution
\cite{coleman:75a,coleman:76}.  For the `pion' mass squared one expects
an expansion of the form
\begin{equation}
  M^2 (m) = 1 + M_1 m + M_2 m^2 + M_3 m^3 + O(m^4) \; ,
  \label{M2EXP}
\end{equation}
where all masses are measured  in units of the basic scale $\mu_0
= e/\sqrt{\pi}$,  the mass of the boson in the massless Schwinger
model, which is thus represented by the `1' in (\ref{M2EXP}). The
first  order  coefficient  $M_1$  was  obtained  analytically  in
\cite{banks:76} via bosonization,
\begin{equation}
  M_1 = 2 e^\gamma = 3.56215 \; . 
  \label{M_1_EX}
\end{equation}
with  $\gamma  =  0.577216$  being  Euler's  constant.    Shortly
afterwards  Bergknoff,  using light-cone  Hamiltonian  techniques
(see below), found a value \cite{bergknoff:77}
\begin{equation}
  M_1 = 2 \pi/\sqrt{3} = 3.62760 \; ,
  \label{M_1_BER}
\end{equation}
which differs from (\ref{M_1_EX})  by 1.8 \%.  One topic of this chapter
will  be the analysis  of this discrepancy  and an
attempt  to  do better  by refining  't~Hooft's  and  Bergknoff's
methods.  This  is  particularly  important  as  the  coefficient
$M_1$  is  directly  related  to  the  vacuum  structure  of  the
Schwinger model by \cite{hamer:82}
\begin{equation}
  M_1 = - 4\pi \cond \cos \theta \; ,
  \label{M_1_COND}
\end{equation}
where $\cond$ is the condensate  of the massless Schwinger  model
\cite{marinari:81,smilga:92a} (in units of $\mu_0$),
\begin{equation}
  \cond = - \frac{1}{2\pi} \e^\gamma = - 0.28347 \; ,
\end{equation}
and $\theta$ the vacuum $\theta$-angle. The result (\ref{M_1_EX}) thus
corresponds to $\theta=0$.  In \cite{heinzl:96b}, the relation
(\ref{M_1_COND}) (for $\theta=0$), which is a 1+1 dimensional analogue
of the Gell-Mann-Oakes-Renner formula (\ref{GOR1}), has been used to
determine the condensate from the `pion' mass squared via
\begin{equation}
  \cond = \left. - \frac{1}{4\pi} \frac{\partial}{\partial m} M^2
  (m) \right\vert_{m=0} = - \frac{1}{4\pi} M_1 \; .
  \label{COND}
\end{equation}
Any inaccuracy in the determination of $M_1$ thus immediately affects
the value of the condensate \cite{prokhvatilov:88,prokhvatilov:89}\footnote{ At
this point it should be mentioned that within light-cone quantization
there have been many attempts to calculate the condensate of the
massless model alternatively by solving for its vacuum structure
\cite{mccartor:91,heinzl:91b,heinzl:92b,mccartor:94a,
  mccartor:94b,mccartor:97a,kalloniatis:96a,gamboa:96}.}.

Recently,  the calculations  of $M^2 (m)$ have  been extended  to
second  order.   Using functional  integral  techniques  and mass
perturbation  theory,  Adam  \cite{adam:96b}  found  the  analytical
expression
\begin{eqnarray}
  M_2 = 4 \pi^2 \cond^2  \, (A \cos 2\theta + B) \, ,
  \label{ADAM1} 
\end{eqnarray}
where $A = -0.6599$ and $B = 1.7277$ are numerical constants, given in
terms of integral expressions containing the `pion' propagator for 
$ m = 0 $.  Inserting the values for $A$, $B$ and the condensate, and
setting $\theta=0$, (\ref{ADAM1}) becomes
\begin{equation}
  M_2 = 3.3874 \; .
  \label{ADAM2}
\end{equation}
This  result  has  been  confirmed  independently  by  Fields  et
al.~\cite{vary:96},  who derived the same integral  expressions  by
summing up all relevant Feynman diagrams in the bosonized  theory
using near-light-cone coordinates.

To first order in $m$, the mass-squared of the 't~Hooft model `pion' has
already been calculated by 't~Hooft \cite{thooft:75} by solving the
associated light-cone bound-state equation.  He derived this equation by
projecting the covariant Bethe-Salpeter equation onto three-dimensional
hypersurfaces of equal light-cone time, $x^\p$.  Soon afterwards, the
equation was rederived using light-cone Hamiltonian techniques
\cite{marinov:74}.  The result for the `pion' mass-squared is
\begin{equation}
  M^2 (m) = 2 \frac{\pi}{\sqrt{3}} m + O(m^2) \; .
  \label{M_1_THO}
\end{equation}
As the light-cone bound state equations of the two models at hand differ
only by an additive contribution due to the anomaly (see
\cite{heinzl:96b,hornbostel:90a} and below), it is not too
surprising that 't~Hooft's and Bergknoff's results coincide.  For the
't~Hooft model, all masses are expressed in units of the basic scale
$\mu_0^2 = g^2 N_C/2\pi$ \footnote{Some authors (including 't~Hooft) use
  a different convention for the coupling, which amounts to the
  replacement $g^2 \to 2 g^2$.}. It is obvious from (\ref{M_1_THO}) that
$M^2$ vanishes in the chiral limit, $m \to 0$. As explained in
\cite{witten:78,zhitnitsky:86}, this is not in contradiction with
Coleman's theorem \cite{coleman:73c} as the `pion' is not a Goldstone
boson and decouples from the $S$-matrix.

The condensate of the 't~Hooft model has first been calculated by
Zhitnitsky \cite{zhitnitsky:86} using basically the identity
(\ref{COND_GEN}),
\begin{equation}
  \cond/N_C = - \frac{1}{4\pi} M_1 = - 0.28868 \; . 
  \label{COND_TH}
\end{equation}
As expected, the condensate is proportional to $N_C$ and can again be
written in terms of $M_1$. The result has been confirmed analytically
\cite{burkardt:89a,lenz:91} and numerically \cite{li:86}. Via
(\ref{COND_TH}), the numerical value \cite{li:86} for the condensate
leads to a numerical estimate for $M_1$,
\begin{equation}
  M_1 = 3.64 \pm 0.05 \; .
  \label{LI}
\end{equation}
Though the numerical calculation did not use light-cone methods, it
seems to favor the 't~Hooft-Bergknoff value (\ref{M_1_BER}), which is
{\it the} standard value for the 't~Hooft model.  Higher order
corrections to (\ref{M_1_THO}) have been discussed in \cite{brower:79}
without explicit calculation of the expansion coefficients.

In recent years, both models have been serving as a testing ground for
new techniques developed in order to solve bound state problems using
light-cone (or, equivalently, light-front) quantization.  These new
methods are `discretized light-cone quantization' (DLCQ)
\cite{maskawa:76,pauli:85a,pauli:85b}, where one works in a finite
volume leading to an equally-spaced momentum grid (cf.~Chapters~4
and~7), and the `light-front Tamm-Dancoff approximation'
\cite{harindranath:90}, which (drastically) truncates the Fock space of
the theory (cf.~Section~7.4), thus limiting the number of constituents a
bound state can have.  The latter approach can be viewed as a
generalization of the techniques of 't~Hooft and Bergknoff.  Both
methods aim at a numerical solution of relativistic bound state
problems. DLCQ has been applied to the massive Schwinger model
\cite{eller:87} and to QCD in 1+1 dimensions (for arbitrary $N_C$)
\cite{hornbostel:90a,burkardt:89b}.  There are analogous LFTD
calculations for both models, QCD$_{1+1}$ \cite{sugihara:94,aoki:95} and
the Schwinger model \cite{mo:93,harada:95}.  These latter works are
closer in spirit to ours than the DLCQ approaches.  We will compare to
all this recent work in more detail later on.

The purpose of this chapter is to obtain the `pion' mass squared of both
the Schwinger and 't~Hooft model to high accuracy including the third
order in $m$.  In addition, we want to calculate the light-cone wave
function of the `pion' with high precision.  We will use the light-front
techniques of 't~Hooft and Bergknoff and extensions thereof.  Our
starting point is a Tamm-Dancoff truncation to the two-particle
(valence) sector.  The low order calculations will be done analytically.
To go beyond, computer algebraic and numerical methods will be applied
and their convergence tested.  The final goal is to shed some light on
the advantages and limitations of this particular approach to
bound-state equations.

The remainder of this chapter is organized as follows.  In
Section~\ref{sec:'tHooft's-Ansatz} we review 't~Hooft's ansatz for the
wave function yielding the lowest order solution of the bound-state
equation. We compare the exact endpoint analysis with a variational
procedure which is introduced at this point.  In
Section~\ref{sec:High-Order-Extension} we extend 't~Hooft's ansatz by
adding more variational parameters in such a way that an analytic
solution can still be obtained. To this end we employ computer algebraic
methods which allow to treat up to five variational parameters. This is
sufficient to achieve rather good convergence.  In
Section~\ref{sec:Comp-With-Numer}, these results are compared with the
outcome of purely numerical calculations.  We conclude in
Section~\ref{sec:Disc-Concl} with some discussion of the presented as
well as related work.

\section{'t~Hooft's Ansatz}
\label{sec:'tHooft's-Ansatz}

Our starting point is the bound state equation of the 't~Hooft and
Schwinger model in the two-particle sector, which, in a unified way, can
be written as \cite{heinzl:96b,hornbostel:90a}
\begin{eqnarray}
  M^2  \phi(x)  &=&  (m^2  -1)  \frac{\phi(x)}{x(1-x)} -  {\cal P}
  \int_0^1 dy \frac{\phi(y)}{(x-y)^2} \nn \\
  &+&  \alpha   \int_0^1   dx  \phi(x)   \;   .
  \label{BSEQ1} 
\end{eqnarray}
This expression corresponds to the first line of (\ref{PI_LCBSE}) with
only the matrix element $W_2 \equiv \bra q \bar q | W | q \bar q \ket$
retained.  We will refer to (\ref{BSEQ1}) as the 't~Hooft-Bergknoff
equation in what follows. $\phi(x)$ denotes the valence part of the
`pion' wave function, $x$ and $y$ the momentum fraction of one of the
two fermions in the meson.  The symbol $ {\cal P}$ indicates that the
integral is defined as a principal value \cite{thooft:74,gelfand:64}. It
regularizes the Coulomb singularity $1/(x-y)^2$ in the matrix element
$W_2$.

The parameter $\alpha$ measures the strength of the anomaly.  In the
't~Hooft model, $\alpha = 0$ (no anomaly), and in the Schwinger model
$\alpha = 1$ (representing the usual chiral anomaly).  The scale
parameters, as already mentioned, are given by $\mu_0^2 = g^2 N_C /2\pi$
and $\mu_0^2 = e^2/\pi$, respectively.  $M$ denotes the mass of the
lowest lying bound state (the `pion').  Our objective is to obtain
(approximate, but accurate) solutions for $M$ and $\phi$.

Upon multiplying (\ref{BSEQ1}) with $\phi (x)$ and integrating over $x$
we obtain for the eigenvalue
\begin{equation}
  M^2 (m) = (m^2  - 1) \frac{I_1}{I_0}  - \frac{I_2}{I_0}  + \alpha
  \frac{I_3^2}{I_0} \; ,
  \label{EIGENVALUE1} 
\end{equation}
where we have defined the integrals
\begin{eqnarray}
  \jot4.5pt
  I_0 &=& \int_0^1 dx \phi^2 (x) \; , \label{I0} \\ 
  I_1 &=& \int_0^1 dx \frac{\phi^2(x)}{x (1-x)} \; , \label{I1} \\
  I_2 &=& {\cal P} \int_0^1  dx dy \frac{\phi  (x) \phi (y)}{(x-y)^2}
  \; , \label{I2} \\
  I_3 &=& \int_0^1 dx \phi (x) \; . \label{I3} 
\end{eqnarray}
$I_0$  is the norm  of the wave  function,  $I_1$  and $I_2$  are
matrix  elements  of  the  mass  and  interaction   term  in  the
light-cone  Hamiltonian  \cite{bergknoff:77,mo:93}  in the  state
$|\phi  \ket$.   The integral  $I_3$ is the wave function  at the
origin.   An independent  formula for the `pion' mass-squared  is
obtained by integrating (\ref{BSEQ1})  over $x$.  This results in
the simple expression
\begin{equation}
  M^2(m) = m^2 \frac{I_4}{I_3} + \alpha \; ,
  \label{EIGENVALUE1A}
\end{equation}
with the additional integral,
\begin{equation}
  I_4 = \int_0^1 dx \frac{\phi(x)}{x(1-x)} \; .
  \label{I4}
\end{equation}
Of course, for the exact wave functions, the right-hand sides of
(\ref{EIGENVALUE1}) and (\ref{EIGENVALUE1A}) have to coincide. As the
wave functions cannot be obtained exactly, we will later use the
agreement between both values for the mass-squared as a measure for the
accuracy of our wave functions.  To determine the latter we will use a
particular class of variational ans\"atze.

In his original work on the subject, \cite{thooft:74,thooft:75},
't~Hooft used the following ansatz for the wave function
\begin{equation}
  \phi (x) = x^\beta (1-x)^\beta \, .
  \label{THO_ANS}
\end{equation}
This  ansatz  is  symmetric  in $x \leftrightarrow  1-x$  (charge
conjugation odd), and $\beta$ is supposed to lie between zero and
one  so  that  the  endpoint  behavior  is  non-analytic.  As  a
non-trivial  boundary condition one has the exact solution of the
massless case,
\begin{equation}
  M^2  = \alpha  \; , \quad  \mbox{and}  \quad  \phi  (x)  = 1 \; ,
  \quad \mbox{{\it i.e.}} \quad \beta = 0 \; . 
  \label{MASSLESS} 
\end{equation}
The main effect  of having  a non-vanishing  fermion  mass is the
vanishing  of the wave  functions  at the  endpoints  implying  a
non-zero  $\beta$.  This suggests the following  series expansion
for $\beta$,
\begin{equation}
  \beta (m) = \beta_1 m + \beta_2 m^2 + \beta_3 m^3 + O(m^4) \; ,
  \label{BETA_EXP}
\end{equation}
and for the `pion' mass squared,
\begin{equation}
  M^2 = \alpha + M_1 m + M_2 m^2 + M_3 m^3 + O(m^4) \; ,
  \label{M2_EXP}
\end{equation}
in accordance with (\ref{M2EXP}).

\subsection{Exact Endpoint Behavior}
\label{sec:Exact-Endp-Behav}

The  exponent   $\beta$  in  (\ref{THO_ANS})   can  actually   be
determined   exactly  by  studying  the  small-$x$  behavior  of
the bound state equation (\ref{BSEQ1}).   To this end we evaluate
the principal value integral for $x \to 0$,
\begin{eqnarray}
  \jot10pt
  {\cal P} \int_0^1 dy \, \frac{y^\beta (1-y)^\beta}{(x-y)^2}
  &=& x^{\beta-1}        \left[{\cal P}         \int_0^\infty  dz
  \frac{z^\beta}{(1-z)^2} + O(x) \right] \nn \\
  &=& - x^{\beta-1} \Big[ \pi\beta \cot \pi\beta + O(x) \Big] \; .
\end{eqnarray}
Plugging this into (\ref{BSEQ1}) and demanding the coefficient of
$x^{\beta-1}$  to  vanish  yields  the  transcendental   equation
\cite{thooft:74}
\begin{equation}
  m^2 - 1 + \pi \beta \cot \pi \beta = 0 \; ,
  \label{COT}
\end{equation}
which  is  independent  of  the  anomaly  $\alpha$.   Using  this
expression  we  can  determine  $\beta$  either  numerically  for
arbitrary $m$ or analytically for small $m$ by expanding
\begin{equation}
  \beta = \frac{\sqrt{3}}{\pi} m \left(1 - \frac{1}{10} m^2 \right)
  + O(m^4) \; . 
  \label{BETA_EXC} 
\end{equation}
Note that the second order coefficient,  $\beta_2$, is vanishing.
Furthermore,  for the exact  $\beta$  one has $\beta_1/\beta_3  =
- 1/10$,  which we will use as a check for our numerical  results
later on.

Our task is now to determine  the coefficients  $M_i$, $i= 1,2,3$
in  (\ref{M2_EXP}).  The  ansatz  (\ref{THO_ANS})  leads  to  the
following   results   for  the  for  the  `pion'   mass   squared
(\ref{EIGENVALUE1}),
\begin{eqnarray}
  M^2 (m, \beta) &=& (m^2 - 1) \Bigg(\frac{1}{\beta} + 4 \Bigg) \nn
  \\
  &+& \Bigg( \frac{1}{4} + \beta \Bigg) \frac{B^2(\beta , \beta)}{B
  (2\beta , 2\beta)} \Bigg[1 + \alpha \frac{\beta}{(2\beta  + 1)^2}
  \Bigg] \; . \nn \\
  \label{EIGENVALUE2}
\end{eqnarray}
In the above,  $B(z_1  , z_2)$  denotes  the Beta function.   The
relevant formulae for double integrals like $I_2$~(\ref{I2})  can
be  found  in  \cite{bardeen:80}  and  \cite{harada:94},  Appendix~C.  For
$\alpha=0$,   (\ref{EIGENVALUE2})   has  also  been  obtained  in
\cite{aoki:95}.   Let us emphasize that this result, which expresses
the `pion'  mass  squared  as a function  of the (exactly  known)
endpoint  exponent  $\beta$,  is only  approximate  as 't~Hooft's
ansatz  (\ref{THO_ANS})  for the wave  function  does  {\em  not}
represent  an exact solution of the bound-state  equation.  It is
only  the  endpoint  behavior  that  is  known  exactly;  in the
intermediate-$x$    region   't~Hooft's   ansatz   is   only   an
approximation that presumably becomes rather bad for large masses
(non-relativistic  limit)  where  the wave  function  is strongly
peaked at $x=1/2$.   The accuracy of the result will be discussed
extensively later on.

In  order  to  find  $M^2$  to  order  $m^3$  we  need  to expand
(\ref{EIGENVALUE2})  to order $\beta^3$,  as $\beta$ itself is of
order $m$. The result is
\begin{eqnarray}
  M^2  (m,  \beta)  &=&  \frac{m^2}{\beta}  +  4  m^2  +  \alpha  +
  \frac{\pi^2}{3} \beta \nonumber \\
  &+& 4 \Big[  \pi^2/3  - 3 \zeta(3)  + \alpha  (\pi^2/12
  - 1) \Big] \beta^2 \nn \\
  &+& \Big[ \frac{3}{5}  \pi^4 - 48 \zeta(3)  + 4\alpha  \Big(4 - 3
  \zeta(3) \Big) \Big] \beta^3 \nn \\
  &+& O(\beta^{\, 4}) \; . 
  \label{EIGENVALUE3} 
\end{eqnarray}
Inserting $\beta$ from (\ref{BETA_EXC}) one finds
\begin{eqnarray}
  M^2  (m)  &=&  \alpha   +  2 \frac{\pi}{\sqrt{3}} m \nonumber \\
  &+& \Bigg[  4 \bigg(2 -  \frac{9}{\pi^2} \zeta(3)  \bigg)  +
  \alpha \bigg(1 - \frac{12}{\pi^2}\bigg) \Bigg] m^2 \nonumber \\
  &+&  \frac{3\sqrt{3}}{\pi^3} \Bigg[\bigg(  \frac{3}{5}  \pi^4  - 48
  \zeta(3) \bigg) + 4 \alpha \Big(4 - 3\zeta(3) \Big) \Bigg] m^3 \nn
  \\
  &+& O(m^4) \; .
  \label{EIGENVALUE4}
\end{eqnarray}
Let us give the explicit  results for the 't~Hooft  and Schwinger
model, {\em i.e.}~for $\alpha$ = 0 and $\alpha$ =1, respectively.
For $\alpha$ = 0, we have
\begin{eqnarray}
  M_1 &=&  2\frac{\pi}{\sqrt{3}} = 3.627599 \; , \label{M21_0} \\
  M_2 &=& 4 \bigg( 2 - \zeta(3)\frac{9}{\pi^2}  \bigg) = 3.615422
  \; , \label{M22_0}\\
  M_3 &=& \frac{3\sqrt{3}}{\pi^3} \Big(  \frac{3}{5}  \pi^4  - 48
  \zeta(3) \Big) = 0.125139 \; , \label{M23_0} 
\end{eqnarray}
and for $\alpha$ = 1, 
\begin{eqnarray}
  M_1 &=&  2\frac{\pi}{\sqrt{3}} = 3.627599 \; , \label{M21_1} \\
  M_2  &=&  4  \bigg(  2 - \zeta(3)\frac{9}{\pi^2}  \bigg)  + 1 -
  \frac{12}{\pi^2} = 3.399568 \; , \label{M22_1} \\
  M_3 &=& \frac{3\sqrt{3}}{\pi^3} \left[ \Big(  \frac{3}{5}  \pi^4  - 48
  \zeta(3) \Big) + 4 \Big( 4 - 3 \zeta(3) \Big)\right] \nonumber \\ 
  &=& 0.389137 \; . 
  \label{M23_1}
\end{eqnarray}
We note that the anomaly does {\em not} contribute to the first order
term which therefore is the same in both the Schwinger and 't~Hooft
model.  This confirms the observation made in Section~8.2, the
coincidence of (\ref{M_1_BER}) and (\ref{M_1_THO}). As $M_1$ is
proportional to the condensate, {\it cf.}~(\ref{COND}) and
(\ref{COND_TH}), the latter is also independent of the anomaly
\cite{heinzl:96b}.

For the Schwinger  model ($\alpha$  = 1), the second order result
can  be  compared  with  the  analytical  calculation   of  Adam,
(\ref{ADAM2}).  Astonishingly, the relative difference is smaller
than for the first order, namely 0.36 \%.  This is accidental, as
we shall see below.

As a cross check, we determine the `pion' mass squared using the
alternative formula (\ref{EIGENVALUE1A}).  Plugging in the ansatz
(\ref{THO_ANS}), we find the fairly simple expression
\begin{equation}
  \tilde  M^2 (m, \beta) = 4 m^2 + \alpha + \frac{2  m^2}{\beta} \; .
  \label{EIGENVALUE5}
\end{equation}
Inserting the expansion (\ref{BETA_EXC}) this becomes
\begin{equation}
  \tilde  M^2 = \alpha + \tilde M_1 m + \tilde M_2 m^2 + \tilde
  M_3 m^3 + O(m^4) \; ,
  \label{EIGENVALUE5A}
\end{equation}
with the coefficients $\tilde M_i$ explicitly given by
\begin{eqnarray}
  \tilde M_1 &=&  2 \pi/\sqrt{3} = 3.627599 \; , \\
  \tilde M_2 &=& 4 \; , \\
  \tilde M_3 &=& \tilde M_1/10 = 0.03627599 \; .
\end{eqnarray}
In (\ref{EIGENVALUE5A})  the anomaly  contributes  only in zeroth
order  (as  $\beta$  is  independent  of $\alpha$),  so that  the
$\tilde  M_i$  are  the  same  for $\alpha=0$  and $\alpha  = 1$.
Comparing with (\ref{M21_0})  and (\ref{M21_1}),  we see that the
first order coefficients  coincide  for the two alternative  mass
formulae.   In addition,  one finds  that  $\tilde  M_2$  roughly
coincides with the $M_2$ of both the 't~Hooft and Schwinger model
to within  10-20 \%.  The value  for $\tilde  M_3$ is smaller  by
approximately an order of magnitude.

We  conclude   that  't~Hooft's   ansatz  works  well  to  lowest
non-trivial order in $m$ but needs improvement if one wants to go
further.   As a preparative  step  for such a development  we now
introduce  a variational  method that can easily  be extended  to
accurately calculate higher orders in $m$.

\subsection{Variational Approach}
\label{sec:Variational-Approach}

At variance  with the above we are now going to regard $\beta$ as
a variational  parameter  to  be  determined  by  minimizing  the
function $M^2 (m ,\beta)$ of (\ref{EIGENVALUE3}).  To this end we
insert  the expansion  (\ref{BETA_EXP})  into (\ref{EIGENVALUE3})
and obtain
\begin{eqnarray}
  M^2  (m)  &=&  \alpha   +  \Bigg(   \frac{\pi^2}{3}   \beta_1   +
  \frac{1}{\beta_1} \Bigg) m \nn \\
  &+& \Bigg[ 4 + \beta_2 \Big(\frac{\pi^2}{3} - \frac{1}{\beta_1^2}
  \Big) \nn \\ 
  &+&  4  \bigg(   \frac{\pi^2}{3}   -  3  \,  \zeta(3)   +  \alpha
  \Big(\frac{\pi^2}{12}  -  1 \Big)  \bigg)  \beta_1^2  \Bigg]  m^2
  \nonumber \\
  &+& \Bigg[ 8 \beta_1 \beta_2  \Big( \frac{\pi^2}{3}  - 3 \zeta(3)
  \Big) + \beta_1^3 \Big( \frac{3}{5} \pi^4 - 48 \zeta(3) \Big) \nn
  \\
  &+& \frac{\beta_2^2}{\beta_1^3}  + \beta_3 \Big(\frac{\pi^2}{3} -
  \frac{1}{\beta_1^2} \Big) \nn \\
  &+& \alpha \bigg(  2 \beta_1  \beta_2  \Big( \frac{\pi^2}{3}  - 4
  \Big) + 4 \beta_1^3  \Big( 4 - 3 \zeta(3) \Big) \bigg) \Bigg] m^3 
  \nonumber \\
  &+& O(m^4) \; .
  \label{EIGENVALUE6}
\end{eqnarray}
Note  that to order  $m$ only the leading  coefficient  $\beta_1$
contributes.  We will furthermore shortly see that the dependence
of $M_2$ on $\beta_2$ and of $M_3$ on $\beta_3$ is only apparent.

Solving the minimization equation, $\partial M^2 / \partial \beta
= 0$, for the coefficients $\beta_i$, leads to
\begin{eqnarray}
  \beta_1 &=& \sqrt{3}/\pi = 0.55133 \; , \label{BETA_1_NUM} \\
  \beta_2 &=& 0.11690 + 0.065612 \alpha \; , \label{BETA_2_NUM} \\
  \beta_3 &=& 0.0049077 - 0.050811 \alpha + 0.019521 \alpha^2 \; . 
  \label{BETA_3_NUM} 
\end{eqnarray}
Comparing with (\ref{BETA_EXC}) we note that the coefficient $\beta_1$
is exact!  Plugging it into (\ref{EIGENVALUE6}) we verify the
statement above that $M_2$ is independent of $\beta_2$ and $M_3$
independent of $\beta_3$.  Therefore, the expressions
(\ref{EIGENVALUE6}) and (\ref{EIGENVALUE4}) coincide up to and
including order $m^2$.  $M_1$ and $M_2$ are thus the same,
irrespective of whether one uses the exact endpoint exponent of
(\ref{BETA_EXC}) or its variational estimate.  The estimates
(\ref{BETA_1_NUM}-\ref{BETA_3_NUM}) for the coefficients $\beta_i$
differ in at least three respects from the exact values of
(\ref{BETA_EXC}): (i) $\beta_2$ and $\beta_3$ depend on $\alpha$, (ii)
$\beta_2 \ne 0$, (iii) $\beta_3 \ne -\beta_1/10$.  However, all these
shortcomings affect at most the third order coefficient $M_3$, and
thus become negligible for small $m$.  Only if one wants to have
reliable numbers for $M_3$, (which we will be going to produce), these
effects have to be taken into account.  The present values of
$\beta_i$ lead to
\begin{eqnarray}
  M_3 (\alpha=0) &=& 0.043597 \; , \nn \\
  M_3 (\alpha=1) &=& 0.190372 \; , 
\end{eqnarray}
which should  be compared  with (\ref{M23_0})  and (\ref{M23_1}),
respectively.  The differences are large so that the agreement is
not particularly  good.  These  discrepancies,  however,  are not
disturbing  at this point,  as we are calculating  a third  order
effect  with just one variational  parameter.   We will do better
later on by enlarging the number of these parameters until we see
satisfactory convergence of the results.

\begin{figure}
  \begin{center}
    \caption{\protect\label{fig-81}\textsl{Comparison   of  the  two
    alternative mass formulae, (\protect\ref{EIGENVALUE2}) for
    $M^2(\beta)$ and (\protect\ref{EIGENVALUE5}) for $\tilde
    M^2(\beta)$, and the second order bosonization results
    \protect\cite{adam:96b,vary:96} for $m$ = 0.1.  The vertical line
    marks the minimum of the curve $M^2(\beta)$ yielding the variational
    estimate for $\beta$ using 't~Hooft's ansatz.  For this value of
    $\beta$ one finds $M^2 (m = 0.1) = 1.3969$ and $\tilde M^2 (m = 0.1)
    = 1.3913$.  The bosonization result (short-dashed horizontal line)
    is $M^2_{\mbox{\scriptsize Bos}} (m = 0.1) = 1.390 \pm 1 \cdot
    10^{-3}$, where the error is basically an estimate of the unknown
    third order coefficient.  Thus, within 't~Hooft's ansatz, the
    alternative mass formula yields a somewhat better result at $m=0.1$.
    In the next sections we will refine our methods so that the
    variational estimates for $M^2$ and $\tilde M^2$ will converge
    towards each other.}}  \vspace{0.5cm}
\includegraphics[width=\graphicwidth]{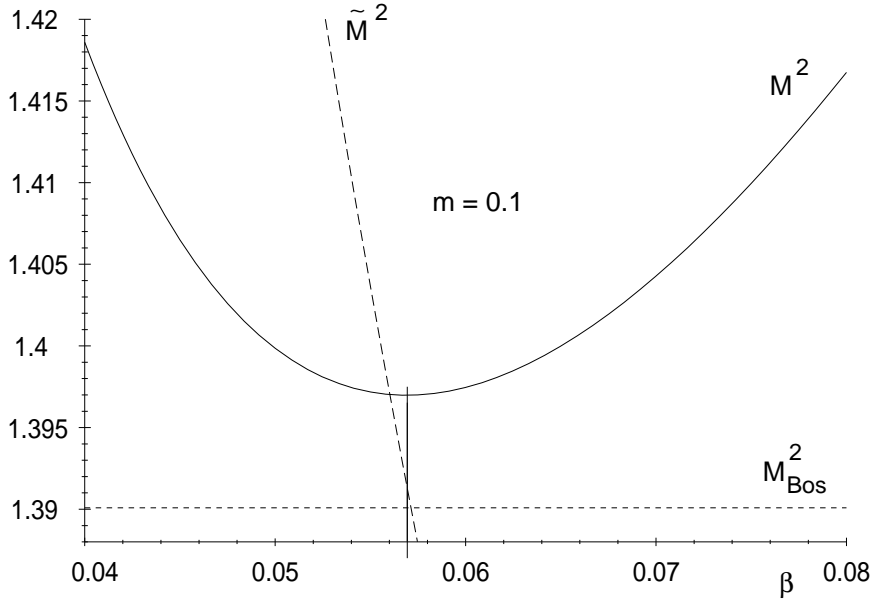}
\end{center} 
\end{figure}

The variational estimate for $\beta$ can also be plugged into the
alternative mass formula, $\tilde M^2 (m, \beta)$, (\ref{EIGENVALUE5}),
which by itself does not constitute a variational problem.  We do not
give the analytical results for the $\tilde M_i$ here but simply refer
to Fig.~\ref{fig-81} for a qualitative comparison of the two alternative
formulae, and to Section~\ref{sec:High-Order-Extension},
Tables~\ref{III} and \ref{VII}, for the actual numbers.

As stated repeatedly,  't~Hooft's ansatz (\ref{THO_ANS})  used so
far has to be extended if one wants to accurately  determine  the
second  and third order coefficients  of $M^2$.  This is our next
issue.

\section{Extension of 't~Hooft's Ansatz}
\label{sec:High-Order-Extension}

For the numerical calculations of \cite{aoki:95,harada:94,harada:95} the
following set of trial functions has been used:
\begin{equation}
  \phi (x) = \sum_{k=0}^N u_k [x(1-x)]^{\beta + k} \; .
  \label{THO_EXT1}
\end{equation}
Obviously, the term with $k=0$ and normalization $u_0 = 1$ corresponds
to 't~Hooft's original ansatz (\ref{THO_ANS}).  The coefficients $u_k$
are treated as additional variational parameters, so that, according
to the variational principle, (\ref{THO_EXT1}) {\em must} yield a
better result than (\ref{THO_ANS}).  The question is, how big the
improvement will be. To this end we will calculate the
coefficients $M_i$ by adding more and more basis functions to
't~Hooft's original ansatz (\ref{THO_ANS}), thus enlarging our
space of variational parameters. We will follow the two different
approaches mentioned above, namely, (i) use the exact $\beta$ from
(\ref{BETA_EXC}), or, (ii) treat $\beta$ as one of the variational
parameters.  For both approaches we will compare the calculated
coefficients $M_i$ with the $\tilde M_i$ obtained from the alternative
mass formula (\ref{EIGENVALUE1A}).  We will continue adding basis
functions until we see our results converge.  The maximum number $N$
of basis functions in this section will be five, {\it i.e.}~'t~Hooft's
original  wave  function  plus  four  corrections.  

A  particular
benefit  of the ansatz  (\ref{THO_EXT1})  is the  fact  that  the
integrals  $I_1$  to $I_4$  can still  be evaluated  analytically
though  the formulae  become rather lengthy.   For this reason we
will  make  heavy  use  of  the  program  package  MAPLE  in what
follows.  As a result we have analytical expressions  for all the
quantities  we calculate.   As these expressions  cover pages and
pages without being very instructive,  we do not display them but
rather the evaluated numbers.  In this sense, our treatment might
be called `semi-analytical'. A major advantage, however, of using
computer  algebraic  manipulations   is  the  high  calculational
accuracy  which is only limited  by the maximum  number of digits
the machine can handle.

\subsection{Purely Variational Approach}
\label{sec:Purely-Vari-Appr}

In  this  subsection,   $\beta$  will  always  be  treated  as  a
variational  parameter  and thus  be obtained  by minimizing  the
`pion' mass squared, $M^2$. In the ansatz (\ref{THO_EXT1}) we use
up to four additional basis functions, so that our maximum $N$ is
four. To avoid an inflation of indices we rename the coefficients
$u_k$, $k= 1, \ldots , 4$ by $a$, $b$, $c$ and $d$, respectively.
Their expansion coefficients, defined via
\begin{eqnarray}
    a   &=&   a_1   m +   a_2   m^2  \, , \nn \\
    b   &=&   b_1   m +   b_2   m^2  \, , \nn \\
    c   &=&   c_1   m +   c_2   m^2  \, , \nn \\
    d   &=&   d_1   m +   d_2   m^2  \, ,
  \label{COEFFS}    
\end{eqnarray}
are determined (recursively)  by minimizing $M^2$ with respect to
them. Let us start with the 't~Hooft model ($\alpha=0$).

\subsubsection{'t~Hooft Model}
\label{sec:'tHooft-Model}

In order to save space we do not display  all the values  for the
coefficients  in (\ref{COEFFS}).   The best values  obtained  are
given  in the last  section  when we discuss  the quality  of the
wave  function.   Here,  we rather  display  the results  for the
coefficients  $\beta_i$ of $\beta$ (see Table \ref{I}), as these can be
compared with the exact values of (\ref{BETA_EXC}).

\begin{table}
\begin{center}
\renewcommand{\arraystretch}{1.2}
\caption{\label{I}\textsl{Expansion   coefficients   of  the  end   point
exponent  $\beta$ for the 't~Hooft model obtained by successively
varying  with respect  to $\beta$  (first line) , $\beta$ and $a$
(second line), etc.  The first column, $\beta_1$,  coincides with
the result  from the exact endpoint  analysis,  which in addition
yields $\beta_2 = 0$, $\beta_3 = - \beta_1/10$.}} 
\vspace{.3cm}

\begin{tabular*}{\textwidth}[t]{l @{\extracolsep\fill} l r r }
  \hline\hline
  \it Ansatz & $\beta_1 = \sqrt{3}/\pi$ & $\beta_2$ & $\beta_3$ \\
  \hline
  't~Hooft & 0.55132889 & 0.11689763  &    0.00490773  \\ 
  a        & 0.55132889 & 0.00976951  & $-$0.04010634   \\ 
  b        & 0.55132889 & 0.00256081  & $-$0.04889172   \\ 
  c        & 0.55132889 & 0.00102086  & $-$0.05209341   \\ 
  d        & 0.55132889 & 0.00050738  & $-$0.05343097   \\ 
\hline\hline
\end{tabular*}
\end{center}
\end{table}

As an important  finding we note that $\beta_1$ remains unchanged
at the exact  value $\sqrt{3}/\pi$  (a `variational  invariant').
$\beta_2$  tends  to zero,  as it should.   The non-vanishing  of
$\beta_2$ only affects the coefficient  $M_3$, as $M_1$ and $M_2$
do not depend on $\beta_2$.  The convergence of $\beta_3$ towards
$-\beta_1/10$  seems somewhat  slow; but as all $M_i$, $i = 1, 2,
3$, are independent of $\beta_3$ this has no observable effect.

In Table~\ref{II}  we list the expansion coefficients of the mass
squared, $M^2$. One notes that the convergence of the results for
$M_2$  and $M_3$  is rather  good.   For $M_2$ we finally  have a
relative accuracy of $8 \cdot 10^{-7}$, and for $M_3$ of $4 \cdot
10^{-5}$.  Furthermore,  the coefficients  are getting smaller if
one adds more basis functions, in accordance with the variational
principle. As the coefficient $M_1$ is entirely determined by the
`variational   invariant'   $\beta_1$  it  remains  unchanged  at
$2\pi/\sqrt{3}$.   This is another important result:  $M_1$ stays
fixed at the standard 't~Hooft-Bergknoff value (\ref{M_1_BER}).

\begin{table}
\renewcommand{\arraystretch}{1.2}
\caption{\label{II} \textsl{Expansion   coefficients   of  $M^2$  for  the
't~Hooft  model obtained by successively  enlarging  the space of
variational parameters.  $M_1$ is the standard 't~Hooft-Bergknoff
result.   Note  the good convergence  towards  the bottom  of the
table.}} 
\vspace{.3cm}

\begin{tabular*}{\textwidth}[t]{l @{\extracolsep\fill} c c c }
  \hline\hline
  \it Ansatz & $M_1 = 2\pi/\sqrt{3}$ & $M_2$ & $M_3$ \\
  \hline
  't~Hooft & 3.62759873 & 3.61542218 & 0.043597197  \\ 
  a        & 3.62759873 & 3.58136872 & 0.061736701  \\ 
  b        & 3.62759873 & 3.58107780 & 0.061805257  \\ 
  c        & 3.62759873 & 3.58105821 & 0.061795547  \\ 
  d        & 3.62759873 & 3.58105532 & 0.061793082  \\ 
  \hline\hline
\end{tabular*}
\end{table}

If we use the alternative  mass formula (\ref{EIGENVALUE1A})  and
evaluate it using the wave function calculated  above we find the
values for $\tilde{M}^2$ listed in Table \ref{III}.

\begin{table}
\renewcommand{\arraystretch}{1.2}
\caption{\label{III} \textsl{Expansion  coefficients  of  the  alternative
mass squared  $\tilde  M^2$ for the 't~Hooft  model  obtained  by
successively enlarging the space of variational parameters. Again,
the   fixed   value   for   $\tilde    M_1$   is   the   standard
't~Hooft-Bergknoff result.}} 
\vspace{.3cm}

\begin{tabular*}{\textwidth}[t]{l @{\extracolsep\fill} c c c }
  \hline\hline
  \it Ansatz & $\tilde{M}_1 = 2\pi/\sqrt{3}$ & $\tilde{M}_2$ 
  & $\tilde{M}_3$ \\
  \hline
  't~Hooft & 3.62759873 & 3.23084437 & 0.130791594   \\ 
  a        & 3.62759873 & 3.54922830 & 0.066592425  \\ 
  b        & 3.62759873 & 3.57265307 & 0.059652468  \\ 
  c        & 3.62759873 & 3.57769969 & 0.060282959  \\ 
  d        & 3.62759873 & 3.57938609 & 0.060854772   \\ 
  \hline\hline
\end{tabular*}
\end{table}

We note that the values of the $\tilde M_i$ are somewhat more sensitive
to the values of the expansion coefficients $\beta_i$, as $\tilde M_2$
does depend on $\beta_2$, and $\tilde M_3$ on $\beta_2$ and $\beta_3$.
To check the quality of our wave functions one should compare the
results for $ M_i $ and $ \tilde M_i$ which are listed in
Tables~\ref{II} and~\ref{III}.  Again, it is gratifying to note that we
are improving our results step by step in the variational procedure.
The values of $\tilde M_i$ converge towards those of $M_i$.  The
relative accuracy is approximately $10^{-3}$.  This error is entirely
due to the difference between the calculated and the real wave function.
It is clear that the accuracy in the single eigenvalue $M^2$
(Table~\ref{II}) is higher than the one for the wave function, where in
principle an infinite number of points has to be calculated.  This
represents an indirect proof that we are indeed improving our wave
functions, and not just the eigenvalues.

\subsubsection{Schwinger Model}
\label{sec:Schwinger-Model}

For the Schwinger  model ($\alpha$ = 1), we perform  exactly  the
same calculations.   The expansion  coefficients  of $\beta$  are
listed in Table~\ref{IV}.
Although  the  coefficients  $\beta_2$,  $\beta_3$  are  somewhat
different  from  those  in Table \ref{I} (which  they  must be as they
lose their dependence  on $\alpha$  only in the strict  limit  of
{\it  exact}  evaluation)  we note the same tendency:   $\beta_2$
converges to zero and $\beta_3$ towards $-\beta_1/10$.

\begin{table}
\renewcommand{\arraystretch}{1.2}
\caption{\label{IV}   \textsl{Expansion   coefficients   of  the  endpoint
exponent $\beta$ for the Schwinger model obtained by successively
enlarging the space of variational parameters.  The first column,
$\beta_1$,  coincides  with  the result  from the exact  endpoint
analysis,  which in addition yields $\beta_2  = 0$, $\beta_3  = -
\beta_1/10$.}} 
\vspace{.3cm}

\begin{tabular*}{\textwidth}[t]{l @{\extracolsep\fill} c c c }
  \hline\hline
  \it Ansatz & $\beta_1$ & $\beta_2$ & $\beta_3$ \\
  \hline
  't~Hooft & 0.55132889 & 0.18250945 & $-$0.026382222  \\ 
  a        & 0.55132889 & 0.01177681 & $-$0.040372413  \\ 
  b        & 0.55132889 & 0.00317437 & $-$0.048081756  \\ 
  c        & 0.55132889 & 0.00127507 & $-$0.051602713  \\ 
  d        & 0.55132889 & 0.00063609 & $-$0.053130398  \\ 
  \hline\hline
\end{tabular*}
\end{table}

The best values for the `pion' mass squared are again provided by
the  variational  results  listed  in  Table~\ref{V}.   The  numerical
accuracy is practically  the same as for the analogous  Table~\ref{II}.
The comparison  with  the alternatively  calculated  coefficients
$\tilde M^2$ is given in Table~\ref{VI}.
Again,  everything  is completely  analogous  to the case  of the
't~Hooft model ($\alpha=0$).

\begin{table}
\renewcommand{\arraystretch}{1.2}
\caption{\label{V} \textsl{Expansion  coefficients of $M^2$ for the Schwinger model
  obtained by successively enlarging the space of variational
  parameters.  Again, the fixed value for $M_1$ is the standard
  't~Hooft-Bergknoff result.  Note the good convergence towards the
  bottom of the table.}}

\vspace{.3cm}
\begin{tabular*}{\textwidth}[t]{l @{\extracolsep\fill} c c c }
  \hline\hline
  \it Ansatz & $M_1 = 2\pi/\sqrt{3}$ & $M_2$ & $M_3$ \\
  \hline
  't~Hooft & 3.62759873 & 3.39956798 & 0.19037224  \\ 
  a        & 3.62759873 & 3.30906326 & 0.34776772  \\ 
  b        & 3.62759873 & 3.30864244 & 0.34820193  \\ 
  c        & 3.62759873 & 3.30861240 & 0.34820513  \\ 
  d        & 3.62759873 & 3.30860791 & 0.34820389  \\ 
  \hline\hline
\end{tabular*}
\end{table}

\begin{table}
\renewcommand{\arraystretch}{1.2}
\caption{\label{VI}  \textsl{Expansion  coefficients  of the alternative
mass squared  $\tilde  M^2$ for the Schwinger  model obtained  by
successively  enlarging  the  space  of  variational  parameters.
Again,  the  fixed  value  for  $\tilde   M_1$  is  the  standard
't~Hooft-Bergknoff result.}}

\vspace{.3cm}  
\begin{tabular*}{\textwidth}[t]{l @{\extracolsep\fill} c  c  c  } 
\hline\hline
\it  Ansatz  & $\tilde{M}_1  =
  2\pi/\sqrt{3}$ & $\tilde{M}_2$ & $\tilde{M}_3$ \\ \hline
  't~Hooft  &  3.62759873  &  2.79913596  & 0.57111672  \\  
  a         &  3.62759873  &  3.27031909  & 0.36868190  \\ 
  b         &  3.62759873  &  3.29819920  & 0.34869835  \\ 
  c         &  3.62759873  &  3.30441759  & 0.34753576  \\ 
  d         &  3.62759873  &  3.30651525  & 0.34762562  \\
  \hline\hline
\end{tabular*} 
\end{table}

\subsection{Variational Approach using the exact $\vcg{\beta}$}
\label{sec:Vari-Appr-using}

In this subsection we calculate $M^2$ and $\tilde M^2$ for both
values of $\alpha$ using the exact value (\ref{BETA_EXC}) for $\beta$.
Thus only $a$, $b$, $c$, $d$ are treated as variational parameters.
This is the procedure employed numerically in \cite{harada:95}. From the
discussion of the preceding subsection, in particular the values for
the $\beta_i$ displayed in Table~\ref{I}, which differ minimally from
the exact values, we expect that the results will be very close to
those from the purely variational approach.

\subsubsection{'t~Hooft Model}
\label{sec:'tHooft-Model-2}

This is exactly what happens as can be seen from Tables~\ref{VII} and
\ref{VIII}.  Comparing Table~\ref{VII} with Table~\ref{II} one finds
that the coefficients $M_1$ and $M_2$ of both tables coincide as these
only depend on $\beta_1$ which is the same in both approaches.  Only for
$M_3$ there are slight differences, due to the dependence on $\beta_2$.
For the $\tilde M_i$, $i=2,3$, the discrepancies are somewhat bigger as
these coefficients do depend on $\beta_2$ and $\beta_3$ (Table~\ref{III}
{\em vs.}~Table~\ref{VIII}). Still the consistency is quite obvious.

\begin{table}
\renewcommand{\arraystretch}{1.2}
\caption{\label{VII}  \textsl{Expansion  coefficients  of $M^2$  ('t~Hooft
model) obtained by using the exact endpoint exponent  $\beta$ and
successively  enlarging  the  space  of variational  parameters..
Again,    the   fixed   value   for   $M_1$   is   the   standard
't~Hooft-Bergknoff result.}}

\vspace{.3cm}

\begin{tabular*}{\textwidth}[t]{l @{\extracolsep\fill} c c c } 
  \hline\hline
  \it Ansatz & $M_1 = 2\pi/\sqrt{3}$
  & $M_2$ & $M_3$ \\ \hline
  't~Hooft & 3.62759873  & 3.61542219  & 0.12513878 \\
  a        & 3.62759873  & 3.58136873  & 0.06230622 \\ 
  b        & 3.62759873  & 3.58107781  & 0.06184441 \\
  c        & 3.62759873  & 3.58105821  & 0.06180178  \\ 
  d        & 3.62759873  & 3.58105533  & 0.06179462 \\ 
  \hline\hline
\end{tabular*} 
\end{table}

\begin{table}
\renewcommand{\arraystretch}{1.2}
\caption{\label{VIII}  \textsl{Expansion  coefficients  of the alternatively
defined  mass squared $\tilde M^2$ ('t~Hooft  model) obtained  by
using  the  exact  endpoint  exponent  $\beta$  and  successively
enlarging the space of variational  parameters.  Again, the fixed
value  for  $\tilde  M_1$  is  the  standard   't~Hooft-Bergknoff
result.}}

\vspace{.3cm}

\begin{tabular*}{\textwidth}[t]{l @{\extracolsep\fill} l l l }
  \hline\hline
  \it Ansatz & $\tilde{M}_1 = 2\pi/\sqrt{3}$ 
  & $\tilde{M}_2$ & $\tilde{M}_3$ \\
  \hline
  't~Hooft & 3.62759873 & 4.0        & 0.36275987   \\ 
         a & 3.62759873 & 3.61350915 & 0.08659575  \\ 
         b & 3.62759873 & 3.58950254 & 0.07106398  \\ 
         c & 3.62759873 & 3.58441672 & 0.06603497  \\ 
         d & 3.62759873 & 3.58272458 & 0.06406179  \\ 
  \hline\hline
\end{tabular*}
\end{table}

Upon comparing the second columns of Tables~\ref{II} and
\ref{III}, respectively Tables~\ref{VII} and \ref{VIII}, one notes a funny
coincidence. In the first case $M_2 $ is slightly bigger than $\tilde
M_2$, and vice versa in the second case.  If one denotes the purely
variational results with a superscript `$v$', and the results obtained
with the exact $\beta$ with a superscript `$e$', one finds that $M_2$,
which is the same in both approaches, is given by the arithmetic mean
\begin{equation}
  M_2 = \frac{1}{2} \left(\tilde M_2^e + \tilde M_2^v \right) \; .
\end{equation}
We have checked this analytically for 't~Hooft's ansatz.  For the
higher orders this is difficult  to do but the numerical evidence
is beyond doubt.

Comparing the third columns of Tables~\ref{II} and \ref{VII}, one
finds that $ M_3^v < M_3^e $.  We thus see a slight tendency that the
results of the purely variational procedure are better than those
obtained using the exact $\beta$.  The same observation has been made
by Mo and Perry \cite{mo:93}, in particular for large values of the
fermion mass $m$ (where the third order coefficient $M_3$ becomes
important).  We interpret this fact as a hint that for larger $m$ the
behavior of the wave function in the intermediate region becomes more
relevant compared to the endpoint behavior.

\subsubsection{Schwinger Model}
\label{sec:Schwinger-Model-2}

For the Schwinger model, the analogous results are listed in
Tables \ref{IX} and \ref{X}.  Exactly the same remarks as above
apply including the size of the errors and the relation between $M_2$
and $\tilde M_2^e$, $\tilde M_2^v$.  In the next section we will
verify our results by purely numerical methods.

\begin{table}
\renewcommand{\arraystretch}{1.2}
\caption{\label{IX}  \textsl{Expansion coefficients  of $M^2$ (Schwinger 
model) obtained by using the exact endpoint exponent  $\beta$ and
successively  enlarging  the  space  of  variational  parameters. 
Again,    the   fixed   value   for   $M_1$   is   the   standard
't~Hooft-Bergknoff result.}}

\vspace{.3cm}
\begin{tabular*}{\textwidth}[t]{l @{\extracolsep\fill} c c c }
  \hline\hline
  \it Ansatz & $M_1 = 2\pi/\sqrt{3}$ & $M_2$ & $M_3$ \\
  \hline
  't~Hooft & 3.62759873 & 3.39956798 & 0.38913656  \\ 
         a & 3.62759873 & 3.30906326 & 0.34859533  \\ 
         b & 3.62759873 & 3.30864244 & 0.34826210  \\ 
         c & 3.62759873 & 3.30861239 & 0.34821485  \\ 
         d & 3.62759873 & 3.30860791 & 0.34820630   \\ 
  \hline\hline
\end{tabular*}
\end{table}

\begin{table}
\renewcommand{\arraystretch}{1.2}
\caption{\label{X} \textsl{Expansion coefficients of the alternatively
defined mass squared $\tilde M^2$ (Schwinger  model) obtained  by
using  the  exact  endpoint  exponent  $\beta$  and  successively
enlarging the space of variational  parameters.  Again, the fixed
value  for  $\tilde  M_1$  is  the  standard   't~Hooft-Bergknoff
result.}}

\vspace{.3cm}
\begin{tabular*}{\textwidth}[t]{l @{\extracolsep\fill} l l l }
  \hline\hline
  \it Ansatz & $\tilde{M}_1 = 2\pi/\sqrt{3}$ 
  & $\tilde{M}_2$ & $\tilde{M}_3$ \\
  \hline
  't~Hooft & 3.627598730 & 4.0         & 0.3627598730  \\ 
         a & 3.627598730 & 3.347807427 & 0.3637551073  \\ 
         b & 3.627598730 & 3.319085690 & 0.3566261465  \\ 
         c & 3.627598730 & 3.312807215 & 0.3523062162  \\ 
         d & 3.627598730 & 3.310700593 & 0.3504584408   \\ 
  \hline\hline
\end{tabular*}
\end{table}

\section{Comparison With Numerical Results}
\label{sec:Comp-With-Numer}

For the numerical calculations the ansatz (\ref{THO_EXT1}) has been
used.  The value for $\beta$ has been determined numerically from
(\ref{COT}).  Again, up to four additional basis functions have been
included (for larger masses even five).  Let us begin with the Schwinger
model ($\alpha=1$).  Here we can use the code developed in \cite{harada:95}.
In Table~\ref{XI} we list $M^2 -1$ as a function of the fermion mass
$m$, calculated within two-particle light-front Tamm-Dancoff
approximation. The notations a,b,c,d are as in the preceding section.

\begin{table}
\renewcommand{\arraystretch}{1.5}
\caption{\label{XI}   \textsl{Numerical   results   using  lowest   order
(two-particle)  light-front Tamm-Dancoff approximation  for $M^2 -1$ as a function of
the fermion mass $m$, for $\alpha = 1$ (Schwinger model).}}

\scriptsize

\vspace{.3cm}
\begin{tabular*}{\textwidth}[t]{l @{\extracolsep\fill} c c c c c c c  }
\hline\hline
\it Ansatz & $m$ = 0.0001 & $m$ = 0.0005 & $m$ = 0.001 & $m$ = 0.005 & $m$ = 0.01 & $m$ = 0.05 & $m$ = 0.1  \\ 
\hline  
't~Hooft  & 0.000362794 & 0.00181465 & 0.00363100 & 0.0182230 & 0.0366163 & 0.189927 & 0.397130   \\
a         & 0.000362858 & 0.00181465 & 0.00363086 & 0.0182207 & 0.0366072 & 0.189695 & 0.396181   \\
b         & $-$         & 0.00181461 & 0.00363100 & 0.0182210 & 0.0366071 & 0.189694 & 0.396176   \\
c         & $-$         & 0.00181459 & $-$        & $-$       & 0.0366071 & 0.189694 & 0.396176   \\
d         & $-$         & $-$        & $-$        & $-$       & 0.0366073 & 0.189694 & 0.396176  \\
\hline\hline
\end{tabular*}
\normalsize
\end{table}

One main difference in comparison with the computer algebraic treatment
of Section~\ref{sec:High-Order-Extension} is the numerical inaccuracy
for small $m$.  This is a general disease of numerical treatments, and
also shared e.g.~by the lattice \cite{hamer:82,marinari:81} or
DLCQ approach \cite{eller:87}.  In these approaches, however, the
numerical errors are typically much larger than ours (see the next
section for an explicit comparison).  Due to the small-$m$ instability,
our numerical calculation does not converge within an arbitrary number
of digits.  As soon as a calculated value for $M^2 - 1$ becomes bigger
as the preceding one (a numerical violation of the variational
principle), we terminate the procedure and pick the smaller value as our
final result.  The difference between these last two values can be used
as an estimate of the numerical inaccuracy.  The errors will be further
discussed below and in the next section when we compare our results with
related work (see Tables~\ref{XV}, \ref{XVI}).

It should  also be pointed  out that  the numerical  analysis  is
conceptually   very   different   from  the  computer   algebraic
treatment.  As one cannot perform a Taylor expansion numerically,
the  expansion  coefficients  of  $M^2$  have  to be obtained  by
fitting polynomials to $M^2 (m)$.  Clearly, this is an additional
source of errors, and one expects the results to be less accurate
than those of the preceding sections.  A cubic fit to the optimum
values  for $M^2 -1$ in Table~\ref{XI}  yields the following  expansion
coefficients of $M^2$,
\begin{eqnarray}
  M_1 &=&  3.62609 \; , \label{EVNUM1_31} \\ 
  M_2 &=&  3.33607 \; , \\
  M_3 &=&  0.22396 \label{EVNUM1_33} \; ,
\end{eqnarray}
which  should  be compared  with the last  line of Table~\ref{V}.   To
estimate  the accuracy  of these values  we also show the results
from a quartic fit, 
\begin{eqnarray}
  M_1 &=&  3.62755 \; , \label{EVNUM1_41} \\ 
  M_2 &=&  3.31029 \; , \\
  M_3 &=&  0.32984 \; . \label{EVNUM1_43}
\end{eqnarray}
Comparing (\ref{EVNUM1_31}-\ref{EVNUM1_33}) and
(\ref{EVNUM1_41}-\ref{EVNUM1_43}) one finds that the stability of the
fits is not too impressive in view of the accuracy we would like to
achieve.  In particular the third order coefficient is numerically
difficult to determine.  We estimate the relative accuracy as being
$10^{-3}$ for $M_1$, $10^{-2}$ for $M_2$ and $10^{-1}$ for $M_3$.  This
is also confirmed by comparing with Table~\ref{V}.

For the Schwinger  model it is possible to check the influence of
higher  particle  sectors  on the `pion'  mass squared  by using  the
machinery developed in \cite{harada:95} for the wave functions of the
higher Fock components.   These are the amplitudes of finding not
only two, but four, six, ...  (anti-)fermions  in the `pion'.  In
Table~\ref{XII},  we list the best values for $M^2-1$  including  up to
six-body  wave  functions.   We also show  the 2-, 4-, and 6-body
content of the total wave function.   It is known that the `pion'
is entirely 2-particle in the chiral limit \cite{mo:93}. For small
mass,  one therefore  expects  only small contributions  from the
higher Fock sectors.  This is confirmed by the numerical results.
Astonishingly,  this feature  persists  up to values  of at least
$m=0.5$ for the fermion mass.  This fact is in agreement with the
observation  of Mo and Perry that the four-particle  component of
the wave function is less than 0.4~\% for {\em all} values of the
fermion mass \cite{mo:93}.  A similar result has been found in the
DLCQ calculations of \cite{eller:87}. In addition we note that there
seems  to  be some  interesting  kind  of hierarchy  between  the
relative strengths of the contributions  from different  particle
sectors.   If we denote the $2k$-particle  amplitude  in the wave
function  by $f_{2k}$,  we find  that  $|f_2|^2  \gg |f_4|^2  \gg
|f_6|^2$,  the individual  proportions  being several  orders  of
magnitude  (see Table~\ref{XII}).   Here,  we explicitly  see the
magic of light-front  field theory at work:  high Fock components
in bound states tend to be largely suppressed,  at variance  with
the situation encountered  in field theory quantized the standard
way.

\begin{table}
\renewcommand{\arraystretch}{1.5}
\caption{\label{XII} \textsl{Numerical Results for $M^2 -1$ as a function
of the fermion  mass  $m$,  for  $\alpha  = 1$ (Schwinger  model)
including 2-, 4- and 6-body wave functions.  The contributions of
the different Fock sectors to the total wave function squared are
given in percent.}}

\scriptsize

\vspace{.3cm}
\begin{tabular*}{\textwidth}[t]{l @{\extracolsep\fill} c c c c c c c  }
  \hline\hline
  \it       & $m$ = 0.0001      & $m$ = 0.0005     & $m$ = 0.001      & $m$ = 0.005     & $m$= 0.01 & $m$ = 0.05 & $m$ = 0.1  \\   \hline
  $M^2 - 1$ & 0.000362793 & 0.00181481 & 0.00363055 & 0.0182179 & 0.0365968 & 0.189468  & 0.395400   \\ 
  \% 2-body & 100.00      & 100.00     & 100.00     & 99.999990 & 99.999962 & 99.999213 & 99.997514  \\
  \% 4-body & $-$         & $-$        & $-$        & 0.000010  & 0.000038  & 0.000782  & 0.002474   \\
  \% 6-body & $-$         & $-$        & $-$        & $-$       & $-$       & 0.000005  & 0.000013   \\ 
\hline\hline  
\end{tabular*} 
\normalsize  
\end{table}

From the point of view of the variational  principle, the results
shown  in Table~\ref{XII}  are a bit better  (i.e.~smaller)  than
those of Table~\ref{XI} (apart from the value for $m=0.0005$),  but the
improvement is rather small. A cubic fit yields
\begin{eqnarray}
  M_1 &=&  3.62667 \; , \label{EVNUM2_31} \\ 
  M_2 &=&  3.23696 \; , \\
  M_3 &=&  0.36235 \; ,\label{EVNUM2_33}
\end{eqnarray}
and a quartic fit
\begin{eqnarray}
  M_1 &=&  3.62747 \; , \label{EVNUM2_41} \\ 
  M_2 &=&  3.20864 \; , \\
  M_3 &=&  0.60365 \; .\label{EVNUM2_43}
\end{eqnarray}
Compared with (\ref{EVNUM1_31}-\ref{EVNUM1_33}) and
(\ref{EVNUM1_41}-\ref{EVNUM1_43}) we do not find any absolute
improvement in $M_1$ that could be distinguished from the numerical
inaccuracy.  For the coefficient $M_2$, which in bosonization schemes is
3.3874, we even get the wrong tendency: it becomes smaller upon
including higher Fock states.  The inaccuracy for $M_3$ is so large that
this coefficient is only determined in its order of magnitude.

Altogether,  we arrive at the very important conclusion  that the
inclusion  of higher particle  sectors  in the light-front  bound
state  equation  of the Schwinger  model  does  not diminish  the
discrepancy    between    the    results    obtained    via   the
't~Hooft-Bergknoff method and those from bosonization techniques.

Let us move to the 't~Hooft  model.  If we put $\alpha=0$  in the
Schwinger  model  code  above  (Code~I)  we find the `pion'  mass
squared  for the 't~Hooft  model.   The best  values  within  the
2-particle  sector  are listed  in Table~\ref{XIII}.   Note that higher
particle  sectors  are  strictly  suppressed  in the  large-$N_C$
limit.  Thus, there is no point in calculating  these, unless one
is  interested  in  $1/N_C$  corrections  to the  't~Hooft  model
results \cite{hornbostel:90a,burkardt:89b,sugihara:94}.

\begin{table}
\renewcommand{\arraystretch}{1.2}
\caption{\label{XIII}  \textsl{$M^2$  as a function  of the fermion  mass
$m$, for $\alpha  = 0$ ('t~Hooft  model),  obtained  with Code I.}}
\vspace{.3cm}

\begin{tabular*}{\textwidth}[t]{l @{\extracolsep\fill} l l l l }
  \hline\hline
  \it $m$ & 0.0001 & 0.001 & 0.01 & 0.1 \\
  \hline
  $  M^2$  & 0.000362795 & 0.00363109 & 0.0366342 & 0.398634 \\ 
  \hline\hline
\end{tabular*}
\end{table}

\begin{table}
\renewcommand{\arraystretch}{1.2}
\caption{\label{XIV}  \textsl{$M^2$ as a function  of the fermion  mass
$m$, for $\alpha = 0$ ('t~Hooft model), obtained with Code II.}} 
\vspace{.3cm}

\begin{tabular*}{\textwidth}[t]{l @{\extracolsep\fill} l l l l }
  \hline\hline
  \it $m$ & 0.0001 & 0.001 & 0.01 & 0.1 \\
  \hline
  $  M^2$  & 0.000362795 & 0.00363118 & 0.0366341 & 0.398634 \\ 
  \hline\hline
\end{tabular*}
\end{table}

In view of the restricted  number of data points only a quadratic
fit makes sense which yields 
\begin{eqnarray}
  M_1 &=&  3.62754 \; , \label{EVNUM31} \\ 
  M_2 &=&  3.58796 \; ,\label{EVNUM32}
\end{eqnarray}
which is consistent with the results of
Section~\ref{sec:High-Order-Extension} (see Table~\ref{II}).  A second
code (Code II), which was independently developed for the 't~Hooft
model \cite{sugihara:94} yields the results displayed in Table~\ref{XIV}.
From a quadratic fit we obtain
\begin{eqnarray}
  M_1 &=&  3.62754 \; , \label{EVNUM41} \\ 
  M_2 &=&  3.58806 \; .\label{EVNUM42}
\end{eqnarray}
Comparing with (\ref{EVNUM31}, \ref{EVNUM32}) we see that both codes
yield the same results within the numerical accuracy.  This is
reassuring, since, as already stated, the codes were developed
independently from each other.

A cubic fit to a total  of 30 data points  produced  with Code~II
yields
\begin{eqnarray}
  M_1 &=& 3.62758 \; , \label{EVNUM51} \\
  M_2 &=& 3.58260 \; , \\
  M_3 &=& 0.06450 \; ,\label{EVNUM52}
\end{eqnarray}
in fair  agreement  with  the variationally  obtained  values  of
Table~\ref{II}.  To check the stability of this fit we can compare with
the extension to fourth order,
\begin{eqnarray}
  M_1 &=& 3.62756 \; , \label{EVNUM61} \\ 
  M_2 &=& 3.58314 \;,  \\
  M_3 &=& 0.06280 \; , \\ 
  M_4 &=& 0.001232 \; .\label{EVNUM64}
\end{eqnarray}
This leads to an estimate  of roughly  $10^{-6}$,  $10^{-4}$  and
$10^{-2}$ for the (absolute) numerical error in the first, second
and third order  coefficient,  respectively.   Furthermore  it is
gratifying  to note that the fourth order coefficient,  $M_4$, is
numerically small.

As is obvious from the discussion above, the numerical errors for
the 't~Hooft  model are much smaller than those for the Schwinger
model.  The basic reason for this is the Schwinger model anomaly.
Note that in the bound state equation the anomaly factor $\alpha$
multiplies an integral over the wave function.  To evaluate this,
one needs the wave function as a whole. In the 't~Hooft model, on
the other  hand,  this  term  is absent  and the eigenvalues  are
dominated entirely by the endpoint behavior of the wave function
which  is known exactly.   This makes the numerical  errors  much
smaller (see also Tables~\ref{XV}, \ref{XVI}).

\section{Discussion}
\label{sec:Disc-Concl}

In  the  preceding  sections  we have  calculated  the  expansion
coefficients $M_i$ in the series
\begin{equation}
  M^2 = \alpha + M_1 m + M_2 m^2 + M_3 m^3
  \label{MEXP}
\end{equation}
for the `pion' mass squared of both the 't~Hooft ($\alpha=0$) and
the Schwinger  model  ($\alpha=1$).   In order  for the expansion
(\ref{MEXP})  to make sense we have considered  only small masses
$m \ll 1$ {\em i.e.}~$m \ll \mu_0$ in the original units. For the
't~Hooft  model we have $\mu_0^2  = g^2 N_C /2\pi$  with $N_C \to
\infty$, $g^2 N_C$ fixed.  Thus, we are working in the {\em weak}
coupling  phase where the limit $N_C \to \infty$  ($g \to 0$), is
taken  {\em before}  the limit $m \to 0$, or, equivalently,  such
that  one  always   has  $m  \gg  g  \sim  1/\sqrt{N_C}   \to  0$
\cite{zhitnitsky:86,zhitnitsky:96b}.   For the Schwinger model, $\mu_0^2 =
e^2 /\pi$, so that small fermion mass corresponds to {\em strong}
coupling.

We have used analytical, computer algebraic and numerical methods
of  different  accuracy.  Nonetheless,  the  overall  picture  is
intrinsically  consistent.  Our main findings are the following:

(i) The first order coefficient,  $M_1$, in the expansion  of the
`pion' mass squared is independent  of the anomaly $\alpha$, {\em
i.e.}~the  same in both the 't~Hooft  and Schwinger  model.  This
confirms  the results of 't~Hooft \cite{thooft:74, thooft:75}   and
Bergknoff \cite{bergknoff:77}.

(ii) The 't~Hooft-Bergknoff  value, $M_1 = 2\pi /\sqrt{3}$,  is a
`variational invariant': it does not get altered by extending the
space   of  variational   parameters.    This  has  been  checked
analytically and numerically. 

(iii) In the Schwinger  model, the 't~Hooft-Bergknoff  value does
not change upon inclusion  of higher particle sectors.   Only the
second  and  third  order  coefficients,  $M_2$  and  $M_3$,  are
affected.

(iv) Thus, for the Schwinger  model, there remains  a few percent
discrepancy  in the coefficients  $M_1$  and  $M_2$  compared  to
bosonization results.

(v) The variational  calculation  yields the `pion' wave function
to a high accuracy.  The endpoint behavior is reproduced exactly
(in leading  order  in $m$).   The behavior  in the intermediate
region,  $0 < x < 1$, gets improved  as can be seen  from  Fig.~\ref{fig-83}
below.

In Table \ref{XV} we summarize  our  optimum  final  results  for the
't~Hooft model ($\alpha$  = 0) and compare with results that have
been obtained previously.

\begin{table}
\renewcommand{\arraystretch}{1.3}
\caption{\label{XV}  \textsl{The expansion  coefficients  $M_i$  for the
  't~Hooft model ($\alpha$ = 0). The errors of our results obtained
  within 2-particle Tamm-Dancoff (2PTD) approximation are estimated by
  comparing the last two lines in Table~\ref{II} (for the variational
  method) and the different polynomial fits (\protect\ref{EVNUM31} -
  \protect\ref{EVNUM64}) (for the numerical results).  The numerical inaccuracies
  given in the last line were estimated  by us using the polynomial
  fit method.}}

\vspace{.3cm}
\small
\begin{tabular*}{\textwidth}[t]{l @{\extracolsep\fill}  l  l  l  } 
\hline\hline
& $M_1$ & $M_2$ & $M_3$ \\ \hline
variational 2PTD & $2\pi/\sqrt{3}$ = 3.627599 & 3.581055 $\pm \, 3
\cdot 10^{-6}$ & 0.061793 $\pm \, 3 \cdot 10^{-6}$ \\ 
numerical  2PTD & 3.62758  $\pm \, 2\cdot 10^{-5}$  & 3.5829 $\pm
\, 3 \cdot 10^{-4}$ & 0.064 $\pm \, 1 \cdot 10^{-3}$ \\
't~Hooft \cite{thooft:75}  & $ 2\pi/\sqrt{3}$  = 3.627599 & $-$ & $-$
\\
Burkardt  \cite{burkardt:89a}  & $2\pi/\sqrt{3}$  = 3.627599 & 3.5812 &
$-$ \\ 
Li \cite{li:86} & 3.64 $\pm$ 0.06 & $-$ & $-$ \\
Li et al. \cite{birse:87} & 3.64 $\pm 0.03$ & 3.60 $\pm 0.06$ & 0.04
$\pm 0.04$ \\
\hline\hline
\end{tabular*} 
\normalsize
\end{table}

Upon inspection of Table~\ref{XV} one finds a very good overall
consistency of the results.  All given values are in good agreement
within error bars.  The scale is set by the variational results which
are the most accurate ones. Within error bars, they are matched by our
numerical results as well as by previous analytical and numerical
calculations.

As far as the latter are concerned,  a few remarks  are in order.
The  coefficients   $M_i$   displayed   in  the  last   line   of
Table~\ref{XV} are obtained by performing polynomial fits to the
values calculated  numerically  in \cite{birse:87}  (which basically
agree with those from 't~Hooft's  original calculation  displayed
in Fig.~5 of \cite{thooft:74}).   The averaged  result of the fits is
shown including  an estimate  of the errors.   It should  also be
stressed that the data in \cite{birse:87} are mainly obtained for $m
> 0.5$ whereas  the bulk of our values for the $M_i$ are obtained
for $m<0.1$.   Nevertheless,  the agreement  with our results  is
satisfactory within error bars.

From Table~\ref{XV} it is obvious that all numerical results favor the
first order 't~Hooft-Bergknoff value of $2\pi/\sqrt{3}$.  This agreement
is particularly gratifying for the data of \cite{birse:87} which were
not obtained via light-cone techniques but within ordinary quantization
and even within a different gauge, namely axial gauge.  The agreement is
therefore highly non-trivial \cite{bars:78a}.

\begin{figure}
\begin{center}
\caption{\protect\label{fig-82} \textsl{The  rescaled `pion' mass $\bar M = M/\protect\sqrt{1  +
    m^2}$ as a function of the fermion mass $m$.  The short-dashed
  curve represents a `phenomenological' parametrization for $\bar
  M$ with $M^2 = 1 + M_1 m + 4 m^2$, which becomes exact for very
  small and very large $m$ and thus smoothly interpolates between the
  strong and weak coupling regions.  The long-dashed curve is our
  second-order and the solid curve our third-order result.  As
  expected, mass perturbation theory breaks down as $m$ becomes of
  order 1.  The crosses are the lattice results \protect\cite{crewther:80},
  the diamonds the lattice results \protect\cite{carson:86}, which go down
  to comparatively small masses, however with large errors.  The
  circles are light-front Tamm-Dancoff results \protect\cite{mo:93}, whereas the squares
  \protect\cite{eller:87} and triangles \protect\cite{elser:94}, as quoted
  in \protect\cite{vary:96}, are DLCQ results.  The values corresponding
  to the triangles are {\em not} included in Table~\ref{XVI}.  The 2\%
  discrepancy in $M_1$ is invisible within the resolution of the
  figure.}}  
\vspace{3cm}
\includegraphics[width=\graphicwidth]{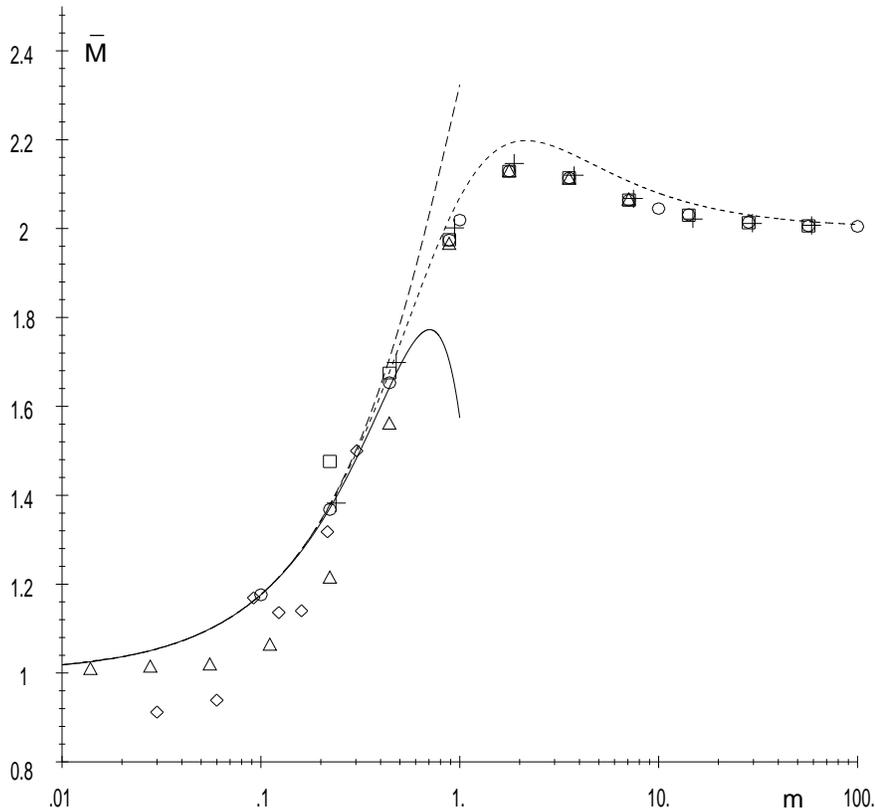}
\end{center}
\end{figure}

For the Schwinger model, ($\alpha$ = 1), we summarize our findings in
Table~\ref{XVI}.  For comparison we have also listed analytical results
obtained via bosonization techniques and data which we extracted from
polynomial fits to numerical light-front Tamm-Dancoff \cite{mo:93}, DLCQ
\cite{eller:87} and lattice results \cite{crewther:80}.  The contents of
Table~\ref{XVI} are graphically displayed in Fig.~\ref{fig-82}, where we have
chosen rescaled units as in \cite{vary:96}.

\begin{table}
\caption{\label{XVI}  \textsl{The  expansion  coefficients  $M_i$  for the
  Schwinger model ($\alpha$ = 1).  The errors for the variational
  2-particle Tamm-Dancoff (2PTD) approximation are estimated by
  comparing the last two lines in Table~\ref{V}.  The errors of our
  numerical calculations are obtained from comparing polynomial fits
  of different order to the numerical results of Tables~\ref{XI} and
  \ref{XII}.  Errors of other results are given where they could be
  estimated analogously.}}

\small

\vspace{.3cm} \begin{tabular*}{\textwidth}[t]{l @{\extracolsep\fill} l l l } 
\hline\hline
& $M_1$ & $M_2$ & $M_3$ \\ \hline 
variational 2PTD & $2\pi/\sqrt{3}$  = 3.627599 & 3.308608 $\pm \,
4 \cdot 10^{-6}$ & 0.348204 $\pm 1 \, \cdot 10^{-6}$ \\
numerical 2PTD & 3.6268 $\pm 8 \cdot 10^{-4}$ & 3.32 $\pm 0.02 $ &
0.28 $\pm 0.06 $ \\
numerical 6PTD & 3.6267 $\pm 4 \cdot 10^{-4}$ & 3.22 $\pm 0.02 $ &
0.5 $\pm 0.1 $ \\
numerical  4PTD  \cite{mo:93}  & 3.62 $\pm 0.07$  & $3.2 \pm 0.3$ &
$0.3 \pm 0.2 $ \\
DLCQ \cite{eller:87}  & 3.7 $\pm 0.2 $ & 3.5 $\pm 0.3 $ & $-$ \\
lattice \cite{crewther:80} & 3.5 $\pm 0.2 $ & 3.7  & 0.02 \\
Adam \cite{adam:96b} & $2 e^\gamma$ = 3.562146 & 3.3874 & $-$ \\
Fields et al. \cite{vary:96} & $2 e^\gamma$ = 3.562146 & 3.387399 &
$-$ \\
\hline\hline
\end{tabular*} 
\normalsize
\end{table}

As already stated, due to the anomaly the numerical errors for the
Schwinger model are at least an order of magnitude larger than for the
't~Hooft model.  In addition, the 2\% discrepancy between our light-cone
results and the analytic bosonization results does not get resolved.
However, we have shown that the discrepancy is not due to (i)
inaccuracies in the wave function or (ii) neglect of contributions from
the next higher Fock sectors (four and six particles).  It should be
mentioned that other numerical methods (DLCQ, lattice) are by far too
inaccurate to distinguish between the 't~Hooft-Bergknoff and the
bosonization value.  We will come back to these issues in a moment.

As is well known, \cite{vary:96, mo:93,vandesande:96}, the DLCQ data are
comparatively inaccurate for small fermion mass $m$ (i.e.~large
coupling).  This is due to the fact that the dominating feature of the
light-cone wave function, its endpoint behavior, is not very accurately
reproduced using an equally-spaced momentum grid.  The poor convergence
of DLCQ for small $m$ has recently been overcome by incorporating the
exact endpoint behavior \cite{vandesande:96}.

The lattice data generally suffer form the same disease of having
rather large error bars for small $m$. The chiral limit, $m = 0$,
can only be reached via extrapolation in a very inaccurate manner
unless  one uses the Schwinger  result  $M^2 (m=0) = 1$ as a bias
\cite{crewther:80}.  A representative  collection of lattice results for
$M_1$ is given in Table~\ref{XVII}.  For the convenience of the reader
we also list the condensate $\cond = - M_1 /4\pi$.

\begin{table}
\renewcommand{\arraystretch}{1.2}
\caption{\label{XVII} \textsl{Lattice   results   for  the  first   order
coefficient $M_1$ and the (negative of the) condensate, $-\cond
= M_1 / 4\pi$.  The results are quoted in chronological  order.
They should be compared with the bosonization  results,  $M_1 =
3.562$,  $-\cond  = 0.283$  and  the  't~Hooft-Bergknoff  values,
$M_1 = 3.628$, $-\cond = 0.289$.}}

\vspace{.3cm} 

\begin{tabular*}{\textwidth}[t]{c @{\extracolsep\fill} c c c c c c} 
\hline\hline
&   \cite{banks:76}  &
 \cite{carroll:76} &  \cite{crewther:80} &
  \cite{marinari:81}   &
 \cite{hamer:82} &   \cite{carson:86} \\
\hline
$M_1$ & 3.42 & 3.644 & 3.48 & 2.97 & 3.33 & 3.77 \\
$-\cond$  &  0.27  &  0.290  &  0.28  & 0.24  & 0.26 & 0.30 \\
\hline\hline
\end{tabular*} 
\end{table}

The values for the Schwinger model condensate in Table~\ref{XVII} should
be  compared  with  our  results.  If  we  assume  that  formulae
(\ref{COND})  and (\ref{ADAM1})  are directly  applicable  to our
calculation we get

\begin{eqnarray}
\cond_1  &\equiv&  -  \frac{M_1}{4\pi}  =  -  \frac{1}{2\sqrt{3}}  = -
0.28868 \; , \label{COND1} \\
\cond_2 &\equiv& - \frac{1}{2\pi}  \sqrt{\frac{M_2}{A+B}}  = - 0.28015
\; , \label{COND2}
\end{eqnarray}

while the standard result is

\begin{equation}
\cond = - \frac{e^\gamma}{2\pi} = - 0.28347 \; .
\label{COND_BOS}
\end{equation}

As already mentioned in Section~8.2, the few percent discrepancies in
$M_1$ and $M_2$ immediately affect the condensate.  Note that the
condensate value obtained from the second order coefficient is closer to
the bosonization result, the relative error being 1.2 \% compared to the
1.8 \% at first order.  Furthermore, the exact result (\ref{COND_BOS})
lies {\em between} our first and second order values, (\ref{COND1}) and
(\ref{COND2}), so that their mean value, $\cond_{1/2} \equiv -0.28442$,
is much closer to (\ref{COND_BOS}), the error being only 0.3\%. A
similar reduction of errors is actually at work if one considers the
`pion' mass-squared of the Schwinger model as a function of $m$.  The
first order overshoots the bosonization value while the second order
contribution is too small.  Adding both one gets closer to the exact
result, at least if the mass $m$ is not too tiny.  For the value of
Fig.~\ref{fig-81}, $m$ = 0.1, one finds to second order $M^2 (0.1) =
1.39585$, while, in this order, $M^2_{\mbox{\scriptsize{Bos}}} (0.1) =
1.39009$. The relative difference is only 0.4\%. Possible resolutions of
the discrepancies between (\ref{COND1}), (\ref{COND2}) and
(\ref{COND_BOS}) will be discussed at the end of this section.

Apart from the `pion' mass squared, the eigenvalue in the
't~Hooft-Bergknoff equation, we have also calculated the associated
eigenfunction, the light-cone wave function.  The latter has been
obtained with high accuracy as can be seen from the good convergence of
the alternatively defined mass squared coefficients, $\tilde M_i$
towards the variational estimates, $M_i$ (see
Section~\ref{sec:High-Order-Extension}).  If we denote $\phi_0 (x) =
x(1-x)$, our most accurate variational ansatz for the wave function can
be written as
\begin{equation}
  \phi  \Big[\beta,  a,  b,  c,  d  \Big]  =  \phi_0^\beta  +  a \,
  \phi_0^{\beta+1}  + b \, \phi_0^{\beta+2} + c \, \phi_0^{\beta+3}
  + d \,  \phi_0^{\beta+4} \; . 
\end{equation}
According   to  our  mass  perturbation   theory,   each  of  the
variational  parameters is expanded in powers of the fermion mass
$m$,  the  coefficients   being  denoted  $\beta_1$,   $\beta_2$,
$\beta_3$, $a_1$, $a_2$, etc.  From 't~Hooft's endpoint analysis,
$\beta$ is known exactly.  It turns out to be independent  of the
anomaly  ($\alpha$)  and thus  is the same  for the 't~Hooft  and
Schwinger  model.  In Table~\ref{XVIII}  we compare  the exact expansion
coefficients  of $\beta$ with their variational  estimates.   The
best estimates for the other variational parameters are listed in
Table~\ref{XIX} for both the 't~Hooft and Schwinger model.

\begin{table}
\renewcommand{\arraystretch}{1.2}
\caption{\label{XVIII} \textsl{Comparison  of the exact expansion  coefficients  of the
  endpoint exponent $\beta$ with their best variational estimates for
  both the 't~Hooft ($\alpha$ = 0) and the Schwinger model ($\alpha$ =
  1).}}

\vspace{.3cm} 

\begin{tabular*}{\textwidth}[t]{l @{\extracolsep\fill} r r r }
\hline\hline
& $\beta_1$ & $\beta_2$ & $\beta_3$ \\ \hline  
exact  &  $\sqrt{3}/\pi$  =  0.55132890  &  0  &  $-\beta_1/10  =
-$0.05513289 \\
variational ($\alpha$ = 0) & $\sqrt{3}/\pi$  = 0.55132890 & 0.00051
& $-$0.053 \\
variational ($\alpha$ = 1) & $\sqrt{3}/\pi$  = 0.55132890 & 0.00063
& $-$0.053 \\ 
\hline\hline
\end{tabular*} 
\end{table}

\begin{table}
\renewcommand{\arraystretch}{1.2}
\caption{\label{XIX}   \textsl{Best   estimates   for   the  variational
parameters  in  the  light-cone  wave  function  for  both  the  't~Hooft
($\alpha$ = 0) and Schwinger model ($\alpha$ = 1).}}

\vspace{.3cm} 

\begin{tabular*}{\textwidth}[t]{l @{\extracolsep\fill} c c c c c c r c }
\hline\hline
& $a_1$  & $a_2$ & $b_1$ & $b_2$  & $c_1$ & $c_2$ & $d_1$ & $d_2$
\\ \hline
$\alpha = 0$ & 0.92 & 1.19 & $-2.2$ & $-3.0$ & 7.0 & 9.2 & $-$9.9
& $-11.8$ \\
$\alpha  = 1$ & 1.42  & 1.03  & $-2.7$  & $-3.1$  & 8.8  & 9.6  &
$-$12.5 & $-12.1$ \\
\hline\hline
\end{tabular*} 
\end{table}

The  errors  in the  coefficients  are  comparatively  large  and
growing from the left to the right in Table~\ref{XIX}. Good numerical
convergence is evident for the coefficients of $a$. For the other
coefficients  our method does not provide enough iteration  steps
to make convergence  explicit.   However, the sensitivity  of the
mass expansion coefficients  $M_i$ and the wave functions  on the
parameters $b$, $c$, $d$ is very weak (see Fig.~\ref{fig-83}).

Let us finally investigate possible sources for the few-percent
discrepancies in the Schwinger model results like for the condensate
values, (\ref{COND1}) and (\ref{COND2}) {\em vs.}~(\ref{COND_BOS}).
First of all, it has to be reemphasized that the light-cone calculations
are conceptually rather different from the usual treatment based on
bosonization within standard equal-time quantization.  We are studying
the lowest lying meson as a relativistic fermion-anti-fermion bound
state.  The fermionic degrees of freedom are {\it explicitly} taken into
account as they define our Fock basis.  In the bosonized theory, which
is a sine-Gordon model with the Lagrangian \cite{coleman:75b, coleman:76}
\begin{equation}
  {\cal  L} = \frac{1}{2}  \partial_\mu  \phi  \partial^\mu  \phi -
  \frac{1}{2} \mu_0^2 \phi^2 - c m \mu_0 \cos \sqrt{4\pi} \phi \; ,
  \label{SINE_GORDON}
\end{equation}
the `pion' is the elementary particle described by the scalar field
$\phi$.  One calculates its mass by perturbation theory in the
`coupling' $c m \mu_0$, $c = e^\gamma / 2\pi$.  In the bosonized theory,
this `coupling' is actually dependent on the normal-ordering scale used
to renormalize the theory.  The standard choice is the most natural one,
namely the Schwinger boson mass $\mu_0$, which explicitly appears in the
Lagrangian (\ref{SINE_GORDON}).  It is, however, not very hard to
determine the general scale dependence of quantities calculated within
perturbation theory.  All one needs is Coleman's re-normal-ordering
prescription \cite{coleman:75b} which for the case at hand is
\begin{equation}
  N_{\mu_0}  \cos \sqrt{4\pi}  \phi = \frac{\mu}{\mu_0}  N_\mu \cos
  \sqrt{4\pi} \phi \; .
  \label{RNO}
\end{equation}
Here, $N_{\mu_0}$ and $N_\mu$ denote normal-ordering with respect
to $\mu_0$ and $\mu$, respectively.  From (\ref{RNO}) one derives
the following identity for the boson mass squared (setting $\mu_0
= 1$),
\begin{eqnarray}
  M^2  &=&  1 + M_1 \, m + M_2 \, m^2 + M_3 \, m^3 + \ldots  \nn \\
  &=& 1 + M_1^\prime  \, \mu m + M_2^\prime  \, \mu^2 m^2 +
  M_3^\prime \, \mu^3 m^3 + \ldots \; . \label{SCALE} 
\end{eqnarray}
\begin{figure}
\begin{center}
\caption{\protect\label{fig-83} \textsl{The light-cone wave function 
    of the 't~Hooft model `pion' for $m$ = 0.1.  The solid curve
    represents the result from 't~Hooft's original ansatz, the dashed
    curve our best result (with maximum number of variational
    parameters).  At the given resolution, however, the curves of {\em
      all} extensions of 't~Hooft's ansatz (a, b, c, d) lie on top of
    each other.}}  
\vspace{1.5cm} 
\includegraphics[scale=0.5,width=12cm]{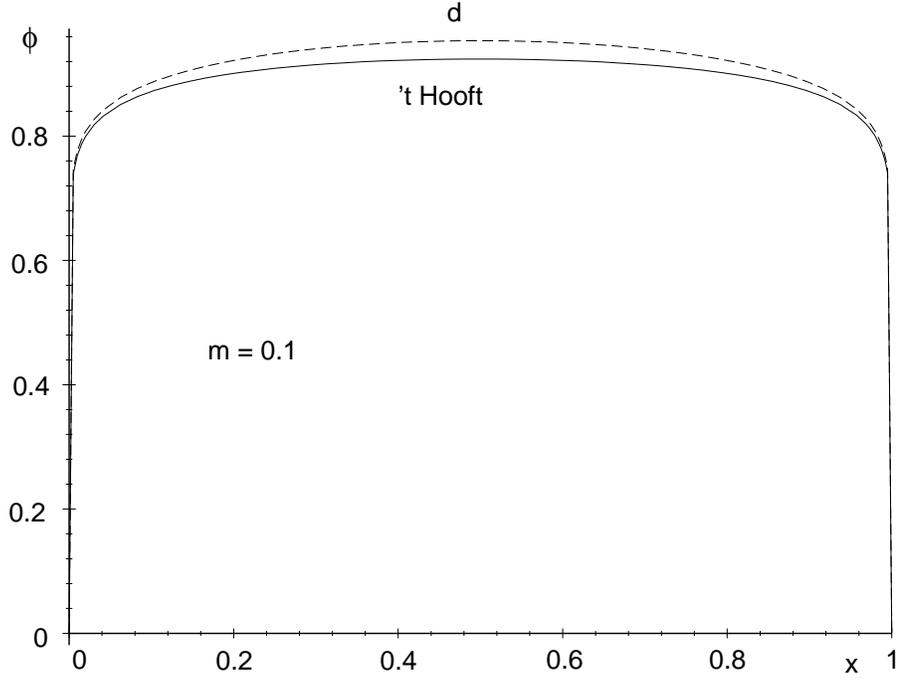}
\end{center}
\end{figure}
In this expression, the $M_i$ denote the coefficients  calculated
with  the  standard  normal-ordering  scale  $\mu_0$,  where  all
tadpoles  can  be  set  to  zero,  while  the  $M^\prime_i$   are
calculated  with  normal-ordering  scale  $\mu$,  where  one  has
non-vanishing  tadpole  contributions.   In order that the `pion'
mass   be  scale   independent   one  has  to  have   that  (upon
reintroducing $\mu_0$)
\begin{equation}
  M_n^\prime = \left( \frac{\mu_0}{\mu} \right)^n M_n \; .
  \label{RESCALE}
\end{equation}
We  have  checked  this  identity  explicitly  in the  first  two
non-trivial   orders   of  mass  perturbation   theory  with  the
Lagrangian (\ref{SINE_GORDON}).   It is also interesting  to note
that the scale dependence of the coefficients  might equivalently
be associated with a multiplicative rescaling of the fermion mass
$m$ by defining  $m^\prime  = \mu m$ and using $m^\prime$  as the
expansion parameter in (\ref{SCALE}).

Now, the upshot of all this is the following.   As we are working
entirely  in the  fermionic  representation,  we do not  a priori
know,  to which  normal-ordering  scale  of the  {\em  bosonized}
theory our results  correspond.   If we assume that it is not the
Schwinger  mass $\mu_0$ but a different scale $\mu$, we can match
the first order coefficients by choosing
\begin{equation}
  \mu      =      \frac{M_1}{M_1^\prime}       \mu_0      =
  \frac{e^\gamma}{\pi/\sqrt{3}} \mu_0 \; ,
  \label{NEWSCALE}
\end{equation}
where $M_1^\prime$  now denotes our light-front  result.  Quite a
similar point of view has been taken by Burkardt  in his paper on
light-cone  quantization  of the sine-Gordon  model \cite{burkardt:93b}.
However, as is obvious from (\ref{RESCALE}),  the change of scale
(\ref{NEWSCALE})    propagates    systematically    through   all
coefficients.   In particular, if $M_1^\prime$ is larger than the
corresponding  bosonization  result,  the same has to be true for
all $M_i^\prime$.  Our coefficient $M^\prime_2$, however, already
fails this requirement.   We thus conclude  that a possible scale
dependence alone cannot explain the discrepancy.

An additional  source of error could still be contributions  from
higher Fock sectors. We have shown numerically that the first few
low orders have very little impact on the expansion  coefficients
$M_i$.   The contribution  from each Fock sector  might  thus  be
exponentially  small.  The whole series, however,  if it could be
summed  up, might  add something  seizable.   Let us assume,  for
example,  that the contribution  of higher Fock sectors  to $M_2$
has  the  following   form\footnote{We   thank  M.~Burkardt   for
suggesting the following example.}
\begin{equation}
  \delta M_2 = \sum_{n=2}^\infty c_{2n}(m) \; ,
  \label{SERIES}
\end{equation}
where $2n$ is the label of the $2n$-particle sector.  If we had $c_{2n}
(m) = e^{-2nm}$, summation of the series would yield $\delta M_2 = 1/2m
+ O(1)$ and thus effectively a contribution to the first order in $m$.
As we are not able to calculate the real high order behavior of the
series in (\ref{SERIES}), we cannot, at the moment, make any definite
conclusion about a possible relevance of such summation effects.  All we
can say is, that if such effects are present, they must be numerically
small, i.e.~at the few percent level.

As a first step towards getting some better control of higher particle
sectors one would like to develop some tools to estimate their
contributions analytically.  This amounts to determine the coefficients
$c_{2n}(m)$, at least for small $n$.  So far, higher Fock states have
only been included numerically.

It might also be interesting to calculate excited states. For the
Schwinger model, this again requires the inclusion of higher Fock
components as pointed out by Mo and Perry \cite{mo:93}.  For the
't~Hooft model, on the other hand, one can stay in the two-particle
sector. One might then try to modify the variational procedure,
basically by taking into account the growing number of nodes, with the
aim to match the low energy spectrum with the high energy behavior
derived by 't~Hooft.  Work in these directions is underway.


\chapter{The Pion Wave Function in the NJL Model}

In this last chapter we will finally study `realistic' light-cone
bound-state equations and wave functions. To this end, we enter the
world of 3+1 dimensions and investigate the pion wave function within a
particularly suited model due to Nambu and Jona-Lasinio (NJL)
\cite{nambu:61a}.

The first derivation, analysis and solution of a light-cone bound-state
equation appeared in 't~Hooft's original paper on what is now called the
't~Hooft model \cite{thooft:74}. We have discussed this model at length in
the last section where we also rederived 't~Hooft's solution.
Interestingly, 't~Hooft did not use the light-cone formalism in the
manner we presented it and which nowadays might be called standard. This
amounts to deriving the canonical light-cone Hamiltonian and setting up
the associated bound-state equation by projecting the Schr\"odinger
equation on the different sectors of Fock space (cf.~Chapter~6).
Instead, he started from covariant equations, namely the Schwinger-Dyson
equations for the quark propagator (or self-energy), and the
Bethe-Salpeter equation for the bound state amplitude, which needs the
quark self-energy as an input. The light-cone bound equation was then
obtained by projecting the Bethe-Salpeter equation onto hypersurfaces of
equal light-cone time thereby spoiling explicit covariance. In this way,
one avoids to explicitly derive the light-cone Hamiltonian, which can be
a tedious enterprise in view of possibly complicated constraints one may
have to solve. This makes it worthwhile to have a closer look at this
way of proceeding.

In a condensed notation, the Schwinger-Dyson equation for the propagator
can be written as
\begin{equation}
  S = S_0 + S_0 \Sigma S \; , 
\end{equation}
where $S$ is the full propagator,  
\begin{equation}
  S (p)  =  \frac{1}{p \!\!\!/ - m_0 - \Sigma} \; ,
\end{equation}
$S_0$ the free propagator,
\begin{equation}
  S_0 (p)  =  \frac{1}{p \!\!\!/ - m_0 } \; ,
\end{equation}
and $\Sigma$ the quark self-energy. The Bethe-Salpeter equation for a
quark-antiquark bound state (a meson), on the other hand, is 
\begin{equation}
  \label{BSE}
  \bsa =  S_1 S_2 K \bsa \; ,
\end{equation}
with $S_1$, $S_2$ the full propagators of quark and anti-quark, $K$ the
Bethe-Salpeter kernel, defined through the interaction one is
considering, and $\bsa$ the Bethe-Salpeter amplitude.  From the latter,
one obtains the light-cone wave function via integration over the energy
variable $p^\m$ \cite{michael:82,brodsky:85,liu:93,mitra:89},
\begin{equation}
  \lca  (\vcg{p})  = \int  \frac{dp^\m}{2\pi}  \bsa  (p) \; ,
  \quad \vcg{p} = (p^\p, \vc{p}_\perp) \; . 
\end{equation}
The program outlined above, deriving the light-cone bound-state equation
from the covariant equations, when applied to QCD, is of course
ambitious but should still be feasible in view of the wealth of results
already obtained in studying the Schwinger-Dyson equations of QCD
\cite{roberts:94,frank:96}.  In a first step, which is understood, say,
as a feasibility study, a simpler approach will be pursued, namely a
calculation within a particularly suited framework, the NJL model. In
the language of the Schwinger-Dyson framework, this amounts to using a
model gluon propagator that is constant in momentum space. Obviously,
such a propagator does not lead to confinement, which however, as
pointed out in the introduction, does not necessarily prohibit a
reasonable description of hadron structure. The NJL model has been
intensively studied within the last decade, and a wealth of results has
been obtained (see the recent reviews
\cite{vogl:91,klevansky:92,hatsuda:94,bijnens:96}). To the best of our
knowledge, however, light-cone wave functions never have been calculated
within the model.

\section{The Condensate}

In its simplest form, the NJL model has a chirally invariant
four-fermion interaction, which can be imagined as the result of
`integrating out' the gluons in the QCD Lagrangian. Within the instanton
model of QCD, this can indeed be done and leads to an NJL type
interaction \cite{diakonov:95}. For two massless ($m_0 = 0$) quark
flavors ($u$ and $d$), the NJL Lagrangian is
\begin{equation}
  \label{NJL_LAG}
  \mathcal{L}_{\mathrm{NJL}} = \bar \psi i \partial \!\!\!/ \psi + G
  \left[ (\bar \psi \psi)^2 + (\bar \psi i \gamma_5 \vcg{\tau}\psi)^2
  \right] \; .
\end{equation}
In what follows we will mostly suppress flavor (and color) degrees of
freedom to keep things as simple as possible.  As the coupling $G$ has
negative mass dimension ($\sim$ mass$^{-2}$), the model is not
renormalizable and needs a cutoff as a parameter that has to be fixed by
phenomenology.  Above a critical coupling, $G > G_c$, a fermion mass is
dynamically generated so that chiral symmetry is spontaneously broken
corresponding to a second order phase transition.  The dynamical fermion
mass is proportional to the fermionic condensate in the vacuum as was
briefly mentioned in the introduction, Section~1.3.

These results are obtained via a self-consistent mean-field solution of
the Schwinger-Dyson equation resulting in the gap equation,
\begin{equation}
  \Sigma  \equiv  m = 2 i G \tr  S(x=0)  = \frac{iGm}{2\pi^4}
  \int \frac{d^4 k}{k^2 - m^2} = - 2 G \cond_m \; .
  \label{GAP}
\end{equation}
This equation determines the dynamically generated mass $m$.  The
integral on the right-hand-side is of course divergent and has to be cut
off (see below). 

The mean-field solution has a very intuitive explanation. One
essentially argues that the main effect of the interaction is to
generate the mass of the quarks which become quasi-particles that
interact only weakly. Neglecting this interaction entirely, one can view
the process of mass generation like in Section 5.2.3 as the transition
of quarks with mass $m_0 = 0$ to mass $m$ resulting in a mass term $m
\bar \psi \psi$ in the Lagrangian (\ref{NJL_LAG}) or in the Hamiltonian.
The condensate is obtained via the Feynman-Hellman theorem by
differentiating the energy density of the quasi-particle Dirac sea, 
\begin{eqnarray}
  \cond_m    &=&    \pad{}{m}    \mathcal{E}    (m)   =   \pad{}{m}
  \int\limits_{-\infty}^0    dk^\p    \int   \frac{d^2
  k_\perp}{16\pi^3}  \frac{m^2  + k_\perp^2}{k^\p}   \nn \\
  &=&   -\frac{m}{8\pi^3}    \int\limits_0^{\infty}
  \frac{dk^\p}{k^\p}  \int  d^2  k_\perp  \; , 
  \label{COND14} 
\end{eqnarray}
Again, as it stands, the integral is divergent and requires
regularization.  In the most straightforward manner one might choose
$m^2/\Lambda \le k^\p \le \Lambda$ and $|\vc{k}_\perp| \le \Lambda$, so
that the condensate becomes
\begin{equation}
  \cond_m  =  - \frac{m}{8\pi^2}  \int\limits_{m^2/\Lambda}^\Lambda
  \frac{dk^\p}{k^\p}  \int\limits_0^{\Lambda^2}  d(k_\perp^2)  =  -
  \frac{m}{8\pi^2} \Lambda^2 \ln \frac{\Lambda^2}{m^2} \; .
  \label{COND15}
\end{equation}
Plugging this result into the gap equation (\ref{GAP}) one finds for the
dynamical mass squared,
\begin{equation}
  m^2  (G)  =  \Lambda^2  \exp  \left(-\frac{4\pi^2}{G  \Lambda^2}
  \right) \; .
  \label{DYN-MASS1}
\end{equation}
The  critical  coupling  is determined  by the vanishing  of this
mass,  $m (G_c)  = 0$,  and from  (\ref{DYN-MASS1})  we find  the
surprising result
\begin{equation}
  G_c = 0 \; .
  \label{GC1}
\end{equation}
This result, however, is wrong since one knows from the conventional
treatment of the model that the critical coupling is finite of the order
$\pi^2/\Lambda^2$, both for covariant and non-covariant cutoff
\cite{nambu:61a}.  In addition, it is quite generally clear that in the
{\em free} theory ($G=0$) chiral symmetry is not broken (as $m_0 = 0$)
and, therefore, this should not happen for arbitrarily small coupling,
either (cf.~Chapter~7).  The remedy is once more to use an information
from the ordinary calculation of the condensate.  We translate the
non-covariant, but rotationally invariant, three-vector cutoff,
$|\vc{k}| \le \Lambda$, into light-cone coordinates \cite{dietmaier:89},
which leads to
\begin{equation}
  0 \le k_\perp^2  \le 2\Lambda  k^\p  - m^2 - (k^\p)^2  \; , \quad
  \frac{m^2}{2\Lambda} \le k^\p \le 2\Lambda \; .  
  \label{CUTOFF}
\end{equation}
Note that the transverse cutoff becomes a polynomial in $k^\p$. The
$k_\perp$-integration thus has to be performed first. Renaming $2
\Lambda \to \Lambda$ and introducing the longitudinal momentum fraction,
$x \equiv k^\p / \Lambda$, (\ref{CUTOFF}) becomes
\begin{equation}
  \label{INVM_CUTOFF}
  0 \le k_\perp^2 \le \Lambda^2 x (1-x) - m^2 \; , \quad m^2 / \Lambda^2
  \le x \le 1 -  m^2 / \Lambda^2 \; .
\end{equation}
Surprisingly, this is just the often used `invariant-mass cutoff', 
\begin{equation}
  M_0^2  \equiv \frac{k_\perp^2 + m^2}{x (1-x)} \le \Lambda^2 \; ,
\end{equation}
with $M_0^2$ the invariant mass-squared (for a two-particle system)
introduced in Chapter~6. Note that the cutoff is also invariant under
the exchange $x \leftrightarrow 1-x$.

For the condensate (\ref{COND14}) the cutoff (\ref{CUTOFF}) yields an
analytic structure different from (\ref{COND15}),
\begin{equation}
  \cond_m  =  -  \frac{m}{8\pi^2}   \left(  2\Lambda^2  -  m^2  \ln
  \frac{\Lambda^2}{m^2} \right) \; ,
  \label{COND16}
\end{equation}
where we have neglected sub-leading terms in the cutoff $\Lambda$.
From (\ref{COND16}),  one infers the correct cutoff dependence of
the critical coupling,
\begin{equation}
  G_c = \frac{2\pi^2}{\Lambda^2} \; .
  \label{GC2}
\end{equation}
The moral of this calculation is that even in a non-renormalizable
theory like the NJL model, the light-cone regularization prescription is
a subtle issue.  In order to get a physically sensible result the
transverse cutoff has to be $k^\p$-dependent.  Clearly, this dependence
cannot be arbitrary but should be constrained from dimensional and
symmetry considerations.  For renormalizable theories, such arguments
have been given by Perry and Wilson \cite{perry:93b}. In the example
above, it was  rotational  invariance that solved
the problem.

The condensate (\ref{COND16}) was originally obtained in
\cite{dietmaier:89}, where, however, a slightly more complicated cut-off
was used. In that work, the condensate was defined covariantly in terms
of the fermion propagator at the origin, $S(x=0)$ like in (\ref{GAP}).
Performing the energy integration over $k^\m$ with the appropriate
cutoff again leads to (\ref{COND16}).

In the NJL model with its second-order phase transition of mean-field
type, the usual analogy with magnetic systems can be made.  Chiral
symmetry corresponds to rotational symmetry, the vacuum energy density
to the Gibbs free energy, and the mass $m$ to an external magnetic
field.  The order parameter measuring the rotational symmetry breaking
is the magnetization.  It is obtained by differentiating the free energy
with respect to the external field.  This is the analogue of expression
(\ref{COND14}) as derived from the Feynman-Hellmann theorem.

\section{The Light-Cone Bound State Equation}

Once the physical fermion mass $m$ is known, it can be plugged into the
Bethe-Salpeter equation (\ref{BSE}) which in ladder approximation reads
\begin{equation}
  \lca  (\vcg{k})  = \int \frac{dk^\m}{2\pi}  i S(k) i S(k-P)
  \int \frac{d^3 \tilde k}{(2\pi)^3} \int \frac{d\tilde k^\m}{2\pi}
  K (k, \tilde k ) \bsa (\tilde k) \; , 
\end{equation}
with $P$ denoting the bound-state four-momentum.  On the left-hand-side,
the projection onto $x^\p = 0$ (i.e.  the $k^\m$-integration) has
already been carried out.  On the right-hand-side, the two integrations
over $k^\m$ and $\tilde k^\m$ still have to be performed. Whether this
can easily be done depends of course crucially on the kernel $K$, which
in principle is a function both energy variables.  For the NJL model,
however, one has (after a Fierz transformation), 
\begin{equation}
  K(k  ,  \tilde  k)  = 2 \gamma_5  \otimes  \gamma_5  - \gamma_\mu
  \gamma_5 \otimes \gamma^\mu \gamma_5 \; , 
\end{equation}
i.e.~the  kernel  is momentum  independent  due to the four-point
contact   interaction!    Thus,  the  $\tilde   k^\m$-integration
immediately  yields  $\lca$,  and  the $k^\m$-integration  can be
performed via residue technique and is completely  determined  by
the poles of the propagators,  $S(k)$ and $S(k-P)$.  As a result,
one finds a non-vanishing  result  only if $0 \le k^\p \le P^\p$,
and {\em one} of the two particles  is put on-shell,  e.g.~$k^2 =
m^2$, as already observed by Gross \cite{gross:88}.

The upshot of all this is nothing but the light-cone bound-state
equation, which reads explicitly
\begin{eqnarray}  
  \label{NJL_LFBSE}
  \lca  (x,  \vc{k}_\perp)  &=&  - \frac{2G}{x(1-x)} \frac{(\hat  k
  \!\!\!/  + m ) \gamma_5  (\hat k \!\!\!/ - P \!\!\!\!/  + m)}{M^2 -
  M_0^2 }  \nn \\
  &\times& \int_0^1 dy \int \frac{d^2 \tilde k_\perp}{8\pi^3} \tr
  \left[   \gamma_5   \lca  (y  ,  \tilde {\vc{k}}_\perp)   \right]
  \theta_\Lambda (y , \tilde{\vc{k}}_\perp) \nn \\
  &+&  \frac{G}{x(1-x)} \frac{(\hat  k  \!\!\!/  +  m  ) \gamma_\mu
  \gamma_5  (\hat  k \!\!\!/  - P \!\!\!\!/  + m)}{M^2  - M_0^2  }
  \nn \\
  &\times& \int_0^1  dy \int \frac{d^2  \tilde k_\perp}{8\pi^3}  \tr \left[
  \gamma^\mu  \gamma_5  \lca  (y  , \tilde{\vc{k}}_\perp)  \right]
  \theta_\Lambda (y , \tilde{\vc{k}}_\perp) \; . \nn \\ 
\end{eqnarray}
Here we have defined the longitudinal momentum fractions $x = k^\p /
P^\p$, $y = \tilde k^\p / P^\p$, the on-shell momentum $\hat k = (\hat
k^\m , \vc{k}_\perp , k^\p)$ with $\hat k^\m = (k_\perp^2 + m^2)/k^\p$,
and the bound state mass squared, $M^2 = P^2$, which is the eigenvalue
to be solved for.  $\theta_\Lambda (x , \vc{k}_\perp )$ denotes the
invariant mass cutoff (\ref{INVM_CUTOFF}).

Now, while (\ref{NJL_LFBSE}) may appear somewhat complicated it is
actually very simple; indeed, it is basically already the solution of
the problem.  The crucial observation is to note that the two integral
expressions are mere normalization constants,
\begin{eqnarray}
  C_\Lambda &\equiv& \int_0^1 dy \int \frac{d^2 \tilde k_\perp}{8\pi^3}
  \tr  \left[  \gamma_5  \lca  (y , \tilde{\vc{k}}_\perp)  \right]
  \theta_\Lambda (y , \tilde{\vc{k}}_\perp) \; , \\
  D_\Lambda   P^\mu   &\equiv&  \int_0^1   dy  \int  \frac{d^2   \tilde
  k_\perp}{8\pi^3}  \tr \left[ \gamma^\mu \gamma_5 \lca (y , \tilde
  {\vc{k}}_\perp) \right] \theta_\Lambda (y , \tilde{\vc{k}}_\perp)
  \; .
\end{eqnarray}
Thus, up to normalization, the solution of the light-cone bound-state
equation (\ref{NJL_LFBSE}) is
\begin{eqnarray}
  \lca   (x,   \vc{k}_\perp)   &=&   -   \frac{2G C_\Lambda}{x(1-x)}
  \frac{(\hat k \!\!\!/ + m ) \gamma_5 (\hat k \!\!\!/ - P \!\!\!\!/ +
  m)}{M^2 - M_0^2 } \nn \\
  &+&  \frac{G  D_\Lambda}{x(1-x)} \frac{(\hat  k  \!\!\!/  + m ) P
  \!\!\!\!/  \gamma_5  (\hat k \!\!\!/ - P \!\!\!\!/  + m)}{M^2 - M_0^2
  } \; .
  \label{NJL_LCA}
\end{eqnarray}
As a first check of our bound state wave function (\ref{NJL_LCA})
we look for a massless  pion in the chiral limit.  To this end we
decompose the light-cone wave function according to \cite{lucha:91},
\begin{equation}
  \lca = \phi_{\mbox{\tiny  S}} + \phi_{\mbox{\tiny  P}} \gamma_5 +
  \phi_{\mbox{\tiny A}}^\mu \gamma_\mu \gamma_5 + \phi_{\mbox{\tiny
  V}}^\mu    \gamma_\mu     +    \phi_{\mbox{\tiny     T}}^{\mu\nu}
  \sigma_{\mu\nu} \; .
\end{equation}
Multiplying (\ref{NJL_LCA}) with $\gamma_5$, taking the trace and
integrating over $\vcg{k}$ we find
\begin{eqnarray}
  C_\Lambda  &=& - \frac{G C_\Lambda}{2  \pi^3} \int_0^1 dx \int d^2
  k_\perp  \frac{M^2  x + M_0^2  (1-x)}{x  (1-x)  (M^2  - M_0^2)  }
  \theta_\Lambda (x , \vc{k}_\perp) \nn \\
  &+& \frac{G D_\Lambda}{2\pi^3}  M^2 \int_0^1  dx \int  d^2 k_\perp
  \theta_\Lambda (x , \vc{k}_\perp) \; .
\end{eqnarray}
In the  chiral  limit  one  expects  a solution  for  $M=0$,  the
Goldstone pion. In this case one obtains
\begin{equation}
  \label{GAP_INT}
  1 = \frac{G}{2\pi^3}  \int_0^1 dx \int d^2 k_\perp \theta_\Lambda
  (x , \vc{k}_\perp) \; .
\end{equation}
This is exactly the gap equation (\ref{GAP}) upon performing the
$k^\m$-integration there (with an invariant mass cutoff
(\ref{INVM_CUTOFF}) understood in both identities). Note once more the 
light-cone peculiarity that the (Fock) measure in (\ref{GAP_INT}) is
entirely mass independent. All the mass dependence, therefore, has to
come from the (invariant mass) cutoff. Otherwise one will get a wrong
behavior of the dynamical mass $m$ as a function of the coupling $G$,
as was the case in (\ref{DYN-MASS1}).
 
With this in mind, we have found that the Goldstone pion is a solution
of the light-cone bound-state equation exactly if the gap equation
holds.  This provides additional evidence for the self-consistency of
the procedure. The same results are well known from the covariant
treatment \cite{nambu:61a} , and it is gratifying to get them also
within the light-cone formalism. The deeper reason for the fact that the
quark self-energy and the bound state amplitude satisfy essentially the
same equation \cite{savkli:97}, is the chiral Ward identity relating the
quark propagator and the pseudo-scalar vertex \cite{dietmaier:88}.

Our next task is to actually evaluate the solution (\ref{NJL_LCA}) of
the bound-state equation. $\lca$ is a Dirac matrix and therefore is not
yet a light-cone wave function as defined in Chapter~6. The relation
between the two quantities has been given by Liu and Soper
\cite{liu:93},
\begin{equation}
  2 P^\p \sqrt{x \bar x} \, \psi (x , \vc{k}_\perp, \lambda , \lambda^\prime) = 
  \bar u (x P^\p , \vc{k}_\perp , \lambda) \gamma^\p \lca (\vcg{k})
  \gamma^\p v (\bar x P^\p , - \vc{k}_\perp , \lambda^\prime) \; ,
\end{equation}
where we have denoted $\bar x \equiv 1-x$ to save space and included a
factor $(x \bar x)^{1/2}$ to match with the conventions of Brodsky and
Lepage \cite{brodsky:89,lepage:80,lepage:81}. A somewhat lengthy
calculation, using the spinor  identities of App.~A, yields the result
\begin{equation}
  \label{PSI_PROV}
  \psi (x , \vc{k}_\perp, \lambda , \lambda^\prime) = \frac{2 G
  P^\p}{M^2 - M_0^2} \left( \frac{2 C_\Lambda}{M} \bar u_\lambda M
  \gamma_5 v_{\lambda^\prime} - D_\Lambda \bar u_\lambda P \!\!\!\!/ \, \gamma_5
  v_{\lambda^\prime} \right)
\end{equation}
with the arguments of the spinors $\bar u$ and $v$ suppressed. At this
point we have to invoke another symmetry principle. Ji et al.~have
pointed out \cite{ji:92} that the spin structure ($\bar u \Gamma v$)
should be consistent with the one obtained  form the instant form
spinors via a subsequent application of a Melosh transformation
\cite{melosh:74} and a boost. Using this recipe, one obtains the
following relation between the constants $C_\Lambda$ and $D_\Lambda$,
\begin{equation}
  \label{CDN}
  2 C_\Lambda / M = - D_\Lambda \equiv N_\Lambda / 2 G P^\p \; .
\end{equation}
As a result, the spin structure in (\ref{PSI_PROV}) coincides with the
standard one used in a number of works
\cite{dziembowski:88,chung:88b,jaus:90,ji:90,ji:92}. The NJL wave function
of the pion finally is
\begin{equation}
  \psi(x, \vc{k}_\perp , \lambda, \lambda^\prime ) =
  \frac{N(\Lambda)}{M^2 - M_0^2} \; \bar u_\lambda (M + P\!\!\!\!/) \, \gamma_5
  \, v_{\lambda^\prime} \, \theta(\Lambda^2 - M_0^2)  \; .
\end{equation}
$N(\Lambda)$ is a cutoff dependent normalization defined in (\ref{CDN}),
and the spin (or helicity) structure is given by
\begin{eqnarray}
  &&\bar u (x P^\p , \vc{k}_\perp , \lambda ) \,  (M + P\!\!\!\!/) \,
  \gamma_5 \, 
  v (\bar x P^\p , - \vc{k}_\perp, \lambda^\prime)  = \nn \\
  && = \frac{1}{\sqrt{x \bar x }} \Big\{ \lambda \Big[ mM + m^2 -
  \vc{k}_\perp^2 + M^2 x \bar x \Big] \delta_{\lambda, -\lambda^\prime}
  - k_{-\lambda} (M + 2m) \delta_{\lambda \lambda^\prime} \Big\}  
\end{eqnarray}
where we have defined $k_\lambda \equiv k^1 + i \lambda k^2$. The first
term with spins anti-parallel corresponds to $L_z =0$, the second one
(with spins parallel) to $L_z = \pm 1$. It has already been pointed out
by Leutwyler that both spin alignments should contribute to the pion
wave function \cite{leutwyler:74a,leutwyler:74c}. Note that the latter is a cutoff
dependent quantity. This is necessary in order to make the wave function
normalizable. In other words, it guarantees the boundary conditions
(\ref{WF_BC}) which state that the wave function drops off sufficiently
fast in $x$ and $k_\perp$.

As we are interested in analyzing the quality of a constituent picture,
we approximate the pion by its valence state,
\begin{equation}
  | \pi: \vcg{P} \ket = \sum_{\lambda, \lambda^\prime} \int_0^1
  \frac{dx}{\sqrt{x (1-x)}} \int \frac{d^2 k_\perp}{16\pi^3} \psi(x,
  \vc{k}_\perp, \lambda, \lambda^\prime ) | q \bar q : x, \vc{k}_\perp, 
  \lambda, \lambda^\prime \ket
\end{equation}
which should be compared with the general expression (\ref{GEN_PI_WF}). 
The normalization of this state is given by (\ref{MOM_NORM}),
\begin{equation}
  \bra \pi: \vcg{P}^\prime | \pi: \vcg{P} \ket = 16 \pi^3
  P^\p \delta^3 (\vcg{P} - \vcg{P}^\prime) \; .
  \label{PI_NORM}
\end{equation}
As usual we work in a frame in which the total transverse momentum
vanishes, i.e.~$\vcg{P} = (P^\p , \vc{P}_\perp = 0)$. Expression
(\ref{PI_NORM}) yields the normalization (\ref{VAL_NORM}) of the wave
function,
\begin{equation}
  \sum_{\lambda \lambda^\prime} \int_0^1 dx \int \frac{d^2 k_\perp}{16
  \pi^3} | \psi (x, \vc{k}_\perp , \lambda , \lambda^\prime ) |^2 = 1 \;
  .
  \label{PSI_NORM}
\end{equation}
It is of course a critical assumption that the probability to find the
pion in its valence state is one. In this way we enforce a constituent
picture {\it by fiat}, and it is clear that such an assumption has to
be explicitly checked. An alternative way of proceeding would be not to
fix the normalization but to use $N(\Lambda)$ as an additional
parameter that has to fixed by phenomenology. We will, however, use
(\ref{PSI_NORM}) and check whether a valence (or constituent) picture
makes sense.

\section{Observables}

With the light-cone wave function at hand, we are in the position to
calculate observables. As a technical prelude, however, we first have to
determine its normalization.  To keep things simple we consider the
chiral limit, $M = 0$. We write the pion wave function as a matrix in
helicity space,
\begin{equation}
  \label{PI_WF_MATRIX}
  \psi (x , \vc{k}_\perp) = - N(\Lambda) \frac{\sqrt{x (1-x)}}{k_\perp^2 +
  m^2} \left( \begin{array}{cc} 
              -2m (k^1 + ik^2) & m^2 - k_\perp^2 \\
              k_\perp^2 - m^2  & - 2m (k^1 - ik^2 )
              \end{array}
       \right) \theta(\Lambda^2 - M_0^2)
\end{equation}
The diagonal terms correspond to parallel spins, the off-diagonal ones to
anti-parallel spins. 

The normalization (\ref{PSI_NORM}) determines $N$ as a function of
$\Lambda$. Using the explicit form of the invariant-mass cutoff,
(\ref{INVM_CUTOFF}), one finds 
\begin{equation}
  \label{INT_N}
  1  = \frac{|N|^2}{8 \pi^2} \int_{\epsilon^2}^{1 - \epsilon^2} dx \, x (1-x)
  \int_0^{\Lambda^2 (x)} d k_\perp^2\; ,
\end{equation}
where we have defined 
\begin{eqnarray}
  \epsilon^2 &\equiv& m^2/\Lambda^2  \\
  \Lambda^2 (x) &\equiv& \Lambda^2 x (1-x) -m^2 = \Lambda^2 \Big[x (1-x) -
  \epsilon^2\Big] \; .
\end{eqnarray}
As an additional approximation we will neglect terms of order
$\epsilon^2$, i.e.~we assume that the cutoff is large compared to the
constituent mass. From the standard values, $\Lambda \simeq 1$ GeV, $m
\simeq 300$ MeV, we expect that this assumption should induce an error
of the order of 10\%. The advantage is that it allows for a simple
analytic evaluation of all the integrals we will encounter. Furthermore,
the leading order will be independent of the actual value of the
constituent mass. From (\ref{INT_N}) one thus finds the normalization, 
\begin{equation}
  \label{NORM}
  |N|^2 = \frac{240 \pi^2}{\Lambda^2} \left[ 1 + 5 \epsilon^2 +
  O(\epsilon^4) \right] \; .
\end{equation}
Choosing $N$ real we obtain to leading order,
\begin{equation}
  N = 2 \sqrt{15} \pi / \Lambda \; .
\end{equation}
This fixes our normalization constant. We proceed by calculating the
pion electromagnetic formfactor. It is defined by the matrix element of
the electromagnetic current $J_{\mathrm{em}}^\mu$ between pion states,
\begin{equation}
  \bra \pi : \vcg{P} | J_{\mathrm{em}}^\mu | \pi : \vcg{P}^\prime \ket = 2 (P
  + P^\prime)^\mu F(Q^2) \; , \quad Q^2 \equiv - (P - P^\prime)^2 \; .
\end{equation}
Considering $\mu = +$ in a frame where $\vcg{P} = (P^\p , \vc{0})$ and
$\vcg{P}^\prime = (P^\p , \vc{q}_\perp)$ one is led to the the Drell-Yan
formula \cite{drell:70c, brodsky:89},
\begin{equation}
  \label{DY}
  F(\vc{q}_\perp^2 ) = \sum_{\lambda \lambda^\prime} \int_0^1 dx \int
  \frac{d^2 k_\perp}{16\pi^3} \psi^* (x , \vc{k}_\perp^\prime , \lambda ,
  \lambda^\prime ) \psi (x , \vc{k}_\perp , \lambda , \lambda^\prime) \; .
\end{equation}
The transverse momentum of the struck quark is $\vc{k}_\perp^\prime =
\vc{k}_\perp + x \vc{q}_\perp$. Note that the formula (\ref{DY}) with
its overlap of two wave functions on the right-hand-side is rather
similar to the non-relativistic result \cite{brodsky:89}. If one sets
the momentum transfer $\vc{q}_\perp$ equal to zero, one is left with the
normalization integral (\ref{PSI_NORM}), so that the form factor is
automatically normalized to one. Plugging the pion wave function
(\ref{PI_WF_MATRIX}) into (\ref{DY}), one finds to first order in
$\vc{q}_\perp^2$,
\begin{equation}
  F(\vc{q}_\perp^2) = 1 - \vc{q}_\perp^2 \frac{|N|^2}{80 \pi^2} \left[
  1 - \frac{20}{3} \epsilon^2 + O(\epsilon^4) \right] \; .
\end{equation}
Inserting the normalization (\ref{NORM}), the formfactor becomes
\begin{equation}
  \label{FORMFAC}
  F(\vc{q}_\perp^2) = 1 - \frac{3}{\Lambda^2} \, \vc{q}_\perp^2 \left[ 1 -
  \frac{5}{3} \epsilon^2 + O(\epsilon^4) \right] \; .
\end{equation}
This leading order result already determines a particular observable,
namely the charge radius of the pion, which is defined via the slope of
the formfactor at zero momentum transfer,
\begin{equation}
  F(\vc{q}_\perp^2) = 1 - \frac{\bra r_\pi^2 \ket}{6} \vc{q}_\perp^2 +
  O(\vc{q}_\perp^4) \; .
\end{equation}
To leading order in $\epsilon^2$ the pion charge radius thus becomes
\begin{equation}
  \label{CHARGE_RAD}
  \bra r_\pi^2 \ket = 18/\Lambda^2 \; ,
\end{equation}
and is thus entirely determined by the cutoff $\Lambda$. We might fix
the latter by using  the experimental result \cite{amendolia:86}
\begin{equation}
  \label{R_PI}
  \bra r_\pi^2 \ket = 0.44 \; \mathrm{fm}^2 \; .
\end{equation}
The cutoff thus becomes (to leading order in $\epsilon^2$),
\begin{equation}
  \label{CUTOFF1}
  \Lambda^2 = \frac{18}{\bra r_\pi^2 \ket} \simeq (1.3 \; \mathrm{GeV})^2 \; .
\end{equation}
Our preferred way, however, to fix the cutoff is to use the pion decay
constant $f_\pi$ = 92.4 MeV \cite{holstein:90}, which is known to a
higher accuracy than the charge radius. Though this does not matter in
view of our present approximations, it seems to be advantageous at least
in principle. $f_\pi$ is given by the `wave function at the origin',
$\psi_0$, defined as
\begin{equation}
  \label{WF_ORIG}
  \psi_0 \equiv \int_0^1 dx \int \frac{d^2 k_\perp}{16 \pi^3} \, \psi
  (x, \vc{k}_\perp, \uparrow , \downarrow) \, \theta(\Lambda^2 - M_0^2)
  = \frac{3 N \Lambda^2}{2048 \pi} + O(\epsilon^2) \; .
\end{equation}
Note that only the $L_z = 0$ component contributes. Using the
normalization (\ref{NORM}) this becomes
\begin{equation}
  \label{PSI_0}
  \psi_0 = \frac{3 \sqrt{15}}{512} \Lambda \simeq 0.022 \, \Lambda \; .
\end{equation}
The pion decay constant has been obtained from weak $\pi$ decay
\cite{brodsky:89} where only the pion valence wave functions contributes
to the decay matrix elements. Therefore, the following result
\cite{brodsky:89,lepage:81} provides an \emph{exact} constraint on the
pion wave function,
\begin{equation}
  \psi_0 = \frac{f_\pi}{2 \sqrt{N_C}} \; .
\end{equation}
Comparing with (\ref{PSI_0}) we find that the  cutoff is fixed through
\begin{equation}
  \label{CUTOFF2}  
  \Lambda = \frac{256}{9 \sqrt{5}} f_\pi \simeq 12.7 f_\pi \simeq 1.2
  \; \mathrm{GeV} \; .
\end{equation}
Having fixed the cutoff, we are in the position to actually predict the
pion charge radius via (\ref{CHARGE_RAD}). The result is 
\begin{equation}
  \label{R_PI_NJL}
  \bra r_\pi^2 \ket = \frac{18}{\Lambda^2} = 0.50 \; \mathrm{fm}^2 = 
  (0.71 \mathrm{fm}  )^2 \; .
\end{equation}
which differs from the experimental value (\ref{R_PI}) by 13\% and
therefore is consistent with the expected accuracy. 

\begin{figure}
  \caption{\protect\label{fig-ff} \textsl{The pion formfactor squared
        vs.~momentum transfer $q^2 \equiv q_\perp^2$. The full line is the
        monopole fit of \protect\cite{amendolia:86}, $|F|^2 = n/(1 +
        q_\perp^2 \bra r_\pi^2 \ket /6)^2$ with $n = 0.991$, $\bra
        r_\pi^2 \ket = 0.431$ fm$^2$; the dashed line is the same fit
        with our values, $n=1$, $\bra r_\pi^2 \ket = 0.50$ fm$^2$. The
        agreement is consistent with the expected accuracy of 10\%.}}
        \vspace{0.5cm}
        \begin{center}
        \includegraphics[scale=0.85]{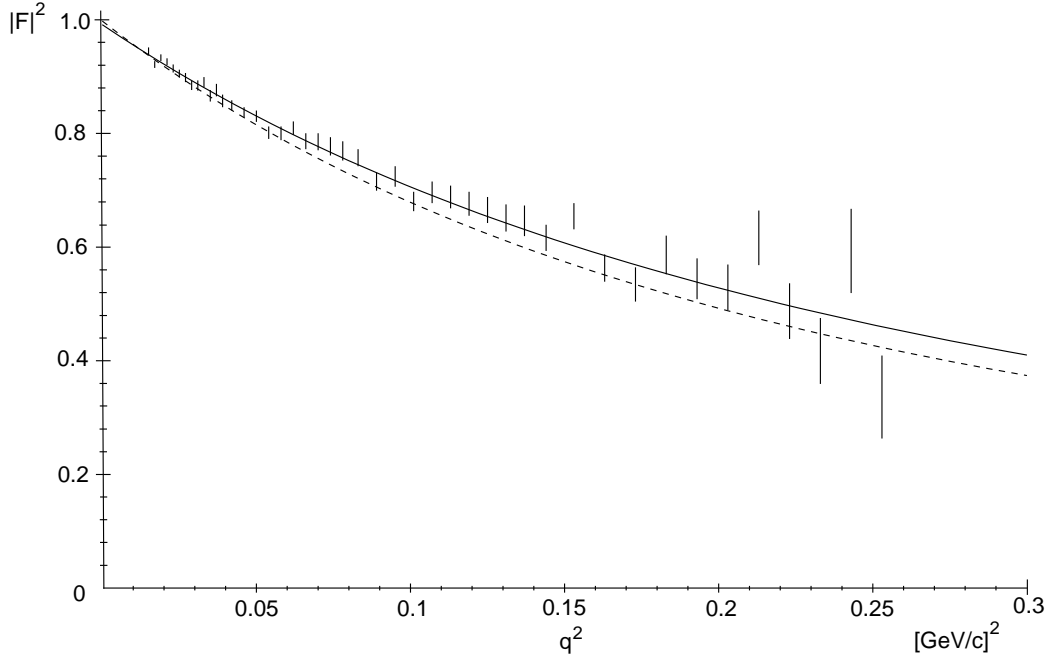}
        \end{center}
\end{figure}
This is also evident from Fig.~\ref{fig-ff} where we compare with the
experimental results of \cite{amendolia:86}. The full and dashed line
correspond to monopole fits of the form
\begin{equation}
  F^2 (\vc{q}_\perp^2) = \frac{n}{(1 + \vc{q}_\perp^2 \bra r_\pi^2 \ket
  /6)^2} \; ,
\end{equation}
with $n=0.991$, $\bra r_\pi^2 \ket = 0.431$ fm$^2$ \cite{amendolia:86}
and our result (\ref{R_PI_NJL}), $n=1$, $\bra r_\pi^2 \ket = 0.50$
fm$^2$, respectively.

If we compare our result (\ref{CUTOFF2}) for the cutoff with the value
for the cutoff $\Lambda_\chi$ below which chiral effective theories make
sense, $\Lambda_\chi = 4 \pi f_\pi \simeq 12.6 f_\pi$ \cite{manohar:84},
we find that the agreement is surprisingly good.  Clearly, this is
somewhat accidental, as we do not expect an accuracy better than 10\%.
This can be seen explicitly again by alternatively using the pion charge
radius to fix $f_\pi$.  Plugging (\ref{CUTOFF1}) into the general
relation (\ref{CUTOFF2}) between cutoff and $f_\pi$, and solving for
$f_\pi$, we predict
\begin{equation}
  f_\pi = \frac{9 \sqrt{5}}{256} \Lambda \simeq 0.077 \, \Lambda \simeq
  100 \mbox{MeV} \; ,
\end{equation}
which indeed is accurate to within 10\%. 

We proceed by calculating the mean transverse momentum,
\begin{equation}
  \bra \vc{k}_\perp^2 \ket = \sum_{\lambda \lambda^\prime} \int_0^1 dx
  \int \frac{d^2 k_\perp}{16\pi^3} \, \vc{k}_\perp^2 \; | \psi(x ,
  \vc{k}_\perp, \lambda , \lambda^\prime )|^2 \, \theta(\Lambda^2 -
  M_0^2) \; .   
\end{equation}
To leading order in $\epsilon^2$ this becomes
\begin{equation}
  \bra \vc{k}_\perp^2 \ket = 
  \frac{|N|^2}{2240 \pi^2}\Lambda^4 \; ,
\end{equation}
or, upon inserting the normalization (\ref{NORM}),
\begin{equation}
  \bra \vc{k}_\perp^2 \ket = \frac{3}{28} \Lambda^2 \simeq (0.43 \;
  \mbox{GeV})^2 \; .
\end{equation}
The r.m.s.~transverse momentum is thus of the order of 400 MeV, which
is a reasonable value \cite{leutwyler:74a}. In particular, this value
shows that the pion is a highly relativistic system as the momenta of
its (valence) constituents exceed the pion mass. 

The inverse of the r.m.s.~momentum can be identified with the
pion core radius, for which one finds
\begin{equation}
  \label{R_PI_CORE}  
  R_\pi \equiv  \bra \vc{k}_\perp^2 \ket^{-1/2} = \sqrt{28/3}
  \; \Lambda^{-1} \simeq 0.47 \; \mbox{fm} \; .
\end{equation}
As expected, it is smaller than the charge  radius which we attribute to the
fact that the charge distribution measured by the charge radius does
not coincide with the distribution of baryon density. The core radius
$R_\pi$ is sometimes related to the decay constant $f_\pi$ via the
dimensionless quantity \cite{weise:84}
\begin{equation}
  C \equiv f_\pi R_\pi  \; .
\end{equation}
In constituent quark models one typically gets 
\begin{equation}
  f_\pi = \frac{\sqrt{3} m}{2\pi} \; ,
\end{equation}
and if the quarks are confined by a harmonic oscillator potential, their
mass is $m=3/2 R_\pi$, so that $ C \simeq 0.4$. This implies the fairly
large value $R_\pi \simeq $ 0.8 fm. Using standard many-body techniques,
Bernard et al.~have calculated this quantity in a model treating the
pion as a collective excitation of the QCD vacuum, and find $C \simeq
0.2$ \cite{bernard:85}. This result is close to what we get from
(\ref{CUTOFF2}) and (\ref{R_PI_CORE}),
\begin{equation}
  C = \frac{9 \sqrt{5}}{256} \sqrt{\frac{28}{3}} \simeq 0.24 \; .
\end{equation}
Thus, though we work within a constituent picture, we do get a
reasonable value. We believe that this is due to the intrinsic
consistency of the light-cone framework with the requirements of
relativity. 

\vfill\eject

\section{The Pion Distribution Amplitude}

\begin{figure}[htb]
\caption{\label{DISAMP_FIG}\textsl{The pion distribution amplitude. The
  full curve represents the asymptotic amplitude, the broken one the
  amplitude calculated within the NJL model. The vertical lines mark the
  constraints (\protect\ref{BJ}) of Belyaev and Johnson.}}  
\vspace{0.5cm}
  \begin{center}
    \includegraphics[scale=0.8]{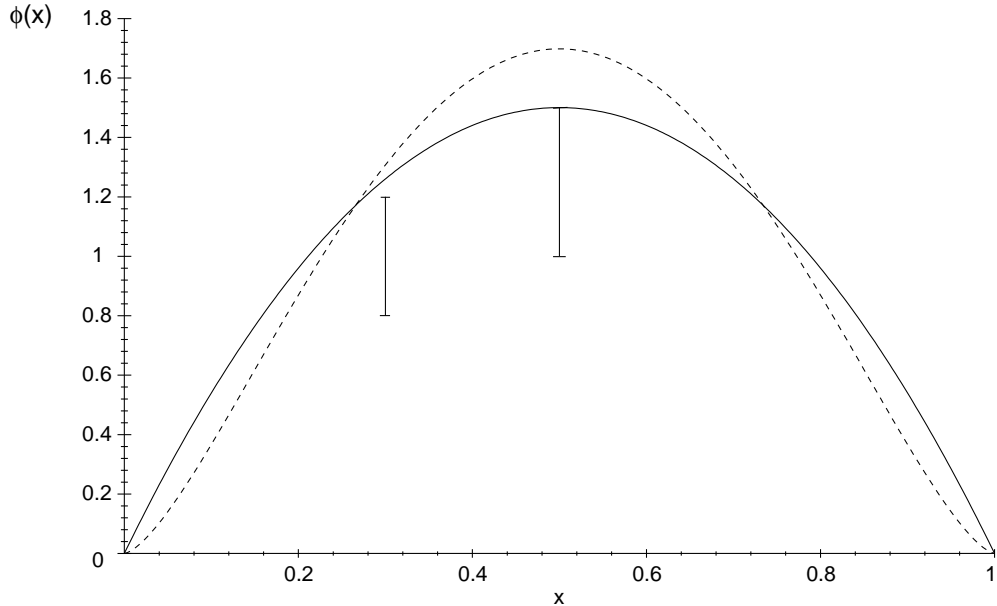}
  \end{center}
\end{figure}

We follow Brodsky and Lepage \cite{brodsky:89,lepage:80} and define the
pion distribution amplitude by integrating the wave function over
$k_\perp$. Due to the cutoff dependence of the wave function, the
integration extends effectively only  up to the cutoff scale 
$\Lambda (x)$,
\begin{equation}
  \phi_\mathrm{BL} (x) \equiv \int \frac{d^2 k_\perp}{16\pi^3} \; \psi_\Lambda (x ,
  \vc{k}\perp, \lambda , \lambda^\prime ) \; . 
\end{equation}
As usual, we evaluate this to leading order in $\epsilon$ so that the
result becomes independent of the constituent mass, 
\begin{equation}
  \phi_\mathrm{BL} (x) = \frac{\sqrt{15}}{4 \pi} \Lambda \, \big[x
    (1-x)\big]^{3/2} = \frac{64 \sqrt{3}}{9 \pi} f_\pi \, \big[x
    (1-x)\big]^{3/2} \; .
\end{equation}
The normalization of $\phi$ is obtained by integrating this  over $x$,
which simply yields the `wave function at the origin', $\psi_0$, defined
in (\ref{WF_ORIG}). We have used the subscript `BL' (for Brodsky-Lepage),
because in what follows we will use a  distribution amplitude which is 
normalized to one, that is,   
\begin{equation}
  \label{DISAMP1}
  \phi (x) \equiv \frac{2 \sqrt{3}}{f_\pi} \, \phi_\mathrm{BL} =
  \frac{128}{3 \pi} \, \big[x     (1-x)\big]^{3/2}  \; . 
\end{equation}
It has been shown by Brodsky and Lepage, that the distribution amplitude
obeys an evolution equation in momentum scale $Q^2$ which can be solved
exactly in the limit $Q^2 \to \infty$. The result is called the
\emph{asymptotic} distribution amplitude, 
\begin{equation}
  \phi_\mathrm{as} (x) = 6 x (1-x) \; .
\end{equation}
The (non-asymptotic) pion distribution amplitude has been a rather
controversial object. For a while people have tended to believe in a
`double-humped' shape of the amplitude (due to a factor $(1 - 2x)^2$),
which was originally found by Chernyak and Zhitnitsky via QCD sum rules
\cite{chernyak:84}. Recent CLEO data, however, \cite{savinov:95} seem to
support an amplitude that is not too different from the asymptotic one
\cite{kroll:96}. It turns out, that our result (\ref{DISAMP1}) is
qualitatively consistent with these findings \cite{belyaev:97b,
  belyaev:97c, petrov:97}. Belyaev and Johnson have recently reported
two constraints which should be satisfied by the distribution amplitude
\cite{belyaev:97b},
\begin{eqnarray}
  \phi (x = 0.3) &=& 1 \pm 0.2 \; , \nn \\
  \phi (x = 0.5) &=& 1.25 \pm 0.25 \; . \label{BJ}
\end{eqnarray}
In Figure \ref{DISAMP_FIG} we have displayed our distribution amplitude
in comparison with the asymptotic one. The vertical lines represent the
constraints (\ref{BJ}). Obviously, these are not satisfied, as we have
\begin{eqnarray}
  \phi (x = 0.3) &=& 1.31  \; , \\
  \phi (x = 0.5) &=& 1.70  \; . 
\end{eqnarray}
As we already overshoot the asymptotic amplitude (for $0.3 < x < 0.7$)
this does not come as a surprise. In order to improve on that we
presumably have to increase our accuracy by taking into account finite
pion and quark masses. Still, we get an amplitude which is at least
qualitatively correct.

It should also be mentioned that it is not obvious what the actual momentum
scale is to which our result corresponds. Our transverse-momentum cutoff
is $x$- dependent,  $\Lambda^2 (x) \simeq \Lambda^2 \, x (1-x)$, so, if we
choose an average $x$ of $\bra x \ket \simeq 1/2$, a natural scale seems
to be\footnote{We thank W.~Schweiger for discussions on this point.}  
\begin{equation}
  \mu \equiv \Big[\bra x \ket (1 -\bra x \ket )\Big]^{1/2} \Lambda \simeq
  \Lambda /2 \simeq 600 \; \mathrm{MeV} \; .
\end{equation}
This is somewhat lower than $\mu \simeq $ 1 GeV, which has been assumed
by Belyaev and Johnson in their analysis of the distribution amplitude
in terms of light-cone quark models \cite{belyaev:97c}.

\chapter*{Conclusions}
\addcontentsline{toc}{chapter}{Conclusions}

In this report we have discussed an alternative approach to relativistic
(quantum) physics based on light-cone dynamics. It makes use of the fact
that for relativistic systems the choice of the time parameter is not
unique. Our particular choice of relativistic dynamics singles out
Dirac's front form so that null-planes tangent to the light-cone become
hypersurfaces of equal time. This apparently trivial change of
coordinates has far-reaching consequences:

\begin{itemize}  
  
\item The number of kinematical (i.e.~interaction independent)
  Poincar\'e generators becomes maximal; there are seven of them instead
  of the usual six, among them the boosts.

\item Lorentz boosts in $z$-direction become diagonal; the light-cone
  time and space coordinates, $x^\p$ and $x^\m$, respectively, do not
  get mixed but rather get rescaled.

\item As a consequence, for many-particle systems one can introduce
  frame-independent relative coordinates, the longitudinal momentum
  fractions, $x_i$, and the relative transverse momenta, $\vc{k}_{\perp
    i}$.    

\item Because of a two-dimensional Galilei invariance, relative and
  center-of-mass motion separate. As a result, many formulae are
  reminiscent of non-relativistic physics and thus very intuitive.
  
\item This is particularly true for light-cone wave functions which,
  because of the last two properties, are boost invariant and do not
  depend on the total momentum of the bound state.  They are therefore
  ideal tools to study relativistic particle systems.
  
\item The last statement even holds for relativistic quantum field
  theory where one combines the unique properties of light-cone
  quantization with a Fock space picture. The central feature making
  this a reasonable idea is the triviality of the light-cone vacuum
  which accordingly is an eigenstate of the fully interacting
  Hamiltonian. This implies that the Fock operators create the \emph{physical}
  particles from the ground state.

\end{itemize}

As expected, however, the principle of `conservation of difficulties' is
at work so that there are problems to overcome. We have focused on the
issue of vacuum properties. In instant-form quantum field theory, in
particular in QCD, many non-perturbative phenomena are attributed to the
non-triviality of the vacuum which shows up via the appearance of
condensates. These are non-vanishing vacuum expectation values which
support the idea that the instant-form vacuum is a complicated many-body
state (like, e.g.~the BCS ground state). Of further physical relevance
is the fact that many of these condensates signal the spontaneous
breakdown of a symmetry. The conceptual problem which arises at this
point is how to reconcile the existence of condensates with the
triviality of the light-cone vacuum.

We have shown that for scalar field theory in 1+1 dimensions, the field 
mode with longitudinal momentum $k^\p$ is the carrier of non-trivial
vacuum properties. Its $c$-number part  coincides with the vacuum
expectation value of the field. The reflection symmetry, mapping the
field onto its negative, is thus seen to be spontaneously broken.

For fermionic fields with a bilinear condensate reflecting the
spontaneous breakdown of chiral symmetry, we had to proceed
differently. We calculated the bound state spectrum of model field
theories and reconstructed  ground state properties like the condensates
from the particle spectrum. The paradigm identity was the
Gell-Mann-Oakes-Renner relation which relates the masses of the bound
states and the fermions to the condensates. Alternatively, one can try
to derive gap equations like in the NJL model, which determine the mass
of the dressed, effective fermions in terms of the condensate.

It is crucial for the latter approach to solve the associated
bound-state problem, or, in other words, the light-cone Schr\"odinger
equation. For a relativistic quantum field theory, this, in principle,
amounts to solving an infinite system of coupled integral equations for
the amplitudes to find an ever increasing number of constituents in the
bound state. Experience, however, shows that the \emph{light-cone}
amplitudes to find more than the valence quanta in the bound state tend
to become rather small. We have confirmed this fact for the theories we
have studied. Note that the same is \emph{not} true within ordinary,
that is, instant-form quantization. The light-cone approach, however,
provides some confidence that also in real QCD a constituent picture may
emerge, so that hadrons are consistently described as light-cone bound
states of a minimal number of two or three constituents.

A first step in this direction was performed using an effective field
theory, the NJL model. We have seen that, though we made a number of
approximations, in particular by \emph{enforcing} a constituent picture,
a number of pionic observables are predicted with reasonable accuracy.

At this point one has to ask whether the ideas just outlined can really
be used in QCD. This is a difficult question. The reason why light-cone
QCD has not been presented in this report is that, in our opinion, there
are still technical problems to overcome. First of all, the quantization
procedure itself is not entirely clear beyond perturbation theory. There
are the usual gauge redundancies to be taken into account
\emph{together} with the peculiar first-order nature of the light-cone
Lagrangian. We have reported on this elsewhere \cite{heinzl:95b}. One
generally uses the light-cone gauge $A^\p = 0$, a variant of an axial
gauge which has its own well-known problems, among them residual gauge
invariance and infrared divergences. However, it is only in this gauge
that the canonical quantization is straightforward. The latter has been
performed quite some time ago \cite{tomboulis:73,casher:76,lepage:81}
and successfully been used as the basis of light-cone perturbation
theory. Some impressive  achievements of the latter method can be found
in the recent review \cite{brodsky:97} which focuses on the light-cone
formulation of gauge theories. 

Still, even if one wants to utilize perturbation theory for bound-state
calculations (in case this is justified) one is confronted with
difficulties. One big problem is to find a systematic renormalization
program in order to get rid of the ubiquitous infinities. As one does
neither have explicit covariance nor rotational invariance there are
very few guiding principles and restrictions.  Accordingly, there is a
rather large variety of divergences which can even become nonlocal,
power counting gets more complicated, and the number of counter-terms
tends to become large \cite{wilson:94}. As a result, to our knowledge,
the light-cone renormalization program has not been pushed beyond one
loop so far.

We therefore consider it worthwhile to first pursue a less ambitious
program, the first steps of which have been undertaken in the last
chapter. One should make use of effective field theories, that are
obtained from QCD in a well-defined manner \cite{kaplan:95,manohar:96}, 
and in general are not renormalizable. The cutoff is a parameter of the
theory which sets the energy scale below which the theory makes sense.
Thus, one only has to deal with the issue of regularization to render
the appearing integrals finite. Still, for the case of the NJL model, we
have seen that already this can be a nontrivial problem. An alternative
candidate suited for quark bound-state studies is the effective quark
theory derived within the QCD instanton model \cite{diakonov:95}. It is
also of the NJL type, however, with a more complicated (and presumably
more realistic) interaction.

Further improvements can  be achieved by using more elaborate kernels in the
covariant Bethe-Salpeter equation. These can be found in the literature
on the Schwinger-Dyson approach to QCD \cite{roberts:94,frank:96} and
basically amount to choosing a model for the gluon propagator. The NJL
model corresponds to an infrared-finite propagator which is constant in
momentum space. Therefore, one does not have confinement. The latter can
be implemented by using a gluon propagator which diverges for small
momenta. Other constraints follow from multiplicative renormalizability.
Clearly, the procedure of deriving the light-cone bound-state equation
will no longer be as simple as in the last chapter but should still be
feasible. Work in this direction is underway.

\chapter*{Acknowledgments}

It is a pleasure to thank my former supervisor and present collaborator
E.~Werner for the continuous support and encouragement which he has
provided over the years. Furthermore, I am grateful to U.~Heinz for some
valuable advice concerning this work. I thank my Regensburg colleagues
for their help, friendship and company, in particular during mountain
hiking: G.~Fischer, T.~Hirschmann, A.~Marquardt, M.~Meister,
M.~Oleszczuk, T.~Pause, S.~Simb\"urger, C.~Skornia, C.~Stern and
B.~Zellermann. Warm thanks go also to the non-scientific staff, G.~Heyer
and C.~Leichtl, for raising my spirits with birthday cakes and the like.

I am indebted to my colleagues in the light-cone community for
discussions and inspiration, the junior ones: M.~Burkardt, K.~Itakura,
A.~Kalloniatis, D.~Rob\-ertson, B.~van de Sande, U.~Trittmann, and
S.~Tsujimaru, as well as the senior ones: A.~Bassetto, P.~Grang{\'e},
J.~Hiller, G.~McCartor, H.-C.~Pauli, R.~Perry, and S.~Pinsky. Special
thanks are due to K.~Harada for his collaboration on the 't~Hooft and
Schwinger model.

\addcontentsline{toc}{chapter}{Acknowledgments}

\begin{appendix}

\chapter{Light-Cone Spinors}

In this appendix we collect some basic formulae representing our
conventions for light-cone spinors in $d$ = 3+1, together with some
frequently occurring identities involving these.  Our conventions for
Dirac matrices are those of Bjorken and Drell \cite{bjorken:64}. 

Let $\psi(x)$ be a solution of the free Dirac equation
\begin{equation}
  \label{DIR_APP}
  (i \partial\!\!\!/  - m) \psi(x) = 0  \quad .
\end{equation}
The Fock expansion of  $\psi$ reads
\begin{equation}
  \label{FOCKEXP_APP}
  \psi(\vcg{x},0) = \sum_\lambda
  \int_0^{\infty}\frac{dk^\p}{ k^\p}\int \frac{d^2k_{\perp}}{16 \pi^3} \,
  \left[b(\vcg{k},\lambda) u(\vcg{k}, \lambda) e^{- i \svcg{k} \cdot \svcg{x}} +
  d^{\dagger} (\vcg{k},\lambda) v(\vcg{k}, \lambda) e^{i \svcg{k} \cdot \svcg{x}}
  \right] \; , 
\end{equation}
where
\begin{equation}
  \vcg{k} \equiv (k^\p,\vc{k}_{\perp}) \; , \quad
  \vcg{x} \equiv  (x^\m , \vc{x}_{\perp}) \; , \quad
  \vcg{k} \cdot \vcg{x} \equiv \sfrac{1}{2} k^\p x^\m -
  \vc{k}_{\perp} \cdot \vc{x}_{\perp} \; .  
\end{equation}
The projections $\psi_{\pm}$ are obtained through the action of the
projectors $\Lambda_{\pm}$, $\psi_{\pm} \equiv \Lambda_{\pm} \psi$, with
(in a two-by-two block notation)
\begin{equation}
  \Lambda_{\pm} \equiv   \frac{1}{4} \gamma^\mp
  \gamma^{\pm}   = \frac{1}{2}
    \left( \begin{array}{c c} 
               \Eins & \pm  \sigma_3  \\
               \pm \sigma_3 & \Eins 
           \end{array} 
    \right) \; . 
\end{equation}
As $\Lambda_{+}$ and $\Lambda_{-}$ are projectors they obey
\begin{equation}
  \Lambda_{\pm}^2 = \Lambda_{\pm} \; , \quad
  \Lambda_{\pm}^{\dagger} = \Lambda_{\pm}  \; ,
\end{equation}
and
\begin{equation}
  \Lambda_{+} + \Lambda_{-} = 1 \; , \quad 
  \Lambda_{+}\Lambda_{-} = \Lambda_{-}\Lambda_{+} = 0
  \;  . 
\end{equation}
The following identities are frequently used ($i = 1,2$),
\begin{eqnarray}
  \Lambda_{\pm} \beta &=& \beta \Lambda_{\mp}  \; , \\
  \Lambda_{\pm} \gamma^i &=& \gamma^i \Lambda_{\pm} \; , \\
  \Lambda_{\pm} \alpha^i &=& \alpha^i \Lambda_{\mp} \; . 
\end{eqnarray}
The spinors  $u$ and $v$ in (\ref{FOCKEXP_APP}) satisfy the equations 
\begin{eqnarray}
  (k\!\!\!/ - m) \, u(\vcg{k}, \lambda) &=& 0  \; , \\
  (k\!\!\!/ + m) \, v(\vcg{k}, \lambda) &=& 0  \; , 
\end{eqnarray}
and are explicitly given by \cite{leutwyler:69,lepage:80,namyslowski:85} 
\begin{eqnarray}
  u(\vcg{k}, \lambda) &=& \frac{1}{\sqrt{k^\p}}(k^\p + \beta m +
  \alpha_i k_i) \, X_\lambda \; , \label{U_SPINOR} \\
  v(\vcg{k}, \lambda) &=&  \frac{1}{\sqrt{k^+}}(k^\p - \beta m +
  \alpha_i k_i) \, X_{-\lambda} \label{V_SPINOR} \; .
\end{eqnarray} 
The $X_{\lambda}$ are eigenspinors of the projector  $\Lambda_{+}$ with
eigenvalue one,

\begin{equation}
  \Lambda_{+} \, X_\lambda  = X_\lambda \; , \quad X_\lambda^\dagger
  X_{\lambda^\prime}^{\,}   = \delta_{\lambda \lambda^\prime} \; , 
\end{equation}
or explicitly,
\begin{equation}
  X_1 \equiv X_{\uparrow} = \frac{1}{\sqrt{2}}
    \left[ \begin{array}{c}
              1 \\ 0 \\ 1 \\ 0
           \end{array} 
    \right] \; , \quad
  X_{-1} \equiv X_{\downarrow} = \frac{1}{\sqrt{2}}
    \left[ \begin{array}{r}
              0 \\ 1 \\ 0 \\ -1
           \end{array} 
    \right] \; .
\end{equation}
If we introduce the two-spinors $\chi_{_\lambda}$, the usual
eigenspinors of $\sigma_3$,
\begin{equation}
  \chi_{_1} \equiv  \chi_{_\uparrow} = 
    \left[ \begin{array}{c}
              1 \\ 0 
           \end{array} 
    \right] \; , \quad 
  \chi_{_{-1}} \equiv  \chi_{_\downarrow} = 
    \left[ \begin{array}{c}
              0 \\ 1 
           \end{array} 
    \right] \; ,
\end{equation}
we can write for $X_{\lambda}$,
\begin{equation}
  X_\lambda = \frac{1}{\sqrt{2}} 
  \left[ \begin{array}{r}
           \chi_{_\lambda}  \\ 
           \lambda \chi_{_\lambda} 
         \end{array} 
  \right] \; .
\end{equation}
Using the identities
\begin{equation}
  \Lambda_{\pm} \spinor{\chi_{_\lambda}}{0} =  
  \frac{1}{2} \spinor{\chi_{_\lambda}}{\pm \lambda\chi_{_\lambda}} \; ,
  \quad \Lambda_{\pm} \spinor{0}{-\lambda \chi_{_\lambda}} = 
  \frac{1}{2} \spinor{\pm \chi_{_{-\lambda}}}{-\lambda 
  \chi_{_{-\lambda}}}  \; ,
\end{equation}
we can rewrite the elementary spinors (\ref{U_SPINOR}, \ref{V_SPINOR})
according to
\begin{eqnarray}
  u(\vcg{k}, \lambda) &=& 
  \sqrt{\frac{2}{k^\p}} \left[ m \Lambda_{-} + (k^\p + \alpha^i k^i)
  \Lambda_{+} \right] \spinor{\chi_{_\lambda}}{0}  \; , \\
  v(\vcg{k}, \lambda) &=& 
  \sqrt{\frac{2}{k^\p}} \left[ m \Lambda_{-} + (k^\p + \alpha^i k^i)
  \Lambda_{+} \right] \spinor{0}{-\lambda \chi_{_{-\lambda}}} \; .
\end{eqnarray}
The projections of the spinors $u$ and $v$ are
\begin{eqnarray}
  u_{+} (\vcg{k}, \lambda) &=&   \sqrt{k^\p} \, X_\lambda \equiv u_+
  (k^\p , \lambda) \; , \\
  v_{+} (\vcg{k}, \lambda) &=& \sqrt{k^\p} \, X_{-\lambda} \equiv v_+
  (k^\p , \lambda) \; , \\
  u_{-} (\vcg{k}, \lambda) &=& \frac{1}{\sqrt{k^\p}} \, 
  (m \beta + \alpha_i k_i) \, X_\lambda  \; , \\
  v_{-} (\vcg{k}, \lambda) &=& \frac{1}{\sqrt{k^\p}} \, 
  (-m \beta + \alpha_i k_i) \, X_{-\lambda} \; .
\end{eqnarray}
One has the completeness relation in the `+'-sector,
\begin{equation}
  \sum_\lambda u_{+}(\vcg{k}, \lambda) u^{\dagger}_{+}
  (\vcg{k}, \lambda) =  \sum_\lambda v_{+}
  (\vcg{k}, \lambda) v^{\dagger}_{+}(\vcg{k}, \lambda)
  = k^\p \Lambda_{+} \; .
\end{equation}
The following provides a (very incomplete) list of matrix elements (for
further matrix elements, see \cite{lepage:80}),
\begin{eqnarray}
  u^{\dagger}_{+} (k^\p, \lambda) \, u_{+}(k^\p,
  \lambda^{\prime}) &=&  v^{\dagger}_{+}(k^\p, \lambda) \,
  v_{+}(k^\p, \lambda^{\prime}) = k^\p \delta_
  {\lambda\lambda^{\prime}} \; , \\
  u^{\dagger}_{\pm} (\vcg{k}, \lambda) \, \beta \, u_{\mp}
  (\vcg{k}, \lambda^{\prime}) &=&  
  m \delta_{\lambda\lambda^{\prime}} \mp \lambda k_{-\lambda}
  \delta_{\lambda,-\lambda^{\prime}} \; , \\
  v^{\dagger}_{\pm} (\vcg{k}, \lambda) \, \beta \, v_{\mp}
  (\vcg{k}, \lambda^{\prime}) &=& 
  -m \delta_{\lambda\lambda^{\prime}} \pm \lambda k_{\lambda}
  \delta_{-\lambda,\lambda^{\prime}} \; , 
\end{eqnarray}  
where we have defined
\begin{equation}
  k_{\lambda} \equiv k^1 + i \lambda k^2 \; , \quad \mbox{so that} \quad
  k_{\lambda} k_{-\lambda} = k_{\perp}^2 \; . 
\end{equation}
Introducing
\begin{equation}
  \gamma_5 = i \gamma^0 \gamma^1 \gamma^2 \gamma^3 = 
  \left( \begin{array}{cc}
            0 & \Eins \\
            \Eins & 0
         \end{array} \right) \; ,
\end{equation}
one finds that $X_\lambda$ is also an eigenspinor of $\gamma_5$,
\begin{equation}
  \gamma_5 X_\lambda = \lambda X_\lambda \; .
\end{equation}
Furthermore, one has
\begin{equation}
  k^i \gamma^i X_\lambda = \lambda k_\lambda X_{-\lambda} \; .
\end{equation}
The last two identities lead to the matrix elements
\begin{eqnarray}
  u_+^\dagger (k^\p , \lambda) \, \gamma_5 \, v_+ (p^\p , \lambda^\prime) &=&
  \sqrt{k^\p p^\p} \, \lambda \, \delta_{\lambda , - \lambda^\prime} \; , \\ 
  u_+^\dagger (k^\p , \lambda) \, k^i \gamma^i \, \gamma_5 \, 
  v_+ (p^\p , \lambda^\prime) &=&
  \sqrt{k^\p p^\p} \, k_{-\lambda} \, \delta_{\lambda \lambda^\prime} \; , 
\end{eqnarray}
which have been used in Chapter~9 to determine the spin structure of the
pion wave function.
\end{appendix}



\addcontentsline{toc}{chapter}{Bibliography}
\baselineskip13pt
\bibliographystyle{h-physrev}


\end{document}